\documentclass[sigconf,10pt]{acmart}
\AtBeginDocument{%
  }

\setcopyright{acmlicensed}
\copyrightyear{2027}
\acmYear{2027}
\acmDOI{XXXXXXX.XXXXXXX}
\acmConference[XXX]{XXX}{XXX}{XXX}
\acmISBN{978-1-4503-XXXX-X/2018/06}

\usepackage{comment}
\usepackage{algorithm}
\usepackage{algpseudocode}
\usepackage{multirow}
\usepackage{float}
\usepackage{subcaption}
\usepackage{graphicx}
\usepackage{dblfloatfix}

\begin{document}

\title{E2LLM: Towards Efficient LLM Serving in
Heterogeneous Edge/Fog Environments}

\author{Truong-Thanh Le}
\affiliation{%
  \institution{Department of Informatics\\Universit of Oslo}
  \city{Oslo}
  \country{Norway}}
\email{truongl@ifi.uio.no}

\author{Hoang-Loc La}
\affiliation{%
  \institution{Department of Computer Science\\UiT The Arctic University of Norway}
  \city{Tromsø}
  \country{Norway}}
\email{hoang.l.la@uit.no}

\author{Amir Taherkordi}
\affiliation{%
  \institution{Department of Informatics\\Universit of Oslo}
  \city{Oslo}
  \country{Norway}}
\email{amirhost@ifi.uio.no}

\author{Frank Eliassen}
\affiliation{%
  \institution{Department of Informatics\\Universit of Oslo}
  \city{Oslo}
  \country{Norway}}
\email{frank@ifi.uio.no}

\author{Phuong Hoai Ha}
\affiliation{%
  \institution{Department of Computer Science\\UiT The Arctic University of Norway}
  \city{Tromsø}
  \country{Norway}}
\email{phuong.hoai.ha@uit.no}

\author{Peiyuan Guan}
\affiliation{%
  \institution{Department of Informatics\\Universit of Oslo}
  \city{Oslo}
  \country{Norway}}
\email{peiyuang@ifi.uio.no}

\renewcommand{\shortauthors}{Truong-Thanh et al.}

\begin{abstract}
  Deploying large language models (LLMs) on resource-constrained infrastructures remains a significant challenge. In addition to satisfying latency requirements, serving systems must efficiently utilize available resources while controlling operational costs. Most existing deployment strategies assume that a complete model can be hosted on a single device, which is often infeasible in edge and fog environments characterized by limited compute and memory capacities.
This paper introduces E2LLM, a distributed serving framework designed for heterogeneous edge infrastructures. Instead of distributing a single model instance across all devices, E2LLM strategically creates multiple model instances and distributes each instance across a subset of devices using model parallelism. To better exploit the asymmetric characteristics of LLM inference, model instances are specialized for either prompt processing (prefill) or token generation (decode), enabling more efficient utilization of heterogeneous resources.
E2LLM combines evolutionary search and optimization techniques to determine an effective deployment configuration. A genetic algorithm groups devices into clusters according to their capabilities and communication characteristics, while a dynamic programming approach identifies model partitions that mitigate bottlenecks arising from model parallel execution under network congestion.
Evaluation across a wide range of deployment settings demonstrates that the proposed framework remains effective under varying model sizes, workload patterns, request rates, and system scales. Compared with alternative deployment strategies, including a completion-time-oriented policy and an implementation inspired by SplitWise, E2LLM consistently delivers lower request queuing delays under congested conditions. Across all 33 congested configurations examined, the framework achieves the smallest maximum waiting time, with improvements reaching up to 45× over competing approaches.
\end{abstract}

\maketitle

\section{Introduction}\label{sec.intro}
\begin{figure*}[!t]
    \centering
    \includegraphics[width=0.8\linewidth]{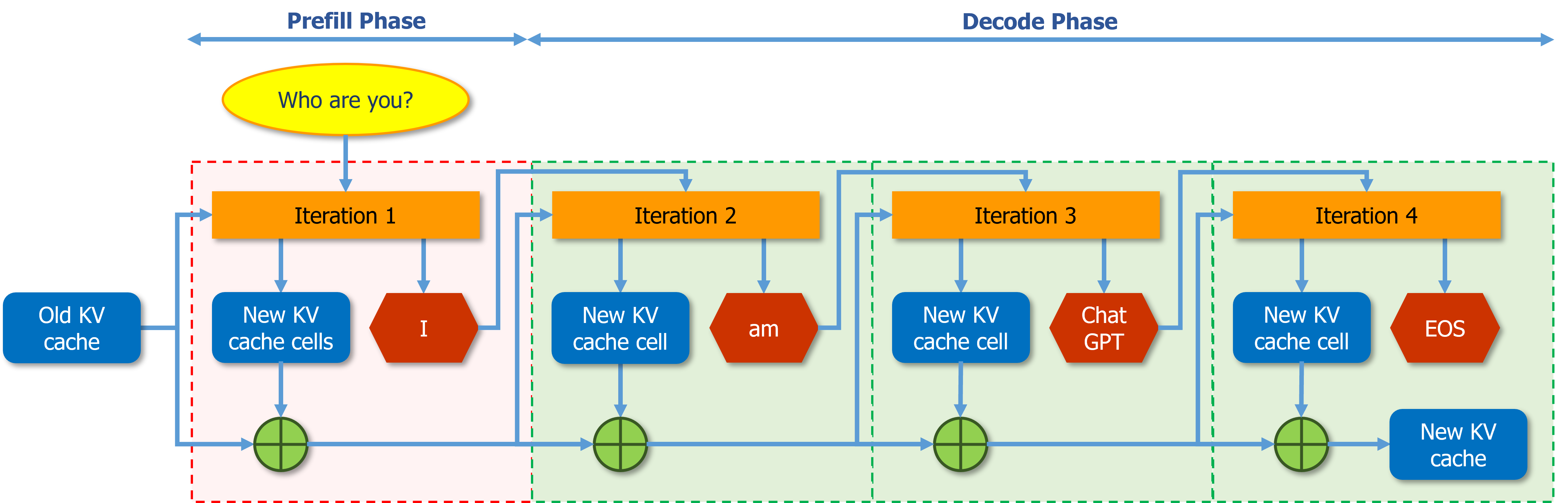}
    \caption{Phases in general LLM Generation Task.}
    \label{fig:Phases}
\end{figure*}

Nowadays, Large Language Models (LLMs) are extremely large, often requiring a few to hundreds of gigabytes of memory and specialized hardware to operate efficiently.
Traditional deployment strategies assume that a single computing device or server can host the entire model \cite{jiang2025thunderserve}, which is feasible in cloud environments but impractical for edge or fog computing scenarios \cite{yang2024efficient}. Usage of such computing platforms is increasingly becoming  important as they enable low-latency processing and improve data privacy. 
Such deployments support, for example, personalized mobile healthcare \cite{ji2025transforming} and real-time decision-making on robotic platforms \cite{gurunathan2025edge}.
Yet, \textit{edge devices typically have limited computational resources, making it impossible to store and run a full LLM on a single device}. Moreover, due to this constraint, \textit{optimizing key performance metrics such as throughput and responsiveness remain a significant challenge when deploying LLMs in edge and fog environments especially in high-demand scenarios.}
Beyond scalability, privacy is a further driver: users handling sensitive or confidential data---such as small companies or individuals fine-tuning LLMs for their own needs~\cite{anisuzzaman2024fine, deng2025hardening}---increasingly prefer to keep computation on their own Edge/Fog infrastructure rather than renting cloud resources, making local deployment both a practical and a cost-effective choice.

Several strategies exist for distributing models across Edge/Fog resources. Hierarchical deployment \cite{li2021appealnet} runs compressed models on end devices and escalates hard inputs upwards, but it routes on prediction accuracy, which is hard to quantify for open-ended text generation. Model (or pipeline) parallelism \cite{qi2025synergistic, le2024optimal, zhang2024edgeshard, laskaridis2020spinn} instead splits the layers across devices and overlaps stages, raising throughput, at the cost of pipeline bubbles and stage imbalance when the partition is chosen poorly.
Moreover, most state-of-the-art methods leveraging model parallelism assume that all devices in the network participate in a single pipeline, meaning that they collectively host \textit{one instance} of the model. This assumption introduces two major drawbacks. First, it significantly increases network demand when the number of devices is large, as every stage depends on inter-device communication. Second, it can lead to resource under-utilization: if a subset of devices is powerful enough to host the model efficiently, forcing all devices into one pipeline wastes potential parallelism. 

Currently, only a few methods deploy multiple replicas, such as HexGen~\cite{jiang2023hexgen}, HexGen-2~\cite{jiang2025hexgen2}, Splitwise~\cite{patel2024splitwise} and ThunderServe~\cite{jiang2025thunderserve}. HexGen-2 is closest to our setting: it forms replicas from \emph{groups} of heterogeneous GPUs and disaggregates prefill from decoding over a network spanning NVLink, PCIe, InfiniBand, RoCE and Ethernet. The common characteristic of these systems is not the assumption that a single device can host the entire model, but rather a deployment environment built around CUDA-class accelerators, datacenter-grade interconnect, and device memory large enough that CPU/RAM offloading serves as a fallback mechanism rather than a core design consideration. In contrast, we investigate a regime in which none of these hold: devices connected through a switched LAN with less than $1$~Gbps bandwidth, and platforms such as Apple Silicon with unified memory, on which CUDA and NCCL are unavailable. Under these constraints, the primary bottlenecks shift from computation to memory capacity and network bandwidth. Consequently, model offloading is no longer an auxiliary optimization but becomes a first-class component of the system cost model.

In this paper, we introduce E2LLM, a method for distributing LLMs across heterogeneous Edge/Fog devices using both \textit{replication} and \textit{model parallelism}. Overall, E2LLM prioritizes latency and bottleneck reduction to ensure responsiveness under resource constraint in Edge/Fog deployment environment. 
We partition the pool of available computing devices into multiple clusters, each hosting one replica of the target model.
Unfortunately, determining the optimal partitioning and clustering strategy is
NP-hard~\cite{jiang2025thunderserve}. The hardness is concentrated in one place: grouping devices into replicas under memory constraints is NP-hard, whereas partitioning layers \emph{within} a fixed replica is solved exactly in $O(M^2N^2)$. This asymmetry motivates our structure --- genetic search over groupings, dynamic programming inside each of them. Our GA design incorporates a two-chromosome representation to capture both clustering and node ordering, along with an elite strategy that preserves a subset of globally best solutions across generations. This ensures that high-quality candidates are not lost during evolution. We also differentiate the role of each replica into Prefill,  which processes all input tokens in a single compute-bound pass, and Decoder, which generates output tokens one at a time and is memory-bound---as illustrated in Figure \ref{fig:Phases}---based on the ratio between input and generated tokens and the computational capability of each device.

Moreover, to distribute the full model across devices within a replica, we use Dynamic Programming (DP) to minimize communication/computation bottlenecks and determine the optimal configuration, such as the number of concurrent client requests, to enhance throughput while still satisfying Quality-of-Service requirements, rather than processing requests sequentially.
Here, we use the term Quality-of-Service (QoS) to subsume the standard Service Level Objectives, namely Time-to-First-Token (TTFT) and Time-per-Output-Token (TPOT), consistent with established distributed-systems terminology.
The results show that E2LLM attains the lowest maximum waiting time in all 33 of the 36 evaluated configurations in which any method experiences congestion. For instance, on eight devices at a $1000$~ms arrival interval the maximum waiting time is $0.7$~s for E2LLM against $31.5$~s for both baselines.

In summary, our contributions are as follows:
\begin{itemize}
    \item We introduce E2LLM, a novel replication-based deployment strategy for LLMs in \textit{heterogeneous Edge/Fog environments} to \textit{maximize throughput and reduce waiting time in high-demand scenarios while targeting a per-request generation-speed floor} (cf. Section \ref{sec.method}).
    \item We design and implement a combined approach using \textit{two-chromosome Genetic Algorithm, Dynamic Programming, and role differentiation} to determine optimal model distribution and resource allocation (cf. Sections \ref{subsec.Method.GA}, \ref{subsec.Method.DP} and \ref{subsec.Method.PrefillDecode} respectively).
    \item We evaluate E2LLM against a finishing-time objective and an adapted SplitWise role-assignment policy on a real heterogeneous cluster, demonstrating \textit{its ability to adapt to changing environment characteristics and to attain the lowest maximum waiting time in every reported configuration} (cf. Section \ref{sec.eval}).
\end{itemize}

\section{Related Work}\label{sec.relatedwork}

Deploying LLMs in resource-constrained environments has become an active area of research due to the growing demand for low-latency, privacy-preserving, and cost-efficient AI services. This challenge involves three main aspects: hierarchical deployment across multiple tiers, parallelization techniques for distributing computation, and replication deployment on heterogeneous devices.

\textbf{Hierarchical deployment across multiple tiers.} A common response to resource limits is to place a lightweight model at the edge and escalate hard inputs to a larger cloud model. AppealNet \cite{li2021appealnet} does this with a learned predictor that trades accuracy against a resource budget. Such schemes still require each tier to hold a complete model, which no device in our setting can.

\begin{figure*}[t]
    \centering
    \includegraphics[width=0.7\linewidth]{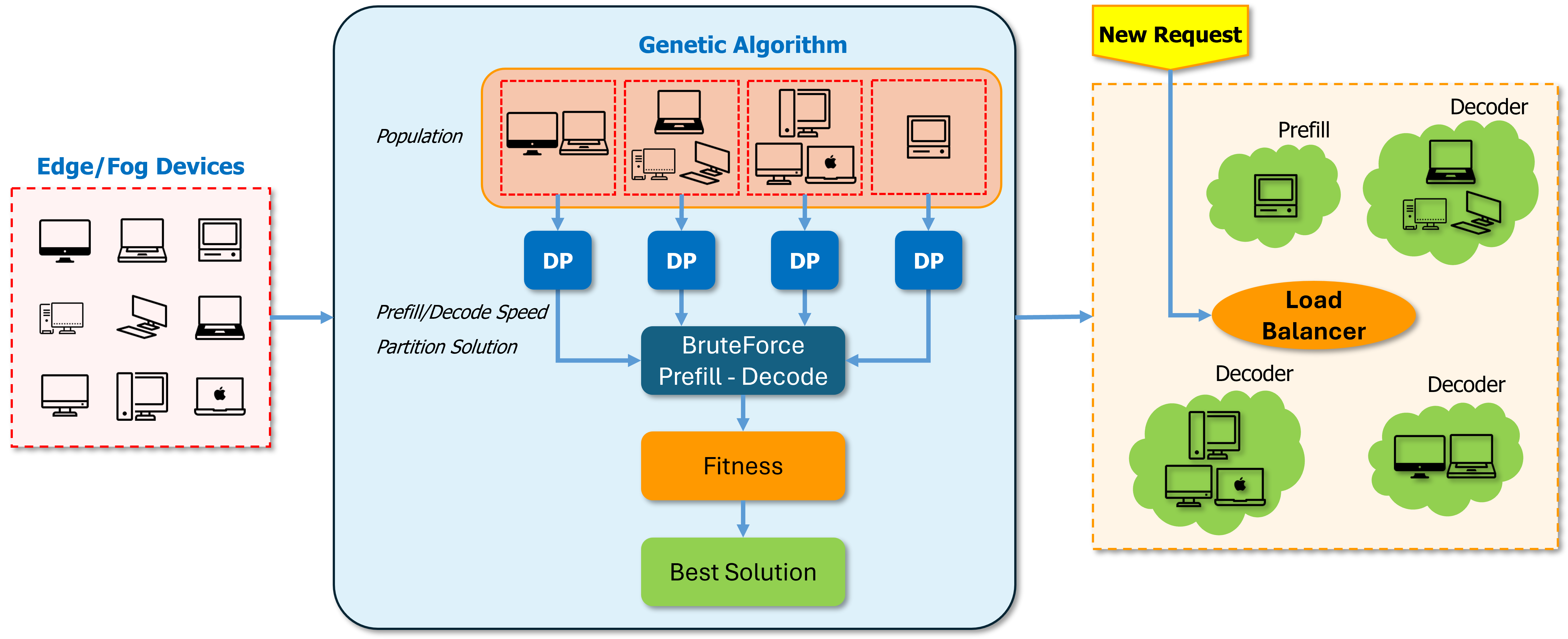}
    \caption{Overview of E2LLM Architecture}
    \label{fig:Overview}
\end{figure*}

\textbf{Parallelization techniques for distributing computation.} To speed up processing and handle large models, researchers use parallelism to split the workload across multiple devices or processors. 
Pipeline parallelism (or model parallelism) splits the model into stages and assigns each stage to a different device. While one stage processes the current input, the next stage can start working on the previous input, creating an assembly-line effect. 
EdgeShard (Zhang et al.) \cite{zhang2024edgeshard} and Le et al. \cite{le2024optimal} highlight a key property of pipeline parallelism in edge-cloud inference: while splitting a DNN into multiple stages and executing them concurrently can reduce latency through overlapping computation and communication, the overall throughput is fundamentally limited by the slowest stage in the pipeline. This bottleneck effect means that even if other stages finish quickly, the end-to-end performance depends on balancing partition sizes and resource allocation to avoid idle time. 
Moreover, graph partitioning methods such as Scotch \cite{pellegrini2012scotch} and METIS \cite{karypis1997metis} can balance workloads and reduce inter-stage communication; however, they optimize a cut objective, which cannot express $\max(\text{stage compute}, \text{link time})$ under the linear ordering a pipeline imposes; a balanced, low-cut partition therefore need not minimize the slowest stage.
PipeEdge~\cite{hu2022pipeedge} partitions layers across heterogeneous edge devices with a dynamic program that minimizes the slowest pipeline stage, placing communication time inside a maximum with computation because the two overlap. Our recurrence follows theirs; the difference is that we fix the device ordering --- admissible because transformer blocks are identical in size and cost --- which reduces the complexity from $O(2^{D}L^{2}D^{2})$ to $O(M^{2}N^{2})$.

\textbf{Replication deployment.} 
Replication improves throughput and latency by routing requests to less-loaded replicas. HexGen \cite{jiang2023hexgen} applies it to LLM inference on heterogeneous cloud GPUs, jointly optimizing replica placement, pipeline partitioning and tensor-parallel configuration to minimize total computation and communication time. HexGen-2~\cite{jiang2025hexgen2} extends this to disaggregated prefill/decode serving on heterogeneous GPUs, forming replica groups by spectral partitioning and sizing them by total memory divided by the replica footprint. It is the closest prior work to ours. ThunderServe~\cite{jiang2025thunderserve} targets KV-cache transfer cost under heterogeneous networks. Both optimize a sum of computation and communication time---a per-request finishing time---which cannot express $\max(\text{stage}, \text{link})$ and therefore does not identify the bottleneck stage that governs throughput once the pipeline is saturated.

Splitwise \cite{patel2024splitwise} separates inference into the Prefill and Decoder phases of Figure \ref{fig:Phases} and provisions each phase separately according to the input-to-generated token ratio. The prefill phase consumes all input tokens in one parallel pass, whereas decoding emits a single token per iteration and therefore dominates end-to-end latency.
This reduces energy consumption and latency relative to uniform execution. However, Splitwise assumes each serving machine holds a complete replica --- possibly through tensor parallelism across eight GPUs --- which no single device in an Edge/Fog cluster can do. Its role assignment is also empirical rather than procedural: the preference for the weaker accelerator in the token phase is reported as a cost- and power-efficiency observation over two homogeneous machine types, not as a constraint, and machines migrate between prefill, decode and mixed pools at runtime. Neither the observation nor the mixed pool transfers directly to a cluster in which a replica is itself composed of several heterogeneous devices and replica strength is undefined until after partitioning.

Furthermore, HexGen, HexGen-2, ThunderServe, and Splitwise all assume CUDA-class accelerators and datacenter-grade interconnect. This raises an important question: \textit{how well can these approaches adapt to edge or fog servers, where resources are far more limited and heterogeneous?} Deploying LLM in such environments introduces new challenges, including constrained memory, lower compute capacity and memory bandwidth. Addressing these challenges is critical for enabling cost-efficient, low-latency, and privacy-preserving AI services beyond the cloud.

\section{Proposed Method}\label{sec.method}

Figure~\ref{fig:Overview} presents the architecture of E2LLM, which aims to enable efficient LLM deployment on centralized heterogeneous Edge/Fog environments. The system first collects device capabilities and network bandwidth information, then applies a Genetic Algorithm to identify clusters that maximize throughput while satisfying QoS constraints. Within each cluster, Dynamic Programming and layer-level latency profiling are used to partition the model and minimize pipeline bottlenecks. Finally, a brute-force search determines the optimal numbers of Prefill and Decoder replicas to minimize waiting time, after which the selected configuration is deployed with a Join Shortest Queue (JSQ) load balancer \cite{patel2024splitwise}.

\subsection{Latency profiling}\label{subsec.Method.profiling}
To accurately predict system performance, we begin by profiling the latency of each layer on each device. Hence, when having a configuration of distributing a pipeline on a cluster, we can identify the time to finish one inference pass, and the bottleneck within pipeline. However, profiling latency at the layer level is challenging because LLM requires enormous memory, making it impractical to load the entire model for detailed analysis. Fortunately, most LLMs share a common architecture consisting of three main components: a language model head, multiple Transformer blocks—which account for the majority of the model’s memory—and an output layer. To simplify profiling, we assume that Transformer blocks of the same type (sliding-window attention, full attention, etc.) have similar latency as transformers of the same type having the same number of FLOPs and GPU kernels. 
We therefore adapt llama.cpp\footnote{\url{https://github.com/ggml-org/llama.cpp}} to load the model with a reduced number of Transformer blocks, which lets us measure per-layer latency without exhausting device memory. It is important to note that in llama.cpp, both the language model head and the output layer must reside on the master device. This constraint ensures correct computation and coordination during inference.

\subsection{Dynamic Programming for model partitioning in pipeline parallelism}\label{subsec.Method.DP}

In this subsection we focus on partitioning a model across the multiple devices of a single replica.
Unlike EdgeShard \cite{zhang2024edgeshard}, which optimizes a single forward pass and then schedules to reduce idle time, we optimize the steady-state bottleneck directly, since with multiple batches in flight the slowest stage governs throughput \cite{le2024optimal}.

\begin{algorithm}[h]
\small
\caption{DP-based Pipeline Allocation algorithm}\label{alg:DP}
\begin{algorithmic}[1]
\Require $M,N$: number of devices/transformer blocks;
\Require $\ell$: profiled compute time of blocks on each device;
\Require $\text{Mem}$: memory constraint of each device;
\Require $B$: profiled link bandwidth between devices;
\Procedure{DP\_Pipe}{$M$,$N$,$\ell$,$\text{Mem}$,$B$}
    \State $\mathrm{DP}_\mu(i,k) \gets +\infty$; $\mathrm{TB}_\mu(i,k)\gets-1~\forall \mu,i,k$
    \For{$\mu \gets 1 \to M$}
        \For{$i \gets 0 \to N-1$}
            \For{$\forall k \gets 1 \to M$}
                \If{$k==1$}
                    \State Eq.\ref{eq:DP2} with mem constraint checking
                    \State $\mathrm{TB}_\mu(i,k) \gets -1$
                \Else
                    \State Eq.\ref{eq:DP1} with mem constraint checking
                    \State store $\mathrm{TB}_\mu(i,k)$ with best $j-1$
                \EndIf
            \EndFor
        \EndFor
    \EndFor

    \State $\mu^{\star} \gets \arg\min_{\mu} \mathrm{DP}_\mu(N-1,M)$

    \State $\text{N\_Layers} \gets [];~i\gets N-1;~k\gets M$
    \While{$i>0$}
        \State $j \gets \mathrm{TB}_{\mu^{\star}}(i,k)$
        \State $\text{N\_Layers.append}(i-j+1)$
        \State $i\gets j-1;~k \gets k-1$
    \EndWhile
    \State \Return N\_Layers.reverse(), $\mu^{\star}$
\EndProcedure
\end{algorithmic}
\end{algorithm}

Our experiments confirm this behavior. With $M$ requests in flight the pipeline emits one token every $T_{slowest}$, so \emph{aggregate} decode throughput is $1/T_{slowest}$. An individual request receives one token per end-to-end forward pass $\Lambda = \sum_k \ell_k + \sum_k \tau(k-1,k)$; per-request TPOT is therefore $\Lambda$, equal to $M\cdot T_{slowest}$ only when the pipeline is balanced. We minimize $T_{slowest}$ for aggregate throughput and impose the per-request quality constraint separately as $1/\Lambda \ge \text{Min\_TPS}$. Additionally, since we are distributing multiple Transformer blocks of identical size, the ordering of nodes in the pipeline is generally irrelevant, except for the node assigned the master role. This master node is responsible for hosting the language model head and output layer. Hence, we just use random or sorted order of the node. For the master node, we simply add the latency of the language model head and the output layer in the dynamic programming process.

Algorithm \ref{alg:DP} provide a pseudo-code of the dynamic programming along with the back tracing process (line 16-22) to determine how many Transformer blocks need to be allocated on each device as well as the master node. The recurrence turns on two decisions. First, which device hosts the master role, enumerated by the outer loop (line 3). Second, given that blocks $j\ldots i$ are placed on device $k$, the best allocation of blocks $0\ldots j-1$ over the first $k-1$ devices is already known, so only the cost of the new stage has to be compared against the running bottleneck (lines 10--11); for $k=1$ the bottleneck is the single stage itself (lines 6--8).

\vspace{-0.5cm}
\begin{equation}
\begin{aligned}
\mathrm{DP}_\mu(i,k) = \min_{0 \le j \le i} \max\Big\{
    &\mathrm{DP}_\mu(j-1,k-1),\; \ell_k(j,i,[k{=}\mu]),\\
    &\tau(k-1,k) \Big\}
\end{aligned}
\label{eq:DP1}
\end{equation}
\begin{equation}
\mathrm{DP}_\mu(i,1) = \ell_1(0,i,[1{=}\mu])
\label{eq:DP2}
\end{equation}

For a fixed master $\mu$ and node ordering, $\mathrm{DP}_\mu(i,k)$ is the minimum achievable bottleneck when the first $i$ blocks are assigned to the first $k$ devices. $\ell_k(j,i,m)$ is the measured compute time of blocks $j\ldots i$ on device $k$, plus the language-model head and output layer when $m$ holds. $\tau(k-1,k)=A/B(k-1,k)$ is the time in seconds to move one microbatch of activations of size $A$ over a link of bandwidth $B$; it sits inside the maximum rather than adding to $\ell_k$ because computation and communication overlap, so a link is itself a pipeline stage. Memory-infeasible candidates take $+\infty$, and the solution is $\min_\mu \mathrm{DP}_\mu(N-1,M)$. Algorithm~\ref{alg:DP} has complexity $O(M^{2}N^{2})$, against $O(M^{2}N^{2}2^{M})$ for EdgeShard~\cite{zhang2024edgeshard} and $O(2^{D}L^{2}D^{2})$ for PipeEdge~\cite{hu2022pipeedge}, both of which search over device subsets; we avoid that by fixing the node ordering. The algorithm returns no solution when the memory constraints cannot be satisfied.

\subsection{Prefill and Decoder phase analysis for phase chosen criteria}\label{subsec.Method.PrefillDecode}
As mentioned earlier, a typical LLM operates in two phases: Prefill and Decoder (Figure \ref{fig:Phases}), each with distinct performance characteristics. Normally, the Prefill phase is generally compute-bound, while the Decoder phase tends to be memory-bound  \cite{patel2024splitwise, jiang2025thunderserve}. 
In addition, analyzing the ratio between input tokens and generated tokens is an important step before making deployment decisions, mainly for task‑specific LLM models such as code generation \cite{gu2023llm}. 
However, general‑purpose LLMs may exhibit highly variable or unseen ratios, making it difficult to perform accurate analysis beforehand. In such cases, a real‑time monitoring and adaptation mechanism would be beneficial, allowing the deployment plan to be dynamically adjusted based on the actual behavior of recent requests. We leave the design and integration of such real‑time analysis techniques for future work.

To study the impact of token distribution, we analyze two datasets from Hugging Face: "lz1bytedance/LongReason" (LZByteDance) \cite{ling2025longreason} and "chillies/DPO\_ielts\_writing" (IELTS) \cite{chillies_dpo_ielts_writing}. For the IELTS dataset, we randomly select five essays per request and ask the model to evaluate them. For LZByteDance, we randomly select a single question for each request. This setup is intended to create workloads with different characteristics and computational demands. Table \ref{tab:RatioPrefillDecoderDataset} summarizes the average ratio between input (INP) and generated (GEN) tokens for these cases in 1000 requests, providing insights into how token distribution impacts Prefill and Decoder workloads.

\begin{table}[]
    \centering
    \caption{Average number of input and generated tokens.}
    \vspace{-0.2cm}
    \label{tab:RatioPrefillDecoderDataset}
    \begin{tabular}{|c|c|c|c|c|}
        \hline
         Model & Dataset &INP & GEN & ratio\\\hline
         Gpt-Oss-20b & LZByteDance  &576 & 588 & 0.98\\\cline{2-5}
         &IELTS & 2291 & 1385 & 1.65\\\hline
         Qwen3-30B-A3B & LZByteDance  &576 & 746 & 0.77\\\cline{2-5}
         Q4\_K\_M&IELTS & 2291 & 1957 & 1.17\\\hline
    \end{tabular}
    \vspace{-0.3cm}
\end{table}

Consider a simple deployment scenario with one Prefill replica and one Decoder replica. Each request must first pass through the Prefill stage and then proceed to the Decoder stage, which naturally forms a two-stage pipeline. In this setup, the system’s throughput is constrained by the slowest stage, a characteristic of pipeline parallelism. Let $PS$ denote the Prefill speed (i.e., tokens per second), $DS$  the Decoder speed, $NP$ the analyzed average number of \textit{input tokens}, and $ND$ the analyzed average number of \textit{generated tokens}. Both $PS$ and $DS$ of a replica are obtained after having the partition plan from Algorithm \ref{alg:DP}.
Hence, when multiple requests arrive at the system, the bottleneck phase
(i.e., the slowest stage) is formulated as:
\begin{equation}
    \mathrm{bottleneck\_phase} = \max\!\left( \frac{NP}{PS},\ \frac{ND}{DS} \right)
    \label{eq:bottleneckPrefillDecoder}
\end{equation}

Consider a workload in which requests arrive at intervals of $T_a$ seconds. The system remains stable only when the bottleneck stage completes service at least as quickly as new requests arrive. Consequently, the formation of a backlog is governed by the excess of the bottleneck stage's service time over the arrival interval. Any positive excess results in sustained queue growth:
\begin{equation}
\mathrm{bottleneck} = \max\!\left(0,\ \mathrm{bottleneck\_phase} - T_a \right)
    \label{eq:bottleneckSystem}
\end{equation}

However, the latency experienced by a request is determined by more than the backlog alone. Even in the absence of queueing, no output token can be generated until the prompt has been fully processed. As a result, the Prefill stage contributes its entire service time, $NP/PS$, to the user-observed latency. Under load, the per-request delay therefore consists of two components: the intrinsic Prefill service time and the queueing delay accumulated at the bottleneck stage. This yields:
\begin{equation}
w =
\begin{cases}
\dfrac{NP}{PS} + \mathrm{bottleneck}, & \mathrm{bottleneck} > 0 \\[6pt]
0, & \text{otherwise}
\end{cases}
\label{eq:perRequestWaiting}
\end{equation}


This type of bottleneck accumulates over time, significantly increasing the waiting time for incoming requests \cite{le2024optimal}. Therefore, minimizing this bottleneck is one of the key advantages of E2LLM, as it ensures better responsiveness and improves user experience under heavy workloads. In contrast, approaches that rely on fixed constraints---where the most powerful device or replica handles the prefill phase while others are assigned to the decoding phase---tend to be less flexible and may exacerbate performance imbalances.

When scaling up the number of Prefill and Decoder replicas, the adjustment is straightforward: we replace the per-replica input and generated token rates with the overall aggregated rates by summing all values across replicas. In other words, $NP$ and $ND$ should reflect the combined throughput of all replicas, ensuring that the bottleneck calculation accounts for the increased parallelism.

\subsection{Genetic Algorithm for deployment LLM on Heterogeneous Edge/Fog Environment}\label{subsec.Method.GA}

\begin{algorithm}[h]
    \small
    \caption{Genetic Algorithm for deployment plan}\label{alg:GA}
\begin{algorithmic}[1]
    \Require $\text{Min\_TPS}$: min generated tokens per second;
    \Require $M,N$: number of devices/transformer blocks;
    \Require $\ell_P$: profiled Prefill compute time of blocks on devices;
    \Require $\ell_D$: profiled Decoder compute time of blocks on devices;
    \Require $\text{Mem}$: memory constraint of each device;
    \Require $B$: profiled link bandwidth between devices;
    \Require $P,G,Q$: number of population/generations/elite citizens;
    \Require $b_{\max}$: maximum KV-cache slots per replica;
    \State Randomly initialize population of $P$ citizens
    \For{$\text{iter} \gets 1 \to G$}
        \For{$\text{citizen} \in \text{population}$}
            \State $\text{Prefill\_Perf} \gets [\,]$;~$\text{Decoder\_Perf} \gets [\,]$
            \For{$rep \in \text{citizen.replicas}$}
                \State $res \gets \text{DP\_Pipe}(|rep|,N,\ell_P,\text{Mem},B)$
                \State $\text{Prefill\_Perf.append}(res)$
                \For{$b \gets 1 \to b_{\max}$}
                    \State $res \gets \text{DP\_Pipe}(|rep|,N,\ell_D(b),\text{Mem},B)$
                    \If{$res$ feasible \textbf{and} $1/\Lambda(res)\ge\text{Min\_TPS}$}
                        \State $\text{Decoder\_Perf.append}(res,b)$
                    \EndIf
                \EndFor
            \EndFor
            \State Brute-force role assignment; keep highest feasible $b$
            \State Obtain fitness (max.\ waiting time, Sec.~\ref{subsec.Method.PrefillDecode})
            \State Store best solution with best fitness
            \State Update the $Q$ elite citizens
        \EndFor
        \State population $\gets$ \{Cross-over; Mutation; Elite citizens\}
    \EndFor
    \State \Return best solution
\end{algorithmic}
\end{algorithm}

Replica formation is the intractable part of the problem (Section \ref{sec.intro}), so we search it with a Genetic Algorithm (Algorithm \ref{alg:GA}). Its effectiveness depends entirely on the gene representation and the crossover and mutation policies, which we therefore state in full. Each chromosome pair encodes node ordering and replica grouping; e.g.\ $[(0,1,2,3,4,5),(3,1,2)]$ places nodes $0$–$2$ in the first replica, node $3$ in the second, and nodes $4,5$ in the third. Crossover recombines the ordering chromosome and inherits the grouping chromosome from one parent, followed by a repair pass that removes duplicate nodes. Mutation fires with probability $0.3$ and, when it fires, swaps two positions in the ordering ($20\%$), regenerates the grouping from a random position ($50\%$), regenerates the grouping entirely ($15\%$), or regenerates both chromosomes ($15\%$). A fixed number of globally best individuals is retained across generations and used in crossover. Appendix \ref{apd.GAComponents} describes these implementations in details.

After generating a chromosome (line 1 and 20), the next step is straightforward. We decompose the chromosome into multiple replicas according to its grouping structure and then apply Algorithm \ref{alg:DP} to determine the optimal deployment plan for each role. For Prefill role, most of the time, only one request can fill up maximum tokens to be processed at once (line 6-7). For Decoder role, we need to find the highest number of parallel requests (line 8-13) that can be processed at the same time to increase throughput while guaranteeing the minimum generated token per second. Moreover, we also cache the result of each replica for reuse, decreasing processing time. This integration ensures that the Genetic Algorithm leverages dynamic programming for fine-grained optimization within each replica, combining global search with local refinement. 

Role assignment is then brute-forced against the congestion model of Section \ref{subsec.Method.PrefillDecode}. Because Edge/Fog clusters hold few replicas, exhaustively enumerating both the role assignment and the number of parallel requests per replica remains cheap, so this step is exact given the grouping the GA supplies.

\subsection{Quality-of-Service Definition}\label{subsec.Method.QoS}
Under heavy demand, requests may queue before processing begins, so a strict waiting-time threshold is impractical; what users notice directly is the rate at which text appears. We therefore express Quality-of-Service as a generation-speed floor. Brysbaert et al.\ \cite{brysbaert2019many} report an average reading speed of about 238 words per minute, and fast readers reach 7--15 words per second; treating one word as one token, \textit{the deployed system should sustain at least 7 tokens per second per request}. Section \ref{sec.eval} describes how this target is translated into the planner's $\text{Min\_TPS}$ threshold.

\section{Evaluation and Discussion}\label{sec.eval}
\providecommand{\figscale}{0.31}


\subsection{Experiment setup}
In the experiments, we use eight devices, as shown in Table \ref{tab:dev}, ranging from embedded to desktop devices. Although the DGX Spark devices are equipped with 128~GB of unified memory, we limit the memory available for model deployment to 12~GB in order to emulate the resource-constrained conditions typical of Edge/Fog environments. Furthermore, our benchmarking results show that DGX Spark provides lower computational performance than the RTX5070, making it a suitable representative of an edge device in our evaluation.

Another key factor that affects system performance is network bandwidth, especially during the decoder phase. For example, transferring an activation of 20kB over a 100Mbps network takes about 1.6ms, not counting network delays. If the system processes 100 tokens per second (about 10ms per token), this extra time reduces performance by roughly 14.6\%. To avoid this bottleneck, each device in a replica should be capable of communicating with each other with a high speed connection. Hence, we use an Edge/Fog environment where all devices are connected through a high-speed Local Area Network (LAN) using a switch, making them as a centralized cluster. Based on our measurements, each device can communicate directly with others at a speed of 920Mbps.

\begin{table}[]
    \centering
    \caption{Device information. (U.M: Unified Memory)}
    \vspace{-0.3cm}
    \label{tab:dev}
    \begin{tabular}{|c|c|c|c|c|}
        \hline
         Device Name & VRAM & RAM & U.M& Dev.ID\\\hline
         RTX5070 & 12GB & 64GB & No & Dev.1\\
         Ryzen 7 9700X & & & & \\\hline
         RTX3060M & 6GB & 64GB & No & Dev.2\\
         Ryzen 5 5800H & & & &\\\hline
         Apple M2 Max & 22GB & 32GB & Yes & Dev.3\\\hline
         Apple M2 Max & 22GB & 32GB & Yes & Dev.4\\\hline
         Apple M1 & 12GB & 16GB & Yes & Dev.5\\\hline
         Apple M1 & 12GB & 16GB & Yes & Dev.6\\\hline
         DGX Spark* & 12GB & 128GB & Yes & Dev.7\\\hline
         DGX Spark* & 12GB & 128GB & Yes & Dev.8\\\hline
    \end{tabular}
    \vspace{-0.5cm}
\end{table}

In addition to the hardware setup, we also define the software and data environment for our experiments. To simulate realistic user requests, we use the datasets LZByteDance and IELTS introduced in Section \ref{sec.method}. We evaluate two state-of-the-art open-source LLMs, Gpt-Oss-20b and Qwen3-30B-A3B-Q4\_K\_M, which contain 24 and 48 Transformer blocks, respectively
Furthermore, we observe that the Apple M2 Max provides approximately 22~GB of memory available for deployment, which is sufficient to host the entire target model. However, it can support only four concurrent requests, as each request requires additional memory for KV-cache storage. 
In this work, the maximum context length is limited to 8,192 tokens; longer contexts would require fewer concurrent requests to avoid OOM failures. This also highlights an additional benefit of pipeline parallelism: besides improving performance, it increases the memory available to a replica and therefore enables higher request concurrency. For example, adding an Apple M1 device with 12~GB of memory increases the capacity from four to eight concurrent requests across two pipeline stages.

\subsection{Baseline chosen}

The most relevant baselines are HexGen, HexGen2, ThunderServe and SplitWise. HexGen minimizes total pipeline execution time rather than the slowest stage, whereas its successor HexGen2 targets per-request finishing time across both phases. We therefore evaluate a \emph{finishing-time} (FT) variant: our planner with the objective replaced by HexGen2's finishing time, holding search space, profiling and runtime fixed. This isolates the objective, but tests only the objective and not HexGen2's max-flow allocator; a faithful port is impossible here, since HexGen2 is NCCL- and CUDA-only and most of our cluster is Apple Silicon.
We also exclude ThunderServe: its main optimization reduces prefill-to-decoder KV-cache transfer, which dominates only when devices as fast as the H100 or A100 generate quickly enough for the network to bind. On our uniform 920~Mbps links that transfer is negligible.

Finally, SplitWise does not use pipeline parallelism, but it offers a more detailed analysis of the prefill and decoder phases and sets the number of replicas per role from the input-to-generated token ratio. As noted in Section~\ref{sec.relatedwork}, it gives no procedure for \emph{which} replicas take which role---only an empirical observation over two homogeneous machine types. We therefore promote that observation into an explicit rule: within each candidate solution, the strongest replicas take the prefill role and the remaining ones decode. This is \emph{our} adaptation rather than SplitWise's design; \emph{adapted SplitWise} is best read as a controlled ablation of our role-assignment policy, holding the genetic search, the dynamic program and the runtime fixed.

\subsection{Experiment results}

\begin{table}[b]
    \centering
    \setlength{\tabcolsep}{2pt}
    \caption{Deployment plans when varying the target model (six devices, IELTS, $1000$~ms).}
    \vspace{-0.3cm}
    \label{tab:dep-model}
    \begin{subtable}[t]{0.48\linewidth}
        \centering
        \caption{Gpt-Oss / HG=ELS=SPW.}
        \begin{tabular}{|c|c|c|c|}
            \hline
            Role & Dev & Sl. & G--C\\\hline
            P & Dev.1$^{\star}$ & 1 & 20--4\\\hline
            \multirow{2}{*}{D} & Dev.2$^{\star}$ & \multirow{2}{*}{14} & 5--1\\
              & Dev.7 &  & 18--0\\\hline
            D & Dev.4$^{\star}$ & 6 & 24--0\\\hline
            \multirow{2}{*}{D} & Dev.5 & \multirow{2}{*}{2} & 15--0\\
              & Dev.6$^{\star}$ &  & 9--0\\\hline
        \end{tabular}
    \end{subtable}\hfill
    \begin{minipage}[t]{0.48\linewidth}
        \centering
        \begin{subtable}[t]{\linewidth}
            \centering
            \caption{Gpt-Oss / Metrics.}
            \begin{tabular}{|l|c|c|c|}
                \hline
                Method & PS & DS & W\\\hline
                HG & 2919 & 12.7 & 318.0\\\hline
                \textbf{ELS} & \textbf{2919} & \textbf{12.7} & \textbf{318.0}\\\hline
                SPW & 2919 & 12.7 & 318.0\\\hline
            \end{tabular}
        \end{subtable}
        \par\vspace{0.3cm}
        \begin{subtable}[t]{\linewidth}
            \centering
            \caption{Qwen3-MoE / Metrics.}
            \begin{tabular}{|l|c|c|c|}
                \hline
                Method & PS & DS & W\\\hline
                HG & 1411 & 11.0 & 794.1\\\hline
                \textbf{ELS} & \textbf{1411} & \textbf{11.0} & \textbf{794.1}\\\hline
                SPW & 1412 & 12.9 & 898.6\\\hline
            \end{tabular}
        \end{subtable}
    \end{minipage}

    \vspace{0.3cm}
    \begin{subtable}[t]{0.48\linewidth}
        \centering
        \caption{Qwen3-MoE / HG=ELS.}
        \begin{tabular}{|c|c|c|c|}
            \hline
            Role & Dev & Sl. & G--C\\\hline
            P & Dev.1$^{\star}$ & 1 & 26--22\\\hline
            \multirow{2}{*}{D} & Dev.2$^{\star}$ & \multirow{2}{*}{4} & 6--22\\
              & Dev.6 &  & 20--0\\\hline
            \multirow{3}{*}{D} & Dev.4$^{\star}$ & \multirow{3}{*}{12} & 23--0\\
              & Dev.5 &  & 7--0\\
              & Dev.7 &  & 18--0\\\hline
        \end{tabular}
    \end{subtable}\hfill
    \begin{subtable}[t]{0.48\linewidth}
        \centering
        \caption{Qwen3-MoE / SPW.}
        \begin{tabular}{|c|c|c|c|}
            \hline
            Role & Dev & Sl. & G--C\\\hline
            P & Dev.1$^{\star}$ & 1 & 26--22\\\hline
            \multirow{3}{*}{D} & Dev.2$^{\star}$ & \multirow{3}{*}{6} & 6--10\\
              & Dev.6 &  & 10--0\\
              & Dev.7 &  & 22--0\\\hline
            \multirow{2}{*}{D} & Dev.4$^{\star}$ & \multirow{2}{*}{8} & 36--0\\
              & Dev.5 &  & 12--0\\\hline
        \end{tabular}
    \end{subtable}
\end{table}

To comprehensively evaluate the benefits of our approach, we conduct experiments across four dimensions: model adaptability, workload characteristics, request arrival rates, and system scalability.
For every experiment below, each figure reports the per-request \emph{prefill speed}, \emph{decoding speed}, and \emph{maximum waiting time} (sorted by finished client index), comparing E2LLM against the HexGen- and adapted SplitWise-based baselines. The accompanying table lists the deployment plan produced by the three methods: HexGen-based (HG), E2LLM (ELS), and adapted SplitWise-based (SPW) for the corresponding configuration.
The ``Role'' column uses P for a prefill replica and D for a decoding replica; within a replica the master device is marked with $^{\star}$; ``Sl.'' is the KV-cache batch capacity (number of slots) of a replica, and ``G--C'' denotes the GPU--CPU split of the assigned Transformer blocks. 
A ``Metrics'' subtable additionally reports, for each configuration and the three methods, the prefill throughput $PS$ and decode throughput $DS$ (in TPS) and the maximum waiting time W (in seconds). It should be noted that the four-device configuration consists of Dev.1, Dev.2, Dev.5, and Dev.6, whereas the six-device configuration consists of Dev.1, Dev.2, Dev.4, Dev.5, Dev.6, and Dev.7. In addition, the request arrival rate are 100ms, 1000ms and 3000ms. 
In total, this results in 36 experimental configurations. However, due to space limitations, most of the detailed results are deferred to Appendix~\ref{apd.fullresults}. The main text focuses on the experiments that best illustrate the key findings and validate the effectiveness of our approach.

\begin{table}[b]
    \centering
    \setlength{\tabcolsep}{2pt}
    \caption{Deployment plans when varying the workload dataset (eight devices, Qwen3-MoE-30B, $3000$~ms).}
    \vspace{-0.3cm}
    \label{tab:dep-dataset}
    \begin{subtable}[t]{.48\linewidth}
        \centering
        \caption{IELTS / HG=ELS.}
        \begin{tabular}{|c|c|c|c|}
            \hline
            Role & Dev & Sl. & G--C\\\hline
            P & Dev.1$^{\star}$ & 1 & 26--22\\\hline
            \multirow{3}{*}{D} & Dev.2 & \multirow{3}{*}{9} & 7--3\\
              & Dev.7 &  & 19--0\\
              & Dev.8$^{\star}$ &  & 19--0\\\hline
            \multirow{2}{*}{D} & Dev.3$^{\star}$ & \multirow{2}{*}{8} & 36--0\\
              & Dev.6 &  & 12--0\\\hline
            \multirow{2}{*}{D} & Dev.4$^{\star}$ & \multirow{2}{*}{8} & 36--0\\
              & Dev.5 &  & 12--0\\\hline
        \end{tabular}
    \end{subtable}\hfill
    \begin{subtable}[t]{.48\linewidth}
        \centering
        \caption{IELTS / SPW.}
        \begin{tabular}{|c|c|c|c|}
            \hline
            Role & Dev & Sl. & G--C\\\hline
            \multirow{2}{*}{P} & Dev.7$^{\star}$ & \multirow{2}{*}{2} & 26--0\\
              & Dev.8 &  & 22--0\\\hline
            \multirow{2}{*}{D} & Dev.1$^{\star}$ & \multirow{2}{*}{10} & 13--19\\
              & Dev.2 &  & 5--11\\\hline
            \multirow{2}{*}{D} & Dev.3$^{\star}$ & \multirow{2}{*}{8} & 36--0\\
              & Dev.6 &  & 12--0\\\hline
            \multirow{2}{*}{D} & Dev.4$^{\star}$ & \multirow{2}{*}{8} & 36--0\\
              & Dev.5 &  & 12--0\\\hline
        \end{tabular}
    \end{subtable}

    \vspace{0.3cm}
    \begin{subtable}[t]{.48\linewidth}
        \centering
        \caption{LZByteDance / HG.}
        \begin{tabular}{|c|c|c|c|}
            \hline
            Role & Dev & Sl. & G--C\\\hline
            P & Dev.1$^{\star}$ & 1 & 26--22\\\hline
            \multirow{3}{*}{D} & Dev.2 & \multirow{3}{*}{9} & 7--3\\
              & Dev.7 &  & 19--0\\
              & Dev.8$^{\star}$ &  & 19--0\\\hline
            \multirow{2}{*}{D} & Dev.3$^{\star}$ & \multirow{2}{*}{8} & 36--0\\
              & Dev.6 &  & 12--0\\\hline
            \multirow{2}{*}{D} & Dev.4$^{\star}$ & \multirow{2}{*}{8} & 36--0\\
              & Dev.5 &  & 12--0\\\hline
        \end{tabular}
    \end{subtable}\hfill
    \begin{minipage}[t]{0.48\linewidth}
        \centering
        \begin{subtable}[t]{\linewidth}
            \centering
            \caption{IELTS / Metrics.}
            \begin{tabular}{|l|c|c|c|}
                \hline
                Method & PS & DS & W\\\hline
                HG & 1412 & 11.9 & 260.5\\\hline
                \textbf{ELS} & \textbf{1412} & \textbf{11.9} & \textbf{260.5}\\\hline
                SPW & 1776 & 11.8 & 342.9\\\hline
            \end{tabular}
        \end{subtable}
        \par\vspace{0.3cm}
        \begin{subtable}[t]{\linewidth}
            \centering
            \caption{LZByteDance / Metrics.}
            \begin{tabular}{|l|c|c|c|}
                \hline
                Method & PS & DS & W\\\hline
                HG & 1147 & 13.9 & 66.3\\\hline
                \textbf{ELS} & \textbf{188} & \textbf{21.2} & \textbf{1.8}\\\hline
                SPW & 954 & 12.8 & 57.7\\\hline
            \end{tabular}
        \end{subtable}
    \end{minipage}

    \vspace{0.3cm}
    \begin{subtable}[t]{.48\linewidth}
        \centering
        \caption{LZByteDance / ELS.}
        \begin{tabular}{|c|c|c|c|}
            \hline
            Role & Dev & Sl. & G--C\\\hline
            \multirow{2}{*}{P} & Dev.2$^{\star}$ & \multirow{2}{*}{2} & 8--20\\
              & Dev.6 &  & 20--0\\\hline
            \multirow{3}{*}{D} & Dev.1$^{\star}$ & \multirow{3}{*}{18} & 13--5\\
              & Dev.7 &  & 15--0\\
              & Dev.8 &  & 15--0\\\hline
            \multirow{2}{*}{D} & Dev.3$^{\star}$ & \multirow{2}{*}{8} & 36--0\\
              & Dev.5 &  & 12--0\\\hline
            D & Dev.4$^{\star}$ & 4 & 48--0\\\hline
        \end{tabular}
    \end{subtable}\hfill
    \begin{subtable}[t]{.48\linewidth}
        \centering
        \caption{LZByteDance / SPW.}
        \begin{tabular}{|c|c|c|c|}
            \hline
            Role & Dev & Sl. & G--C\\\hline
            \multirow{2}{*}{P} & Dev.1$^{\star}$ & \multirow{2}{*}{2} & 25--12\\
              & Dev.2 &  & 11--0\\\hline
            \multirow{3}{*}{D} & Dev.3$^{\star}$ & \multirow{3}{*}{12} & 12--0\\
              & Dev.7 &  & 18--0\\
              & Dev.8 &  & 18--0\\\hline
            \multirow{3}{*}{D} & Dev.4$^{\star}$ & \multirow{3}{*}{6} & 30--0\\
              & Dev.5 &  & 9--0\\
              & Dev.6 &  & 9--0\\\hline
        \end{tabular}
    \end{subtable}
\end{table}

\subsubsection{Model adaptability}
\label{sec:exp-model}
We first fix the cluster configuration (six devices), the workload (IELTS), and the request arrival interval ($1000$~ms), and vary the target model between Gpt-Oss-20b and Qwen3-MoE-30B (Q4\_K\_M). Table~\ref{tab:dep-model} presents the resulting deployment plans and aggregate performance metrics. All three methods adapt their deployment strategies when the target model changes. HexGen produces the same deployment plans as E2LLM for both models, whereas Adapted Splitwise selects different configurations.
Notably, Adapted Splitwise achieves a higher per-request decoding speed for Qwen3-MoE-30B, but supports fewer concurrent decoding slots. Consequently, although all methods achieve similar overall decoding throughput, HexGen and E2LLM can serve more requests simultaneously, resulting in lower queueing delays and smaller maximum waiting times.
This observation reveals a limitation of the analytical model in Section~\ref{subsec.Method.PrefillDecode}, which does not explicitly consider the effect of concurrency on waiting time. Incorporating this factor could lead to more accurate deployment decisions, which we leave for future work.

\subsubsection{Impact of the workload characteristics}
\label{sec:exp-dataset}

\begin{figure*}[t]
    \centering
    \begin{subfigure}{\figscale\textwidth}
        \includegraphics[width=\linewidth]{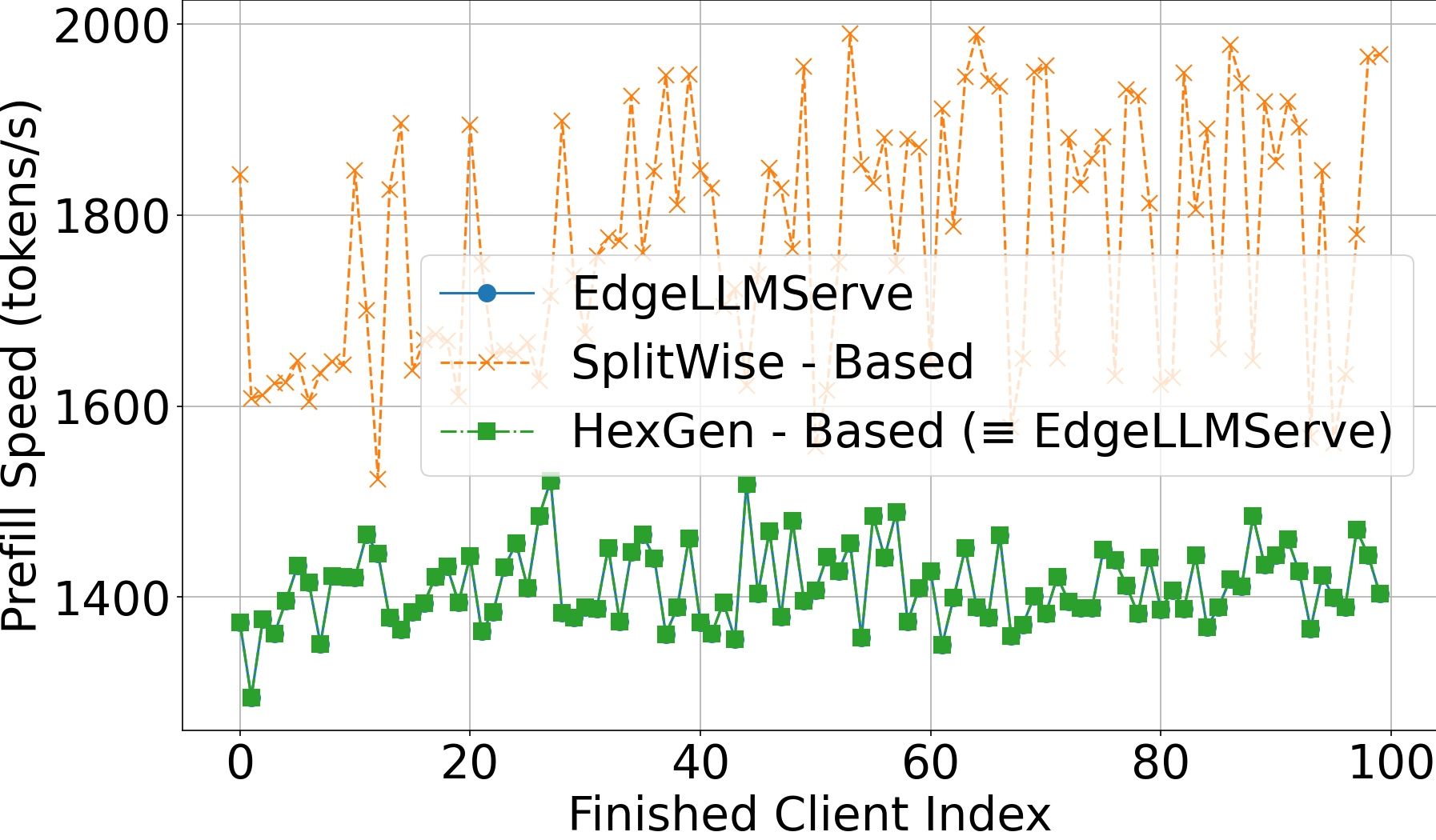}
        \vspace{-0.5cm}
        \caption{IELTS: prefill speed}
    \end{subfigure}\hfill
    \begin{subfigure}{\figscale\textwidth}
        \includegraphics[width=\linewidth]{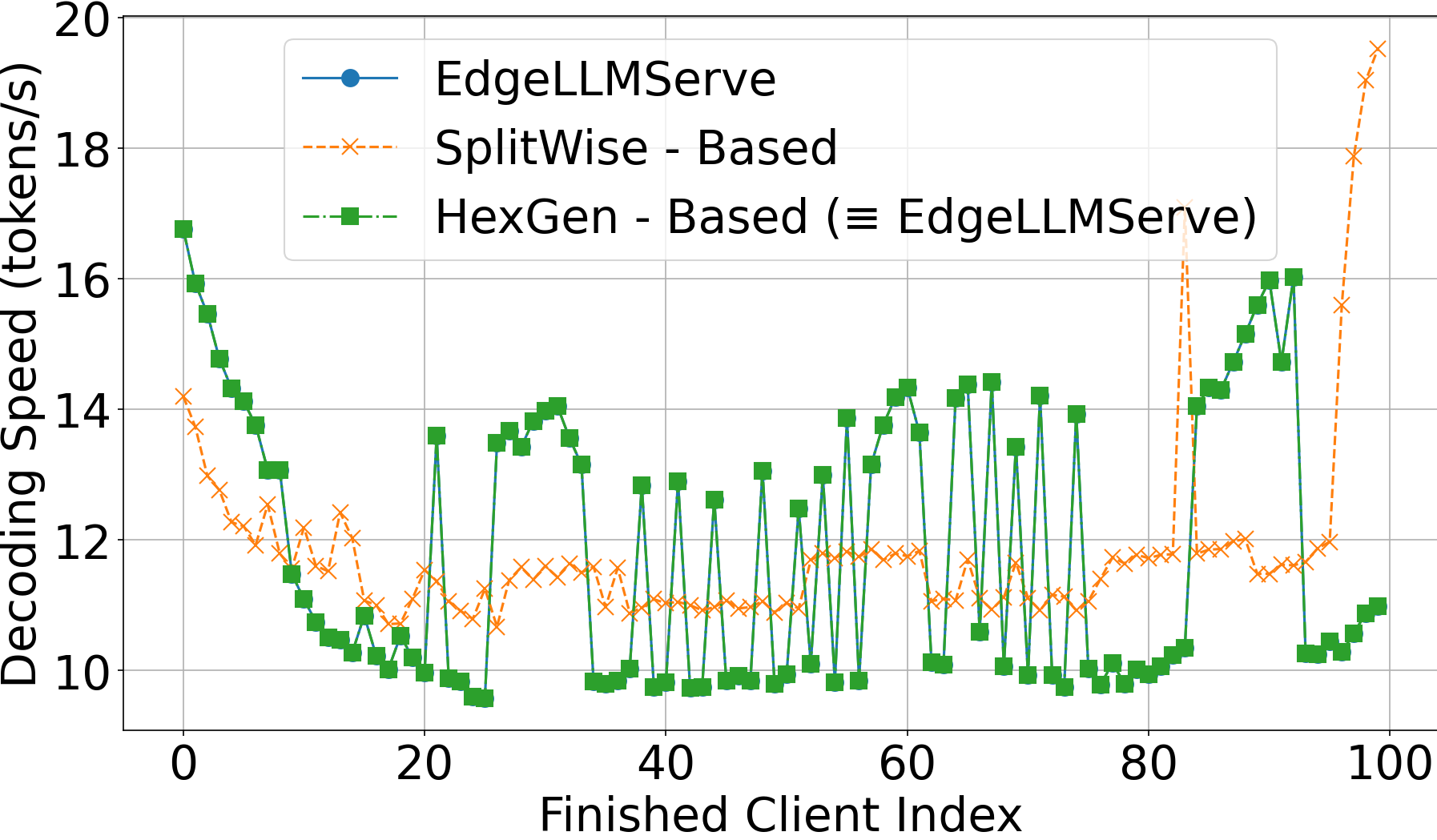}
        \vspace{-0.5cm}
        \caption{IELTS: decoding speed}
    \end{subfigure}\hfill
    \begin{subfigure}{\figscale\textwidth}
        \includegraphics[width=\linewidth]{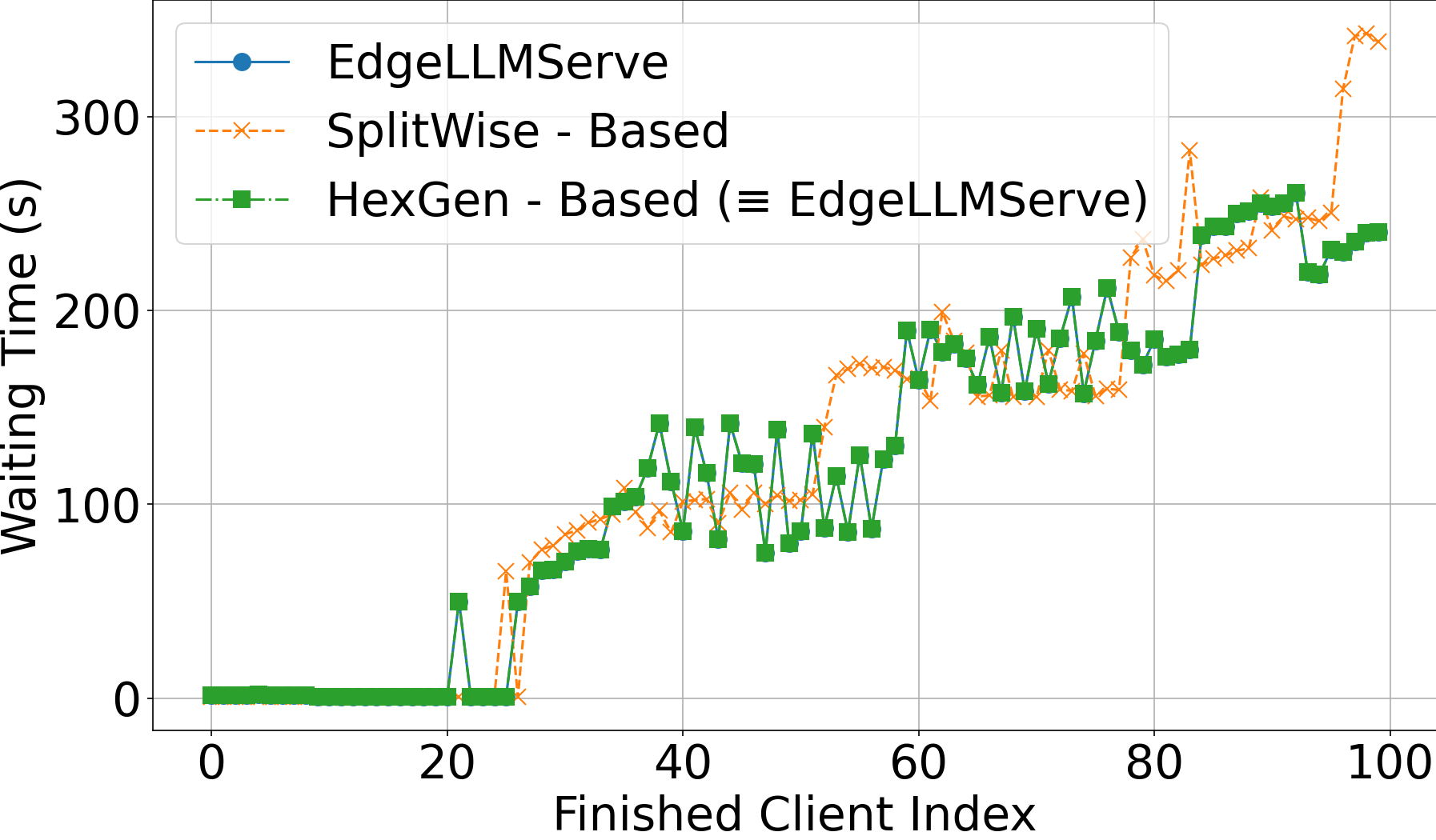}
        \vspace{-0.5cm}
        \caption{IELTS: waiting time}
    \end{subfigure}

    \vspace{0.1cm}
    \begin{subfigure}{\figscale\textwidth}
        \includegraphics[width=\linewidth]{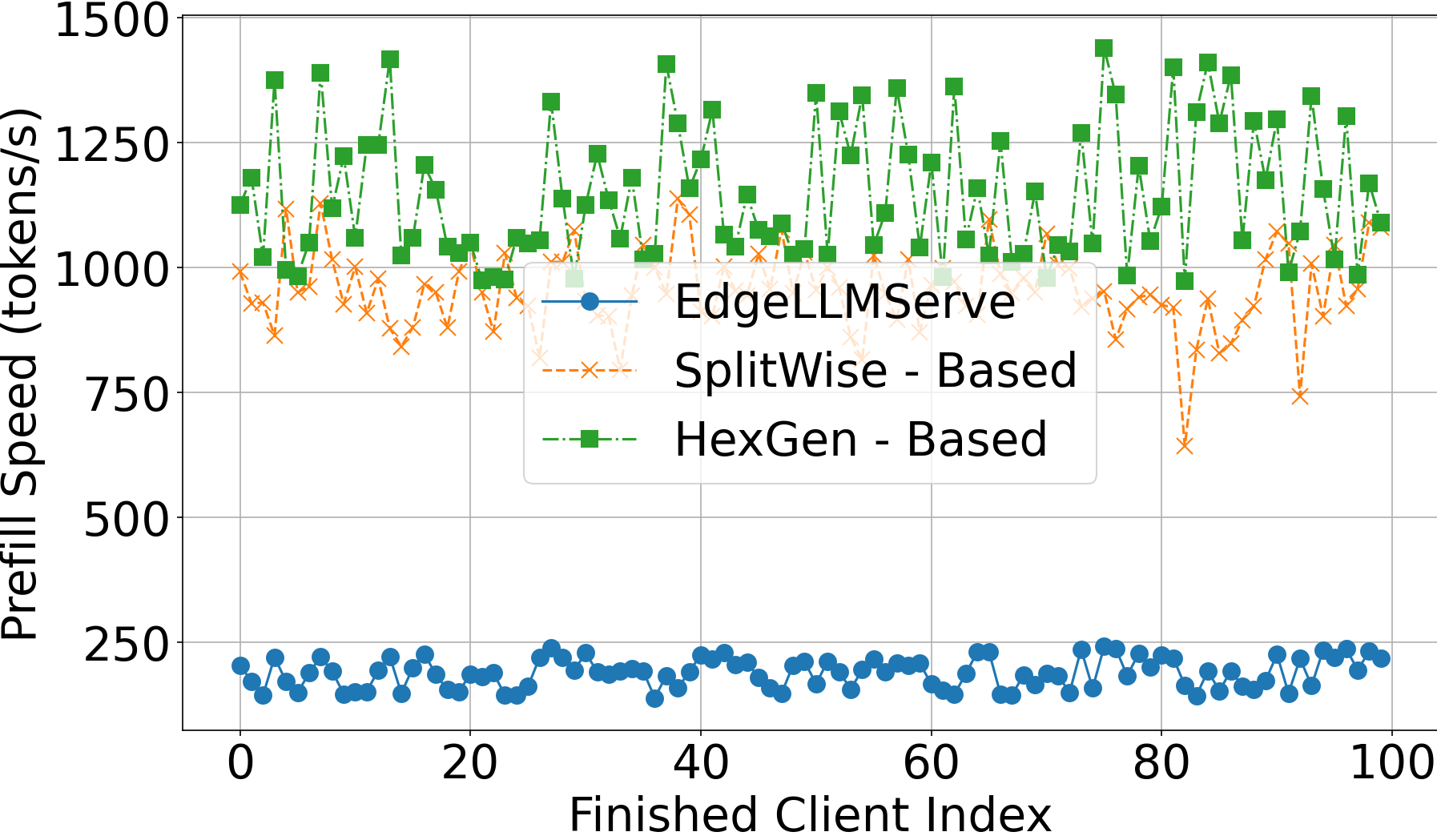}
        \vspace{-0.5cm}
        \caption{LZByteDance: prefill speed}
    \end{subfigure}\hfill
    \begin{subfigure}{\figscale\textwidth}
        \includegraphics[width=\linewidth]{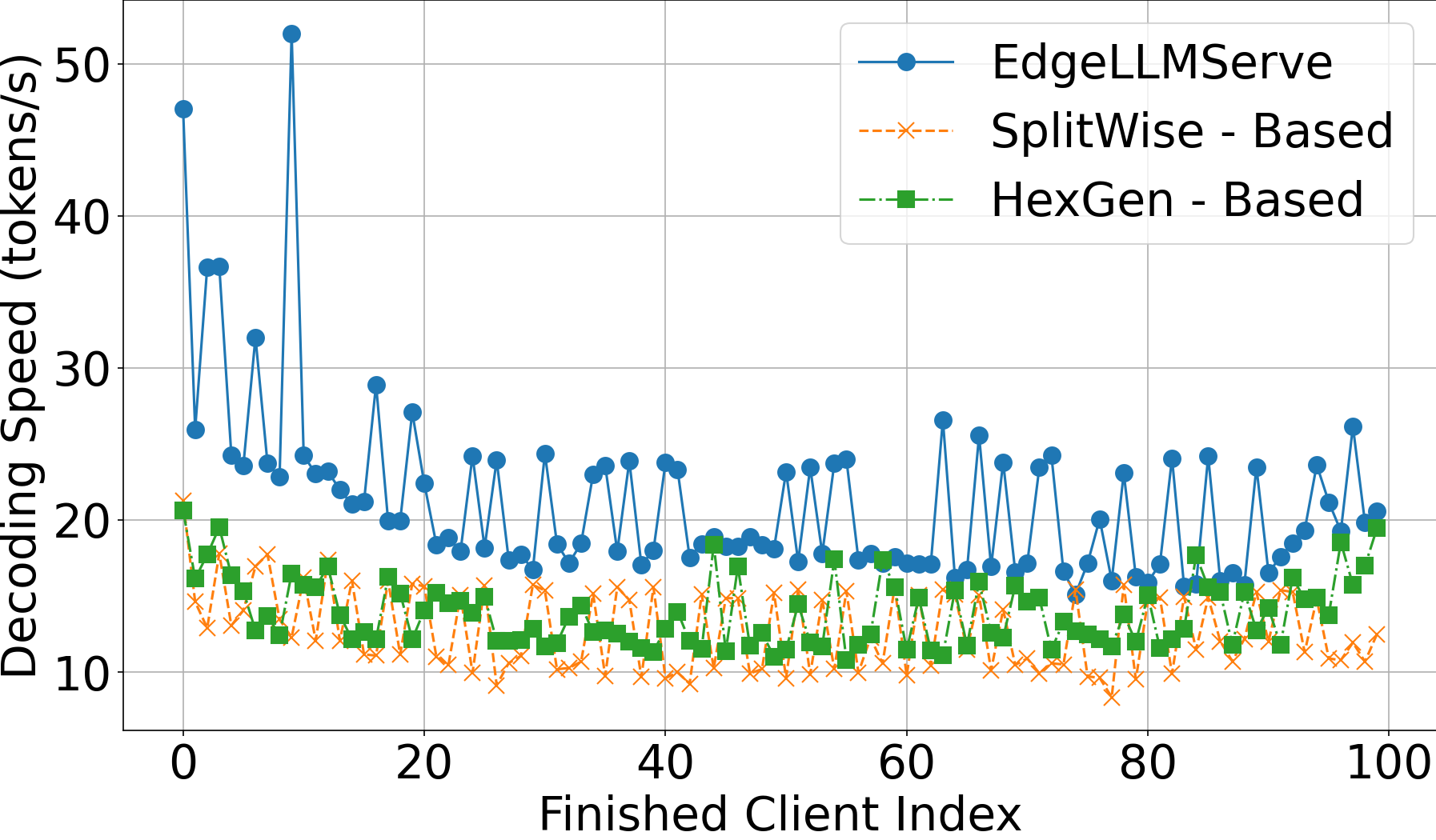}
        \vspace{-0.5cm}
        \caption{LZByteDance: decoding speed}
    \end{subfigure}\hfill
    \begin{subfigure}{\figscale\textwidth}
        \includegraphics[width=\linewidth]{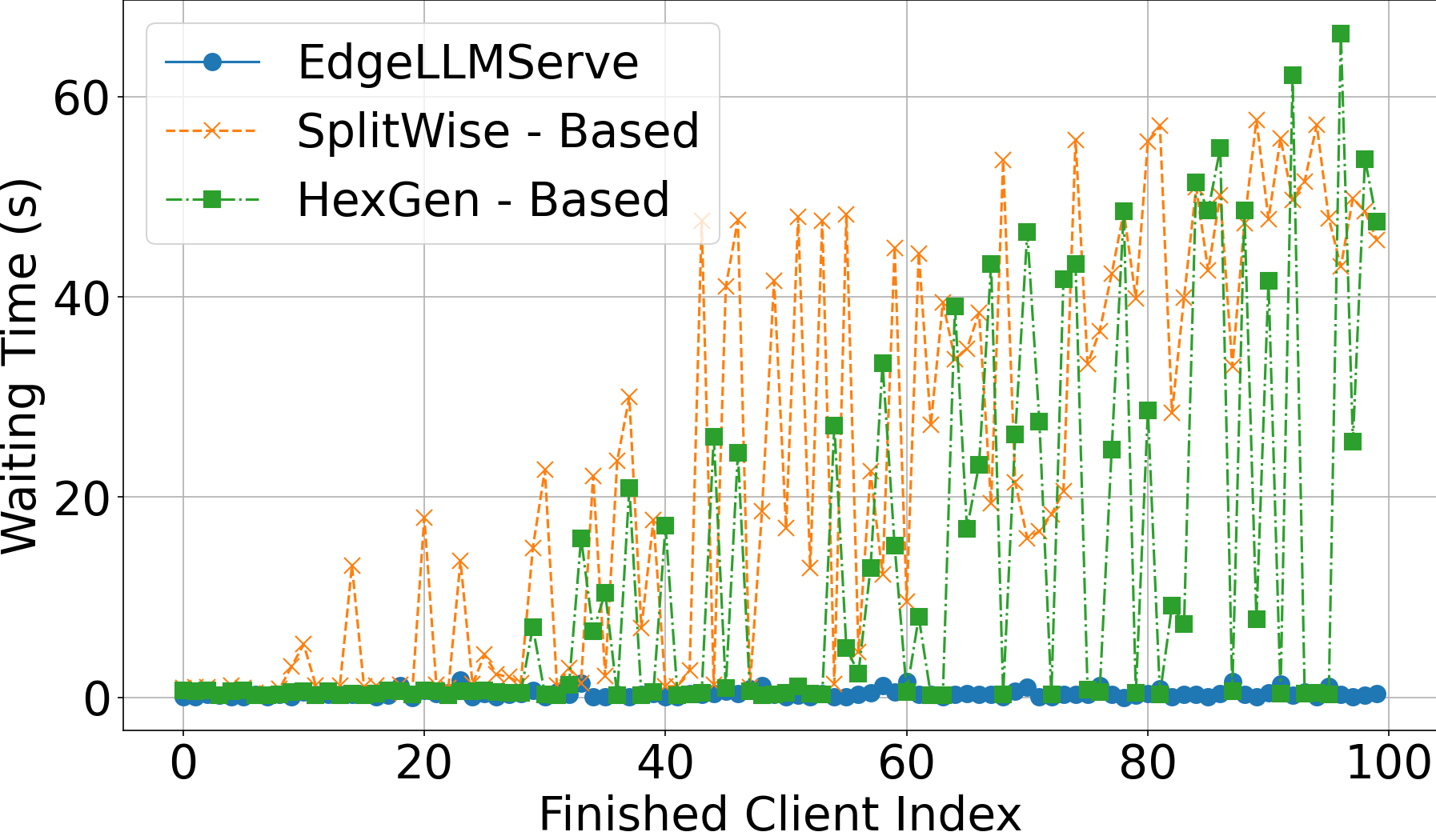}
        \vspace{-0.5cm}
        \caption{LZByteDance: waiting time}
    \end{subfigure}
    \vspace{-0.2cm}
    \caption{Impact of the workload dataset (eight devices, Qwen3-MoE-30B, $3000$~ms).}
    \label{fig:exp-dataset}
\end{figure*}

Next, we fix the cluster configuration (eight devices), the model (Qwen3-MoE-30B), and the request arrival interval (3,000 ms), while varying the workload between the IELTS and LZByteDance datasets. Table~\ref{tab:dep-dataset}, which reports the aggregate metrics, shows that E2LLM effectively adapts to changes in the input-to-output token ratio by rebalancing Prefill and Decoder capacities. As the workload shifts from IELTS to LZByteDance, the system correspondingly shifts the bottleneck from the Decoder phase to the Prefill phase by adjusting the number of Prefill replicas and allocating more computational resources to Decoder replicas when necessary.
Furthermore, Figure~\ref{fig:exp-dataset} shows that the waiting time of E2LLM remains close to zero for both datasets, indicating the absence of significant congestion in either phase. Consequently, the observed latency mainly consists of preprocessing overhead and KV-cache transmission time. In contrast, HexGen maintains the same deployment configuration across both workloads. As a result, its performance is comparable to E2LLM on the IELTS dataset but degrades significantly on the LZByteDance dataset, where the workload characteristics differ substantially.
Adapted Splitwise also demonstrates some degree of adaptability when the workload changes. However, because its resource allocation policy between Prefill and Decoder stages is implicit rather than workload-aware, it is unable to achieve the same level of performance as E2LLM. These results highlight the importance of dynamically balancing Prefill and Decoder resources according to workload characteristics in order to minimize queueing delays and improve overall system performance.

\begin{figure*}[b]
    \centering
    \begin{subfigure}{\figscale\textwidth}
        \includegraphics[width=\linewidth]{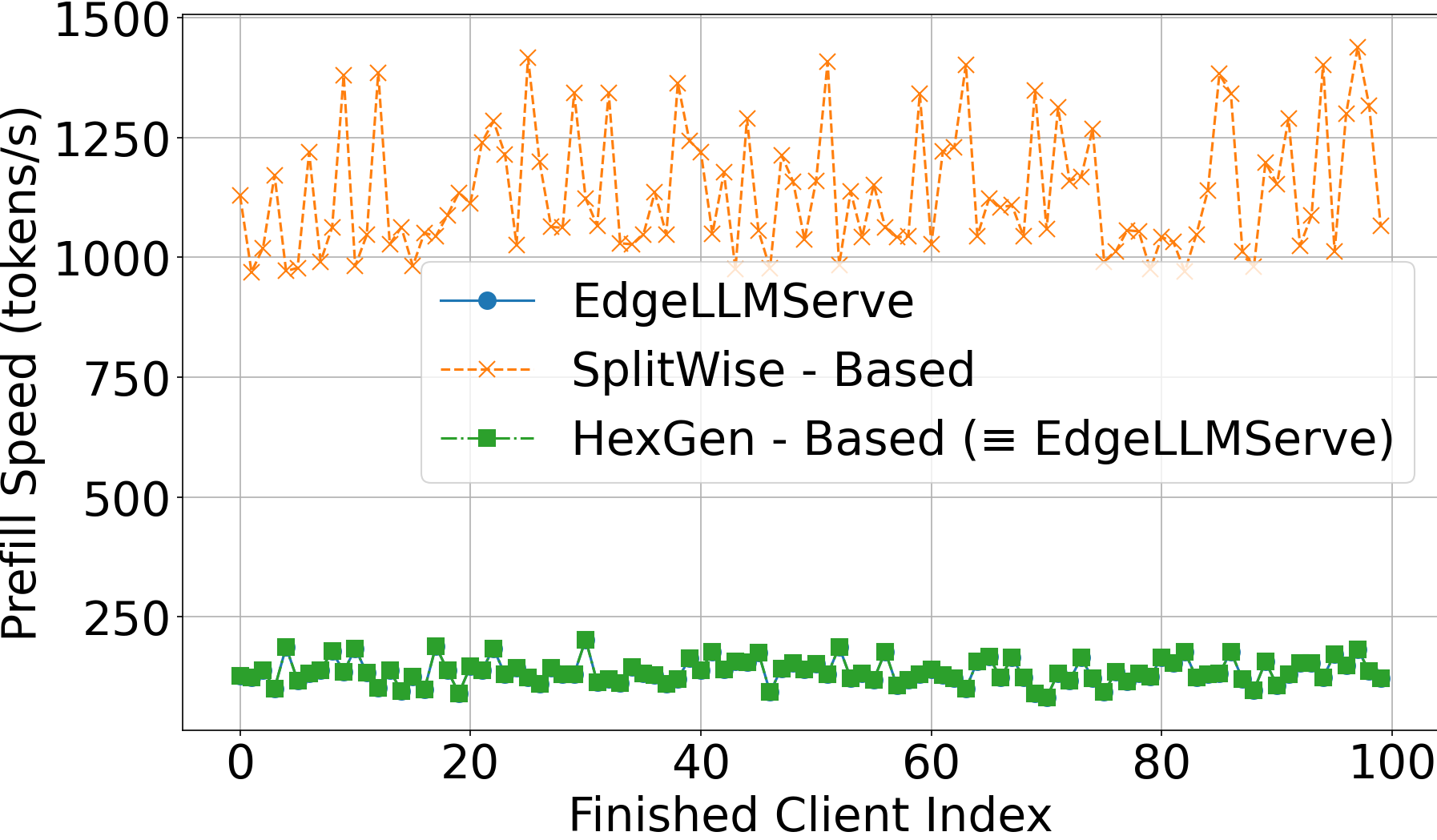}
        \vspace{-0.5cm}
        \caption{4 devices: prefill speed}
    \end{subfigure}\hfill
    \begin{subfigure}{\figscale\textwidth}
        \includegraphics[width=\linewidth]{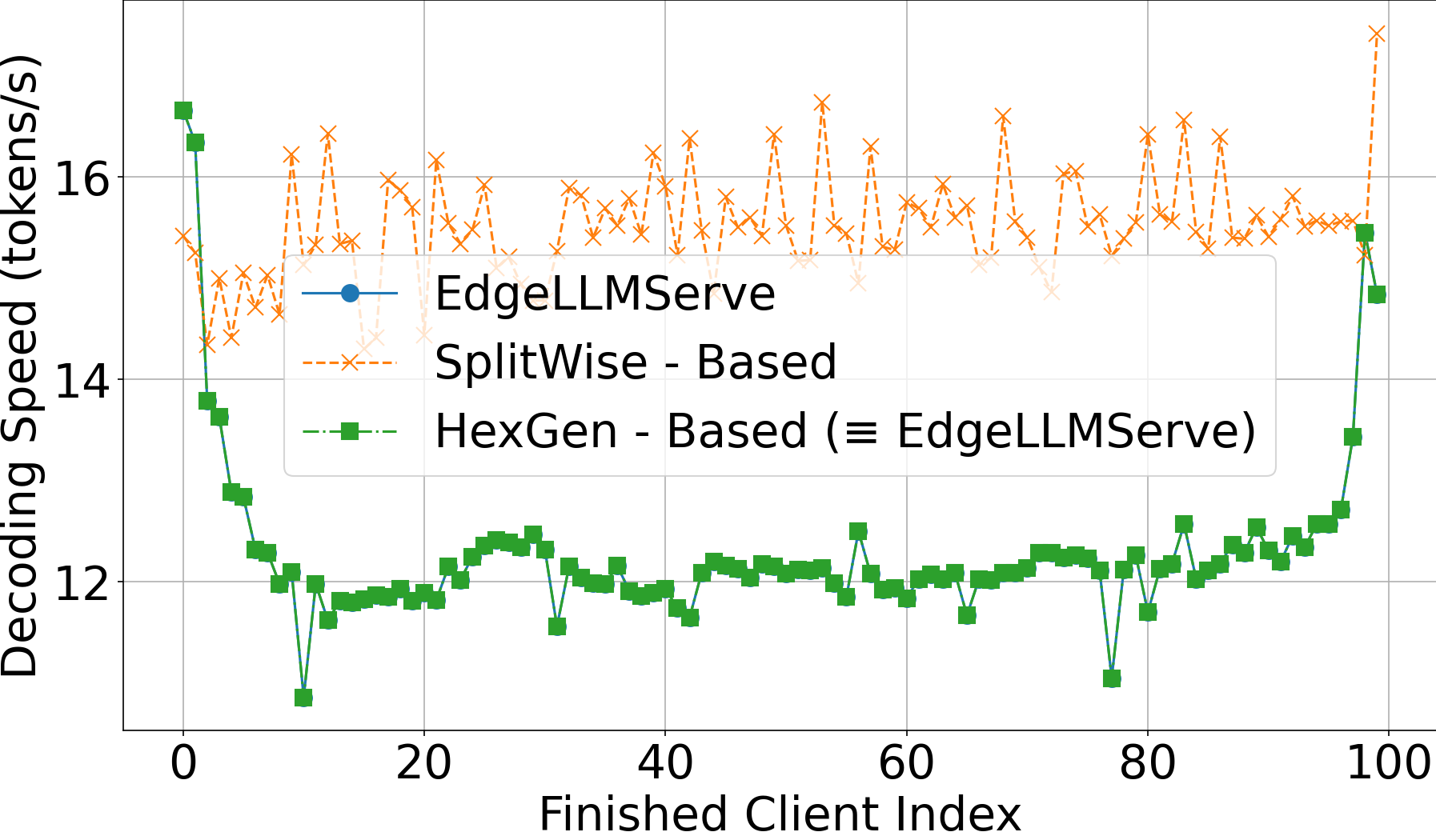}
        \vspace{-0.5cm}
        \caption{4 devices: decoding speed}
    \end{subfigure}\hfill
    \begin{subfigure}{\figscale\textwidth}
        \includegraphics[width=\linewidth]{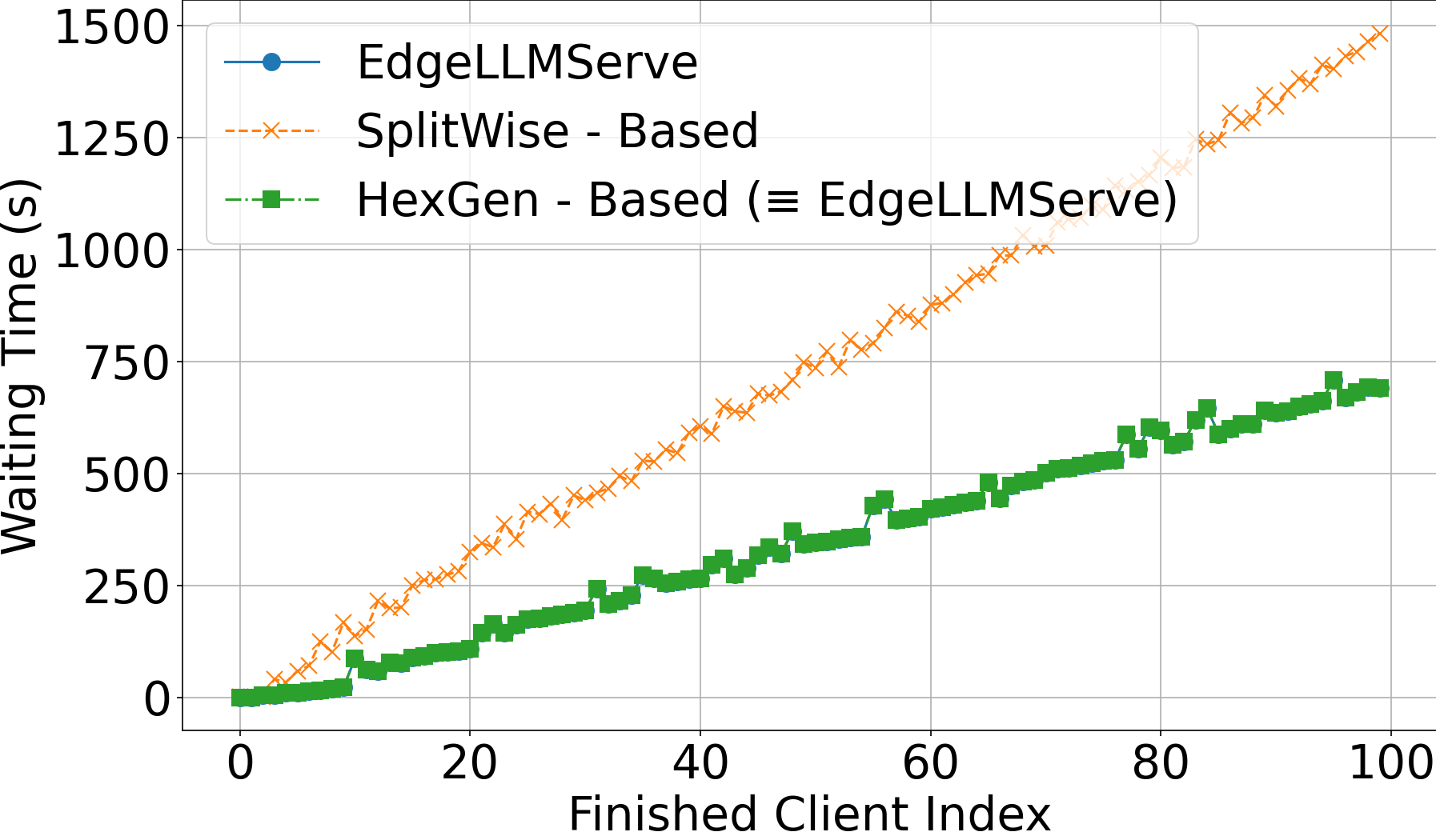}
        \vspace{-0.5cm}
        \caption{4 devices: waiting time}
    \end{subfigure}

    \vspace{0.1cm}
    \begin{subfigure}{\figscale\textwidth}
        \includegraphics[width=\linewidth]{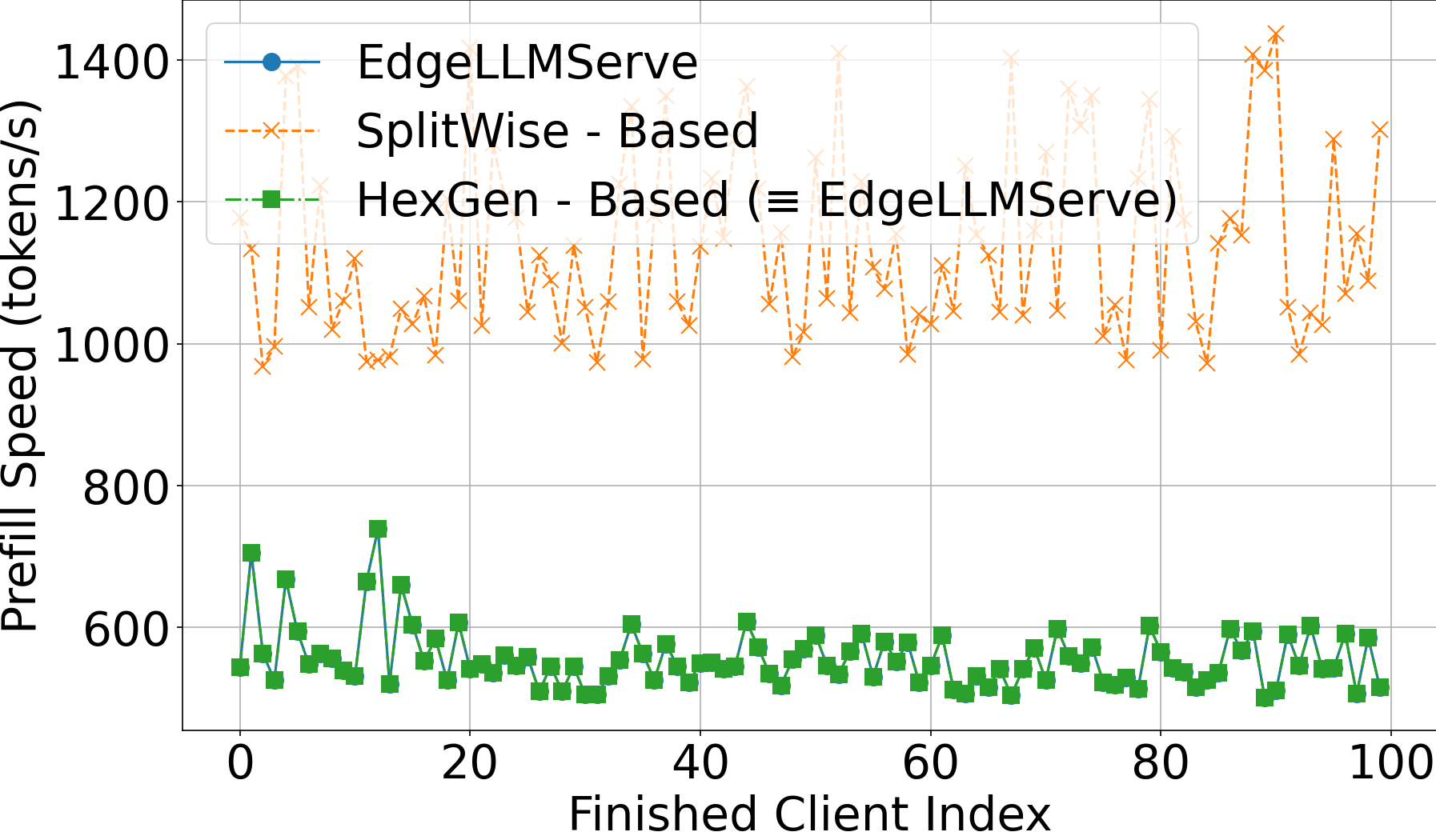}
        \vspace{-0.5cm}
        \caption{6 devices: prefill speed}
    \end{subfigure}\hfill
    \begin{subfigure}{\figscale\textwidth}
        \includegraphics[width=\linewidth]{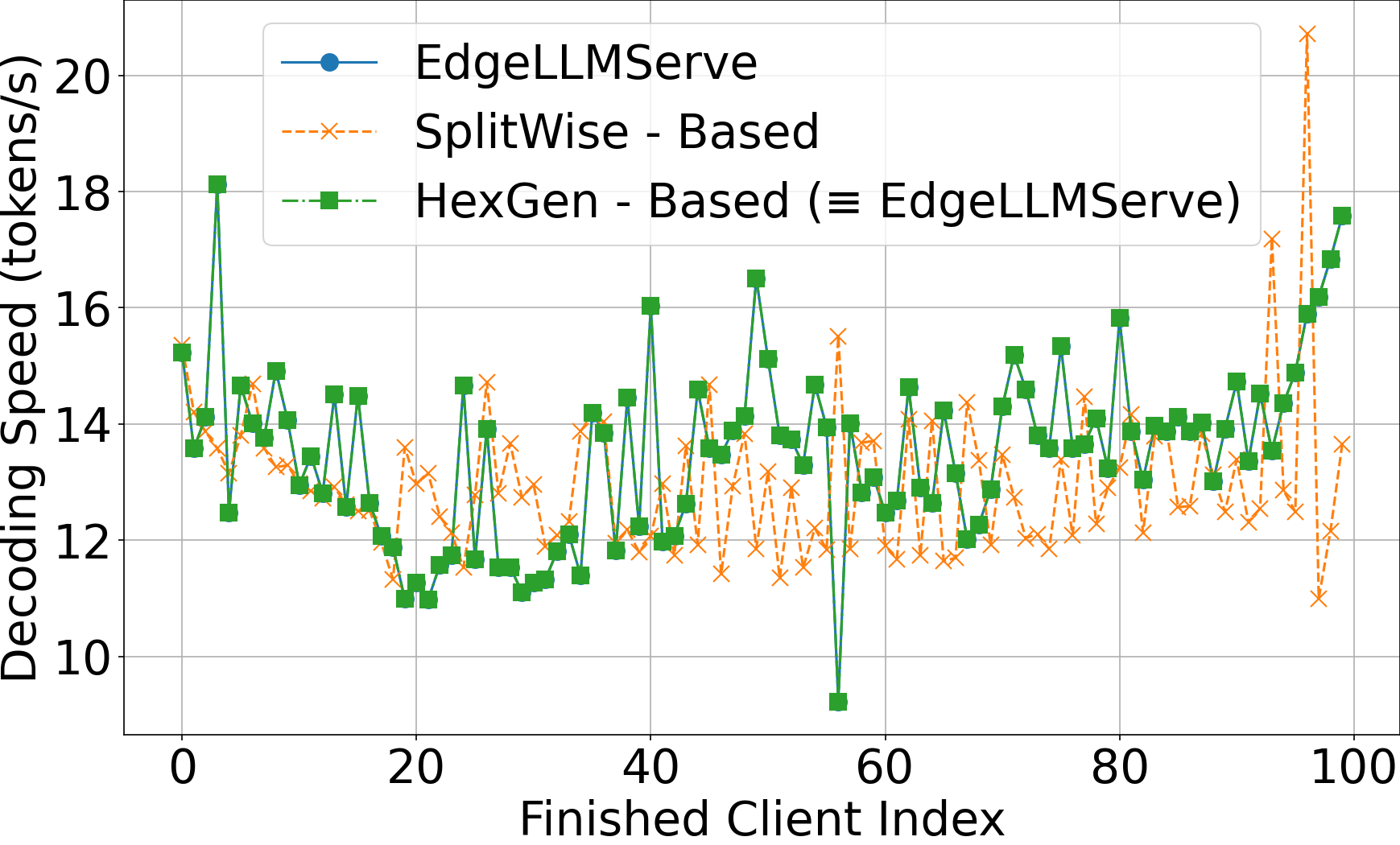}
        \vspace{-0.5cm}
        \caption{6 devices: decoding speed}
    \end{subfigure}\hfill
    \begin{subfigure}{\figscale\textwidth}
        \includegraphics[width=\linewidth]{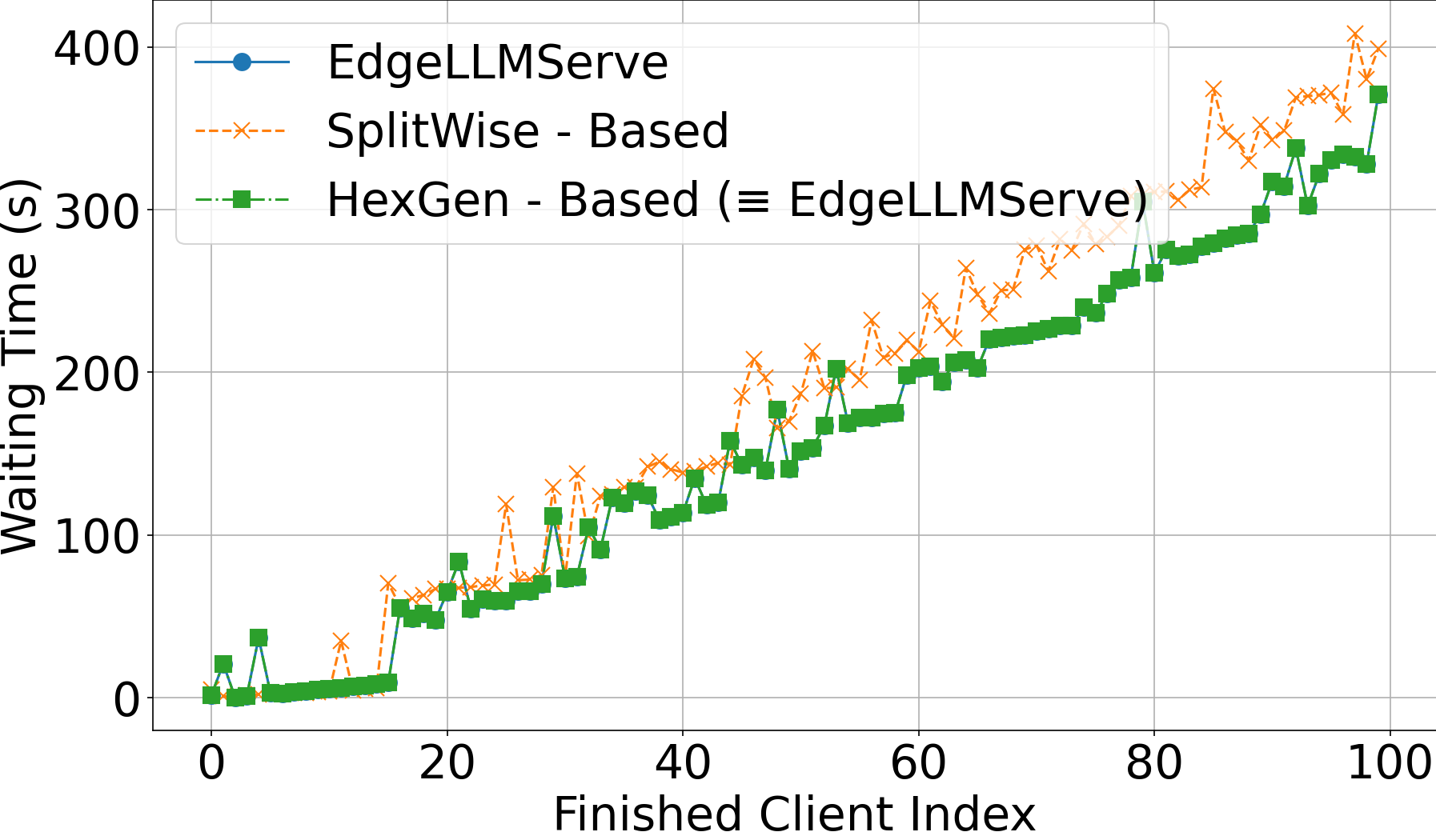}
        \vspace{-0.5cm}
        \caption{6 devices: waiting time}
    \end{subfigure}

    \vspace{0.1cm}
    \begin{subfigure}{\figscale\textwidth}
        \includegraphics[width=\linewidth]{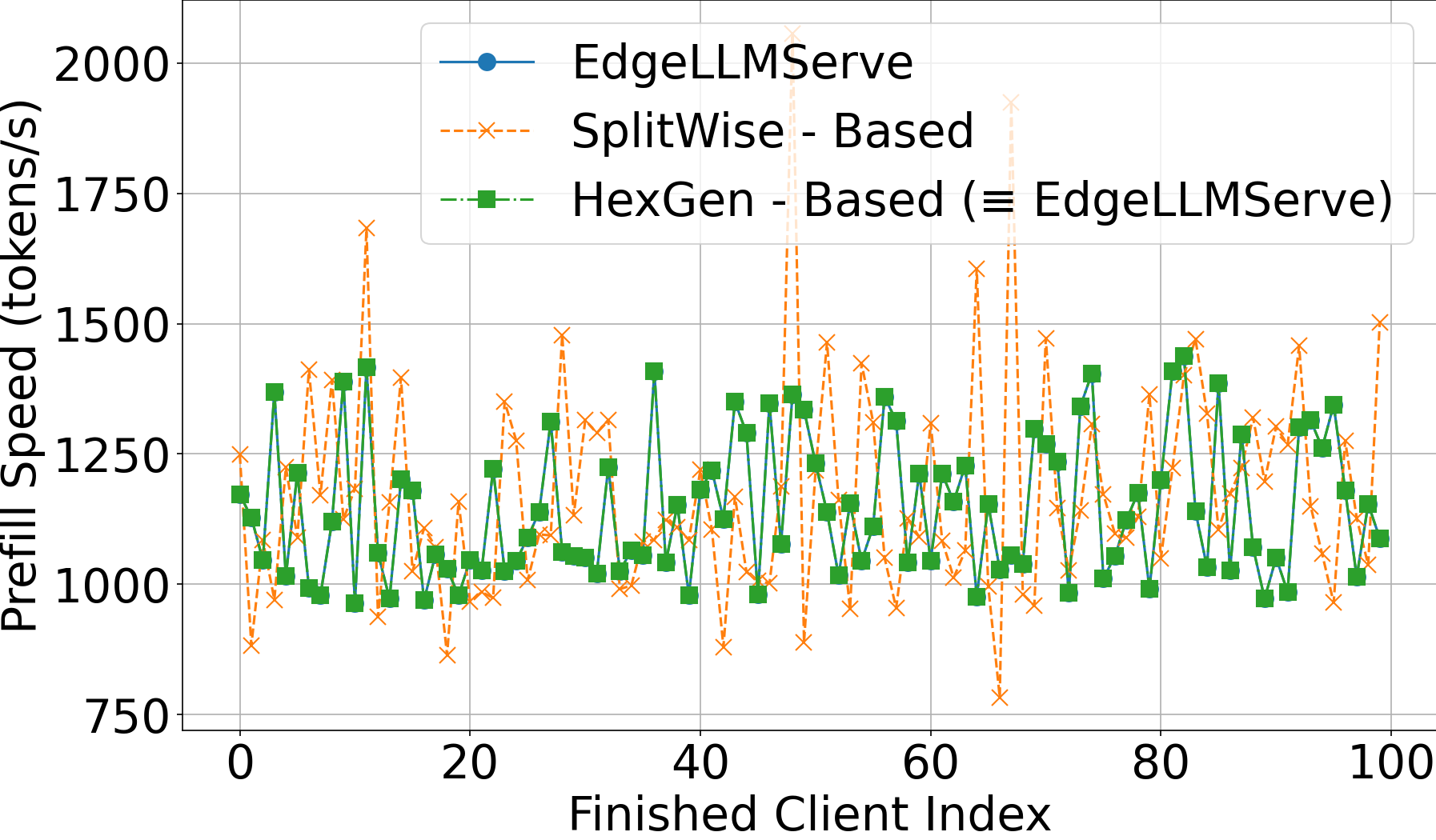}
        \vspace{-0.5cm}
        \caption{8 devices: prefill speed}
    \end{subfigure}\hfill
    \begin{subfigure}{\figscale\textwidth}
        \includegraphics[width=\linewidth]{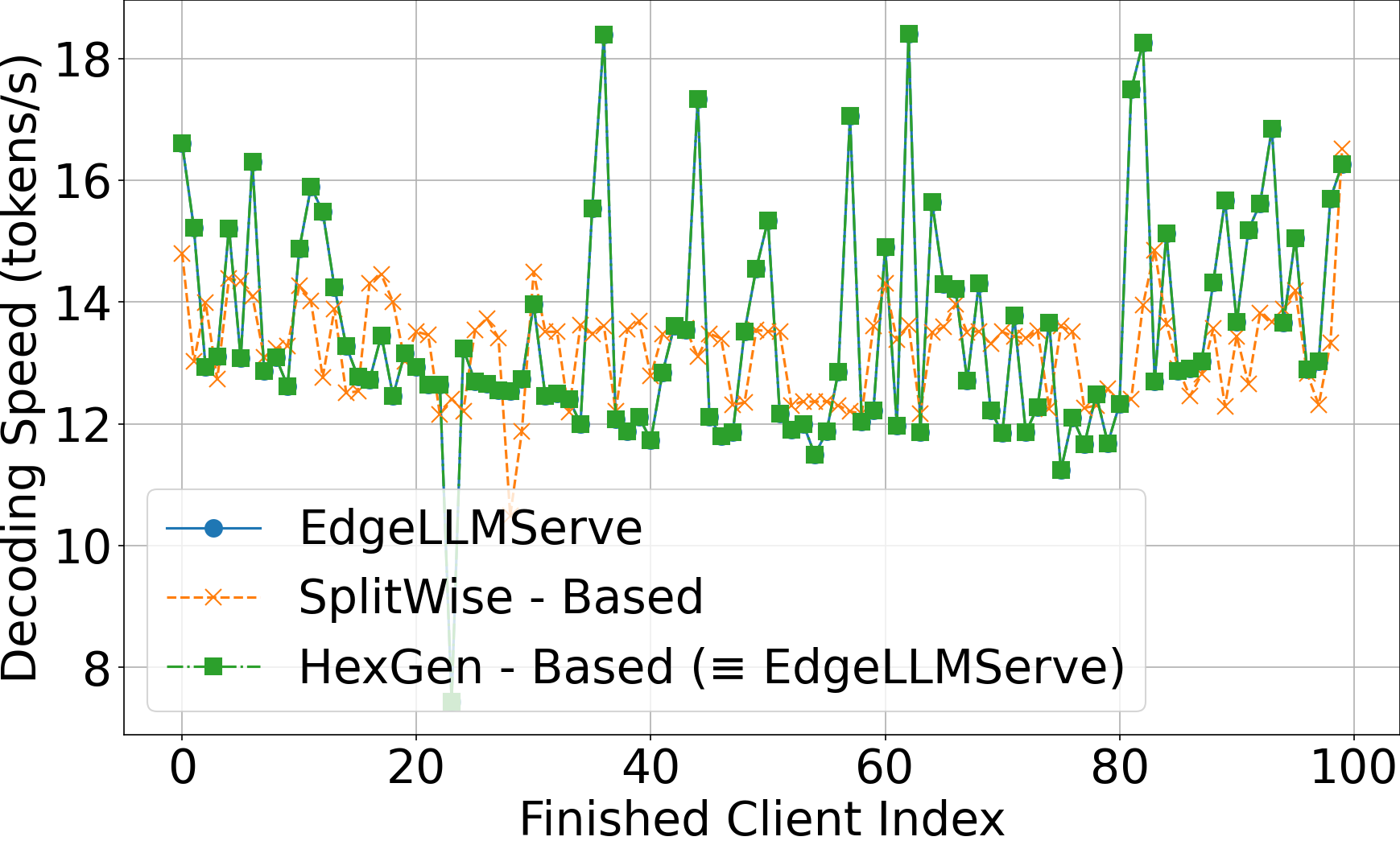}
        \vspace{-0.5cm}
        \caption{8 devices: decoding speed}
    \end{subfigure}\hfill
    \begin{subfigure}{\figscale\textwidth}
        \includegraphics[width=\linewidth]{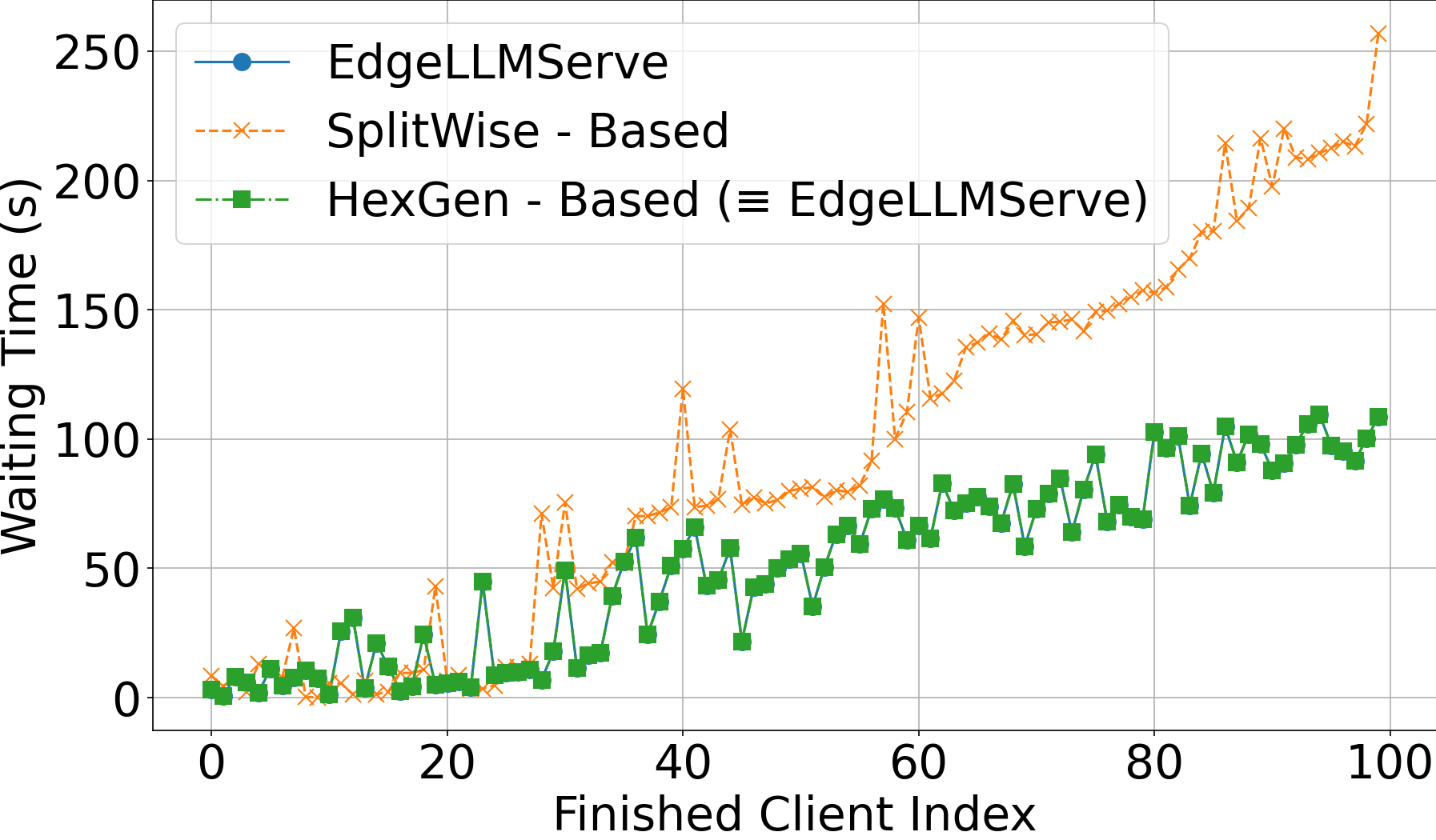}
        \vspace{-0.5cm}
        \caption{8 devices: waiting time}
    \end{subfigure}
    \vspace{-0.2cm}
    \caption{Scalability with cluster size (Qwen3-MoE-30B, LZByteDance, $100$~ms).}
    \label{fig:exp-scale}
\end{figure*}

\subsubsection{Impact of the request arrival rate}
\label{sec:exp-arrival}
We next fix the cluster configuration (eight devices), the model (Gpt-Oss-20b), and the workload (LZByteDance), while varying the mean request arrival interval from $100$~ms (heavy load) to $1000$~ms and $3000$~ms (light load). As the offered load decreases, E2LLM adapts its deployment strategy by assigning the Prefill replica to a less powerful device and reallocating stronger devices to the Decoder phase. This increases the available decoding capacity while maintaining sufficient Prefill throughput. As shown in Table~\ref{tab:dep-arrival}, such adaptation reduces the maximum waiting time from approximately $67$~s at a $100$~ms arrival interval to less than $1$~s at both $1000$~ms and $3000$~ms.
Notably, the decoding throughput of E2LLM increases as the arrival interval becomes larger, resulting in improved responsiveness under lighter workloads. In contrast, both HexGen and Adapted Splitwise maintain the same deployment configuration regardless of the arrival rate. While the decoding throughput of HexGen and Adapted Splitwise also increases under the lightest workload because their Decoder replicas are no longer fully saturated, their static deployment strategies prevent them from fully utilizing the heterogeneous resources available in the cluster. Consequently, they exhibit higher waiting times and lower overall efficiency than E2LLM across different load conditions.

\begin{table}[b]
    \centering
    \setlength{\tabcolsep}{2pt}
    \caption{Deployment plans when varying the arrival interval (eight devices, Gpt-Oss-20b, LZByteDance).}
    \vspace{-0.3cm}
    \label{tab:dep-arrival}
    \begin{subtable}[t]{.48\linewidth}
        \centering
        \caption{HG=SPW (all rates), ELS @$100$~ms.}
        \begin{tabular}{|c|c|c|c|}
            \hline
            Role & Dev & Sl. & G--C\\\hline
            P & Dev.1$^{\star}$ & 1 & 20--4\\\hline
            \multirow{2}{*}{D} & Dev.2$^{\star}$ & \multirow{2}{*}{8} & 5--4\\
              & Dev.4 &  & 15--0\\\hline
            D & Dev.3$^{\star}$ & 5 & 24--0\\\hline
            \multirow{2}{*}{D} & Dev.5 & \multirow{2}{*}{2} & 15--0\\
              & Dev.6$^{\star}$ &  & 9--0\\\hline
            \multirow{2}{*}{D} & Dev.7 & \multirow{2}{*}{30} & 13--0\\
              & Dev.8$^{\star}$ &  & 11--0\\\hline
        \end{tabular}
    \end{subtable}\hfill
    \begin{minipage}[t]{0.48\linewidth}
        \centering
        \begin{subtable}[t]{\linewidth}
            \centering
            \caption{$100$~ms / Metrics.}
            \begin{tabular}{|l|c|c|c|}
                \hline
                Method & PS & DS & W\\\hline
                HG & 2464 & 13.1 & 66.6\\\hline
                \textbf{ELS} & \textbf{2464} & \textbf{13.1} & \textbf{66.6}\\\hline
                SPW & 2464 & 13.1 & 66.6\\\hline
            \end{tabular}
        \end{subtable}
        \par\vspace{0.3cm}
        \begin{subtable}[t]{\linewidth}
            \centering
            \caption{$1000$~ms / Metrics.}
            \begin{tabular}{|l|c|c|c|}
                \hline
                Method & PS & DS & W\\\hline
                HG & 2480 & 13.0 & 31.5\\\hline
                \textbf{ELS} & \textbf{754} & \textbf{16.2} & \textbf{0.7}\\\hline
                SPW & 2480 & 13.0 & 31.5\\\hline
            \end{tabular}
        \end{subtable}
        \par\vspace{0.3cm}
        \begin{subtable}[t]{\linewidth}
            \centering
            \caption{$3000$~ms / Metrics.}
            \begin{tabular}{|l|c|c|c|}
                \hline
                Method & PS & DS & W\\\hline
                HG & 2481 & 18.9 & 0.6\\\hline
                \textbf{ELS} & \textbf{294} & \textbf{21.5} & \textbf{0.6}\\\hline
                SPW & 2481 & 18.9 & 0.6\\\hline
            \end{tabular}
        \end{subtable}
    \end{minipage}

    \vspace{0.3cm}
    \begin{subtable}[t]{.48\linewidth}
        \centering
        \caption{ELS @$1000$~ms.}
        \begin{tabular}{|c|c|c|c|}
            \hline
            Role & Dev & Sl. & G--C\\\hline
            P & Dev.3$^{\star}$ & 1 & 24--0\\\hline
            \multirow{2}{*}{D} & Dev.1 & \multirow{2}{*}{14} & 17--1\\
              & Dev.2$^{\star}$ &  & 5--1\\\hline
            D & Dev.4$^{\star}$ & 6 & 24--0\\\hline
            \multirow{2}{*}{D} & Dev.5 & \multirow{2}{*}{2} & 15--0\\
              & Dev.6$^{\star}$ &  & 9--0\\\hline
            \multirow{2}{*}{D} & Dev.7 & \multirow{2}{*}{30} & 13--0\\
              & Dev.8$^{\star}$ &  & 11--0\\\hline
        \end{tabular}
    \end{subtable}\hfill
    \begin{subtable}[t]{.48\linewidth}
        \centering
        \caption{ELS @$3000$~ms.}
        \begin{tabular}{|c|c|c|c|}
            \hline
            Role & Dev & Sl. & G--C\\\hline
            P & Dev.2$^{\star}$ & 1 & 5--19\\\hline
            \multirow{2}{*}{D} & Dev.1 & \multirow{2}{*}{32} & 14--0\\
              & Dev.7$^{\star}$ &  & 10--0\\\hline
            \multirow{3}{*}{D} & Dev.3$^{\star}$ & \multirow{3}{*}{6} & 14--0\\
              & Dev.5 &  & 5--0\\
              & Dev.6 &  & 5--0\\\hline
            \multirow{2}{*}{D} & Dev.4$^{\star}$ & \multirow{2}{*}{18} & 8--0\\
              & Dev.8 &  & 16--0\\\hline
        \end{tabular}
    \end{subtable}
\end{table}

\subsubsection{Scalability with cluster size}
\label{sec:exp-scale}
Finally, we fix the model (Qwen3-MoE-30B), the workload (LZByteDance), and the request arrival interval ($100$~ms), and scale the cluster from four to six and then eight devices. E2LLM effectively exploits the additional hardware resources by increasing the number of Decoder replicas, reducing the maximum waiting time from $709$~s to $371$~s and finally to $110$~s as the cluster size grows from four to six and eight devices, respectively.
In most bursty-workload scenarios (i.e., a $100$~ms arrival interval), including those reported in the Appendix, HexGen produces the same deployment plan as E2LLM and therefore achieves similar performance. However, as observed in the previous subsection, HexGen may generate different deployment plans in some cases, which can result in lower performance. Consequently, matching the performance of E2LLM is not guaranteed across all workloads and system configurations.
Adapted Splitwise also adjusts its deployment strategy as additional devices become available. However, due to its implicit resource-allocation policy between the Prefill and Decoder phases, the reduction in maximum waiting time is considerably smaller than that achieved by E2LLM. 
It should be noted that the throughput values reported in Table~\ref{tab:dep-scale} are measured on a per-request basis. Therefore, aggregate throughput depends on slot utilization. For instance, in the eight-device configuration, E2LLM provides a Prefill throughput of $1146$ tokens/s per request, yielding a maximum aggregate throughput of $2292$ tokens/s when both slots are occupied. Under low-load conditions, where only one slot is active, the effective throughput falls between these values because pipeline parallelism cannot be fully exploited, as shown in Table \ref{tab:dep-dataset}.

\begin{table}[!hb]
    \centering
    \setlength{\tabcolsep}{2pt}
    \caption{Deployment plans when scaling the cluster size (Qwen3-MoE-30B, LZByteDance, $100$~ms).}
    \vspace{-0.3cm}
    \label{tab:dep-scale}
    \begin{subtable}[t]{.48\linewidth}
        \centering
        \caption{4 dev. / HG=ELS.}
        \begin{tabular}{|c|c|c|c|}
            \hline
            Role & Dev & Sl. & G--C\\\hline
            \multirow{2}{*}{P} & Dev.5$^{\star}$ & \multirow{2}{*}{1} & 22--0\\
              & Dev.6 &  & 26--0\\\hline
            \multirow{2}{*}{D} & Dev.1$^{\star}$ & \multirow{2}{*}{10} & 13--19\\
              & Dev.2 &  & 5--11\\\hline
        \end{tabular}
    \end{subtable}\hfill
    \begin{subtable}[t]{.48\linewidth}
        \centering
        \caption{4 dev. / SPW.}
        \begin{tabular}{|c|c|c|c|}
            \hline
            Role & Dev & Sl. & G--C\\\hline
            P & Dev.1$^{\star}$ & 1 & 26--22\\\hline
            \multirow{3}{*}{D} & Dev.2$^{\star}$ & \multirow{3}{*}{3} & 7--15\\
              & Dev.5 &  & 13--0\\
              & Dev.6 &  & 13--0\\\hline
        \end{tabular}
    \end{subtable}

    \vspace{0.3cm}
    \begin{subtable}[t]{.48\linewidth}
        \centering
        \caption{6 dev. / HG=ELS.}
        \begin{tabular}{|c|c|c|c|}
            \hline
            Role & Dev & Sl. & G--C\\\hline
            P & Dev.4$^{\star}$ & 1 & 48--0\\\hline
            \multirow{3}{*}{D} & Dev.1$^{\star}$ & \multirow{3}{*}{12} & 16--5\\
              & Dev.2 &  & 6--3\\
              & Dev.7 &  & 18--0\\\hline
            \multirow{2}{*}{D} & Dev.5$^{\star}$ & \multirow{2}{*}{2} & 22--0\\
              & Dev.6 &  & 26--0\\\hline
        \end{tabular}
    \end{subtable}\hfill
    \begin{subtable}[t]{.48\linewidth}
        \centering
        \caption{6 dev. / SPW.}
        \begin{tabular}{|c|c|c|c|}
            \hline
            Role & Dev & Sl. & G--C\\\hline
            P & Dev.1$^{\star}$ & 1 & 26--22\\\hline
            \multirow{3}{*}{D} & Dev.2$^{\star}$ & \multirow{3}{*}{6} & 6--10\\
              & Dev.6 &  & 10--0\\
              & Dev.7 &  & 22--0\\\hline
            \multirow{2}{*}{D} & Dev.4$^{\star}$ & \multirow{2}{*}{8} & 36--0\\
              & Dev.5 &  & 12--0\\\hline
        \end{tabular}
    \end{subtable}

    \vspace{0.3cm}
    \begin{subtable}[t]{.48\linewidth}
        \centering
        \caption{8 dev. / HG=ELS.}
        \begin{tabular}{|c|c|c|c|}
            \hline
            Role & Dev & Sl. & G--C\\\hline
            P & Dev.1$^{\star}$ & 1 & 26--22\\\hline
            \multirow{3}{*}{D} & Dev.2 & \multirow{3}{*}{9} & 7--3\\
              & Dev.7 &  & 19--0\\
              & Dev.8$^{\star}$ &  & 19--0\\\hline
            \multirow{2}{*}{D} & Dev.3$^{\star}$ & \multirow{2}{*}{8} & 36--0\\
              & Dev.6 &  & 12--0\\\hline
            \multirow{2}{*}{D} & Dev.4$^{\star}$ & \multirow{2}{*}{8} & 36--0\\
              & Dev.5 &  & 12--0\\\hline
        \end{tabular}
    \end{subtable}\hfill
    \begin{subtable}[t]{.48\linewidth}
        \centering
        \caption{8 dev. / SPW.}
        \begin{tabular}{|c|c|c|c|}
            \hline
            Role & Dev & Sl. & G--C\\\hline
            \multirow{2}{*}{P} & Dev.7$^{\star}$ & \multirow{2}{*}{2} & 26--0\\
              & Dev.8 &  & 22--0\\\hline
            \multirow{2}{*}{D} & Dev.1$^{\star}$ & \multirow{2}{*}{10} & 13--19\\
              & Dev.2 &  & 5--11\\\hline
            \multirow{2}{*}{D} & Dev.3$^{\star}$ & \multirow{2}{*}{8} & 36--0\\
              & Dev.6 &  & 12--0\\\hline
            \multirow{2}{*}{D} & Dev.4$^{\star}$ & \multirow{2}{*}{8} & 36--0\\
              & Dev.5 &  & 12--0\\\hline
        \end{tabular}
    \end{subtable}

    \vspace{0.3cm}
    \begin{minipage}[t]{0.48\linewidth}
        \centering
        \begin{subtable}[t]{\linewidth}
            \centering
            \caption{4 dev. / Metrics.}
            \begin{tabular}{|l|c|c|c|}
                \hline
                Method & PS & DS & W\\\hline
                HG & 136 & 12.3 & 709.4\\\hline
                \textbf{ELS} & \textbf{136} & \textbf{12.3} & \textbf{709.4}\\\hline
                SPW & 1135 & 15.5 & 1482.8\\\hline
            \end{tabular}
        \end{subtable}
        \par\vspace{0.3cm}
        \begin{subtable}[t]{\linewidth}
            \centering
            \caption{6 dev. / Metrics.}
            \begin{tabular}{|l|c|c|c|}
                \hline
                Method & PS & DS & W\\\hline
                HG & 557 & 13.5 & 370.9\\\hline
                \textbf{ELS} & \textbf{557} & \textbf{13.5} & \textbf{370.9}\\\hline
                SPW & 1141 & 13.0 & 408.4\\\hline
            \end{tabular}
        \end{subtable}
    \end{minipage}\hfill
    \begin{minipage}[t]{0.48\linewidth}
        \centering
        \begin{subtable}[t]{\linewidth}
            \centering
            \caption{8 dev. / Metrics.}
            \begin{tabular}{|l|c|c|c|}
                \hline
                Method & PS & DS & W\\\hline
                HG & 1146 & 13.5 & 109.6\\\hline
                \textbf{ELS} & \textbf{1146} & \textbf{13.5} & \textbf{109.6}\\\hline
                SPW & 1175 & 13.3 & 256.9\\\hline
            \end{tabular}
        \end{subtable}
    \end{minipage}
\end{table}

Figure~\ref{fig:exp-scale} further illustrates these observations, highlighting the substantial gap in waiting time between E2LLM and adapted Splitwise. Moreover, this gap becomes increasingly pronounced over time as requests continue to accumulate, demonstrating the superior ability of E2LLM to mitigate congestion under sustained bursty workloads.
Overall, these results show that E2LLM scales more effectively with increasing cluster size by allocating additional resources to the phase that most limits performance, thereby achieving significantly lower queueing delays and better utilization of available hardware resources.

\subsubsection{Discussion.} Overall, E2LLM achieves the best performance across all evaluated scenarios, reducing waiting time by up to 50\% under high-load conditions while adapting effectively to changes in models, workloads, arrival rates, and cluster size.
However, several limitations remain. First, the method assumes that device performance during latency profiling is representative of pipeline-parallel execution. In practice, factors such as power and frequency management \cite{le2026empiricalanalysisgpufrequency}, asynchronous execution, thermal conditions, and GPU throughput characteristics \cite{le2026pm2lat} can affect pipeline latency. Incorporating these factors could improve profiling accuracy and deployment decisions.
Second, our implementation relies on llama.cpp, which uses a star-network architecture in which intermediate outputs are routed through a master node. This introduces additional communication overhead and requires both the first and last pipeline stages to reside on the master, effectively creating an $(N+1)$-stage pipeline on $N$ devices. For simplicity, our model treats the first and last stages as a single stage, which does not fully reflect the actual system behavior.
Addressing these limitations, particularly through more accurate performance modeling and peer-to-peer communication, is left for future work. The source code of E2LLM will be released upon publication of this article. 
Finally, as discussed in Section~\ref{subsec.Method.PrefillDecode}, incorporating dynamic workload analysis and adaptive deployment mechanisms is essential for capturing evolving user behavior and providing a more realistic representation of service performance.

\section{Conclusion}\label{sec.conclusion}
We presented EdgeLLMServe, a replication-based deployment framework for serving LLMs in heterogeneous Edge/Fog environments. By combining full-model replication, Prefill–Decoder role separation, congestion-aware optimization, Dynamic Programming, and Genetic Algorithms, EdgeLLMServe effectively mitigates pipeline bottlenecks and reduces response latency under resource constraints. Experiments on a real-world cluster demonstrate up to 2× higher decoding throughput than HexGen and Splitwise, and more than 50\% lower waiting time than Splitwise in high-demand scenarios. These results demonstrate the effectiveness of workload-aware resource allocation for delivering cost-efficient, low-latency, and privacy-preserving LLM services beyond the cloud.

\section*{Acknowledgment}
This work was supported in part by the Norwegian Research Council under Grant 322473 (AirQMan project) and the European Commission under Grant 101086541 (MISO project).

\clearpage
\bibliographystyle{ACM-Reference-Format}
\bibliography{references}

\clearpage
\appendix
\section{Genetic Algorithm components}\label{apd.GAComponents}

\textbf{Gene definition.} In this work, we represent each gene using two chromosomes. The first chromosome encodes the ordering of nodes/devices, while the second chromosome specifies the grouping of nodes into replicas. For example, the gene: $[(0,1,2,3,4,5), (3,1,2)]$ indicates that there are three replicas. The first replica contains nodes 0, 1, and 2; the second replica consists of node 3; and the remaining nodes (4 and 5) form the third replica. This representation allows the Genetic Algorithm to explore both device ordering and replica composition during optimization.

\textbf{Cross-over policy.} During crossover, we randomly select two candidate genes based on their fitness scores (effectiveness). The first chromosome, which represents the node ordering, is crossed over between the two parents, while the second chromosome, representing replica grouping, is randomly chosen from one of them. However, this process can produce invalid genes because node duplication may occur in the first chromosome. To address this, a repairing procedure is applied to ensure that each node appears exactly once, restoring the validity of the gene before proceeding to the next generation.

\textbf{Mutation strategy.} Mutation introduces diversity into the population and prevents premature convergence. In our approach, mutation operates on both chromosomes of a gene. 
We set the probability of a gene be mutated to about 30\%. 
When being mutated, there are:
\begin{itemize}
    \item 20\% chance that we exchange two random position in the first chromosome.
    \item 50\% chance that we destroy all replica size from a random position in the second chromosome and then regenerate again.
    \item 15\% chance that we completely destroy the second chromosome and regenerate again.
    \item 15\% chance that we destroy both chromosome and regenerate again.
\end{itemize}

\textbf{Elite citizen.} In addition to traditional crossover and mutation, we implement an elite preservation strategy to maintain high-quality solutions across generations. Specifically, we define a maximum number of globally best-performing individuals (elite citizens) to retain after each iteration. These elite individuals are then used in crossover operations with the current population, ensuring that strong genetic material propagates through subsequent generations. This strategy accelerates convergence toward optimal deployment plans while reducing the risk of losing high-quality solutions during random mutations.

\section{Detailed evaluation results}\label{apd.fullresults}

For completeness, this appendix reports every evaluated configuration, organised by cluster size, target model, workload dataset, and request arrival interval. For each configuration the deployment plan of each method---HexGen-based (HG), EdgeLLMServe (ELS), and adapted SplitWise-based (SPW)---is shown alongside the aggregate Metrics, followed by the per-request figures. In the ``Mode'' column, P and D denote prefill and decoding replicas; the master device of each replica is marked with $^{\star}$. In the Metrics table, PS and DS are the prefill/decode throughput (TPS) and W the maximum waiting time (s).

\onecolumn
\providecommand{\appfigscale}{0.3}

\subsection{Four devices}
\subsubsection{Gpt-Oss-20b}
\paragraph{1. IELTS dataset. Arrival rate of 100ms}\mbox{}
\begin{table}[H]
    \centering
    \setlength{\tabcolsep}{3pt}
    \caption{Deployment plans and metrics --- Four devices, Gpt-Oss-20b, IELTS, arrival $100$~ms.}
    \label{apd:t1}
    \begin{subtable}[t]{0.24\textwidth}
        \centering
        \caption{HG}
        \begin{tabular}{|c|c|c|c|}
            \hline
            Mode & Dev & Slots & G--C\\\hline
            P& Dev.2$^{\star}$ & 1 & 5--19\\\hline
            D& Dev.1$^{\star}$ & 7 & 17--7\\\hline
            \multirow{2}{*}{D} & Dev.5 & \multirow{2}{*}{2} & 15--0\\
              & Dev.6$^{\star}$ &  & 9--0\\\hline
        \end{tabular}
    \end{subtable}\hfill
    \begin{subtable}[t]{0.24\textwidth}
        \centering
        \caption{ELS}
        \begin{tabular}{|c|c|c|c|}
            \hline
            Mode & Dev & Slots & G--C\\\hline
            P& Dev.2$^{\star}$ & 1 & 5--19\\\hline
            D& Dev.1$^{\star}$ & 7 & 17--7\\\hline
            \multirow{2}{*}{D} & Dev.5 & \multirow{2}{*}{2} & 15--0\\
              & Dev.6$^{\star}$ &  & 9--0\\\hline
        \end{tabular}
    \end{subtable}\hfill
    \begin{subtable}[t]{0.24\textwidth}
        \centering
        \caption{SPW}
        \begin{tabular}{|c|c|c|c|}
            \hline
            Mode & Dev & Slots & G--C\\\hline
            P& Dev.1$^{\star}$ & 1 & 20--4\\\hline
            \multirow{3}{*}{D} & Dev.2$^{\star}$ & \multirow{3}{*}{6} & 4--6\\
              & Dev.5 &  & 7--0\\
              & Dev.6 &  & 7--0\\\hline
        \end{tabular}
    \end{subtable}\hfill
    \begin{subtable}[t]{0.24\textwidth}
        \centering
        \caption{Metrics.}
        \begin{tabular}{|l|c|c|c|}
            \hline
            Method & PS & DS & W\\\hline
            HG & 375 & 22.5 & 673.6\\\hline
            \textbf{ELS} & \textbf{375} & \textbf{22.5} & \textbf{673.6}\\\hline
            SPW & 2913 & 12.1 & 1718.1\\\hline
        \end{tabular}
    \end{subtable}
\end{table}
\begin{figure}[H]
    \centering
    \begin{subfigure}{\appfigscale\textwidth}
        \includegraphics[width=\linewidth]{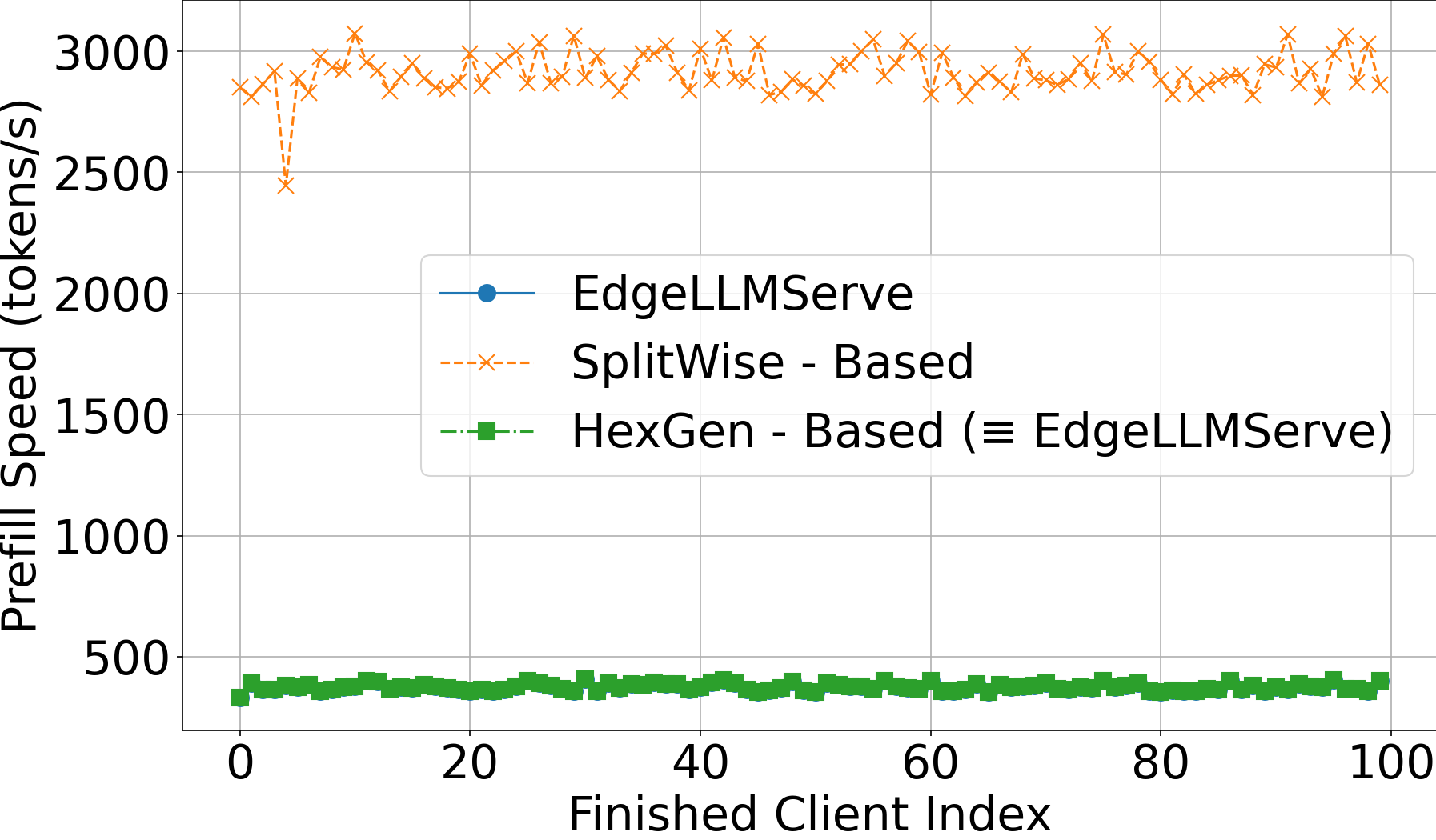}
        \caption{Prefill speed}
    \end{subfigure}\hfill
    \begin{subfigure}{\appfigscale\textwidth}
        \includegraphics[width=\linewidth]{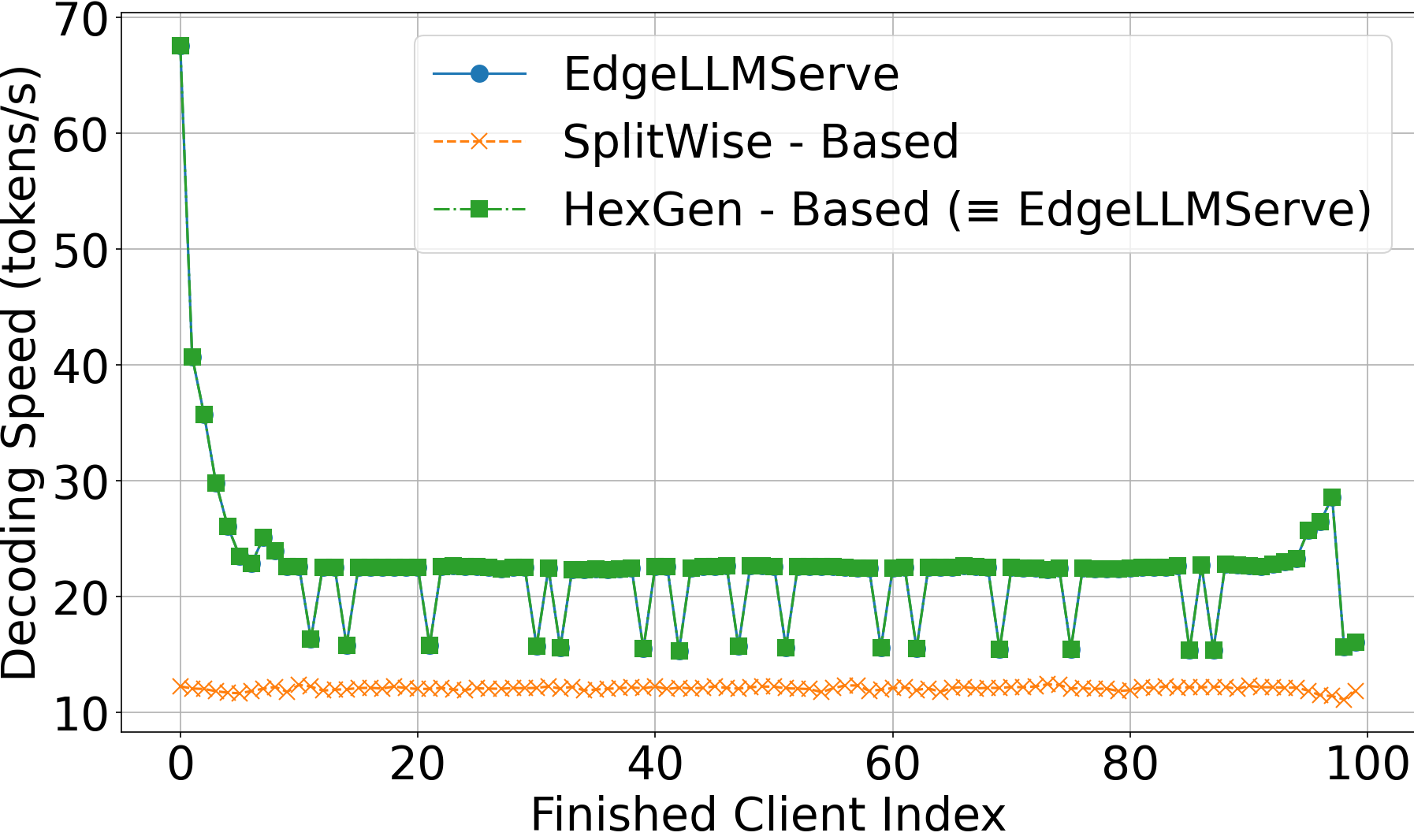}
        \caption{Decoding speed}
    \end{subfigure}\hfill
    \begin{subfigure}{\appfigscale\textwidth}
        \includegraphics[width=\linewidth]{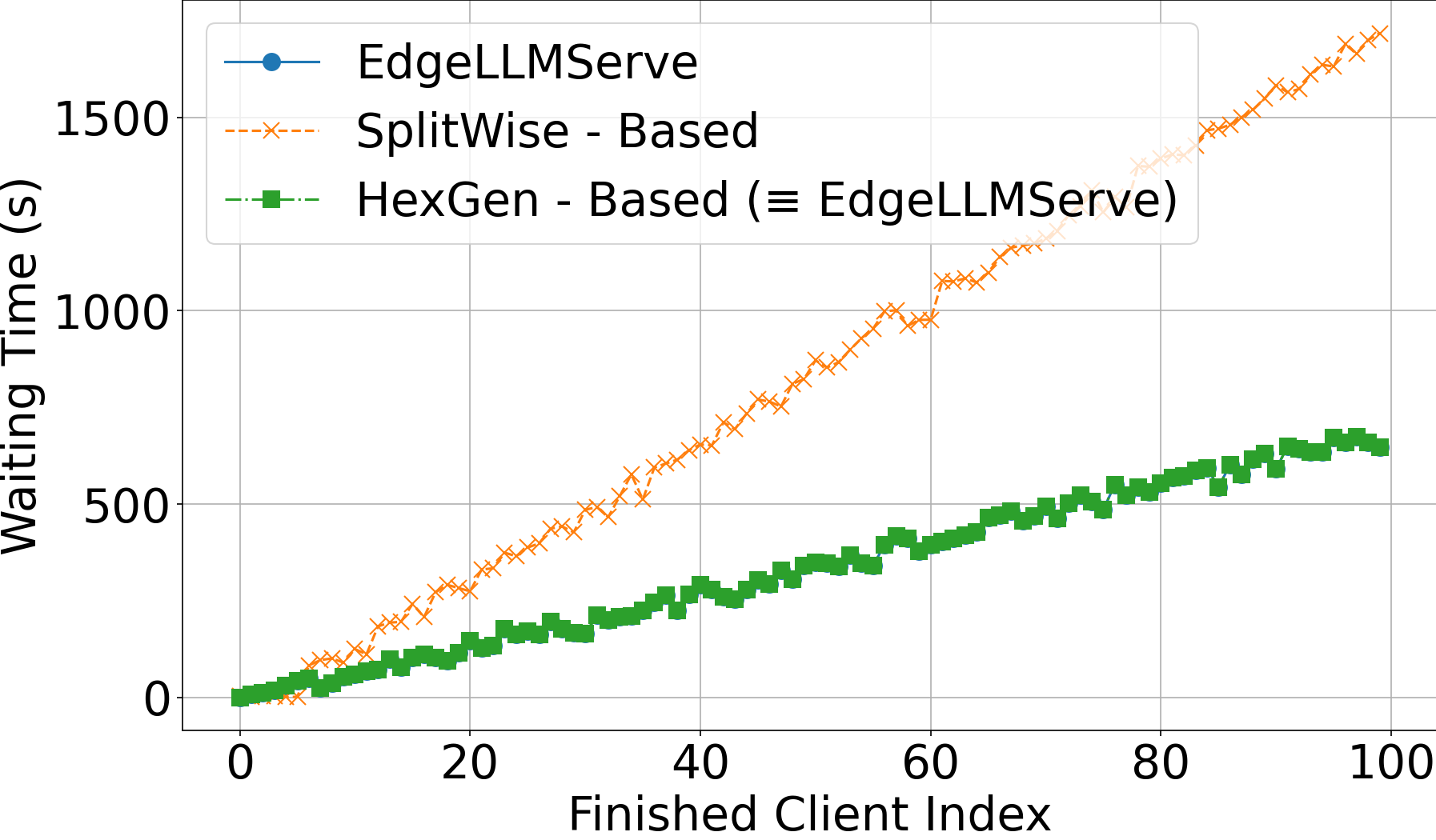}
        \caption{Waiting time}
    \end{subfigure}
    \caption{Per-request metrics --- Four devices, Gpt-Oss-20b, IELTS, arrival $100$~ms.}
    \label{apd:f1}
\end{figure}

\paragraph{2. IELTS dataset. Arrival rate of 1000ms}\mbox{}
\begin{table}[H]
    \centering
    \setlength{\tabcolsep}{3pt}
    \caption{Deployment plans and metrics --- Four devices, Gpt-Oss-20b, IELTS, arrival $1000$~ms.}
    \label{apd:t2}
    \begin{subtable}[t]{0.24\textwidth}
        \centering
        \caption{HG}
        \begin{tabular}{|c|c|c|c|}
            \hline
            Mode & Dev & Slots & G--C\\\hline
            P& Dev.2$^{\star}$ & 1 & 5--19\\\hline
            D& Dev.1$^{\star}$ & 7 & 17--7\\\hline
            \multirow{2}{*}{D} & Dev.5 & \multirow{2}{*}{2} & 15--0\\
              & Dev.6$^{\star}$ &  & 9--0\\\hline
        \end{tabular}
    \end{subtable}\hfill
    \begin{subtable}[t]{0.24\textwidth}
        \centering
        \caption{ELS}
        \begin{tabular}{|c|c|c|c|}
            \hline
            Mode & Dev & Slots & G--C\\\hline
            P& Dev.2$^{\star}$ & 1 & 5--19\\\hline
            D& Dev.1$^{\star}$ & 7 & 17--7\\\hline
            \multirow{2}{*}{D} & Dev.5 & \multirow{2}{*}{2} & 15--0\\
              & Dev.6$^{\star}$ &  & 9--0\\\hline
        \end{tabular}
    \end{subtable}\hfill
    \begin{subtable}[t]{0.24\textwidth}
        \centering
        \caption{SPW}
        \begin{tabular}{|c|c|c|c|}
            \hline
            Mode & Dev & Slots & G--C\\\hline
            P& Dev.1$^{\star}$ & 1 & 20--4\\\hline
            \multirow{3}{*}{D} & Dev.2$^{\star}$ & \multirow{3}{*}{6} & 4--6\\
              & Dev.5 &  & 7--0\\
              & Dev.6 &  & 7--0\\\hline
        \end{tabular}
    \end{subtable}\hfill
    \begin{subtable}[t]{0.24\textwidth}
        \centering
        \caption{Metrics.}
        \begin{tabular}{|l|c|c|c|}
            \hline
            Method & PS & DS & W\\\hline
            HG & 375 & 22.2 & 590.5\\\hline
            \textbf{ELS} & \textbf{375} & \textbf{22.2} & \textbf{590.5}\\\hline
            SPW & 2919 & 12.1 & 1687.1\\\hline
        \end{tabular}
    \end{subtable}
\end{table}
\begin{figure}[H]
    \centering
    \begin{subfigure}{\appfigscale\textwidth}
        \includegraphics[width=\linewidth]{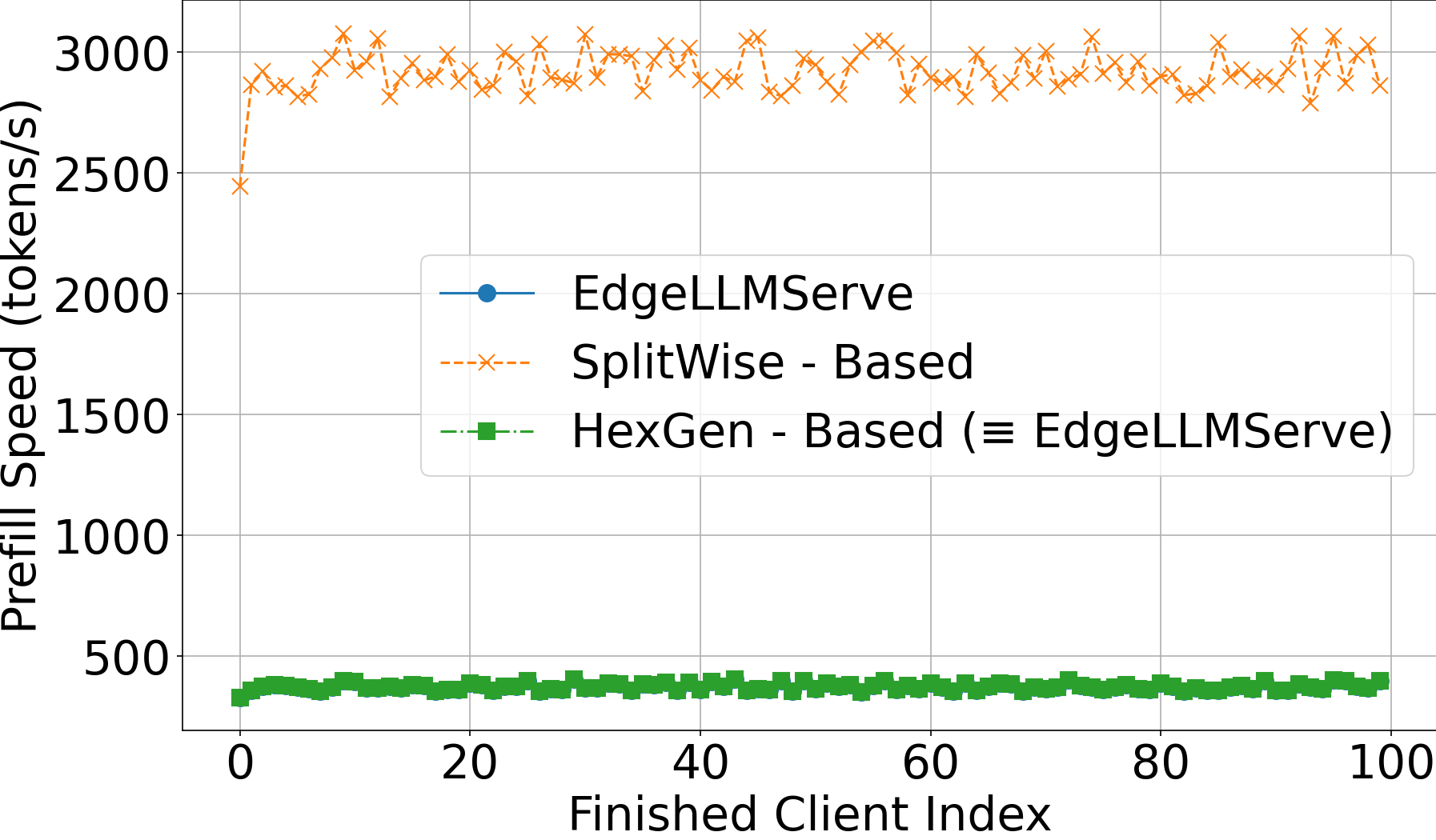}
        \caption{Prefill speed}
    \end{subfigure}\hfill
    \begin{subfigure}{\appfigscale\textwidth}
        \includegraphics[width=\linewidth]{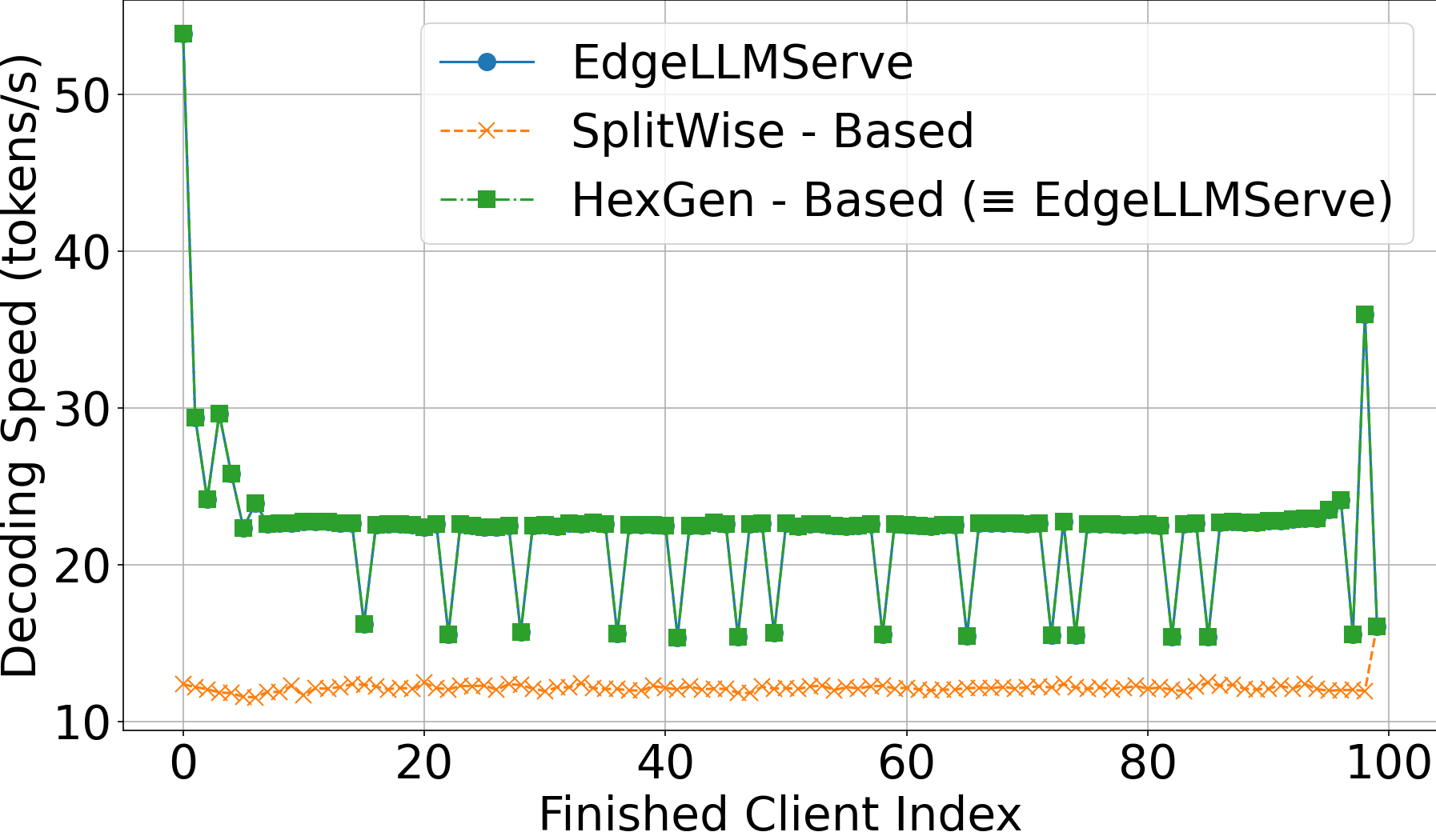}
        \caption{Decoding speed}
    \end{subfigure}\hfill
    \begin{subfigure}{\appfigscale\textwidth}
        \includegraphics[width=\linewidth]{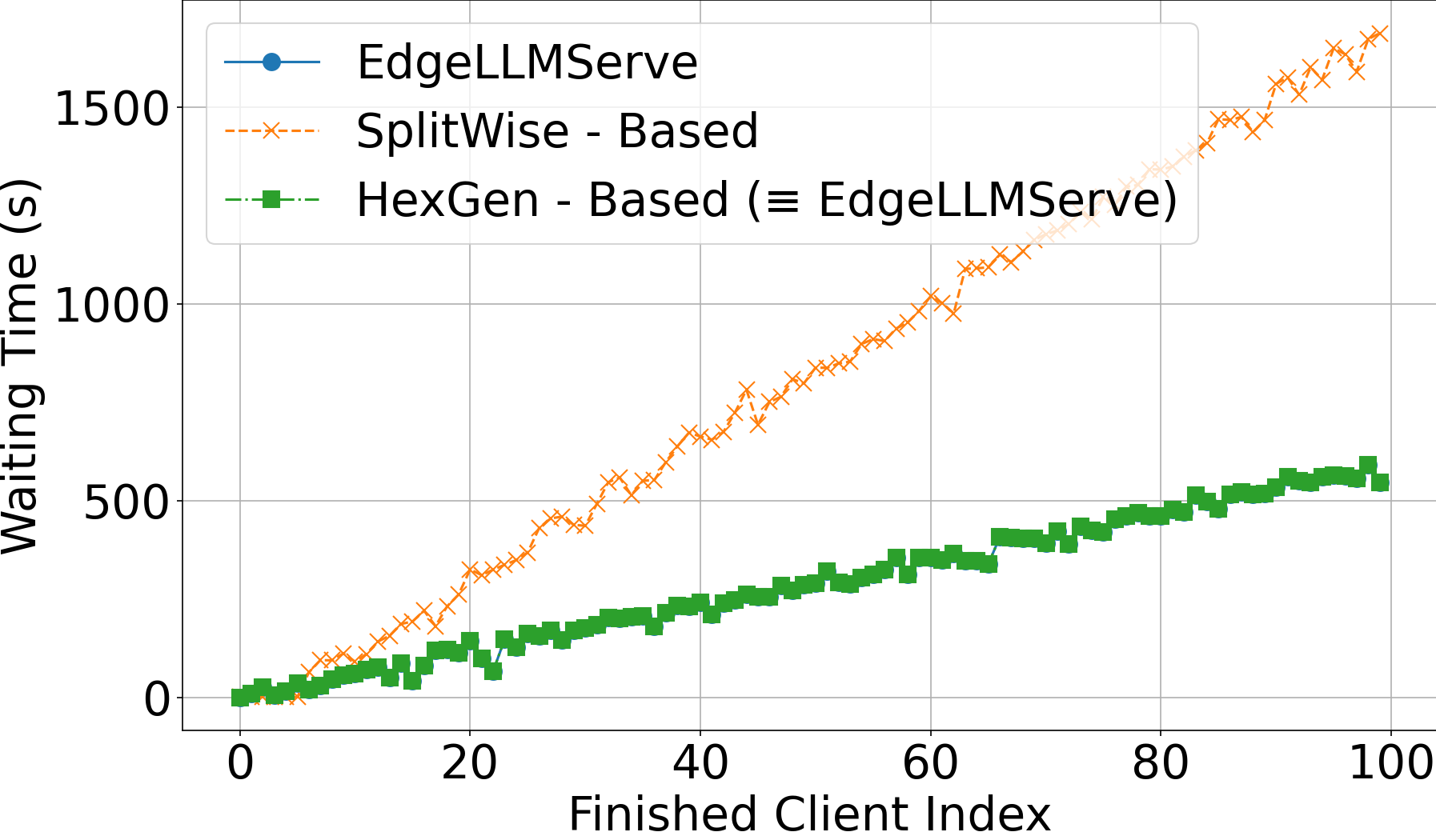}
        \caption{Waiting time}
    \end{subfigure}
    \caption{Per-request metrics --- Four devices, Gpt-Oss-20b, IELTS, arrival $1000$~ms.}
    \label{apd:f2}
\end{figure}

\paragraph{3. IELTS dataset. Arrival rate of 3000ms}\mbox{}
\begin{table}[H]
    \centering
    \setlength{\tabcolsep}{3pt}
    \caption{Deployment plans and metrics --- Four devices, Gpt-Oss-20b, IELTS, arrival $3000$~ms.}
    \label{apd:t3}
    \begin{subtable}[t]{0.24\textwidth}
        \centering
        \caption{HG}
        \begin{tabular}{|c|c|c|c|}
            \hline
            Mode & Dev & Slots & G--C\\\hline
            P& Dev.2$^{\star}$ & 1 & 5--19\\\hline
            D& Dev.1$^{\star}$ & 7 & 17--7\\\hline
            \multirow{2}{*}{D} & Dev.5 & \multirow{2}{*}{2} & 15--0\\
              & Dev.6$^{\star}$ &  & 9--0\\\hline
        \end{tabular}
    \end{subtable}\hfill
    \begin{subtable}[t]{0.24\textwidth}
        \centering
        \caption{ELS}
        \begin{tabular}{|c|c|c|c|}
            \hline
            Mode & Dev & Slots & G--C\\\hline
            P& Dev.2$^{\star}$ & 1 & 5--19\\\hline
            D& Dev.1$^{\star}$ & 7 & 17--7\\\hline
            \multirow{2}{*}{D} & Dev.5 & \multirow{2}{*}{2} & 15--0\\
              & Dev.6$^{\star}$ &  & 9--0\\\hline
        \end{tabular}
    \end{subtable}\hfill
    \begin{subtable}[t]{0.24\textwidth}
        \centering
        \caption{SPW}
        \begin{tabular}{|c|c|c|c|}
            \hline
            Mode & Dev & Slots & G--C\\\hline
            P& Dev.1$^{\star}$ & 1 & 20--4\\\hline
            \multirow{3}{*}{D} & Dev.2$^{\star}$ & \multirow{3}{*}{6} & 4--6\\
              & Dev.5 &  & 7--0\\
              & Dev.6 &  & 7--0\\\hline
        \end{tabular}
    \end{subtable}\hfill
    \begin{subtable}[t]{0.24\textwidth}
        \centering
        \caption{Metrics.}
        \begin{tabular}{|l|c|c|c|}
            \hline
            Method & PS & DS & W\\\hline
            HG & 375 & 22.4 & 404.3\\\hline
            \textbf{ELS} & \textbf{375} & \textbf{22.4} & \textbf{404.3}\\\hline
            SPW & 2928 & 12.1 & 1497.7\\\hline
        \end{tabular}
    \end{subtable}
\end{table}
\begin{figure}[H]
    \centering
    \begin{subfigure}{\appfigscale\textwidth}
        \includegraphics[width=\linewidth]{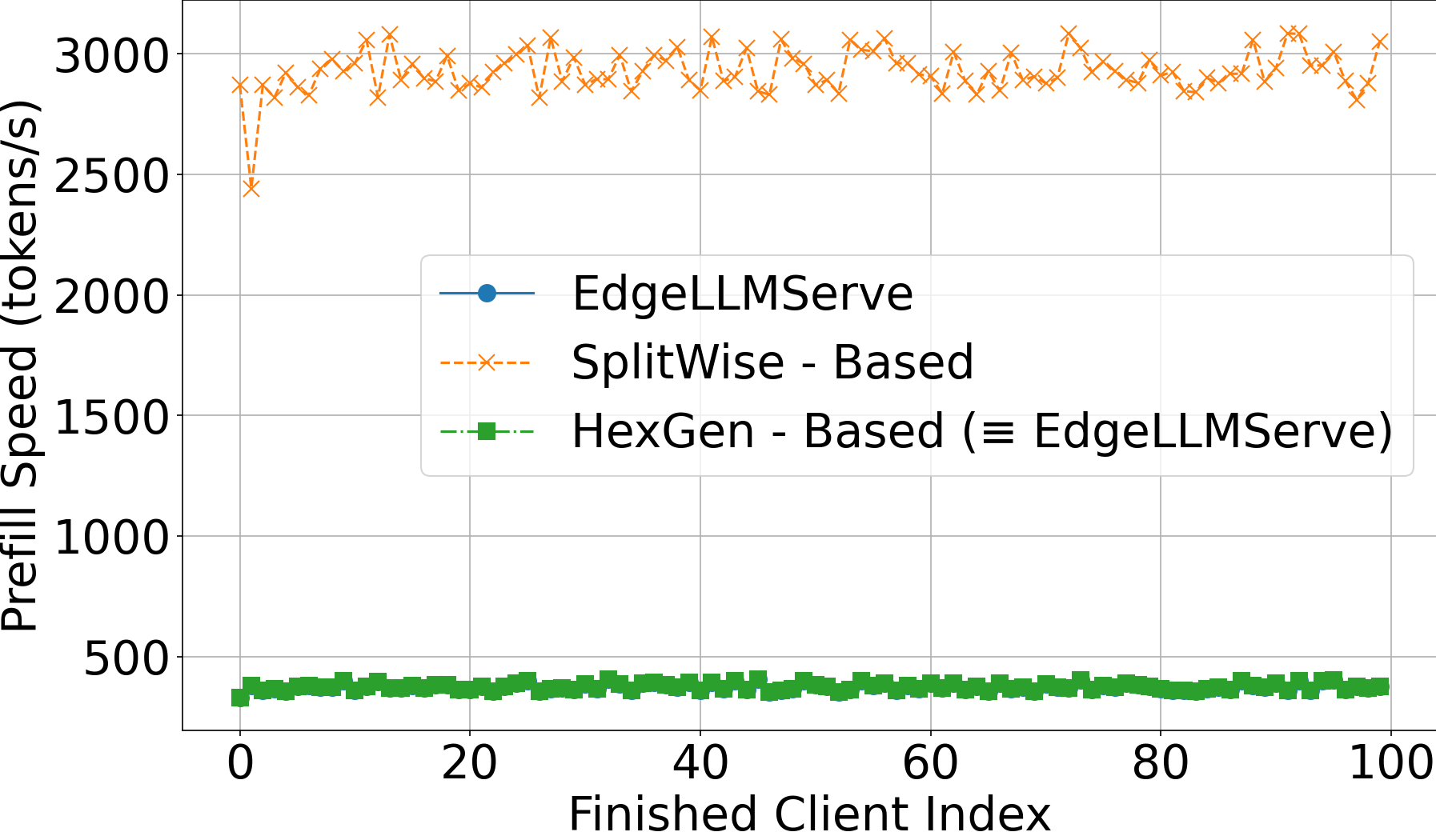}
        \caption{Prefill speed}
    \end{subfigure}\hfill
    \begin{subfigure}{\appfigscale\textwidth}
        \includegraphics[width=\linewidth]{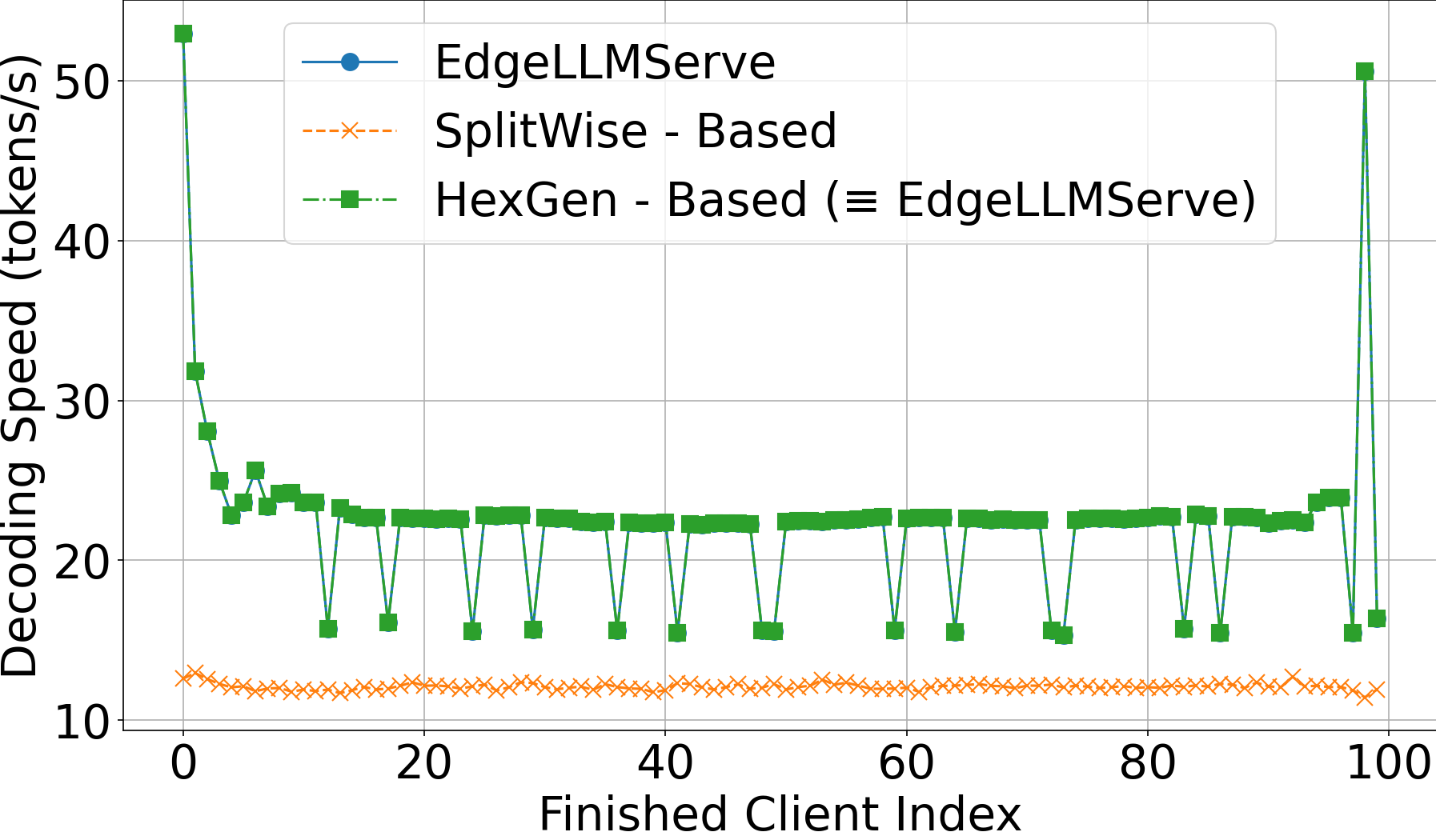}
        \caption{Decoding speed}
    \end{subfigure}\hfill
    \begin{subfigure}{\appfigscale\textwidth}
        \includegraphics[width=\linewidth]{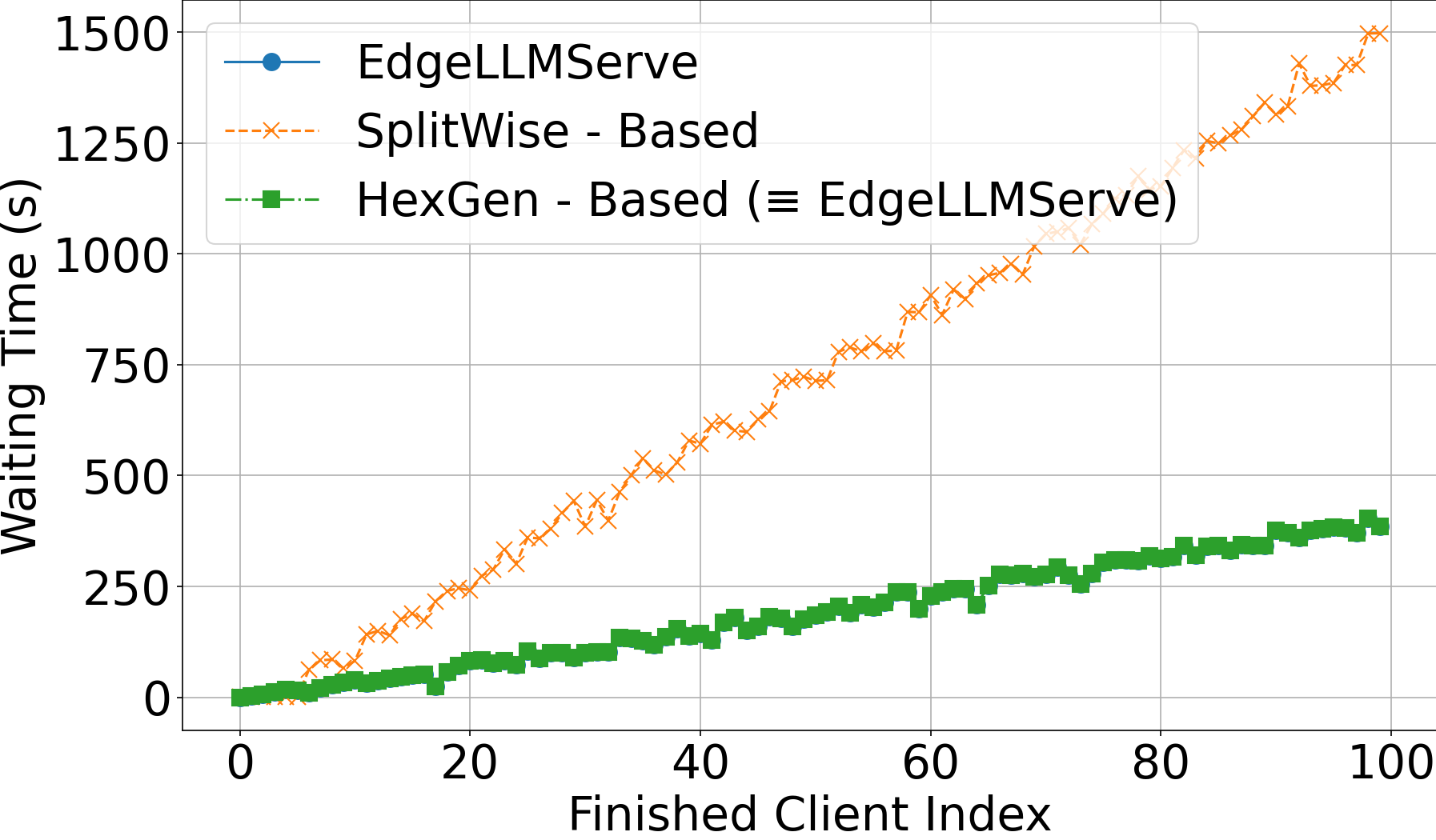}
        \caption{Waiting time}
    \end{subfigure}
    \caption{Per-request metrics --- Four devices, Gpt-Oss-20b, IELTS, arrival $3000$~ms.}
    \label{apd:f3}
\end{figure}

\paragraph{4. LZByteDance dataset. Arrival rate of 100ms}\mbox{}
\begin{table}[H]
    \centering
    \setlength{\tabcolsep}{3pt}
    \caption{Deployment plans and metrics --- Four devices, Gpt-Oss-20b, LZByteDance, arrival $100$~ms.}
    \label{apd:t4}
    \begin{subtable}[t]{0.24\textwidth}
        \centering
        \caption{HG}
        \begin{tabular}{|c|c|c|c|}
            \hline
            Mode & Dev & Slots & G--C\\\hline
            P& Dev.2$^{\star}$ & 1 & 5--19\\\hline
            D& Dev.1$^{\star}$ & 7 & 17--7\\\hline
            \multirow{2}{*}{D} & Dev.5 & \multirow{2}{*}{2} & 15--0\\
              & Dev.6$^{\star}$ &  & 9--0\\\hline
        \end{tabular}
    \end{subtable}\hfill
    \begin{subtable}[t]{0.24\textwidth}
        \centering
        \caption{ELS}
        \begin{tabular}{|c|c|c|c|}
            \hline
            Mode & Dev & Slots & G--C\\\hline
            P& Dev.2$^{\star}$ & 1 & 5--19\\\hline
            D& Dev.1$^{\star}$ & 7 & 17--7\\\hline
            \multirow{2}{*}{D} & Dev.5 & \multirow{2}{*}{2} & 15--0\\
              & Dev.6$^{\star}$ &  & 9--0\\\hline
        \end{tabular}
    \end{subtable}\hfill
    \begin{subtable}[t]{0.24\textwidth}
        \centering
        \caption{SPW}
        \begin{tabular}{|c|c|c|c|}
            \hline
            Mode & Dev & Slots & G--C\\\hline
            P& Dev.1$^{\star}$ & 1 & 20--4\\\hline
            \multirow{3}{*}{D} & Dev.2$^{\star}$ & \multirow{3}{*}{6} & 4--6\\
              & Dev.5 &  & 7--0\\
              & Dev.6 &  & 7--0\\\hline
        \end{tabular}
    \end{subtable}\hfill
    \begin{subtable}[t]{0.24\textwidth}
        \centering
        \caption{Metrics.}
        \begin{tabular}{|l|c|c|c|}
            \hline
            Method & PS & DS & W\\\hline
            HG & 293 & 23.2 & 266.0\\\hline
            \textbf{ELS} & \textbf{293} & \textbf{23.2} & \textbf{266.0}\\\hline
            SPW & 2465 & 12.5 & 647.7\\\hline
        \end{tabular}
    \end{subtable}
\end{table}
\begin{figure}[H]
    \centering
    \begin{subfigure}{\appfigscale\textwidth}
        \includegraphics[width=\linewidth]{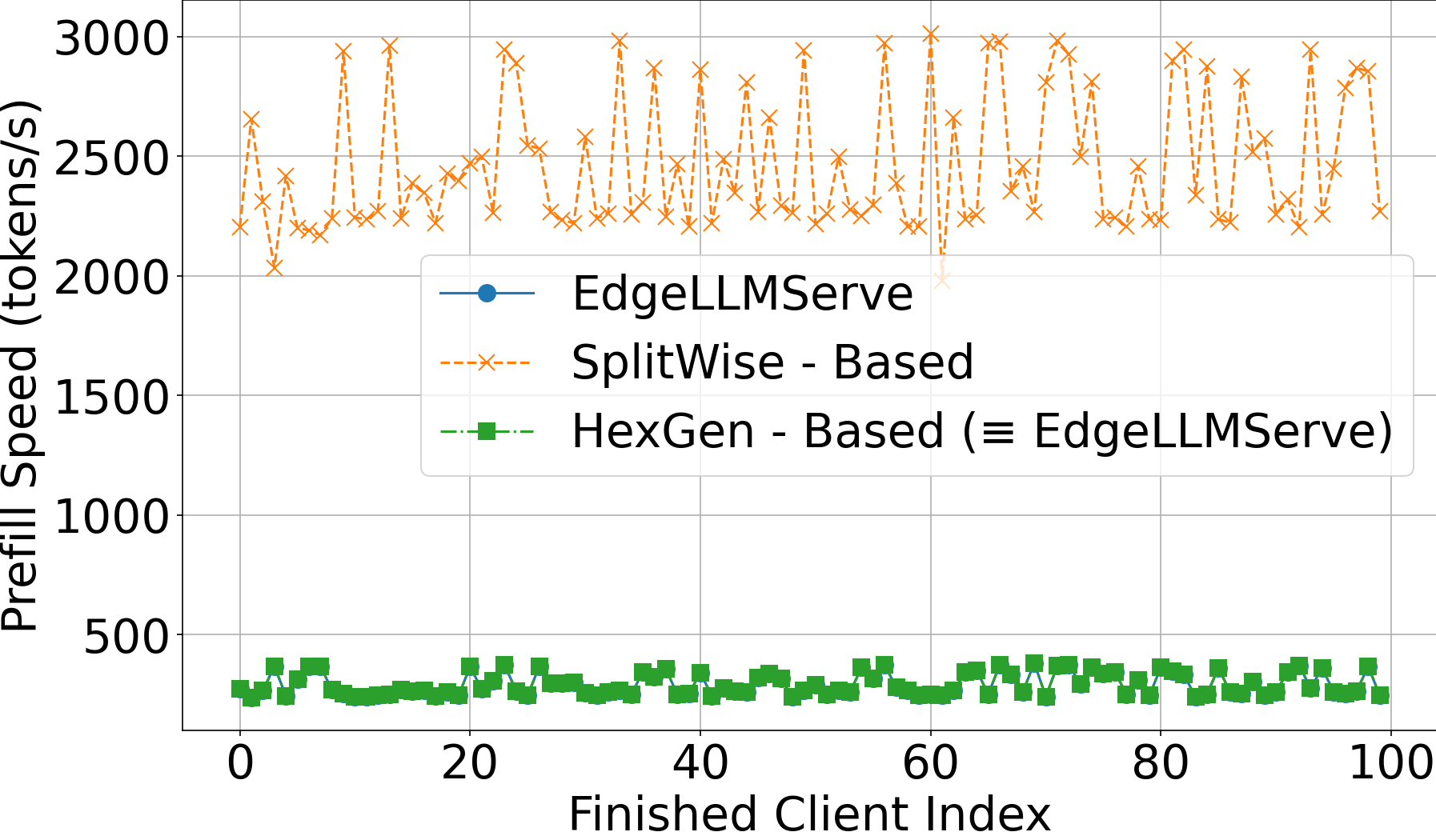}
        \caption{Prefill speed}
    \end{subfigure}\hfill
    \begin{subfigure}{\appfigscale\textwidth}
        \includegraphics[width=\linewidth]{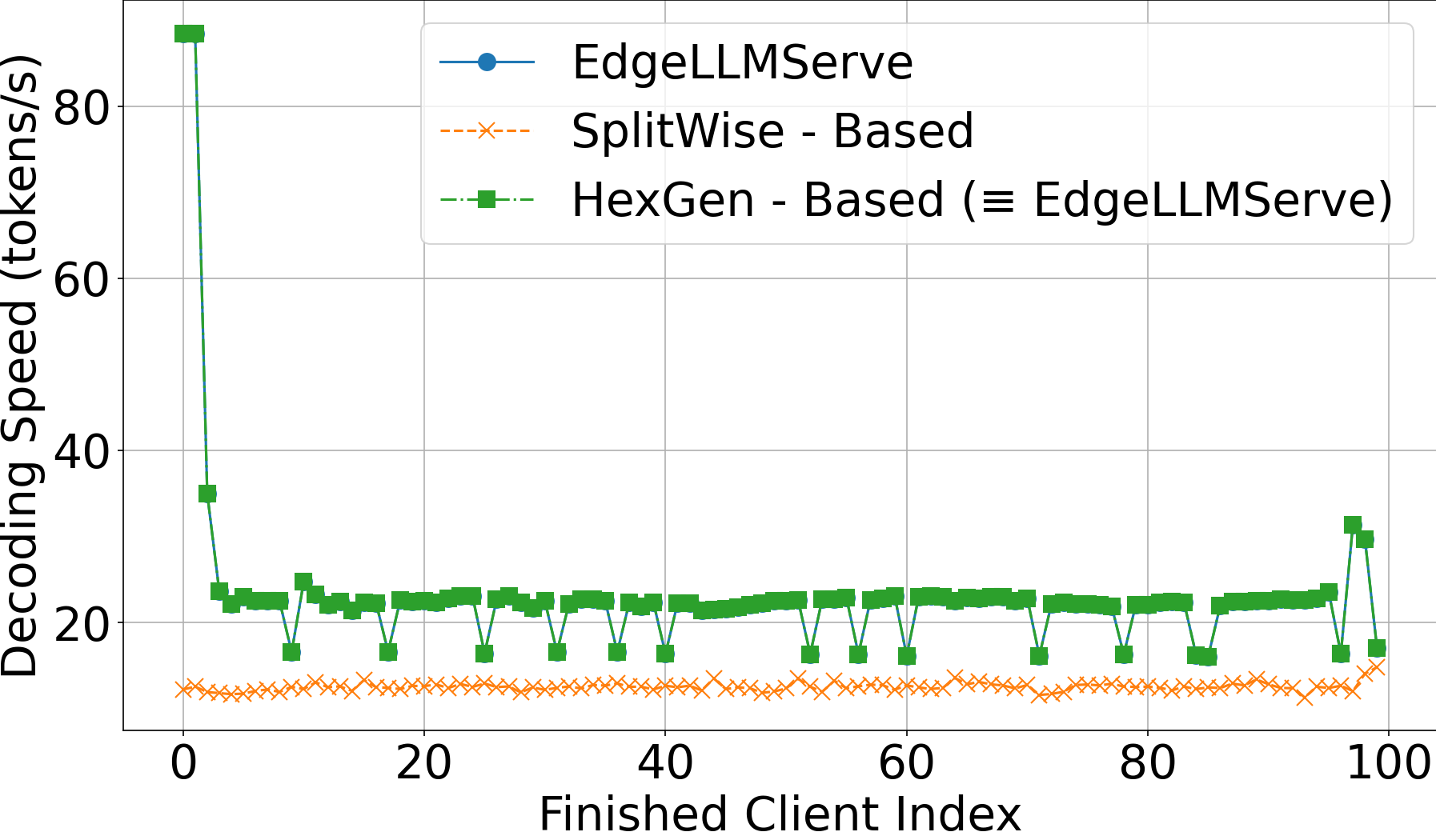}
        \caption{Decoding speed}
    \end{subfigure}\hfill
    \begin{subfigure}{\appfigscale\textwidth}
        \includegraphics[width=\linewidth]{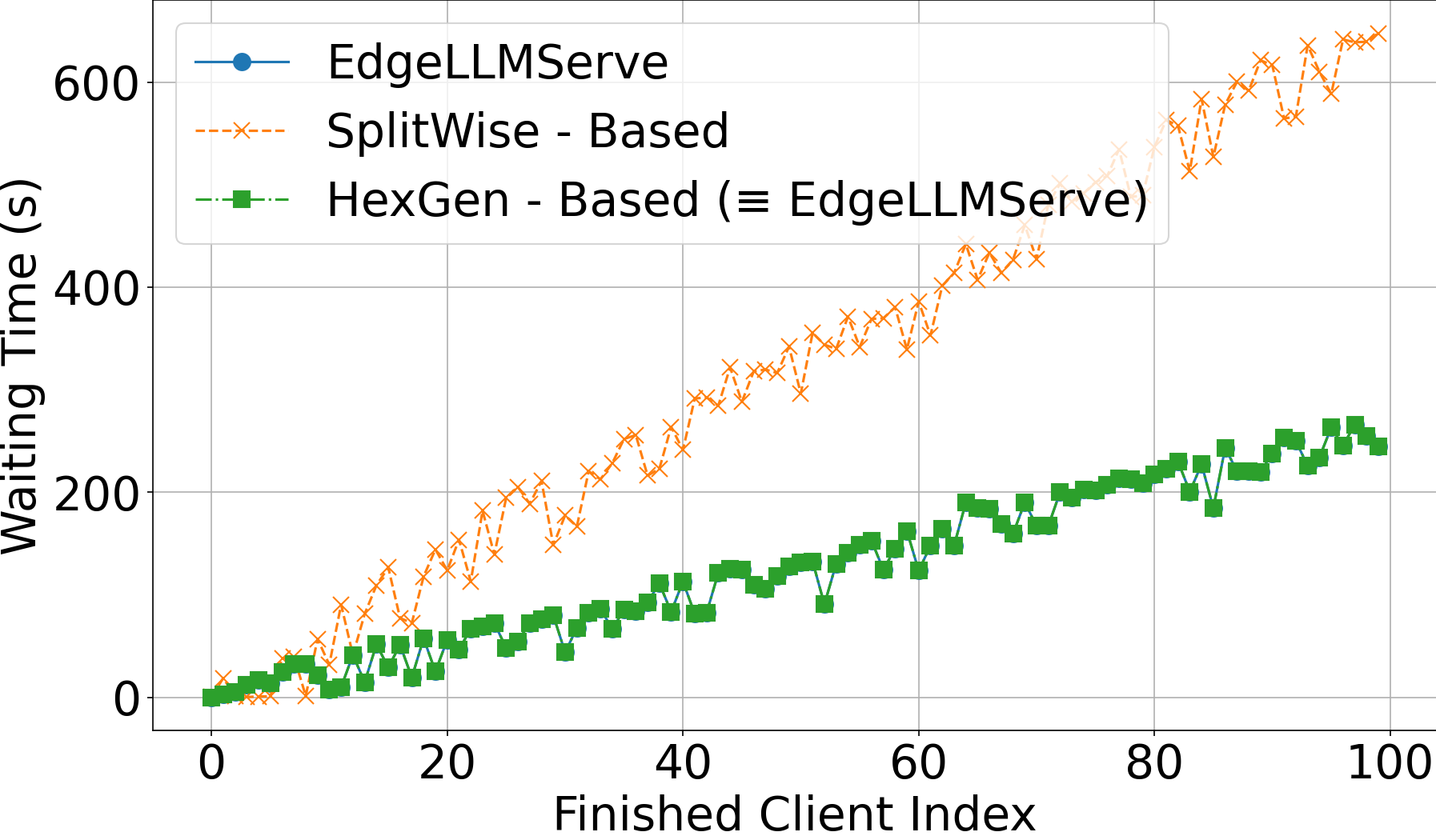}
        \caption{Waiting time}
    \end{subfigure}
    \caption{Per-request metrics --- Four devices, Gpt-Oss-20b, LZByteDance, arrival $100$~ms.}
    \label{apd:f4}
\end{figure}

\paragraph{5. LZByteDance dataset. Arrival rate of 1000ms}\mbox{}
\begin{table}[H]
    \centering
    \setlength{\tabcolsep}{3pt}
    \caption{Deployment plans and metrics --- Four devices, Gpt-Oss-20b, LZByteDance, arrival $1000$~ms.}
    \label{apd:t5}
    \begin{subtable}[t]{0.24\textwidth}
        \centering
        \caption{HG}
        \begin{tabular}{|c|c|c|c|}
            \hline
            Mode & Dev & Slots & G--C\\\hline
            P& Dev.2$^{\star}$ & 1 & 5--19\\\hline
            D& Dev.1$^{\star}$ & 7 & 17--7\\\hline
            \multirow{2}{*}{D} & Dev.5 & \multirow{2}{*}{2} & 15--0\\
              & Dev.6$^{\star}$ &  & 9--0\\\hline
        \end{tabular}
    \end{subtable}\hfill
    \begin{subtable}[t]{0.24\textwidth}
        \centering
        \caption{ELS}
        \begin{tabular}{|c|c|c|c|}
            \hline
            Mode & Dev & Slots & G--C\\\hline
            P& Dev.2$^{\star}$ & 1 & 5--19\\\hline
            D& Dev.1$^{\star}$ & 7 & 17--7\\\hline
            \multirow{2}{*}{D} & Dev.5 & \multirow{2}{*}{2} & 15--0\\
              & Dev.6$^{\star}$ &  & 9--0\\\hline
        \end{tabular}
    \end{subtable}\hfill
    \begin{subtable}[t]{0.24\textwidth}
        \centering
        \caption{SPW}
        \begin{tabular}{|c|c|c|c|}
            \hline
            Mode & Dev & Slots & G--C\\\hline
            P& Dev.1$^{\star}$ & 1 & 20--4\\\hline
            \multirow{3}{*}{D} & Dev.2$^{\star}$ & \multirow{3}{*}{6} & 4--6\\
              & Dev.5 &  & 7--0\\
              & Dev.6 &  & 7--0\\\hline
        \end{tabular}
    \end{subtable}\hfill
    \begin{subtable}[t]{0.24\textwidth}
        \centering
        \caption{Metrics.}
        \begin{tabular}{|l|c|c|c|}
            \hline
            Method & PS & DS & W\\\hline
            HG & 293 & 22.4 & 198.7\\\hline
            \textbf{ELS} & \textbf{293} & \textbf{22.4} & \textbf{198.7}\\\hline
            SPW & 2482 & 12.5 & 599.3\\\hline
        \end{tabular}
    \end{subtable}
\end{table}
\begin{figure}[H]
    \centering
    \begin{subfigure}{\appfigscale\textwidth}
        \includegraphics[width=\linewidth]{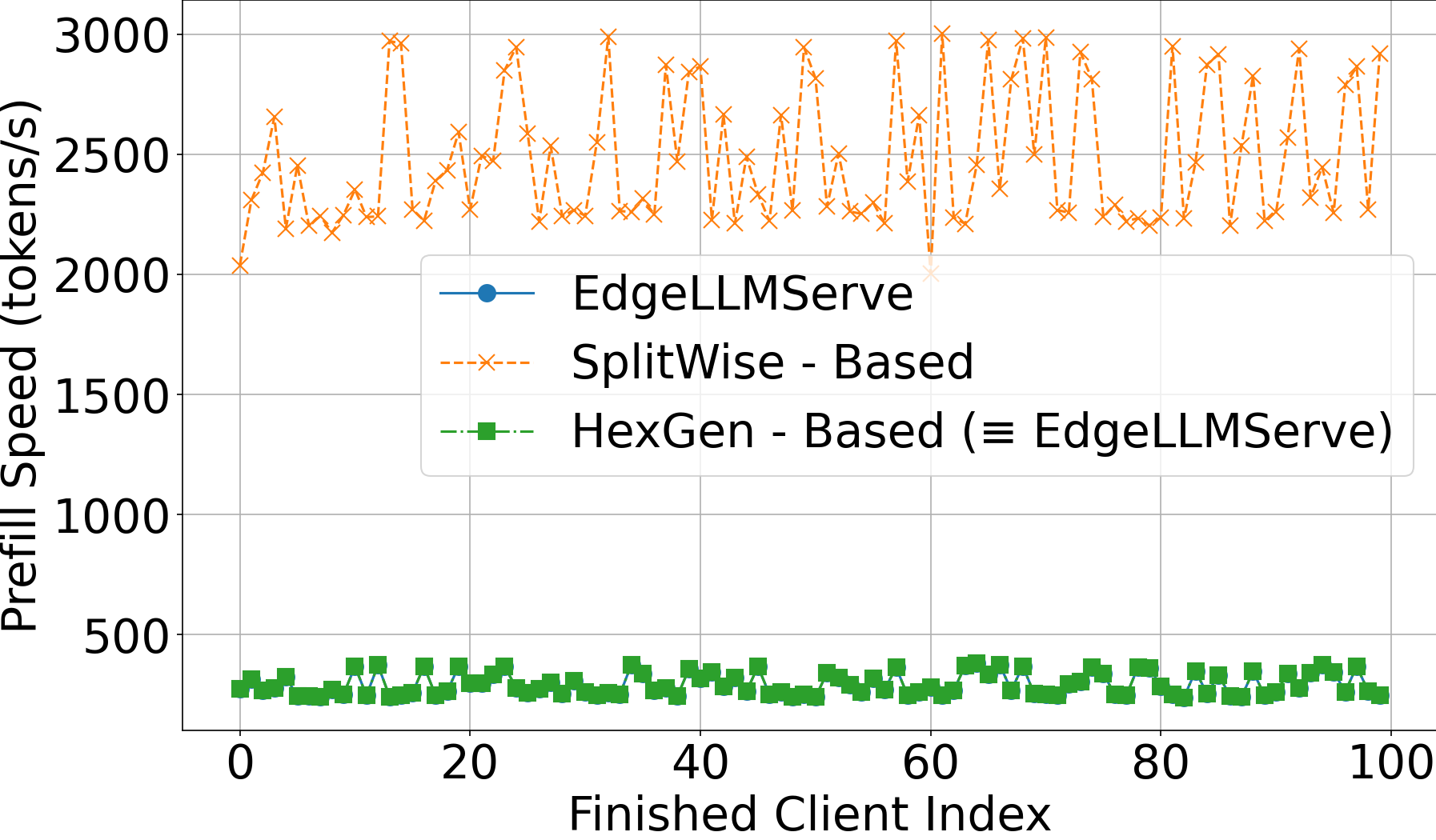}
        \caption{Prefill speed}
    \end{subfigure}\hfill
    \begin{subfigure}{\appfigscale\textwidth}
        \includegraphics[width=\linewidth]{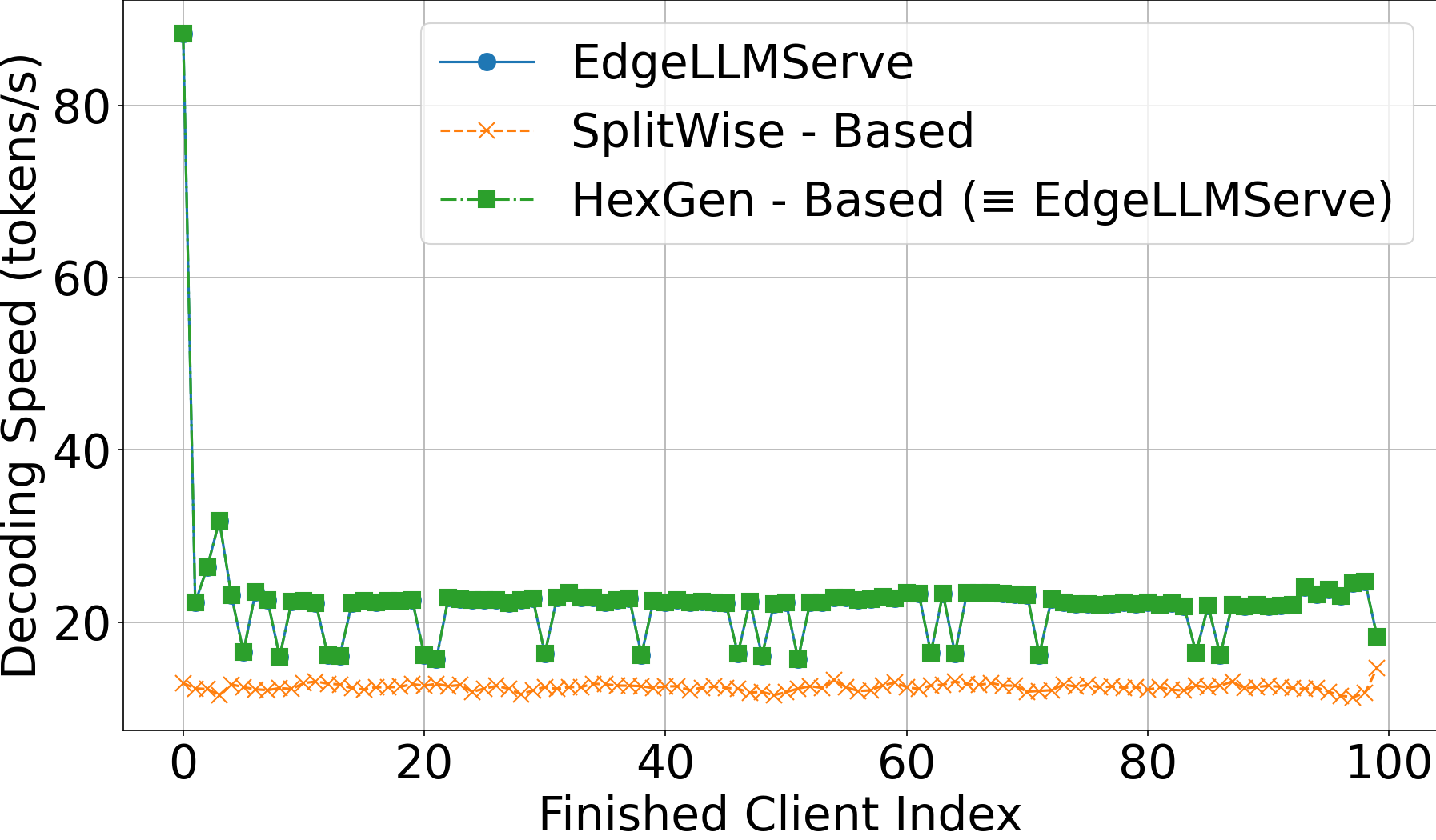}
        \caption{Decoding speed}
    \end{subfigure}\hfill
    \begin{subfigure}{\appfigscale\textwidth}
        \includegraphics[width=\linewidth]{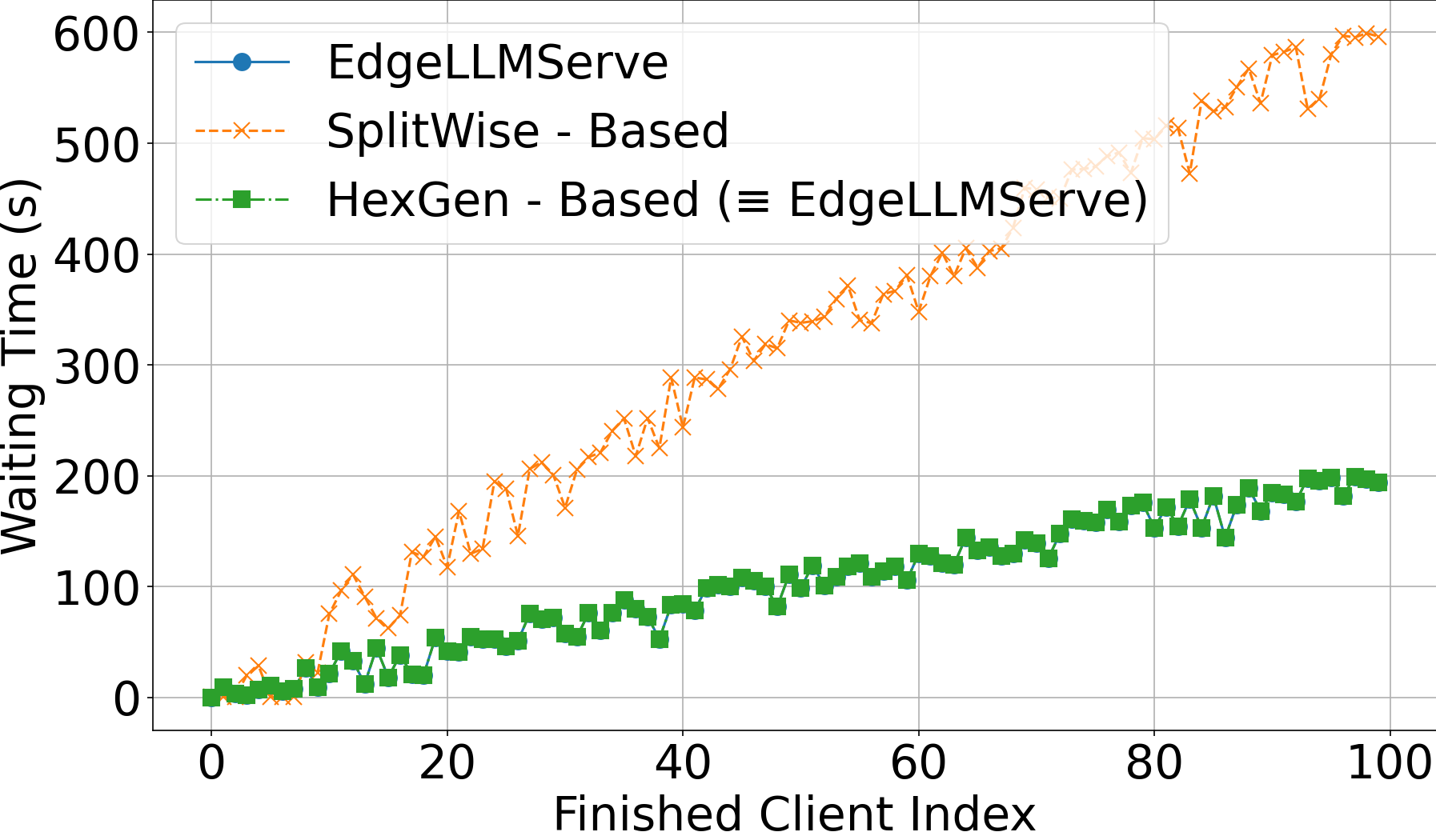}
        \caption{Waiting time}
    \end{subfigure}
    \caption{Per-request metrics --- Four devices, Gpt-Oss-20b, LZByteDance, arrival $1000$~ms.}
    \label{apd:f5}
\end{figure}

\paragraph{6. LZByteDance dataset. Arrival rate of 3000ms}\mbox{}
\begin{table}[H]
    \centering
    \setlength{\tabcolsep}{3pt}
    \caption{Deployment plans and metrics --- Four devices, Gpt-Oss-20b, LZByteDance, arrival $3000$~ms.}
    \label{apd:t6}
    \begin{subtable}[t]{0.24\textwidth}
        \centering
        \caption{HG}
        \begin{tabular}{|c|c|c|c|}
            \hline
            Mode & Dev & Slots & G--C\\\hline
            P& Dev.2$^{\star}$ & 1 & 5--19\\\hline
            D& Dev.1$^{\star}$ & 7 & 17--7\\\hline
            \multirow{2}{*}{D} & Dev.5 & \multirow{2}{*}{2} & 15--0\\
              & Dev.6$^{\star}$ &  & 9--0\\\hline
        \end{tabular}
    \end{subtable}\hfill
    \begin{subtable}[t]{0.24\textwidth}
        \centering
        \caption{ELS}
        \begin{tabular}{|c|c|c|c|}
            \hline
            Mode & Dev & Slots & G--C\\\hline
            P& Dev.2$^{\star}$ & 1 & 5--19\\\hline
            D& Dev.1$^{\star}$ & 7 & 17--7\\\hline
            \multirow{2}{*}{D} & Dev.5 & \multirow{2}{*}{2} & 15--0\\
              & Dev.6$^{\star}$ &  & 9--0\\\hline
        \end{tabular}
    \end{subtable}\hfill
    \begin{subtable}[t]{0.24\textwidth}
        \centering
        \caption{SPW}
        \begin{tabular}{|c|c|c|c|}
            \hline
            Mode & Dev & Slots & G--C\\\hline
            P& Dev.1$^{\star}$ & 1 & 20--4\\\hline
            \multirow{3}{*}{D} & Dev.2$^{\star}$ & \multirow{3}{*}{6} & 4--6\\
              & Dev.5 &  & 7--0\\
              & Dev.6 &  & 7--0\\\hline
        \end{tabular}
    \end{subtable}\hfill
    \begin{subtable}[t]{0.24\textwidth}
        \centering
        \caption{Metrics.}
        \begin{tabular}{|l|c|c|c|}
            \hline
            Method & PS & DS & W\\\hline
            HG & 294 & 22.7 & 18.1\\\hline
            \textbf{ELS} & \textbf{294} & \textbf{22.7} & \textbf{18.1}\\\hline
            SPW & 2480 & 12.5 & 454.1\\\hline
        \end{tabular}
    \end{subtable}
\end{table}
\begin{figure}[H]
    \centering
    \begin{subfigure}{\appfigscale\textwidth}
        \includegraphics[width=\linewidth]{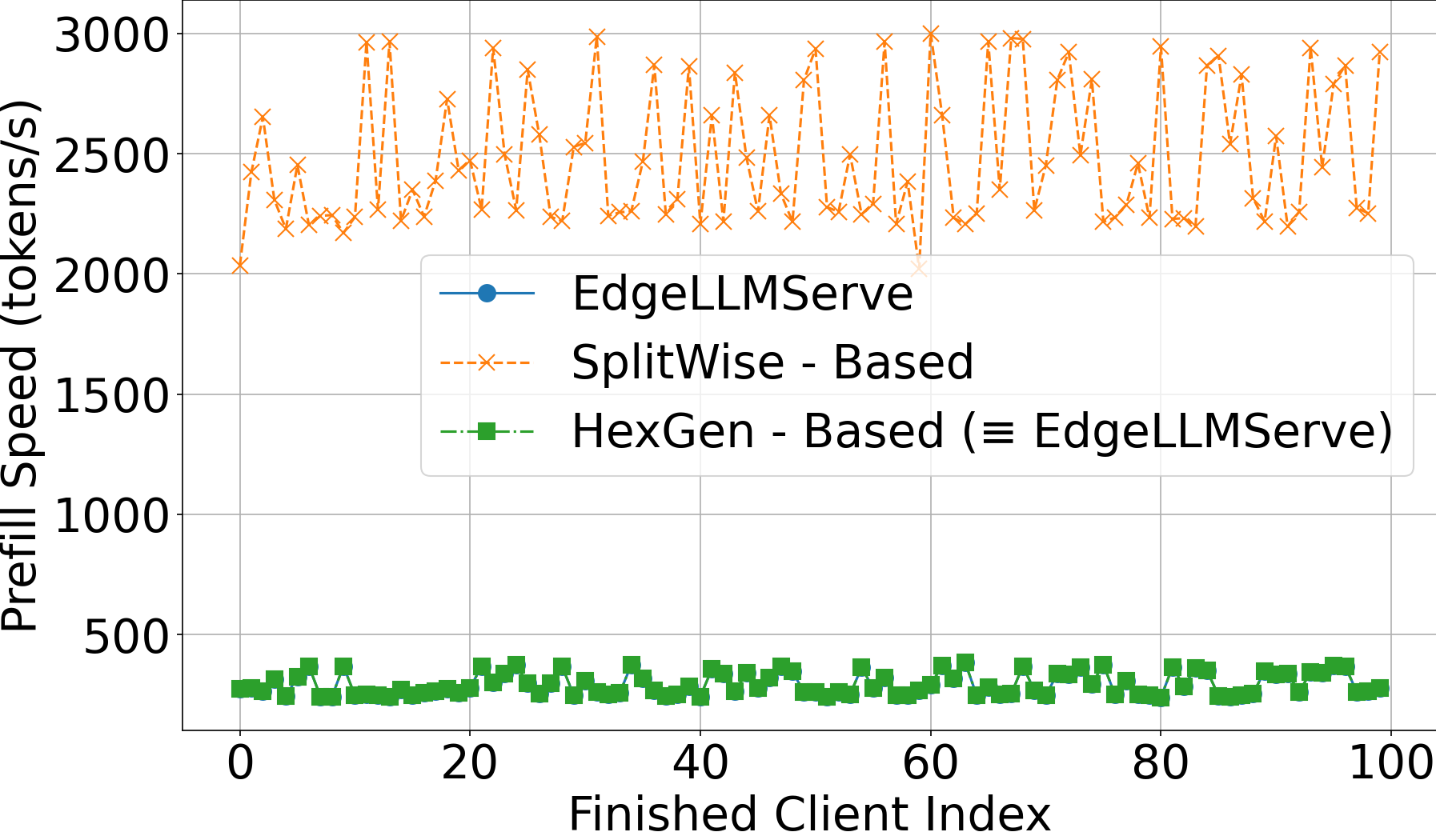}
        \caption{Prefill speed}
    \end{subfigure}\hfill
    \begin{subfigure}{\appfigscale\textwidth}
        \includegraphics[width=\linewidth]{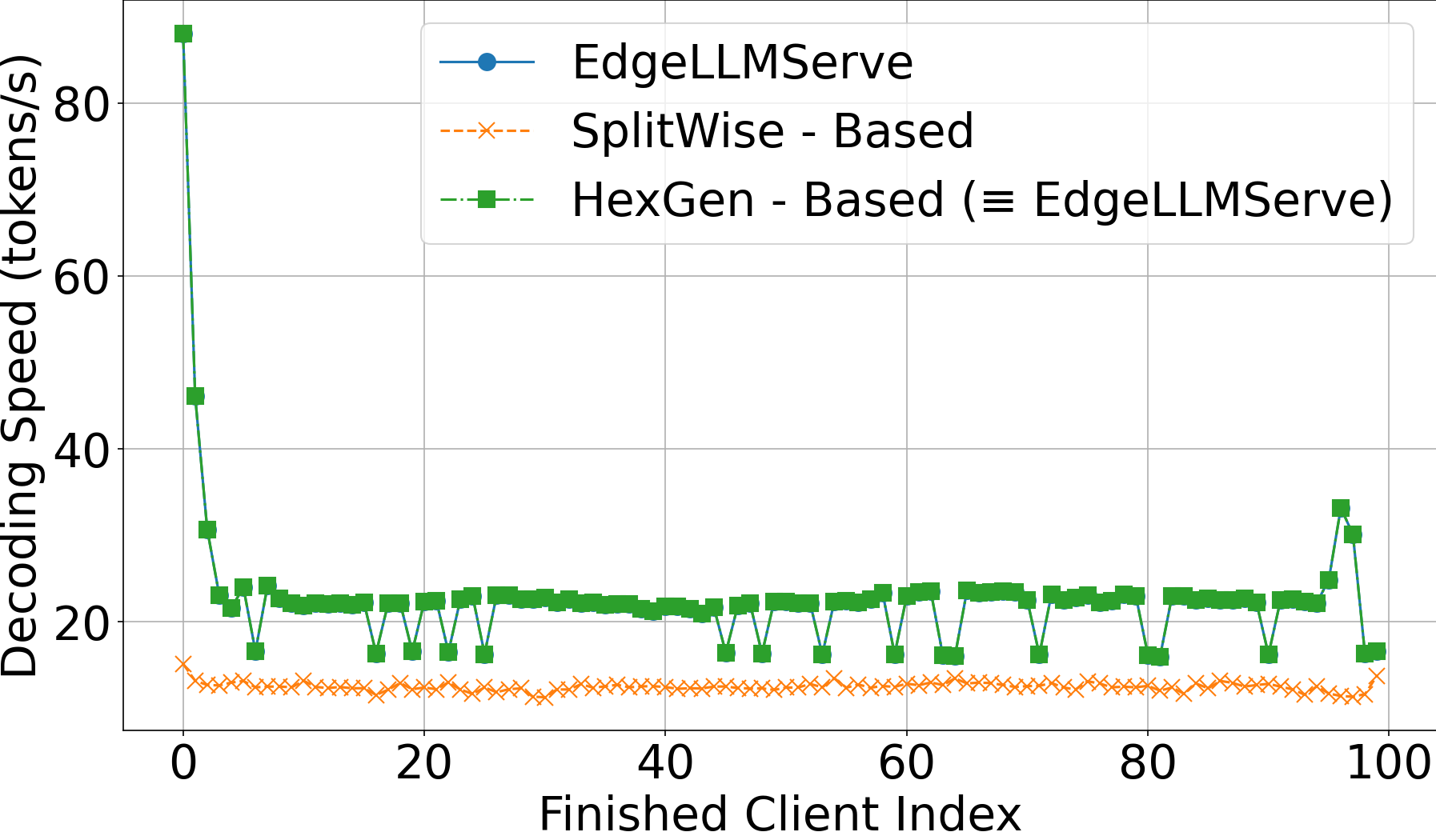}
        \caption{Decoding speed}
    \end{subfigure}\hfill
    \begin{subfigure}{\appfigscale\textwidth}
        \includegraphics[width=\linewidth]{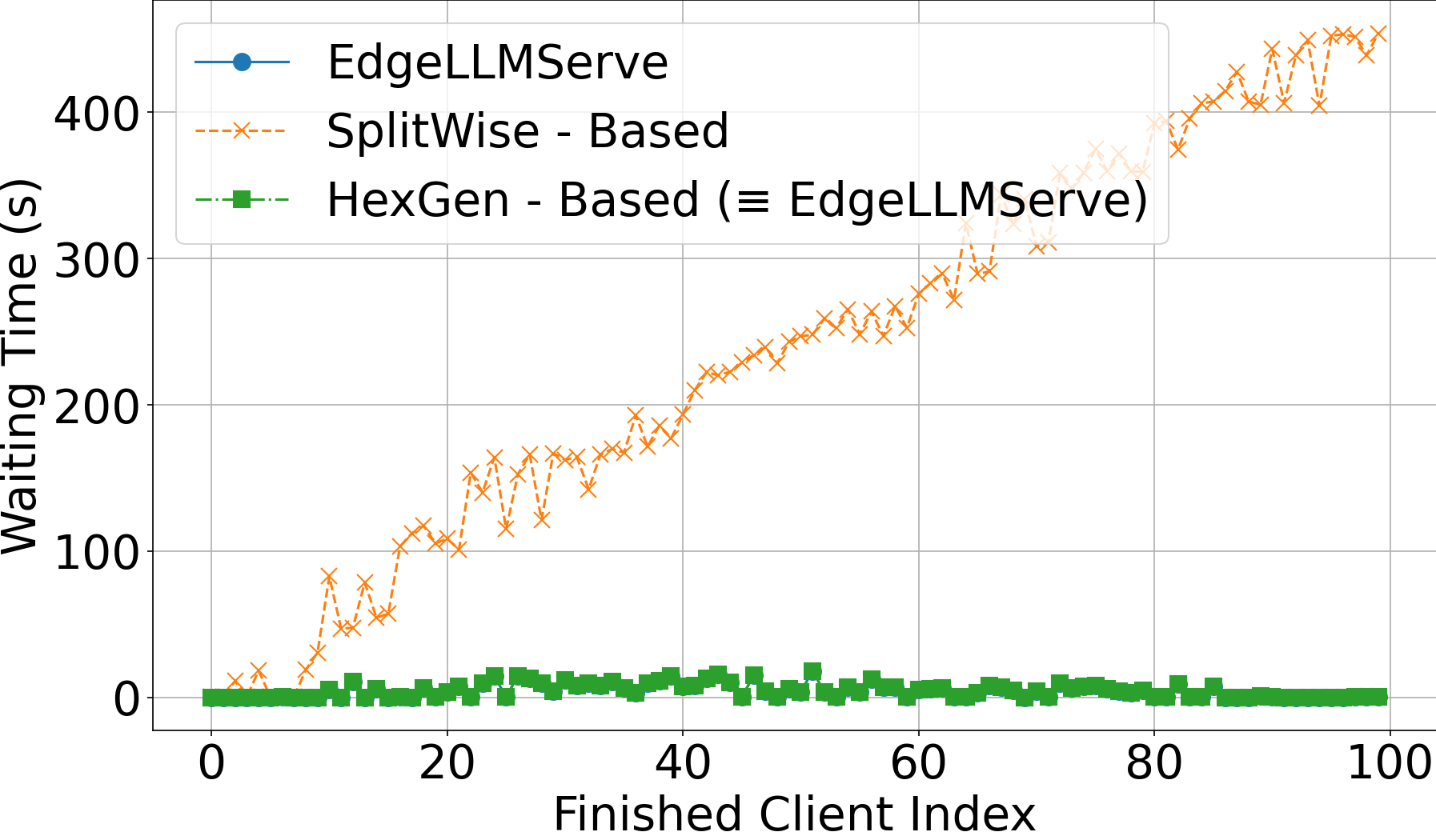}
        \caption{Waiting time}
    \end{subfigure}
    \caption{Per-request metrics --- Four devices, Gpt-Oss-20b, LZByteDance, arrival $3000$~ms.}
    \label{apd:f6}
\end{figure}

\clearpage

\subsubsection{Qwen3-MoE-30B}
\paragraph{1. IELTS dataset. Arrival rate of 100ms}\mbox{}
\begin{table}[H]
    \centering
    \setlength{\tabcolsep}{3pt}
    \caption{Deployment plans and metrics --- Four devices, Qwen3-MoE-30B, IELTS, arrival $100$~ms.}
    \label{apd:t7}
    \begin{subtable}[t]{0.24\textwidth}
        \centering
        \caption{HG}
        \begin{tabular}{|c|c|c|c|}
            \hline
            Mode & Dev & Slots & G--C\\\hline
            P& Dev.2$^{\star}$ & 1 & 9--39\\\hline
            \multirow{3}{*}{D} & Dev.1$^{\star}$ & \multirow{3}{*}{9} & 15--15\\
              & Dev.5 &  & 9--0\\
              & Dev.6 &  & 9--0\\\hline
        \end{tabular}
    \end{subtable}\hfill
    \begin{subtable}[t]{0.24\textwidth}
        \centering
        \caption{ELS}
        \begin{tabular}{|c|c|c|c|}
            \hline
            Mode & Dev & Slots & G--C\\\hline
            P& Dev.2$^{\star}$ & 1 & 9--39\\\hline
            \multirow{3}{*}{D} & Dev.1$^{\star}$ & \multirow{3}{*}{9} & 15--15\\
              & Dev.5 &  & 9--0\\
              & Dev.6 &  & 9--0\\\hline
        \end{tabular}
    \end{subtable}\hfill
    \begin{subtable}[t]{0.24\textwidth}
        \centering
        \caption{SPW}
        \begin{tabular}{|c|c|c|c|}
            \hline
            Mode & Dev & Slots & G--C\\\hline
            P& Dev.1$^{\star}$ & 1 & 26--22\\\hline
            \multirow{3}{*}{D} & Dev.2$^{\star}$ & \multirow{3}{*}{3} & 7--15\\
              & Dev.5 &  & 13--0\\
              & Dev.6 &  & 13--0\\\hline
        \end{tabular}
    \end{subtable}\hfill
    \begin{subtable}[t]{0.24\textwidth}
        \centering
        \caption{Metrics.}
        \begin{tabular}{|l|c|c|c|}
            \hline
            Method & PS & DS & W\\\hline
            HG & 259 & 10.4 & 2021.4\\\hline
            \textbf{ELS} & \textbf{259} & \textbf{10.4} & \textbf{2021.4}\\\hline
            SPW & 1405 & 15.6 & 4032.4\\\hline
        \end{tabular}
    \end{subtable}
\end{table}
\begin{figure}[H]
    \centering
    \begin{subfigure}{\appfigscale\textwidth}
        \includegraphics[width=\linewidth]{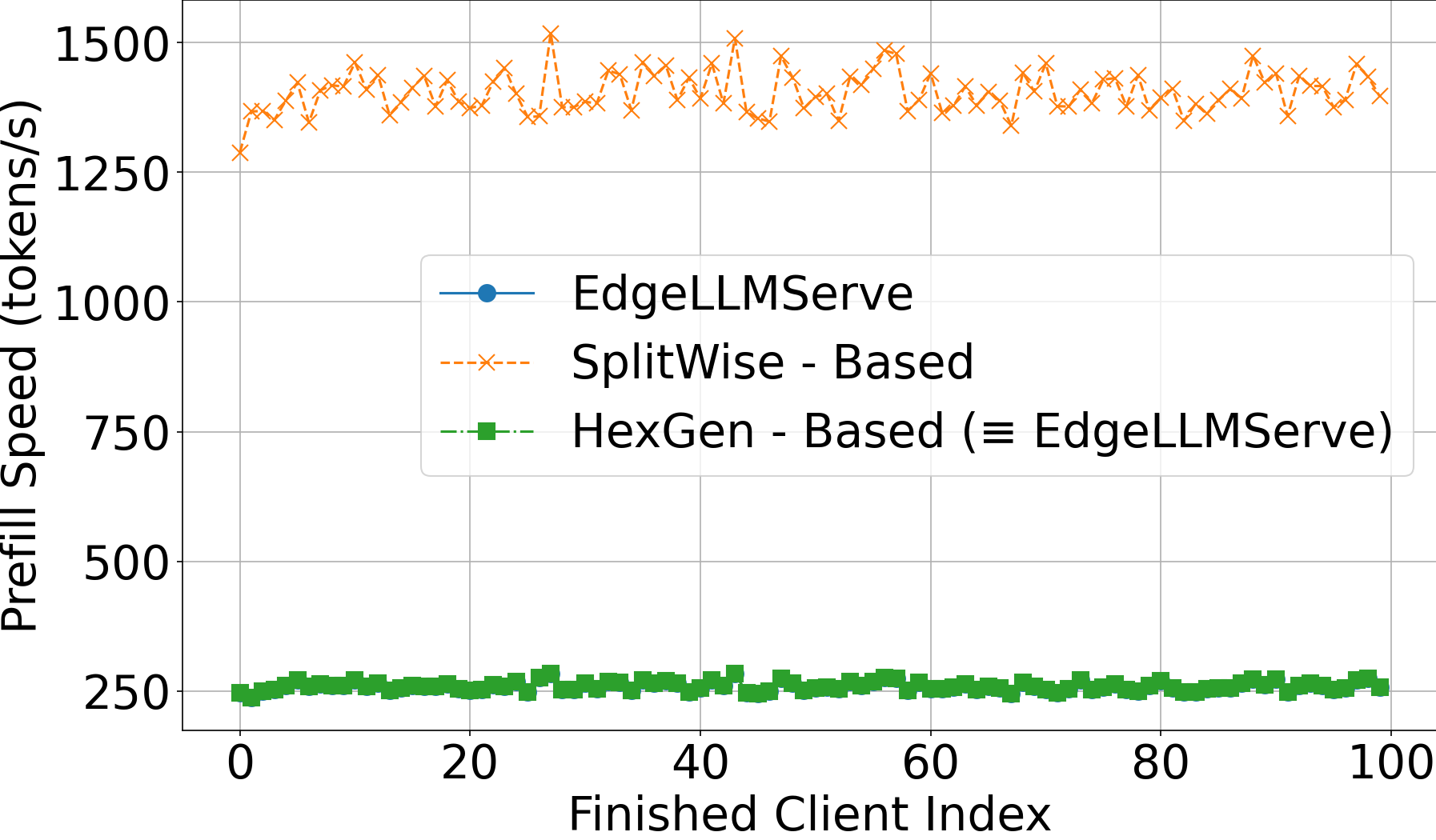}
        \caption{Prefill speed}
    \end{subfigure}\hfill
    \begin{subfigure}{\appfigscale\textwidth}
        \includegraphics[width=\linewidth]{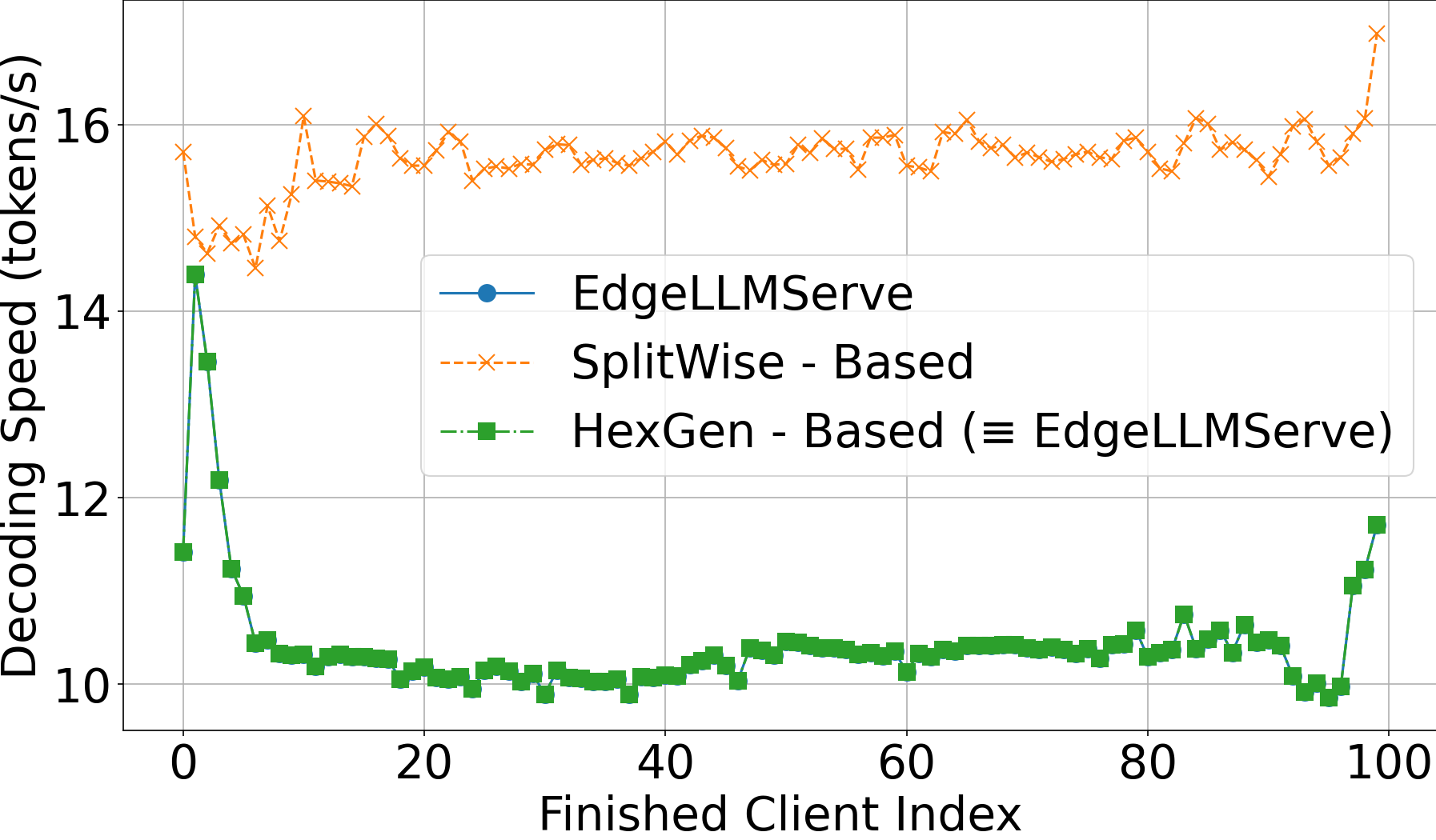}
        \caption{Decoding speed}
    \end{subfigure}\hfill
    \begin{subfigure}{\appfigscale\textwidth}
        \includegraphics[width=\linewidth]{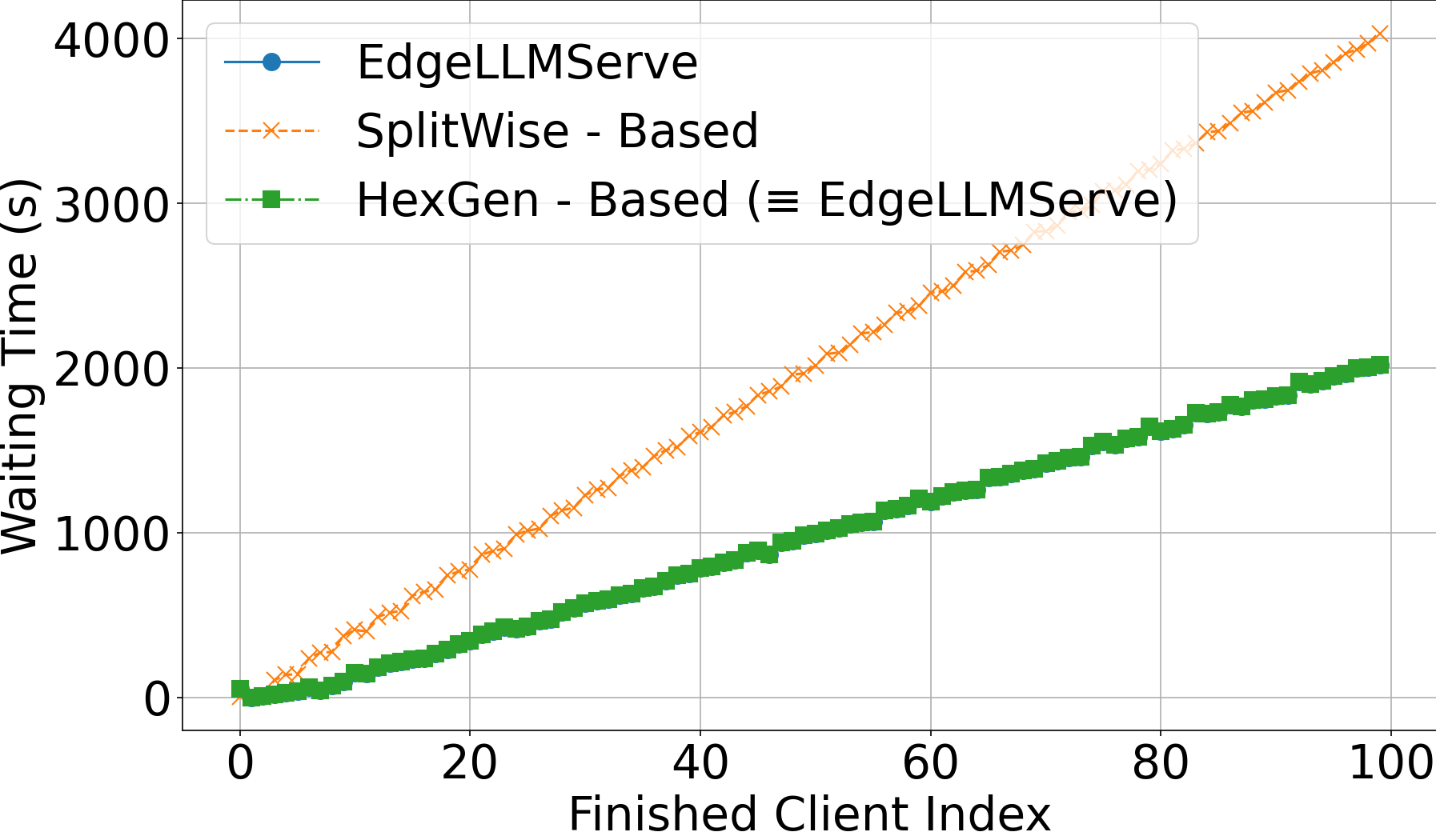}
        \caption{Waiting time}
    \end{subfigure}
    \caption{Per-request metrics --- Four devices, Qwen3-MoE-30B, IELTS, arrival $100$~ms.}
    \label{apd:f7}
\end{figure}

\paragraph{2. IELTS dataset. Arrival rate of 1000ms}\mbox{}
\begin{table}[H]
    \centering
    \setlength{\tabcolsep}{3pt}
    \caption{Deployment plans and metrics --- Four devices, Qwen3-MoE-30B, IELTS, arrival $1000$~ms.}
    \label{apd:t8}
    \begin{subtable}[t]{0.24\textwidth}
        \centering
        \caption{HG}
        \begin{tabular}{|c|c|c|c|}
            \hline
            Mode & Dev & Slots & G--C\\\hline
            P& Dev.2$^{\star}$ & 1 & 9--39\\\hline
            \multirow{3}{*}{D} & Dev.1$^{\star}$ & \multirow{3}{*}{9} & 15--15\\
              & Dev.5 &  & 9--0\\
              & Dev.6 &  & 9--0\\\hline
        \end{tabular}
    \end{subtable}\hfill
    \begin{subtable}[t]{0.24\textwidth}
        \centering
        \caption{ELS}
        \begin{tabular}{|c|c|c|c|}
            \hline
            Mode & Dev & Slots & G--C\\\hline
            P& Dev.2$^{\star}$ & 1 & 9--39\\\hline
            \multirow{3}{*}{D} & Dev.1$^{\star}$ & \multirow{3}{*}{9} & 15--15\\
              & Dev.5 &  & 9--0\\
              & Dev.6 &  & 9--0\\\hline
        \end{tabular}
    \end{subtable}\hfill
    \begin{subtable}[t]{0.24\textwidth}
        \centering
        \caption{SPW}
        \begin{tabular}{|c|c|c|c|}
            \hline
            Mode & Dev & Slots & G--C\\\hline
            P& Dev.1$^{\star}$ & 1 & 26--22\\\hline
            \multirow{3}{*}{D} & Dev.2$^{\star}$ & \multirow{3}{*}{3} & 7--15\\
              & Dev.5 &  & 13--0\\
              & Dev.6 &  & 13--0\\\hline
        \end{tabular}
    \end{subtable}\hfill
    \begin{subtable}[t]{0.24\textwidth}
        \centering
        \caption{Metrics.}
        \begin{tabular}{|l|c|c|c|}
            \hline
            Method & PS & DS & W\\\hline
            HG & 259 & 10.5 & 1997.8\\\hline
            \textbf{ELS} & \textbf{259} & \textbf{10.5} & \textbf{1997.8}\\\hline
            SPW & 1405 & 15.5 & 3964.0\\\hline
        \end{tabular}
    \end{subtable}
\end{table}
\begin{figure}[H]
    \centering
    \begin{subfigure}{\appfigscale\textwidth}
        \includegraphics[width=\linewidth]{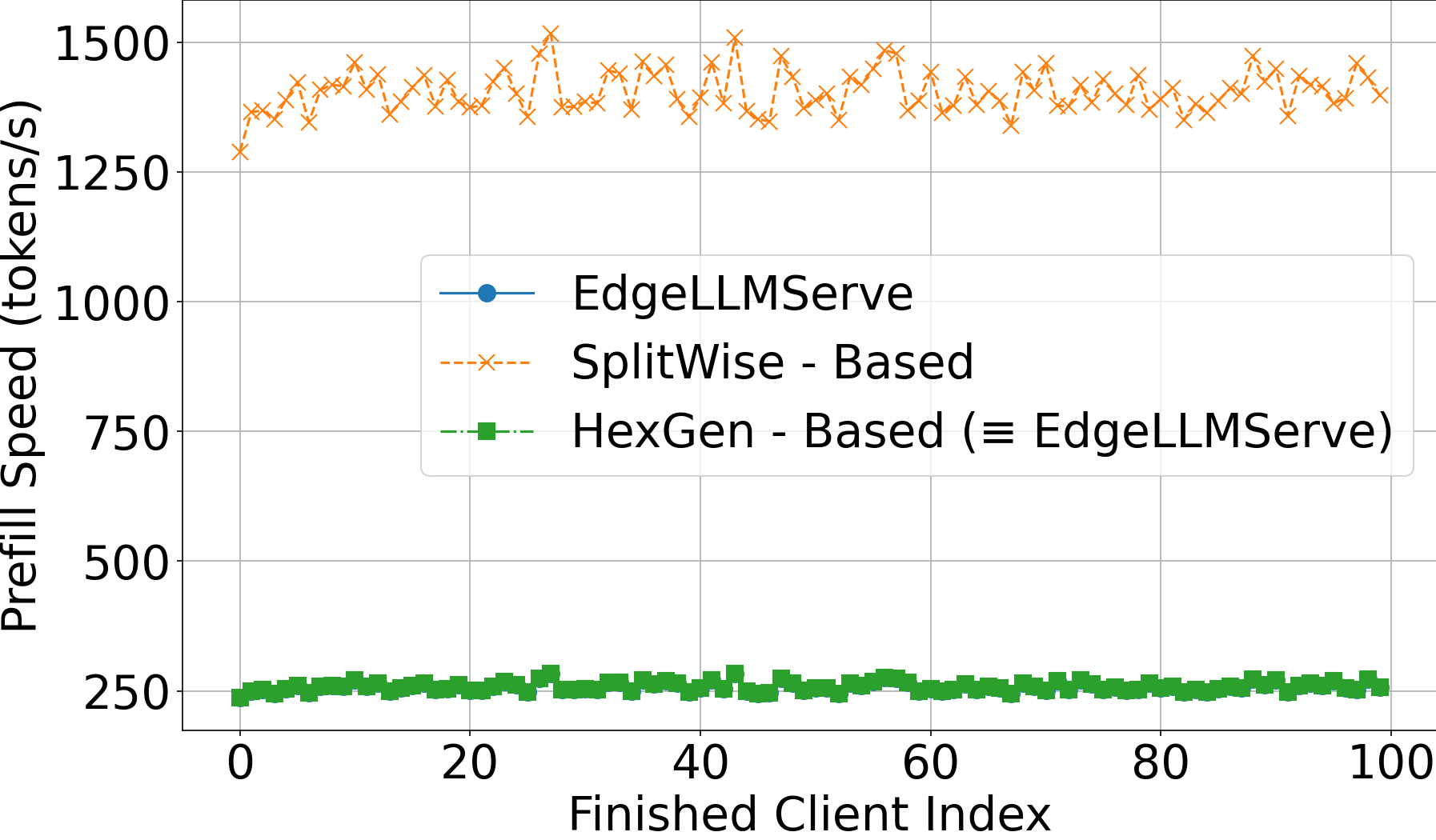}
        \caption{Prefill speed}
    \end{subfigure}\hfill
    \begin{subfigure}{\appfigscale\textwidth}
        \includegraphics[width=\linewidth]{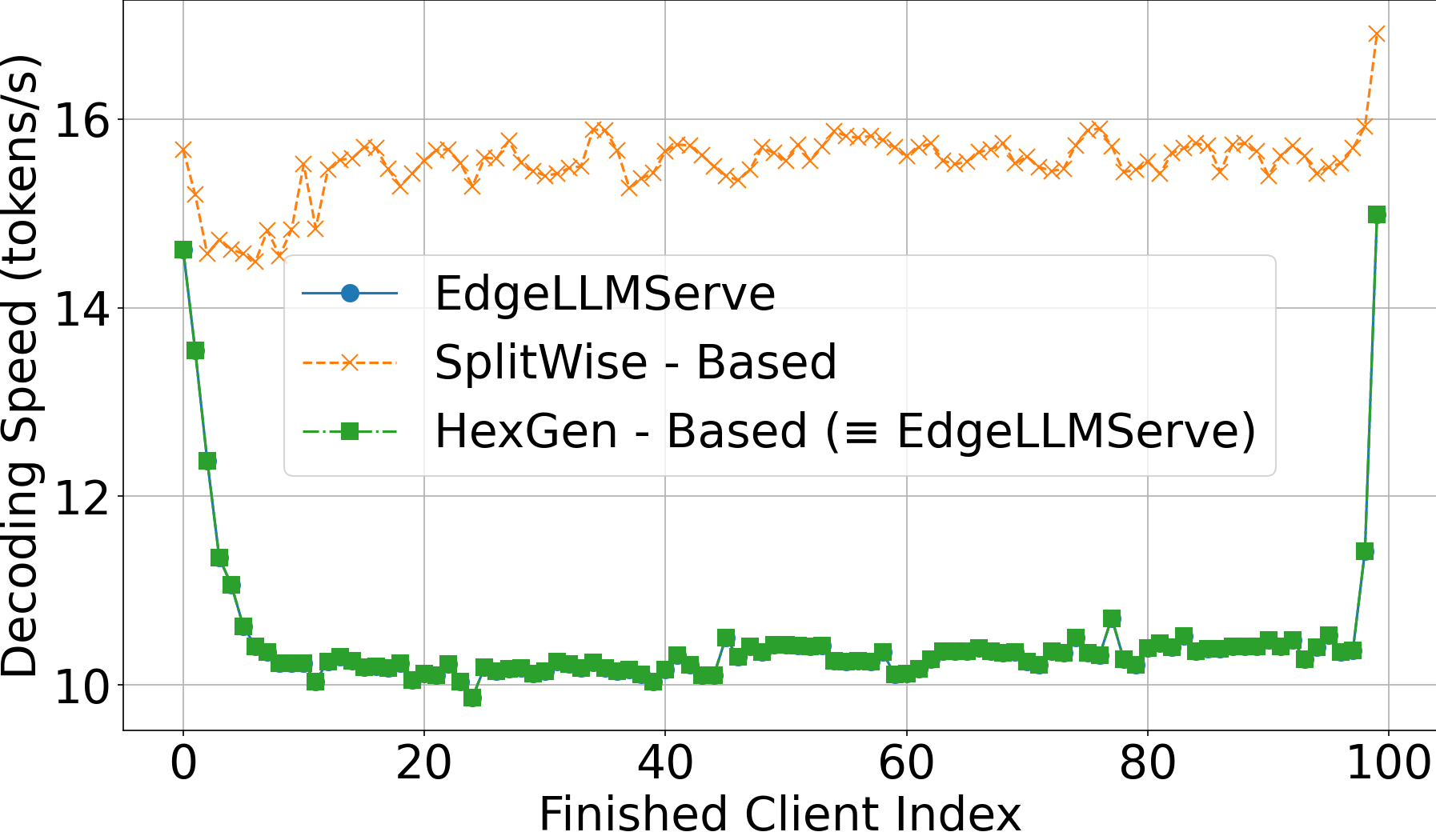}
        \caption{Decoding speed}
    \end{subfigure}\hfill
    \begin{subfigure}{\appfigscale\textwidth}
        \includegraphics[width=\linewidth]{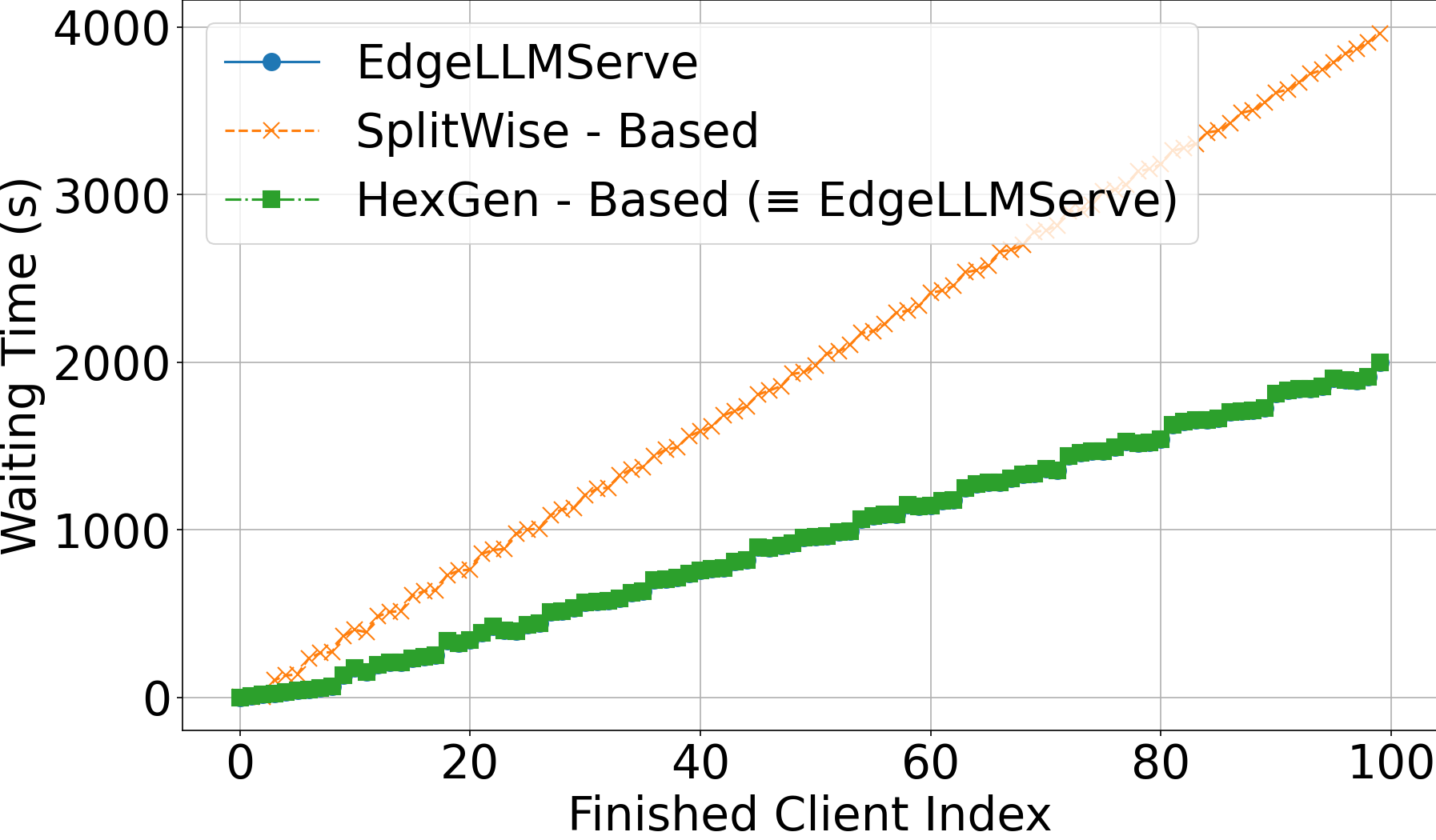}
        \caption{Waiting time}
    \end{subfigure}
    \caption{Per-request metrics --- Four devices, Qwen3-MoE-30B, IELTS, arrival $1000$~ms.}
    \label{apd:f8}
\end{figure}

\paragraph{3. IELTS dataset. Arrival rate of 3000ms}\mbox{}
\begin{table}[H]
    \centering
    \setlength{\tabcolsep}{3pt}
    \caption{Deployment plans and metrics --- Four devices, Qwen3-MoE-30B, IELTS, arrival $3000$~ms.}
    \label{apd:t9}
    \begin{subtable}[t]{0.24\textwidth}
        \centering
        \caption{HG}
        \begin{tabular}{|c|c|c|c|}
            \hline
            Mode & Dev & Slots & G--C\\\hline
            P& Dev.2$^{\star}$ & 1 & 9--39\\\hline
            \multirow{3}{*}{D} & Dev.1$^{\star}$ & \multirow{3}{*}{9} & 15--15\\
              & Dev.5 &  & 9--0\\
              & Dev.6 &  & 9--0\\\hline
        \end{tabular}
    \end{subtable}\hfill
    \begin{subtable}[t]{0.24\textwidth}
        \centering
        \caption{ELS}
        \begin{tabular}{|c|c|c|c|}
            \hline
            Mode & Dev & Slots & G--C\\\hline
            P& Dev.2$^{\star}$ & 1 & 9--39\\\hline
            \multirow{3}{*}{D} & Dev.1$^{\star}$ & \multirow{3}{*}{9} & 15--15\\
              & Dev.5 &  & 9--0\\
              & Dev.6 &  & 9--0\\\hline
        \end{tabular}
    \end{subtable}\hfill
    \begin{subtable}[t]{0.24\textwidth}
        \centering
        \caption{SPW}
        \begin{tabular}{|c|c|c|c|}
            \hline
            Mode & Dev & Slots & G--C\\\hline
            P& Dev.1$^{\star}$ & 1 & 26--22\\\hline
            \multirow{3}{*}{D} & Dev.2$^{\star}$ & \multirow{3}{*}{3} & 7--15\\
              & Dev.5 &  & 13--0\\
              & Dev.6 &  & 13--0\\\hline
        \end{tabular}
    \end{subtable}\hfill
    \begin{subtable}[t]{0.24\textwidth}
        \centering
        \caption{Metrics.}
        \begin{tabular}{|l|c|c|c|}
            \hline
            Method & PS & DS & W\\\hline
            HG & 258 & 10.4 & 1789.3\\\hline
            \textbf{ELS} & \textbf{258} & \textbf{10.4} & \textbf{1789.3}\\\hline
            SPW & 1406 & 15.6 & 3887.4\\\hline
        \end{tabular}
    \end{subtable}
\end{table}
\begin{figure}[H]
    \centering
    \begin{subfigure}{\appfigscale\textwidth}
        \includegraphics[width=\linewidth]{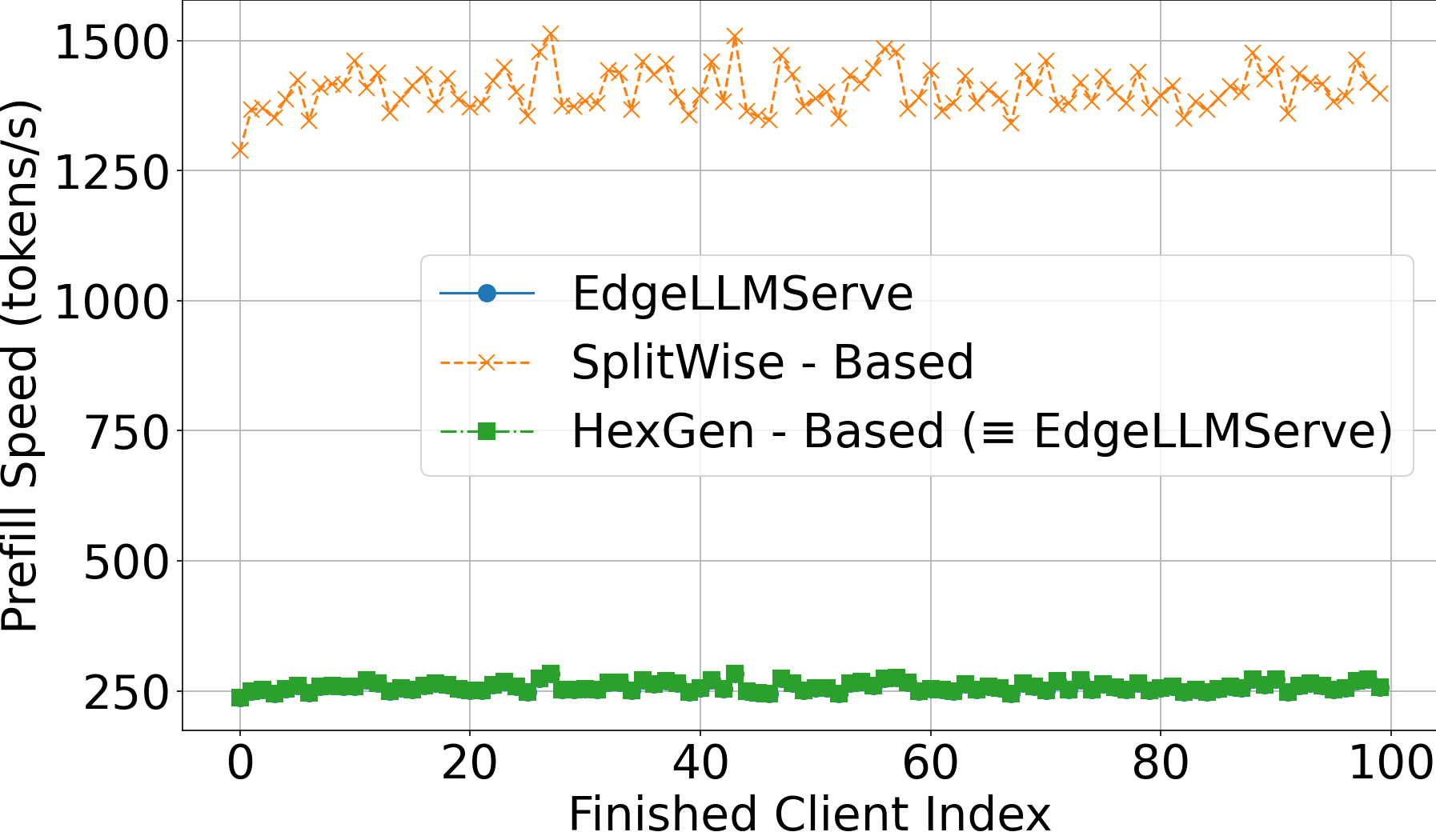}
        \caption{Prefill speed}
    \end{subfigure}\hfill
    \begin{subfigure}{\appfigscale\textwidth}
        \includegraphics[width=\linewidth]{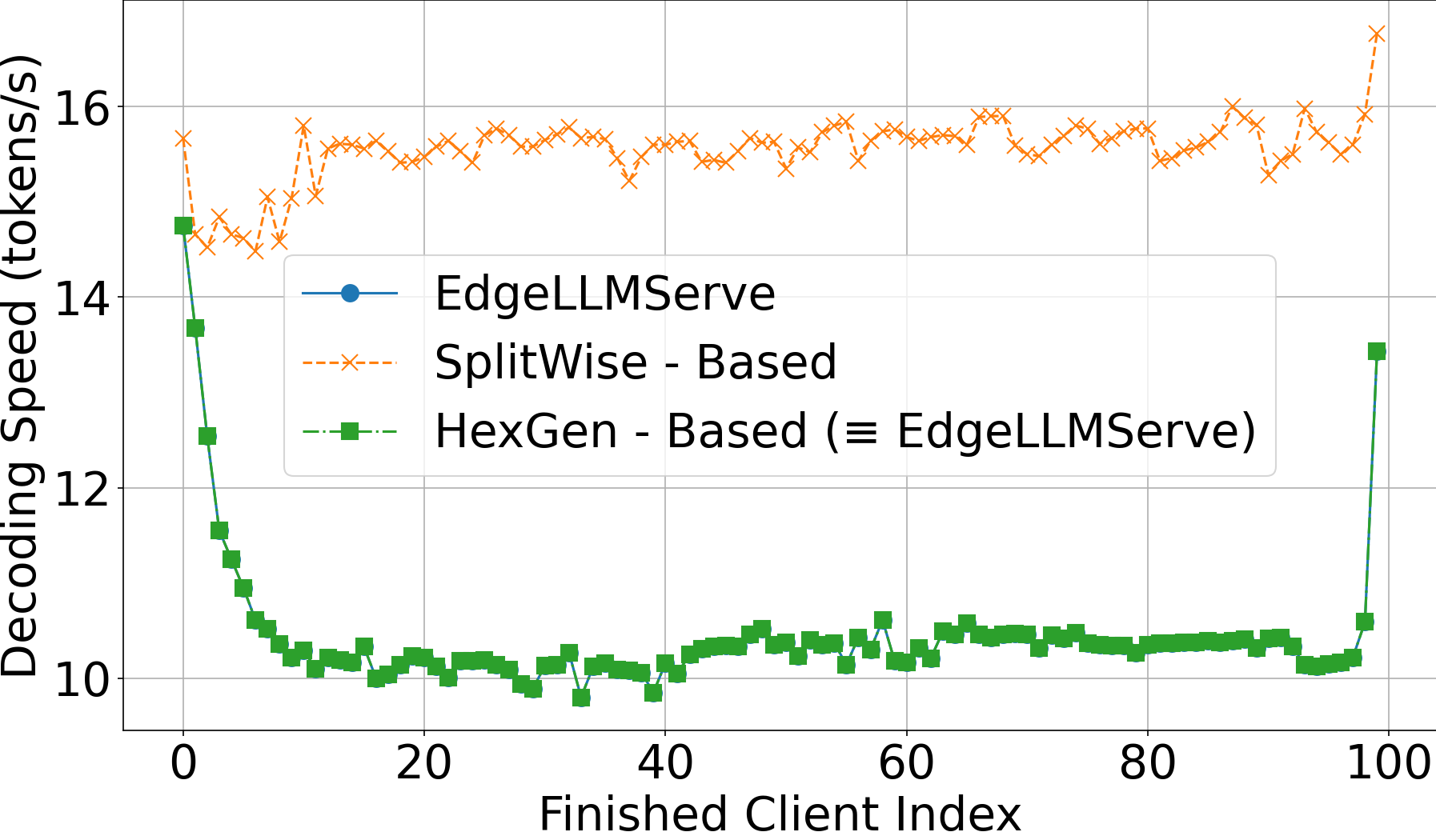}
        \caption{Decoding speed}
    \end{subfigure}\hfill
    \begin{subfigure}{\appfigscale\textwidth}
        \includegraphics[width=\linewidth]{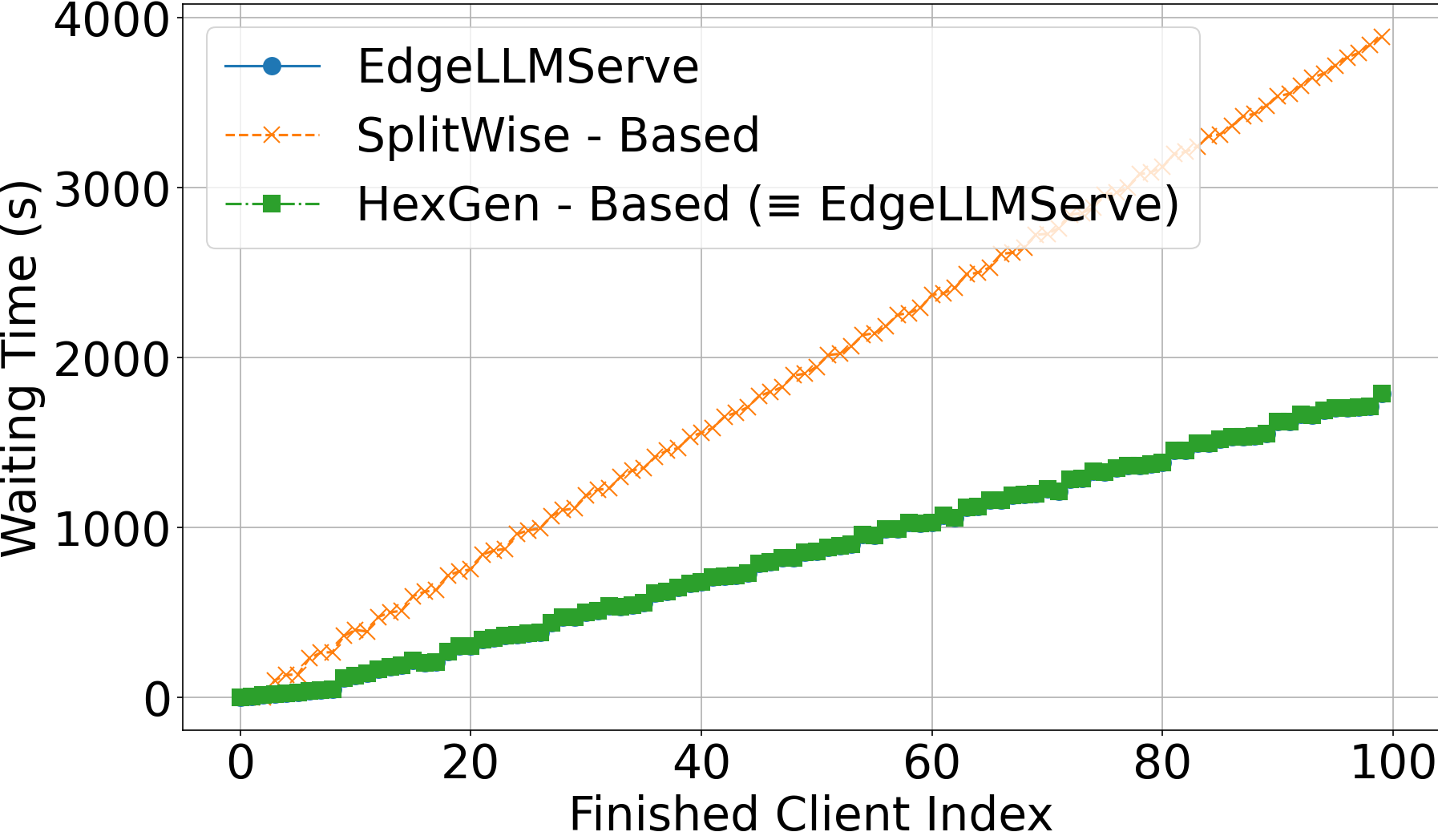}
        \caption{Waiting time}
    \end{subfigure}
    \caption{Per-request metrics --- Four devices, Qwen3-MoE-30B, IELTS, arrival $3000$~ms.}
    \label{apd:f9}
\end{figure}

\paragraph{4. LZByteDance dataset. Arrival rate of 100ms}\mbox{}
\begin{table}[H]
    \centering
    \setlength{\tabcolsep}{3pt}
    \caption{Deployment plans and metrics --- Four devices, Qwen3-MoE-30B, LZByteDance, arrival $100$~ms.}
    \label{apd:t10}
    \begin{subtable}[t]{0.24\textwidth}
        \centering
        \caption{HG}
        \begin{tabular}{|c|c|c|c|}
            \hline
            Mode & Dev & Slots & G--C\\\hline
            \multirow{2}{*}{P} & Dev.5$^{\star}$ & \multirow{2}{*}{1} & 22--0\\
              & Dev.6 &  & 26--0\\\hline
            \multirow{2}{*}{D} & Dev.1$^{\star}$ & \multirow{2}{*}{10} & 13--19\\
              & Dev.2 &  & 5--11\\\hline
        \end{tabular}
    \end{subtable}\hfill
    \begin{subtable}[t]{0.24\textwidth}
        \centering
        \caption{ELS}
        \begin{tabular}{|c|c|c|c|}
            \hline
            Mode & Dev & Slots & G--C\\\hline
            \multirow{2}{*}{P} & Dev.5$^{\star}$ & \multirow{2}{*}{1} & 22--0\\
              & Dev.6 &  & 26--0\\\hline
            \multirow{2}{*}{D} & Dev.1$^{\star}$ & \multirow{2}{*}{10} & 13--19\\
              & Dev.2 &  & 5--11\\\hline
        \end{tabular}
    \end{subtable}\hfill
    \begin{subtable}[t]{0.24\textwidth}
        \centering
        \caption{SPW}
        \begin{tabular}{|c|c|c|c|}
            \hline
            Mode & Dev & Slots & G--C\\\hline
            P& Dev.1$^{\star}$ & 1 & 26--22\\\hline
            \multirow{3}{*}{D} & Dev.2$^{\star}$ & \multirow{3}{*}{3} & 7--15\\
              & Dev.5 &  & 13--0\\
              & Dev.6 &  & 13--0\\\hline
        \end{tabular}
    \end{subtable}\hfill
    \begin{subtable}[t]{0.24\textwidth}
        \centering
        \caption{Metrics.}
        \begin{tabular}{|l|c|c|c|}
            \hline
            Method & PS & DS & W\\\hline
            HG & 136 & 12.3 & 709.4\\\hline
            \textbf{ELS} & \textbf{136} & \textbf{12.3} & \textbf{709.4}\\\hline
            SPW & 1135 & 15.5 & 1482.8\\\hline
        \end{tabular}
    \end{subtable}
\end{table}
\begin{figure}[H]
    \centering
    \begin{subfigure}{\appfigscale\textwidth}
        \includegraphics[width=\linewidth]{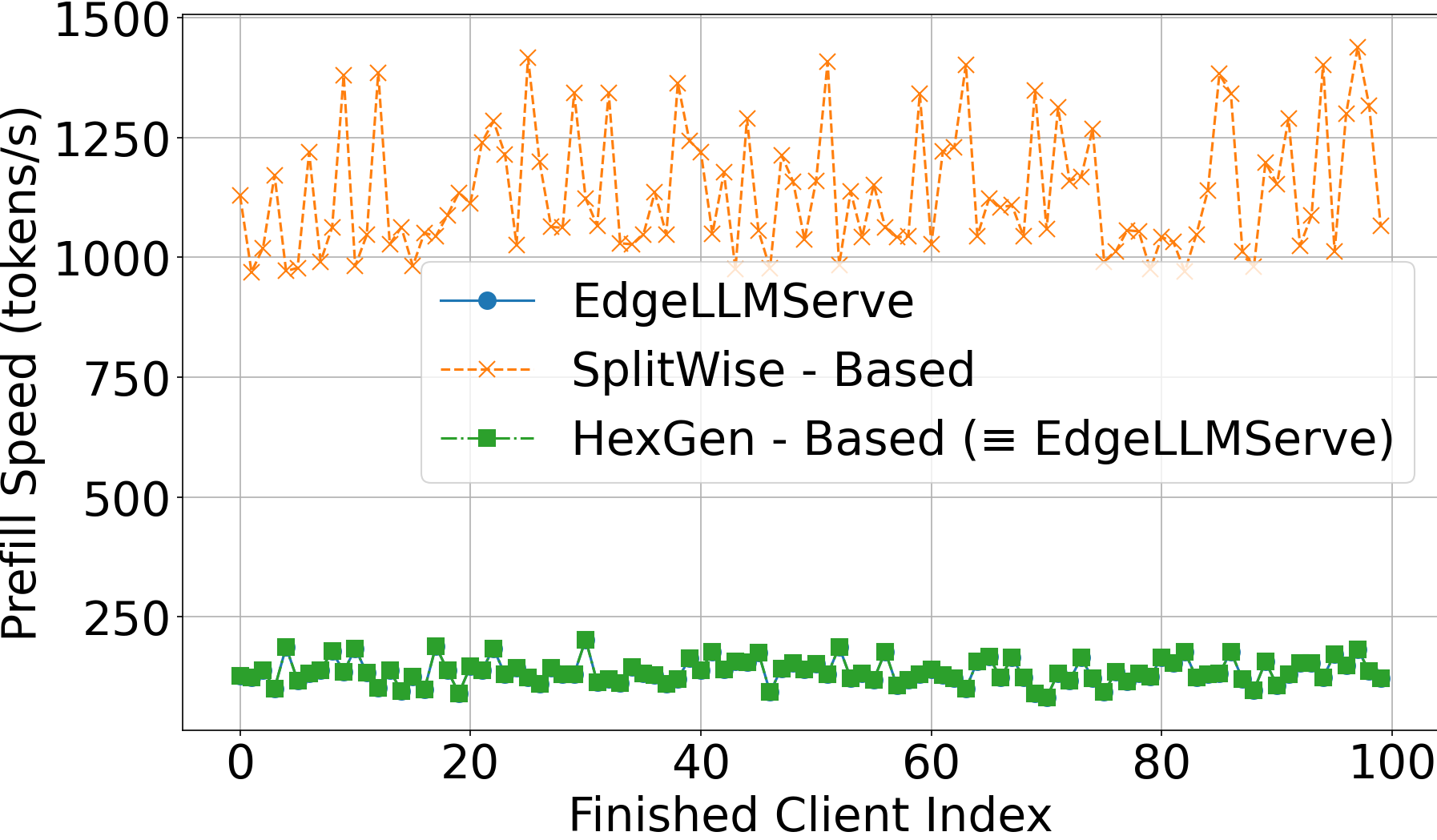}
        \caption{Prefill speed}
    \end{subfigure}\hfill
    \begin{subfigure}{\appfigscale\textwidth}
        \includegraphics[width=\linewidth]{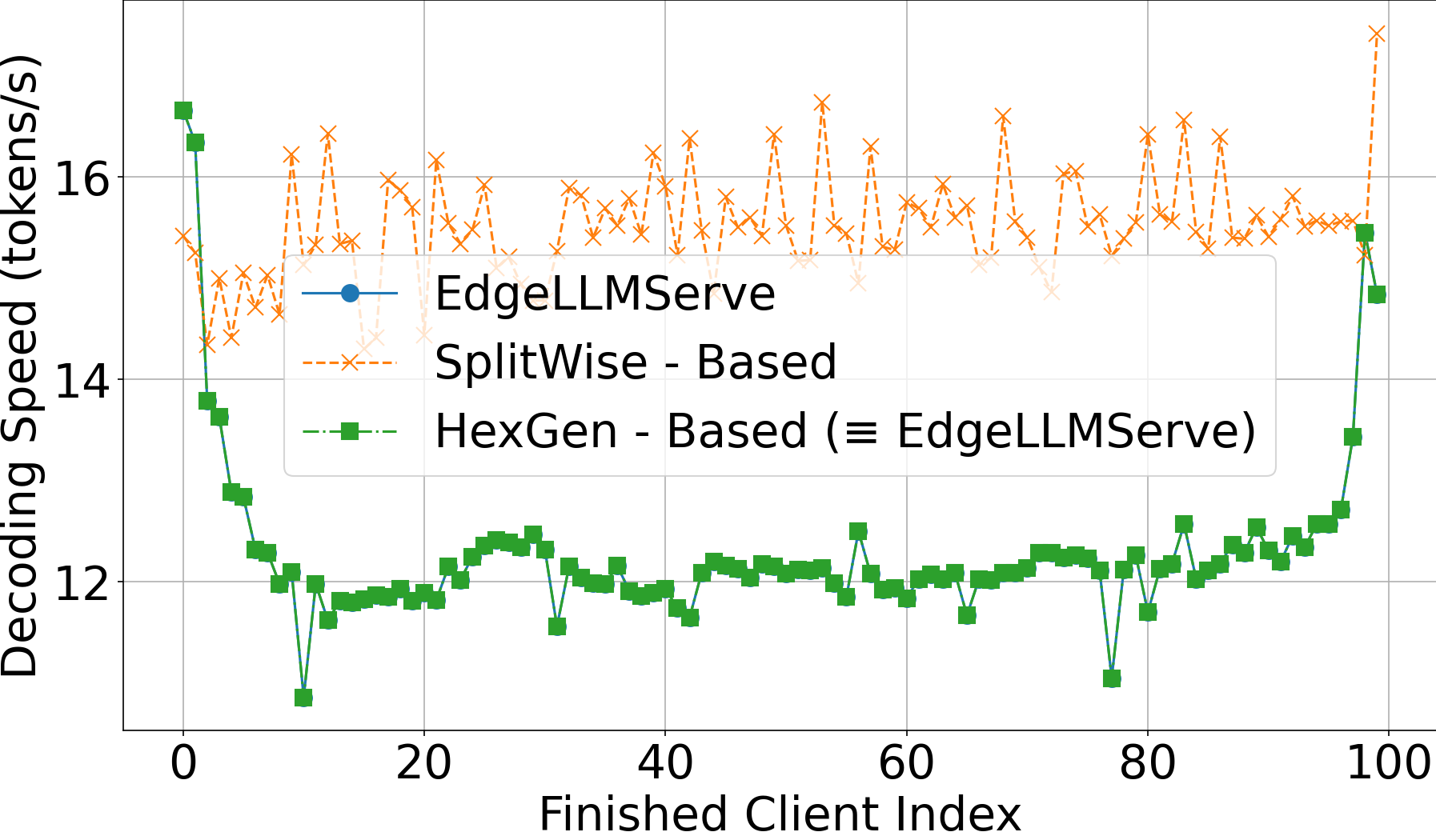}
        \caption{Decoding speed}
    \end{subfigure}\hfill
    \begin{subfigure}{\appfigscale\textwidth}
        \includegraphics[width=\linewidth]{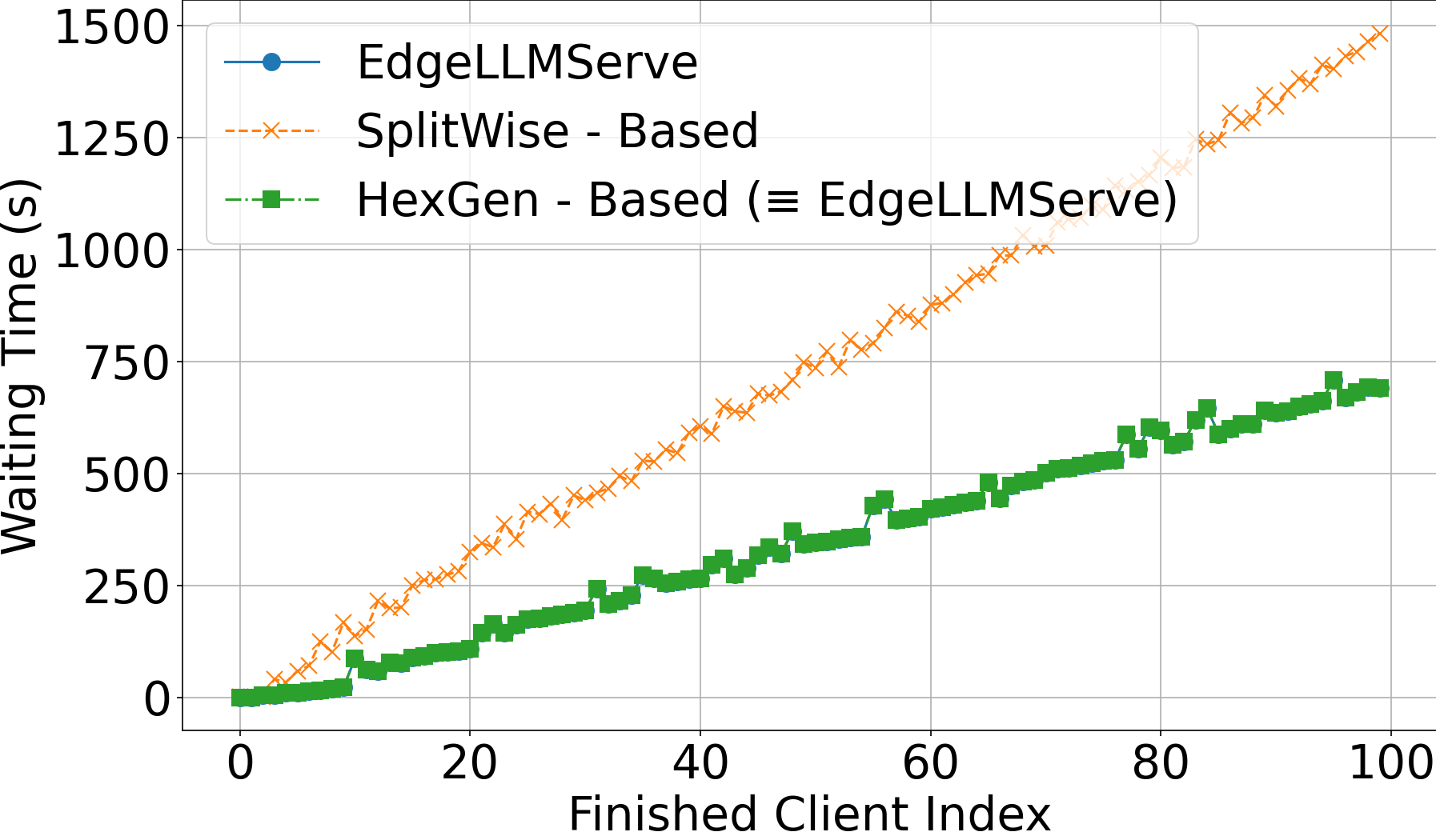}
        \caption{Waiting time}
    \end{subfigure}
    \caption{Per-request metrics --- Four devices, Qwen3-MoE-30B, LZByteDance, arrival $100$~ms.}
    \label{apd:f10}
\end{figure}

\paragraph{5. LZByteDance dataset. Arrival rate of 1000ms}\mbox{}
\begin{table}[H]
    \centering
    \setlength{\tabcolsep}{3pt}
    \caption{Deployment plans and metrics --- Four devices, Qwen3-MoE-30B, LZByteDance, arrival $1000$~ms.}
    \label{apd:t11}
    \begin{subtable}[t]{0.24\textwidth}
        \centering
        \caption{HG}
        \begin{tabular}{|c|c|c|c|}
            \hline
            Mode & Dev & Slots & G--C\\\hline
            \multirow{2}{*}{P} & Dev.5$^{\star}$ & \multirow{2}{*}{1} & 22--0\\
              & Dev.6 &  & 26--0\\\hline
            \multirow{2}{*}{D} & Dev.1$^{\star}$ & \multirow{2}{*}{10} & 13--19\\
              & Dev.2 &  & 5--11\\\hline
        \end{tabular}
    \end{subtable}\hfill
    \begin{subtable}[t]{0.24\textwidth}
        \centering
        \caption{ELS}
        \begin{tabular}{|c|c|c|c|}
            \hline
            Mode & Dev & Slots & G--C\\\hline
            \multirow{2}{*}{P} & Dev.5$^{\star}$ & \multirow{2}{*}{1} & 22--0\\
              & Dev.6 &  & 26--0\\\hline
            \multirow{2}{*}{D} & Dev.1$^{\star}$ & \multirow{2}{*}{10} & 13--19\\
              & Dev.2 &  & 5--11\\\hline
        \end{tabular}
    \end{subtable}\hfill
    \begin{subtable}[t]{0.24\textwidth}
        \centering
        \caption{SPW}
        \begin{tabular}{|c|c|c|c|}
            \hline
            Mode & Dev & Slots & G--C\\\hline
            P& Dev.1$^{\star}$ & 1 & 26--22\\\hline
            \multirow{3}{*}{D} & Dev.2$^{\star}$ & \multirow{3}{*}{3} & 7--15\\
              & Dev.5 &  & 13--0\\
              & Dev.6 &  & 13--0\\\hline
        \end{tabular}
    \end{subtable}\hfill
    \begin{subtable}[t]{0.24\textwidth}
        \centering
        \caption{Metrics.}
        \begin{tabular}{|l|c|c|c|}
            \hline
            Method & PS & DS & W\\\hline
            HG & 140 & 12.5 & 607.0\\\hline
            \textbf{ELS} & \textbf{140} & \textbf{12.5} & \textbf{607.0}\\\hline
            SPW & 1145 & 15.5 & 1454.4\\\hline
        \end{tabular}
    \end{subtable}
\end{table}
\begin{figure}[H]
    \centering
    \begin{subfigure}{\appfigscale\textwidth}
        \includegraphics[width=\linewidth]{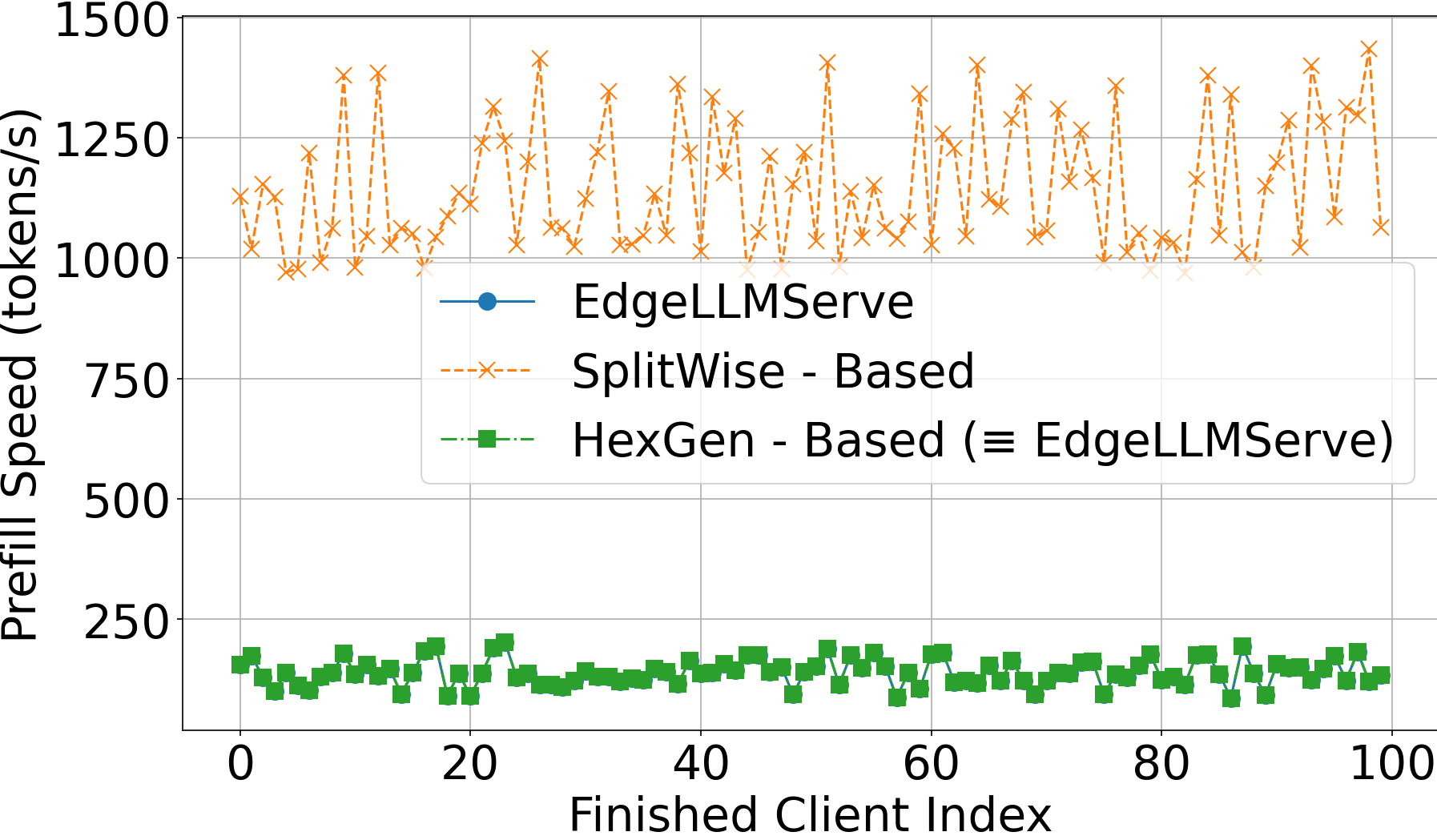}
        \caption{Prefill speed}
    \end{subfigure}\hfill
    \begin{subfigure}{\appfigscale\textwidth}
        \includegraphics[width=\linewidth]{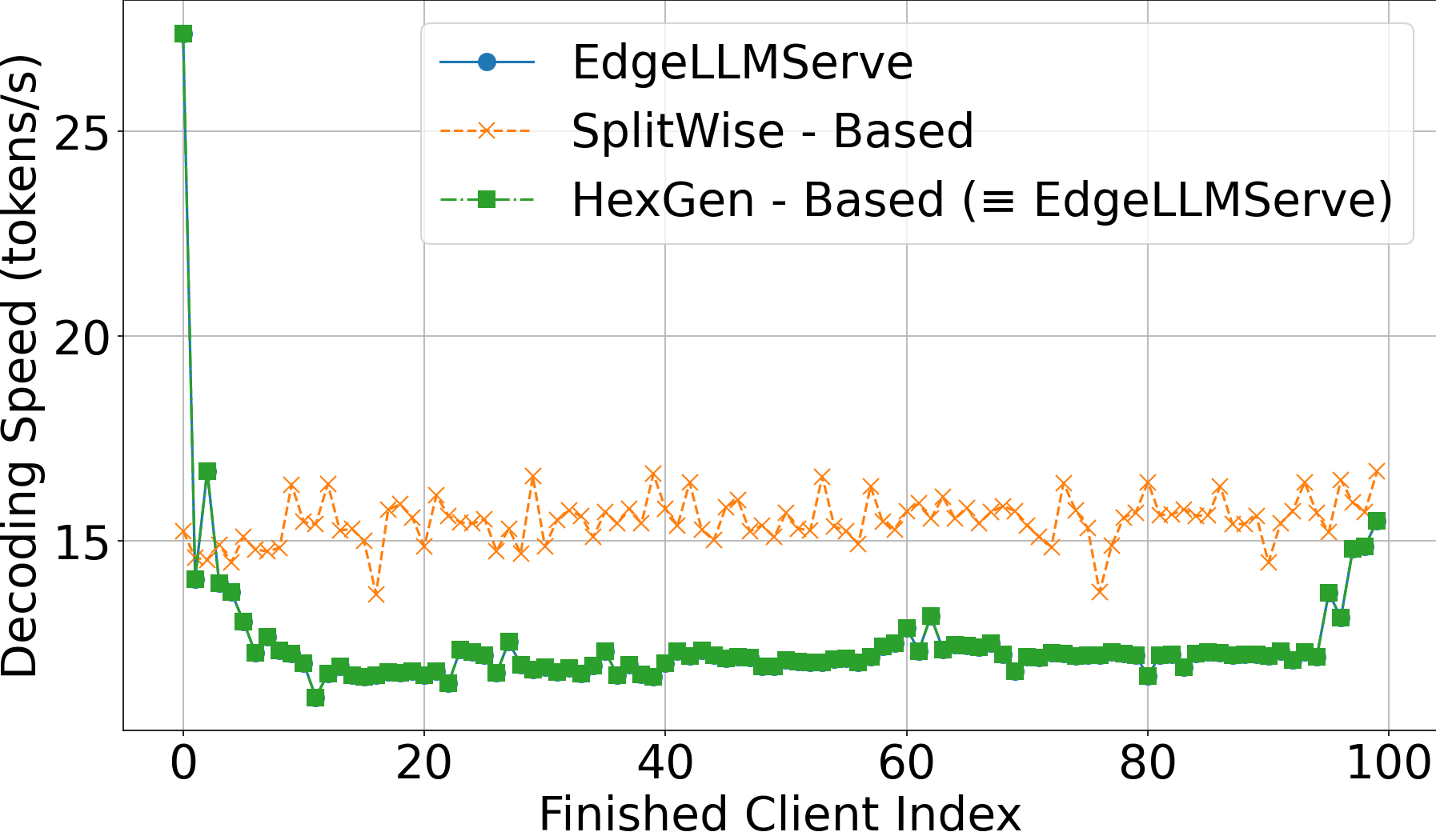}
        \caption{Decoding speed}
    \end{subfigure}\hfill
    \begin{subfigure}{\appfigscale\textwidth}
        \includegraphics[width=\linewidth]{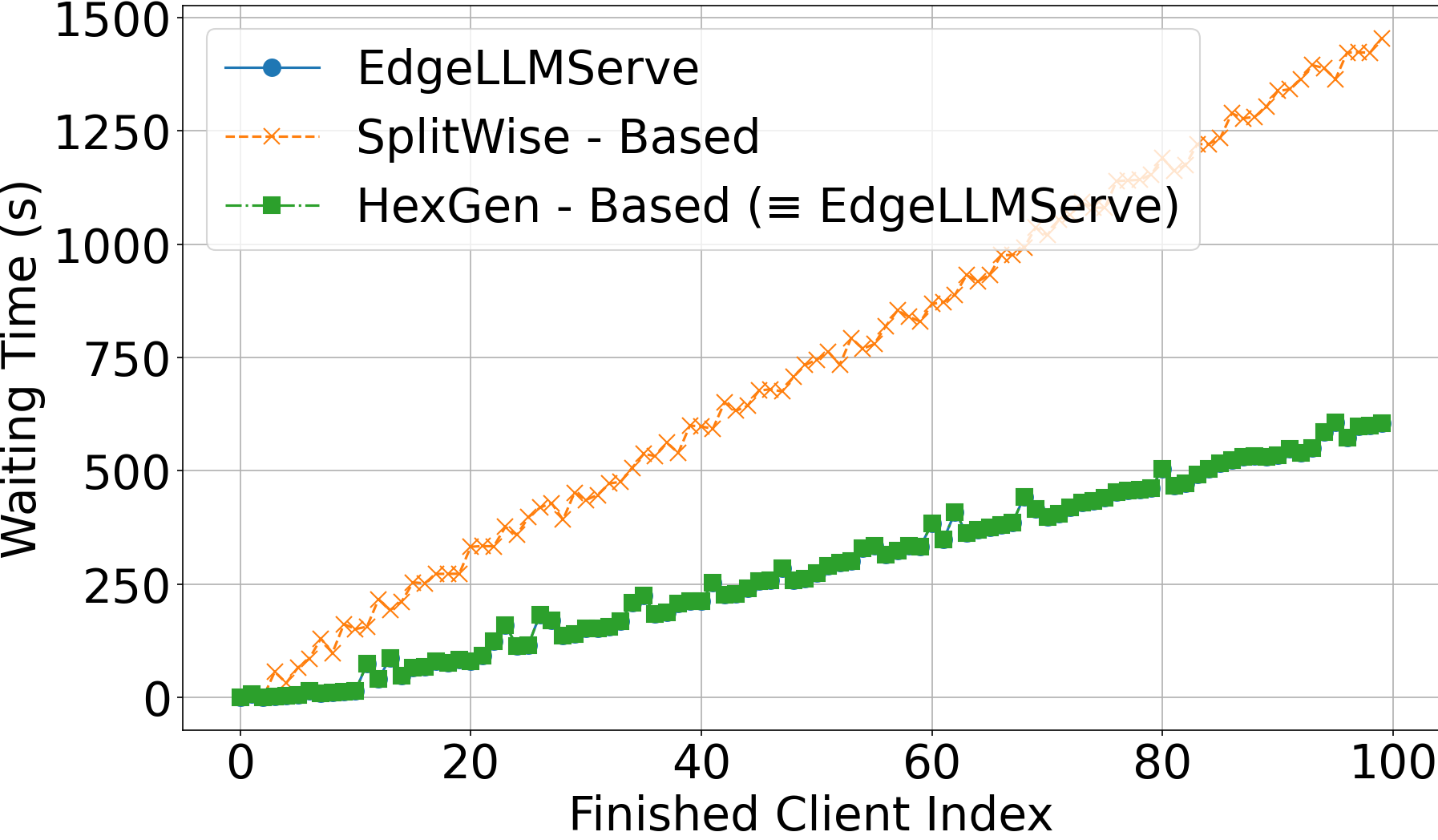}
        \caption{Waiting time}
    \end{subfigure}
    \caption{Per-request metrics --- Four devices, Qwen3-MoE-30B, LZByteDance, arrival $1000$~ms.}
    \label{apd:f11}
\end{figure}

\paragraph{6. LZByteDance dataset. Arrival rate of 3000ms}\mbox{}
\begin{table}[H]
    \centering
    \setlength{\tabcolsep}{3pt}
    \caption{Deployment plans and metrics --- Four devices, Qwen3-MoE-30B, LZByteDance, arrival $3000$~ms.}
    \label{apd:t12}
    \begin{subtable}[t]{0.24\textwidth}
        \centering
        \caption{HG}
        \begin{tabular}{|c|c|c|c|}
            \hline
            Mode & Dev & Slots & G--C\\\hline
            \multirow{2}{*}{P} & Dev.5$^{\star}$ & \multirow{2}{*}{1} & 22--0\\
              & Dev.6 &  & 26--0\\\hline
            \multirow{2}{*}{D} & Dev.1$^{\star}$ & \multirow{2}{*}{10} & 13--19\\
              & Dev.2 &  & 5--11\\\hline
        \end{tabular}
    \end{subtable}\hfill
    \begin{subtable}[t]{0.24\textwidth}
        \centering
        \caption{ELS}
        \begin{tabular}{|c|c|c|c|}
            \hline
            Mode & Dev & Slots & G--C\\\hline
            \multirow{2}{*}{P} & Dev.5$^{\star}$ & \multirow{2}{*}{1} & 22--0\\
              & Dev.6 &  & 26--0\\\hline
            \multirow{2}{*}{D} & Dev.1$^{\star}$ & \multirow{2}{*}{10} & 13--19\\
              & Dev.2 &  & 5--11\\\hline
        \end{tabular}
    \end{subtable}\hfill
    \begin{subtable}[t]{0.24\textwidth}
        \centering
        \caption{SPW}
        \begin{tabular}{|c|c|c|c|}
            \hline
            Mode & Dev & Slots & G--C\\\hline
            P& Dev.1$^{\star}$ & 1 & 26--22\\\hline
            \multirow{3}{*}{D} & Dev.2$^{\star}$ & \multirow{3}{*}{3} & 7--15\\
              & Dev.5 &  & 13--0\\
              & Dev.6 &  & 13--0\\\hline
        \end{tabular}
    \end{subtable}\hfill
    \begin{subtable}[t]{0.24\textwidth}
        \centering
        \caption{Metrics.}
        \begin{tabular}{|l|c|c|c|}
            \hline
            Method & PS & DS & W\\\hline
            HG & 145 & 12.5 & 431.3\\\hline
            \textbf{ELS} & \textbf{145} & \textbf{12.5} & \textbf{431.3}\\\hline
            SPW & 1145 & 15.6 & 1323.0\\\hline
        \end{tabular}
    \end{subtable}
\end{table}
\begin{figure}[H]
    \centering
    \begin{subfigure}{\appfigscale\textwidth}
        \includegraphics[width=\linewidth]{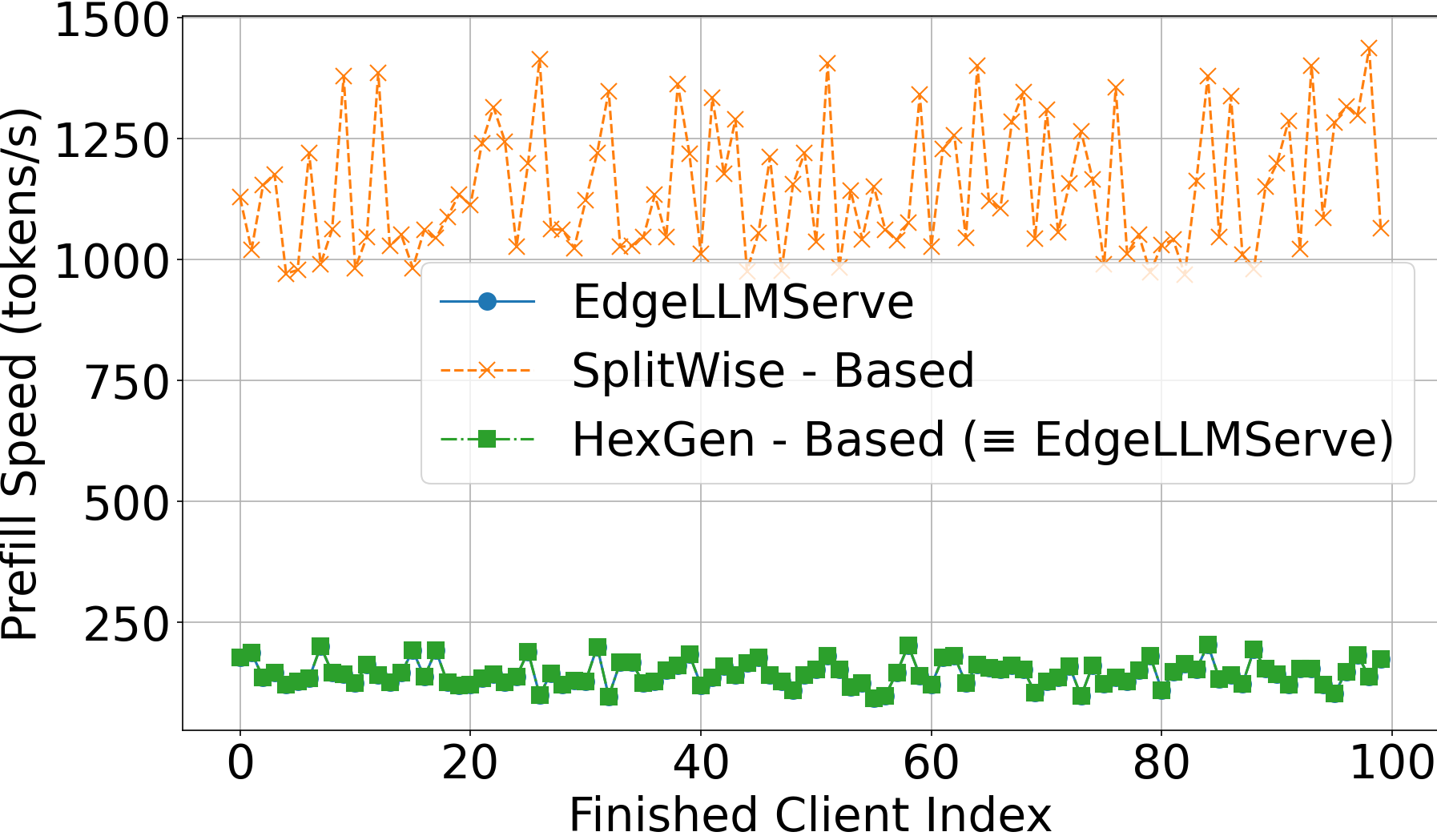}
        \caption{Prefill speed}
    \end{subfigure}\hfill
    \begin{subfigure}{\appfigscale\textwidth}
        \includegraphics[width=\linewidth]{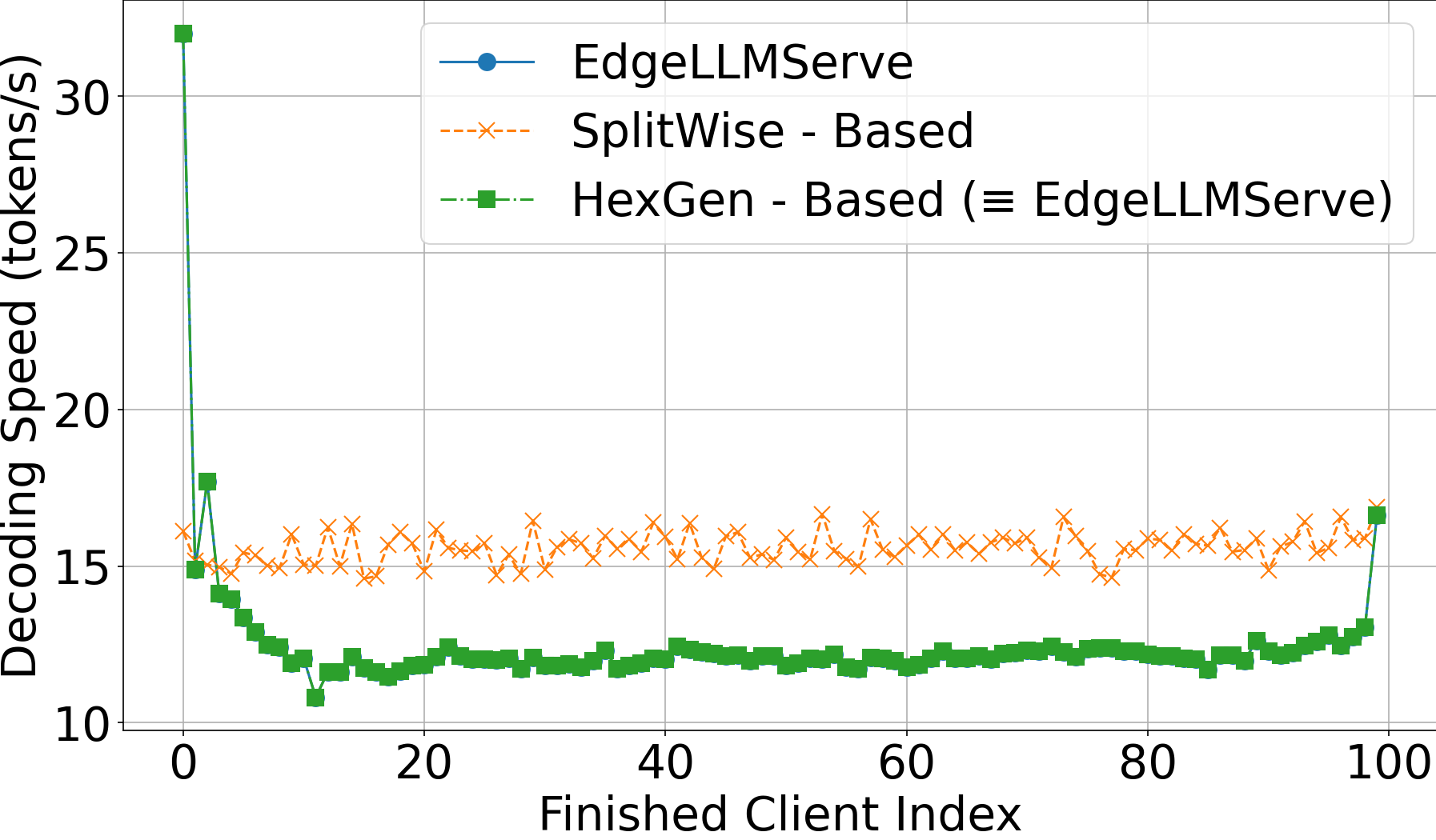}
        \caption{Decoding speed}
    \end{subfigure}\hfill
    \begin{subfigure}{\appfigscale\textwidth}
        \includegraphics[width=\linewidth]{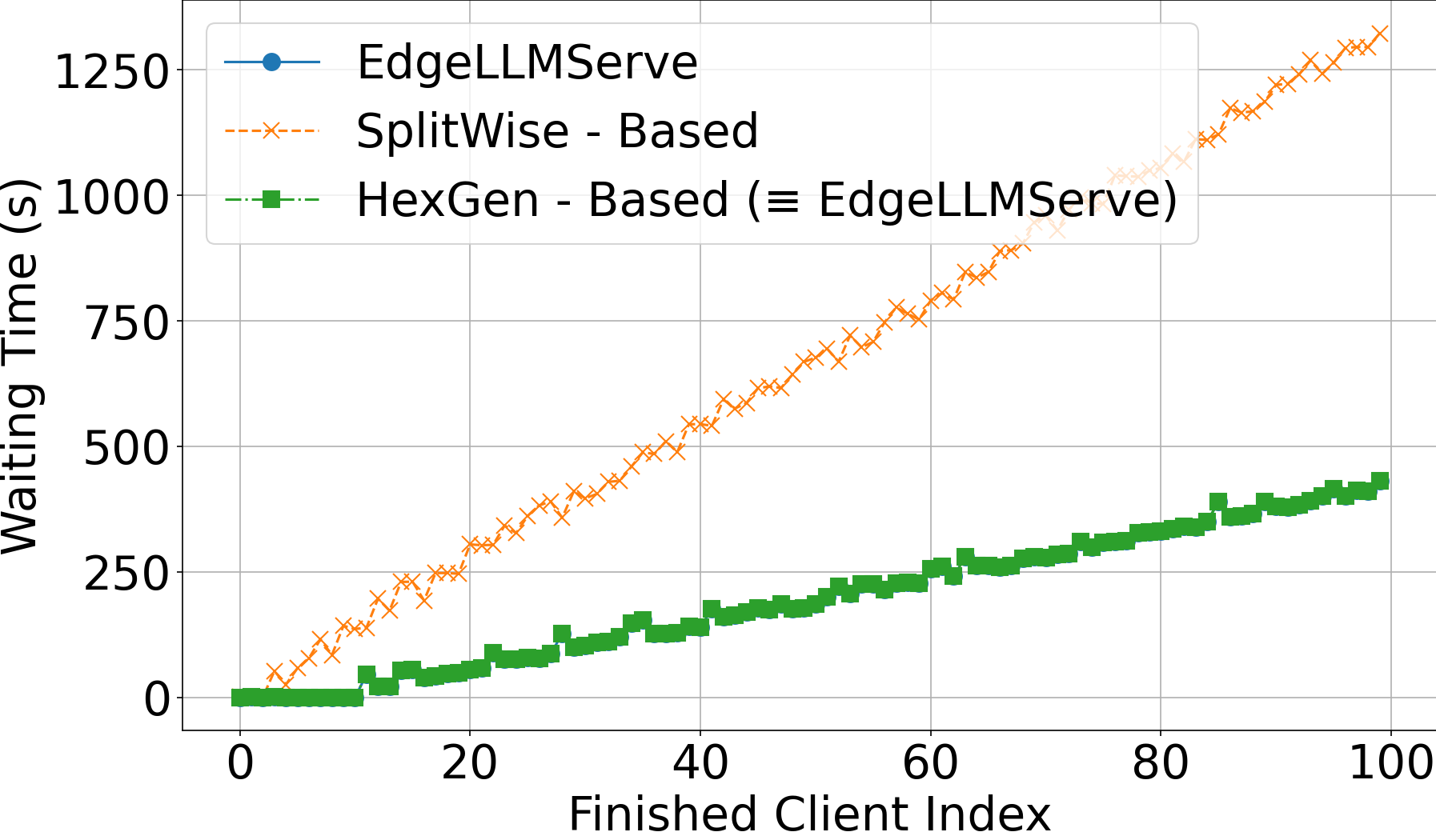}
        \caption{Waiting time}
    \end{subfigure}
    \caption{Per-request metrics --- Four devices, Qwen3-MoE-30B, LZByteDance, arrival $3000$~ms.}
    \label{apd:f12}
\end{figure}

\clearpage
\subsection{Six devices}
\subsubsection{Gpt-Oss-20b}
\paragraph{1. IELTS dataset. Arrival rate of 100ms}\mbox{}
\begin{table}[H]
    \centering
    \setlength{\tabcolsep}{3pt}
    \caption{Deployment plans and metrics --- Six devices, Gpt-Oss-20b, IELTS, arrival $100$~ms.}
    \vspace{-0.3cm}
    \label{apd:t13}
    \begin{subtable}[t]{0.24\textwidth}
        \centering
        \caption{HG}
        \begin{tabular}{|c|c|c|c|}
            \hline
            Mode & Dev & Slots & G--C\\\hline
            P& Dev.1$^{\star}$ & 1 & 20--4\\\hline
            \multirow{2}{*}{D} & Dev.2$^{\star}$ & \multirow{2}{*}{14} & 5--1\\
              & Dev.7 &  & 18--0\\\hline
            D& Dev.4$^{\star}$ & 6 & 24--0\\\hline
            \multirow{2}{*}{D} & Dev.5 & \multirow{2}{*}{2} & 15--0\\
              & Dev.6$^{\star}$ &  & 9--0\\\hline
        \end{tabular}
    \end{subtable}\hfill
    \begin{subtable}[t]{0.24\textwidth}
        \centering
        \caption{ELS}
        \begin{tabular}{|c|c|c|c|}
            \hline
            Mode & Dev & Slots & G--C\\\hline
            P& Dev.1$^{\star}$ & 1 & 20--4\\\hline
            \multirow{2}{*}{D} & Dev.2$^{\star}$ & \multirow{2}{*}{14} & 5--1\\
              & Dev.7 &  & 18--0\\\hline
            D& Dev.4$^{\star}$ & 6 & 24--0\\\hline
            \multirow{2}{*}{D} & Dev.5 & \multirow{2}{*}{2} & 15--0\\
              & Dev.6$^{\star}$ &  & 9--0\\\hline
        \end{tabular}
    \end{subtable}\hfill
    \begin{subtable}[t]{0.24\textwidth}
        \centering
        \caption{SPW}
        \begin{tabular}{|c|c|c|c|}
            \hline
            Mode & Dev & Slots & G--C\\\hline
            P& Dev.1$^{\star}$ & 1 & 20--4\\\hline
            \multirow{2}{*}{D} & Dev.2$^{\star}$ & \multirow{2}{*}{14} & 5--1\\
              & Dev.7 &  & 18--0\\\hline
            D& Dev.4$^{\star}$ & 6 & 24--0\\\hline
            \multirow{2}{*}{D} & Dev.5 & \multirow{2}{*}{2} & 15--0\\
              & Dev.6$^{\star}$ &  & 9--0\\\hline
        \end{tabular}
    \end{subtable}\hfill
    \begin{subtable}[t]{0.24\textwidth}
        \centering
        \caption{Metrics.}
        \begin{tabular}{|l|c|c|c|}
            \hline
            Method & PS & DS & W\\\hline
            HG & 2917 & 12.8 & 302.8\\\hline
            \textbf{ELS} & \textbf{2917} & \textbf{12.8} & \textbf{302.8}\\\hline
            SPW & 2917 & 12.8 & 302.8\\\hline
        \end{tabular}
    \end{subtable}
\end{table}
\begin{figure}[H]
    \centering
    \begin{subfigure}{\appfigscale\textwidth}
        \includegraphics[width=\linewidth]{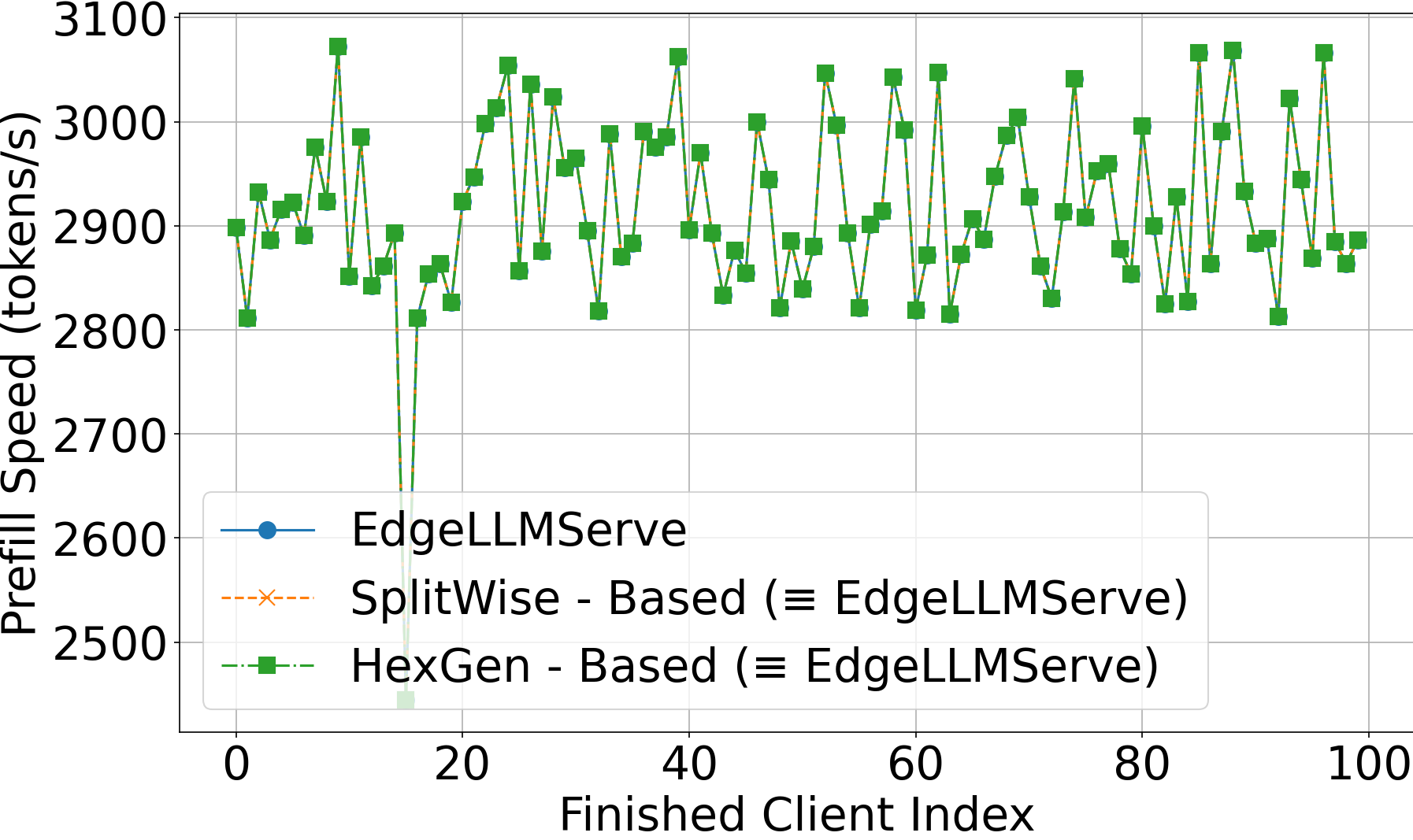}
        \caption{Prefill speed}
    \end{subfigure}\hfill
    \begin{subfigure}{\appfigscale\textwidth}
        \includegraphics[width=\linewidth]{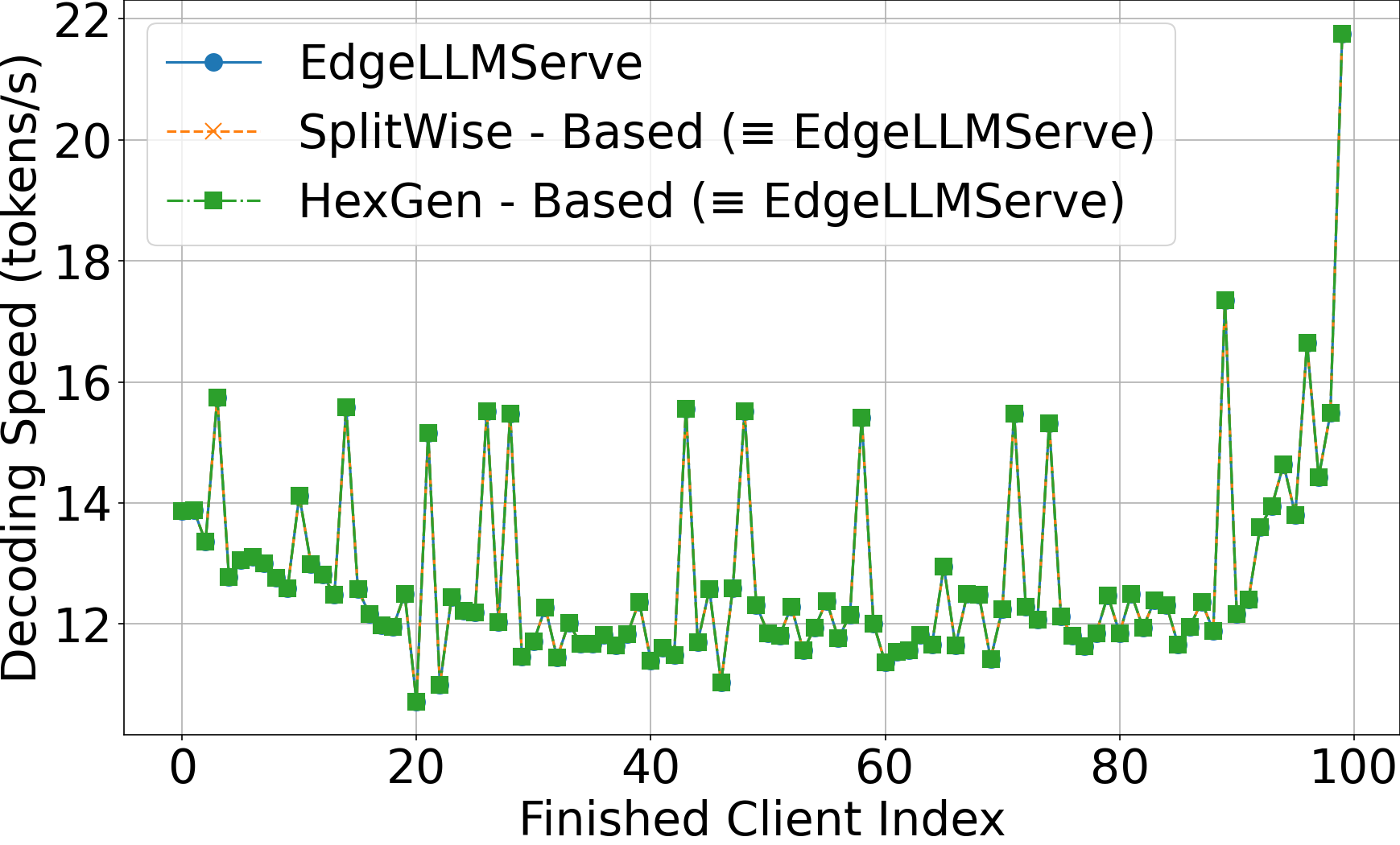}
        \caption{Decoding speed}
    \end{subfigure}\hfill
    \begin{subfigure}{\appfigscale\textwidth}
        \includegraphics[width=\linewidth]{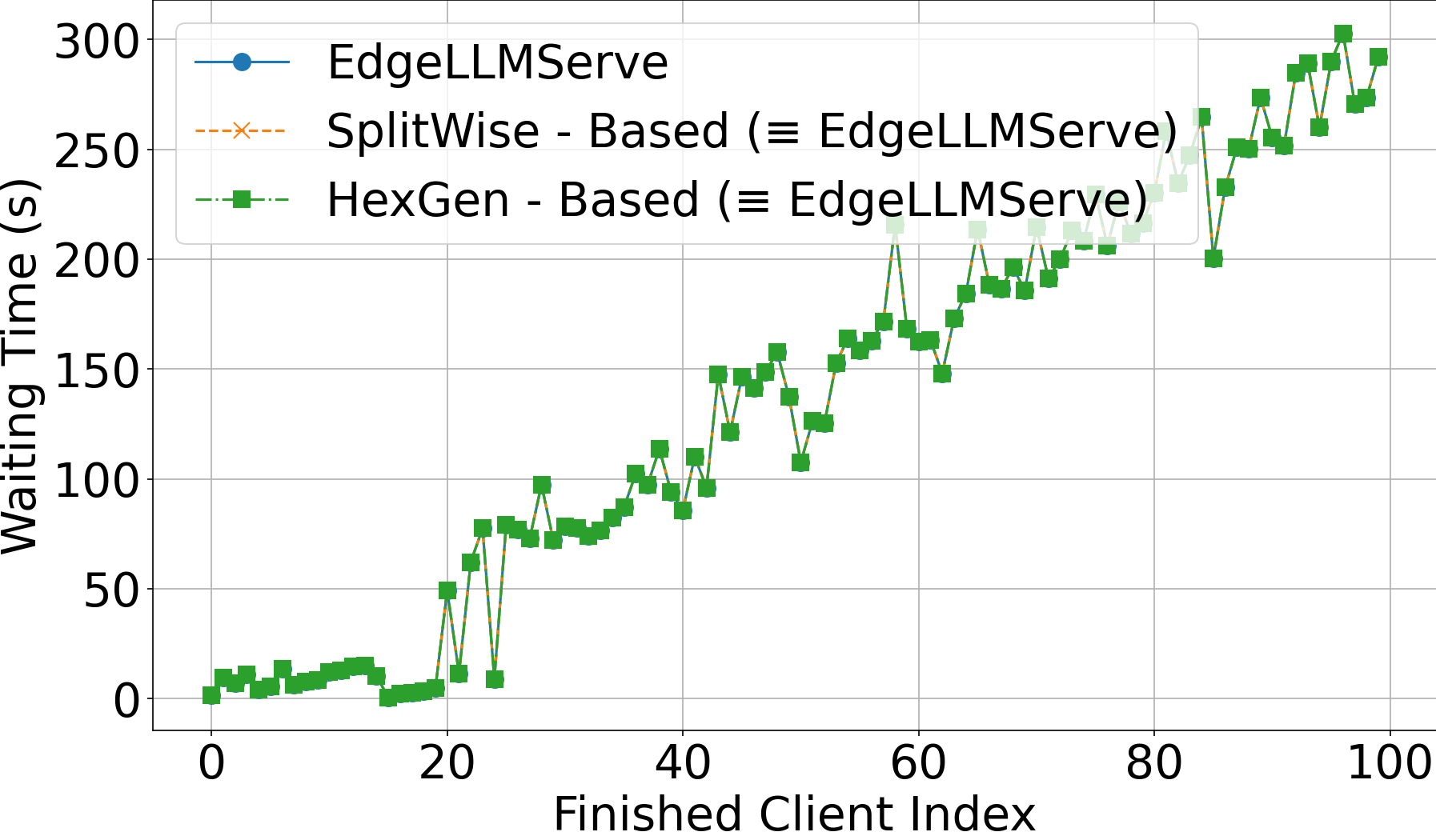}
        \caption{Waiting time}
    \end{subfigure}
    \vspace{-0.3cm}
    \caption{Per-request metrics --- Six devices, Gpt-Oss-20b, IELTS, arrival $100$~ms.}
    \label{apd:f13}
\end{figure}

\paragraph{2. IELTS dataset. Arrival rate of 1000ms}\mbox{}
\begin{table}[H]
    \centering
    \setlength{\tabcolsep}{3pt}
    \caption{Deployment plans and metrics --- Six devices, Gpt-Oss-20b, IELTS, arrival $1000$~ms.}
    \vspace{-0.3cm}
    \label{apd:t14}
    \begin{subtable}[t]{0.24\textwidth}
        \centering
        \caption{HG}
        \begin{tabular}{|c|c|c|c|}
            \hline
            Mode & Dev & Slots & G--C\\\hline
            P& Dev.1$^{\star}$ & 1 & 20--4\\\hline
            \multirow{2}{*}{D} & Dev.2$^{\star}$ & \multirow{2}{*}{14} & 5--1\\
              & Dev.7 &  & 18--0\\\hline
            D& Dev.4$^{\star}$ & 6 & 24--0\\\hline
            \multirow{2}{*}{D} & Dev.5 & \multirow{2}{*}{2} & 15--0\\
              & Dev.6$^{\star}$ &  & 9--0\\\hline
        \end{tabular}
    \end{subtable}\hfill
    \begin{subtable}[t]{0.24\textwidth}
        \centering
        \caption{ELS}
        \begin{tabular}{|c|c|c|c|}
            \hline
            Mode & Dev & Slots & G--C\\\hline
            P& Dev.1$^{\star}$ & 1 & 20--4\\\hline
            \multirow{2}{*}{D} & Dev.2$^{\star}$ & \multirow{2}{*}{14} & 5--1\\
              & Dev.7 &  & 18--0\\\hline
            D& Dev.4$^{\star}$ & 6 & 24--0\\\hline
            \multirow{2}{*}{D} & Dev.5 & \multirow{2}{*}{2} & 15--0\\
              & Dev.6$^{\star}$ &  & 9--0\\\hline
        \end{tabular}
    \end{subtable}\hfill
    \begin{subtable}[t]{0.24\textwidth}
        \centering
        \caption{SPW}
        \begin{tabular}{|c|c|c|c|}
            \hline
            Mode & Dev & Slots & G--C\\\hline
            P& Dev.1$^{\star}$ & 1 & 20--4\\\hline
            \multirow{2}{*}{D} & Dev.2$^{\star}$ & \multirow{2}{*}{14} & 5--1\\
              & Dev.7 &  & 18--0\\\hline
            D& Dev.4$^{\star}$ & 6 & 24--0\\\hline
            \multirow{2}{*}{D} & Dev.5 & \multirow{2}{*}{2} & 15--0\\
              & Dev.6$^{\star}$ &  & 9--0\\\hline
        \end{tabular}
    \end{subtable}\hfill
    \begin{subtable}[t]{0.24\textwidth}
        \centering
        \caption{Metrics.}
        \begin{tabular}{|l|c|c|c|}
            \hline
            Method & PS & DS & W\\\hline
            HG & 2919 & 12.7 & 318.0\\\hline
            \textbf{ELS} & \textbf{2919} & \textbf{12.7} & \textbf{318.0}\\\hline
            SPW & 2919 & 12.7 & 318.0\\\hline
        \end{tabular}
    \end{subtable}
\end{table}
\begin{figure}[H]
    \centering
    \begin{subfigure}{\appfigscale\textwidth}
        \includegraphics[width=\linewidth]{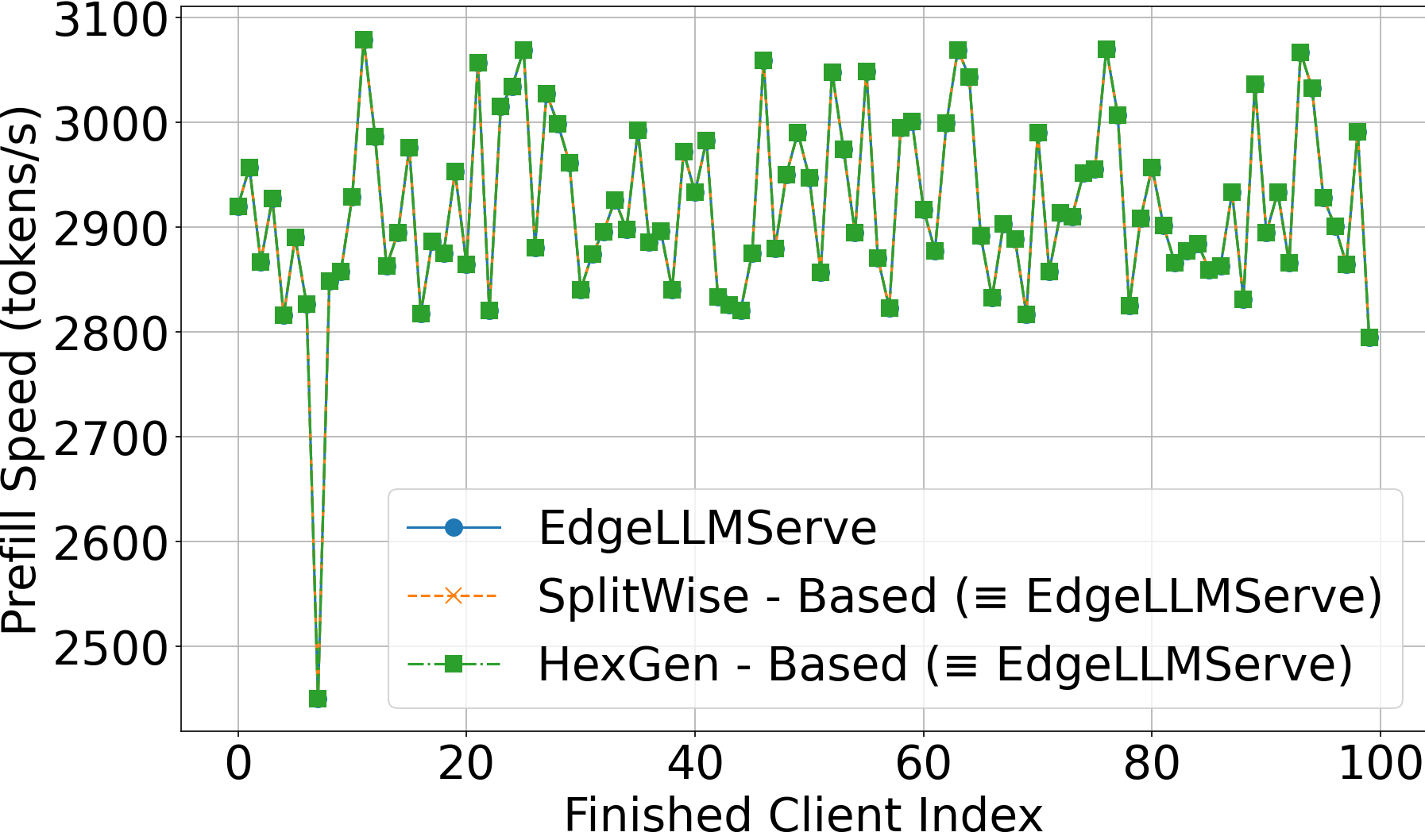}
        \caption{Prefill speed}
    \end{subfigure}\hfill
    \begin{subfigure}{\appfigscale\textwidth}
        \includegraphics[width=\linewidth]{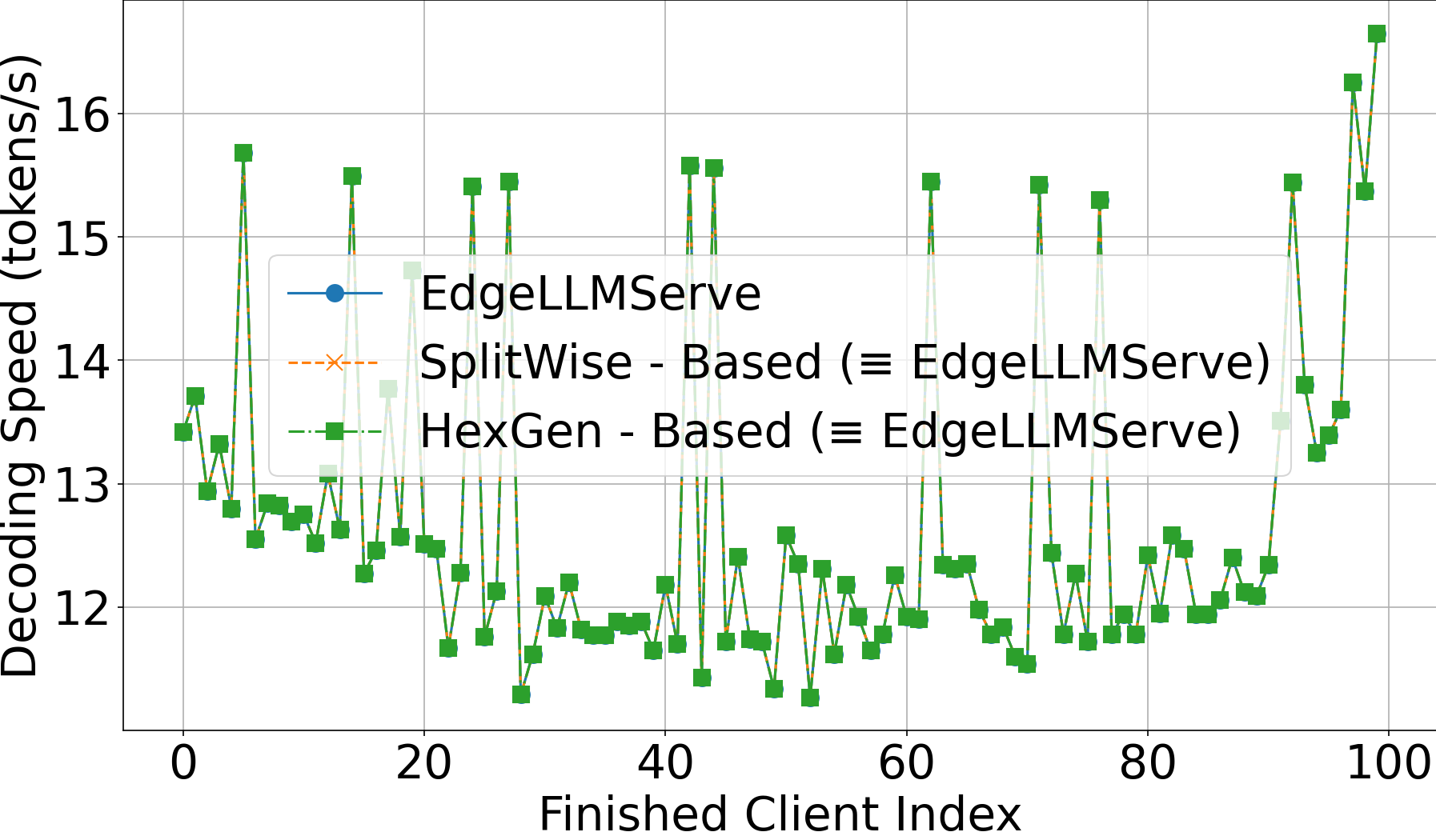}
        \caption{Decoding speed}
    \end{subfigure}\hfill
    \begin{subfigure}{\appfigscale\textwidth}
        \includegraphics[width=\linewidth]{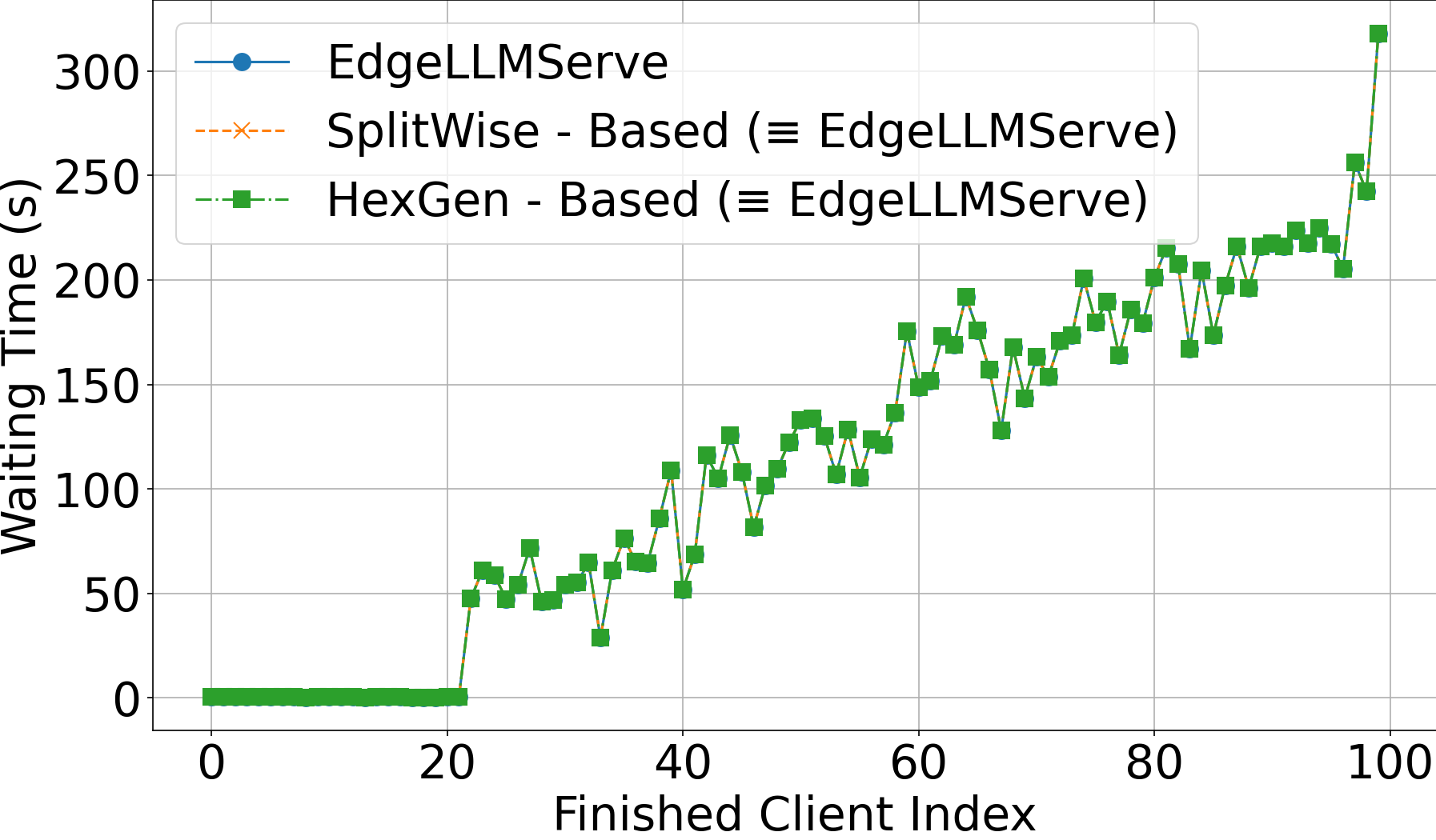}
        \caption{Waiting time}
    \end{subfigure}
    \vspace{-0.3cm}
    \caption{Per-request metrics --- Six devices, Gpt-Oss-20b, IELTS, arrival $1000$~ms.}
    \label{apd:f14}
\end{figure}

\paragraph{3. IELTS dataset. Arrival rate of 3000ms}\mbox{}
\begin{table}[H]
    \centering
    \setlength{\tabcolsep}{3pt}
    \caption{Deployment plans and metrics --- Six devices, Gpt-Oss-20b, IELTS, arrival $3000$~ms.}
    \label{apd:t15}
    \begin{subtable}[t]{0.24\textwidth}
        \centering
        \caption{HG}
        \begin{tabular}{|c|c|c|c|}
            \hline
            Mode & Dev & Slots & G--C\\\hline
            P& Dev.1$^{\star}$ & 1 & 20--4\\\hline
            \multirow{2}{*}{D} & Dev.2$^{\star}$ & \multirow{2}{*}{14} & 5--1\\
              & Dev.7 &  & 18--0\\\hline
            D& Dev.4$^{\star}$ & 6 & 24--0\\\hline
            \multirow{2}{*}{D} & Dev.5 & \multirow{2}{*}{2} & 15--0\\
              & Dev.6$^{\star}$ &  & 9--0\\\hline
        \end{tabular}
    \end{subtable}\hfill
    \begin{subtable}[t]{0.24\textwidth}
        \centering
        \caption{ELS}
        \begin{tabular}{|c|c|c|c|}
            \hline
            Mode & Dev & Slots & G--C\\\hline
            P& Dev.4$^{\star}$ & 1 & 24--0\\\hline
            \multirow{2}{*}{D} & Dev.1 & \multirow{2}{*}{32} & 14--0\\
              & Dev.7$^{\star}$ &  & 10--0\\\hline
            \multirow{3}{*}{D} & Dev.2$^{\star}$ & \multirow{3}{*}{6} & 4--6\\
              & Dev.5 &  & 7--0\\
              & Dev.6 &  & 7--0\\\hline
        \end{tabular}
    \end{subtable}\hfill
    \begin{subtable}[t]{0.24\textwidth}
        \centering
        \caption{SPW}
        \begin{tabular}{|c|c|c|c|}
            \hline
            Mode & Dev & Slots & G--C\\\hline
            P& Dev.1$^{\star}$ & 1 & 20--4\\\hline
            \multirow{2}{*}{D} & Dev.2$^{\star}$ & \multirow{2}{*}{14} & 5--1\\
              & Dev.7 &  & 18--0\\\hline
            D& Dev.4$^{\star}$ & 6 & 24--0\\\hline
            \multirow{2}{*}{D} & Dev.5 & \multirow{2}{*}{2} & 15--0\\
              & Dev.6$^{\star}$ &  & 9--0\\\hline
        \end{tabular}
    \end{subtable}\hfill
    \begin{subtable}[t]{0.24\textwidth}
        \centering
        \caption{Metrics.}
        \begin{tabular}{|l|c|c|c|}
            \hline
            Method & PS & DS & W\\\hline
            HG & 2926 & 12.7 & 209.1\\\hline
            \textbf{ELS} & \textbf{709} & \textbf{14.5} & \textbf{59.0}\\\hline
            SPW & 2926 & 12.7 & 209.1\\\hline
        \end{tabular}
    \end{subtable}
\end{table}
\begin{figure}[H]
    \centering
    \begin{subfigure}{\appfigscale\textwidth}
        \includegraphics[width=\linewidth]{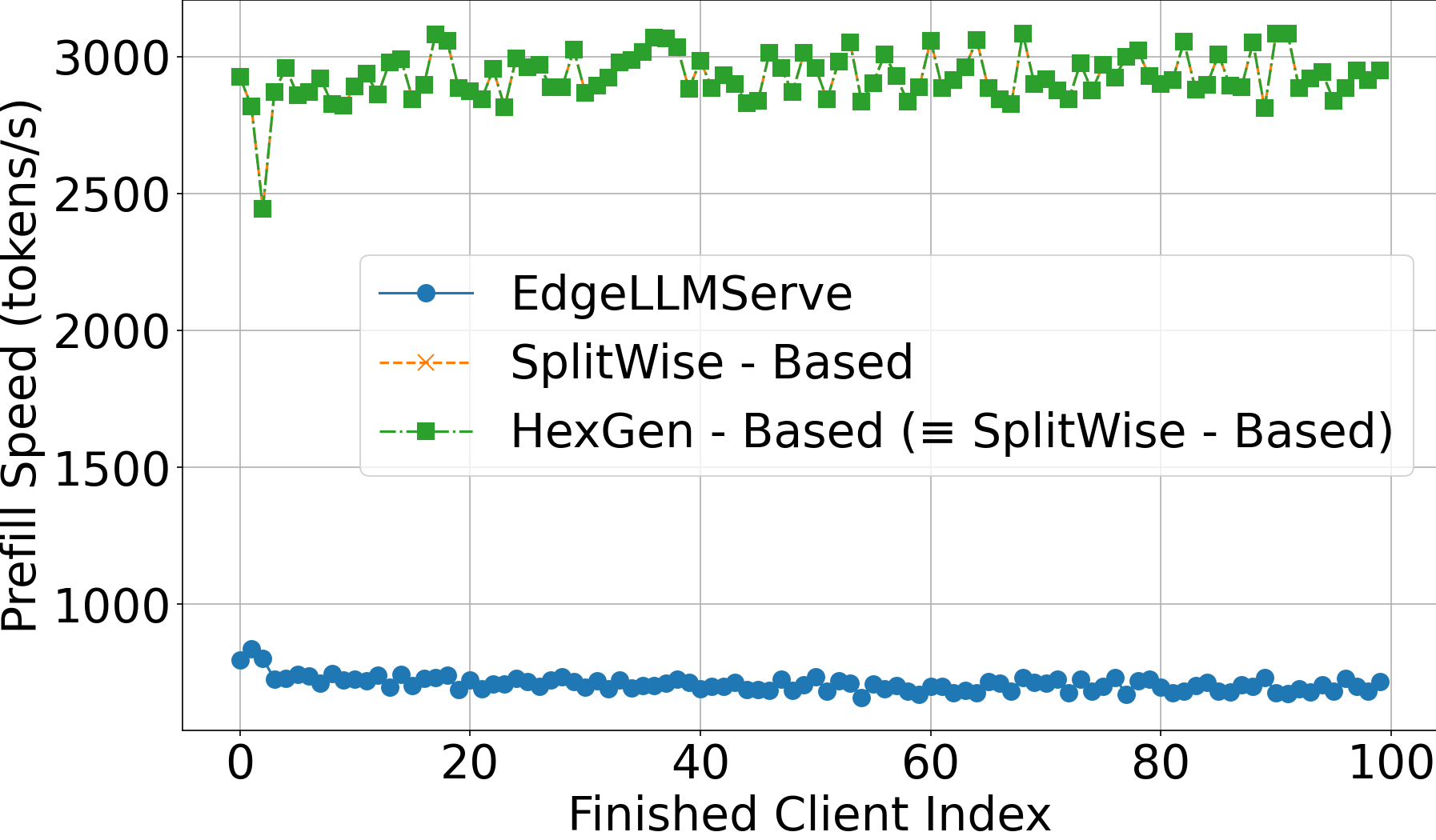}
        \caption{Prefill speed}
    \end{subfigure}\hfill
    \begin{subfigure}{\appfigscale\textwidth}
        \includegraphics[width=\linewidth]{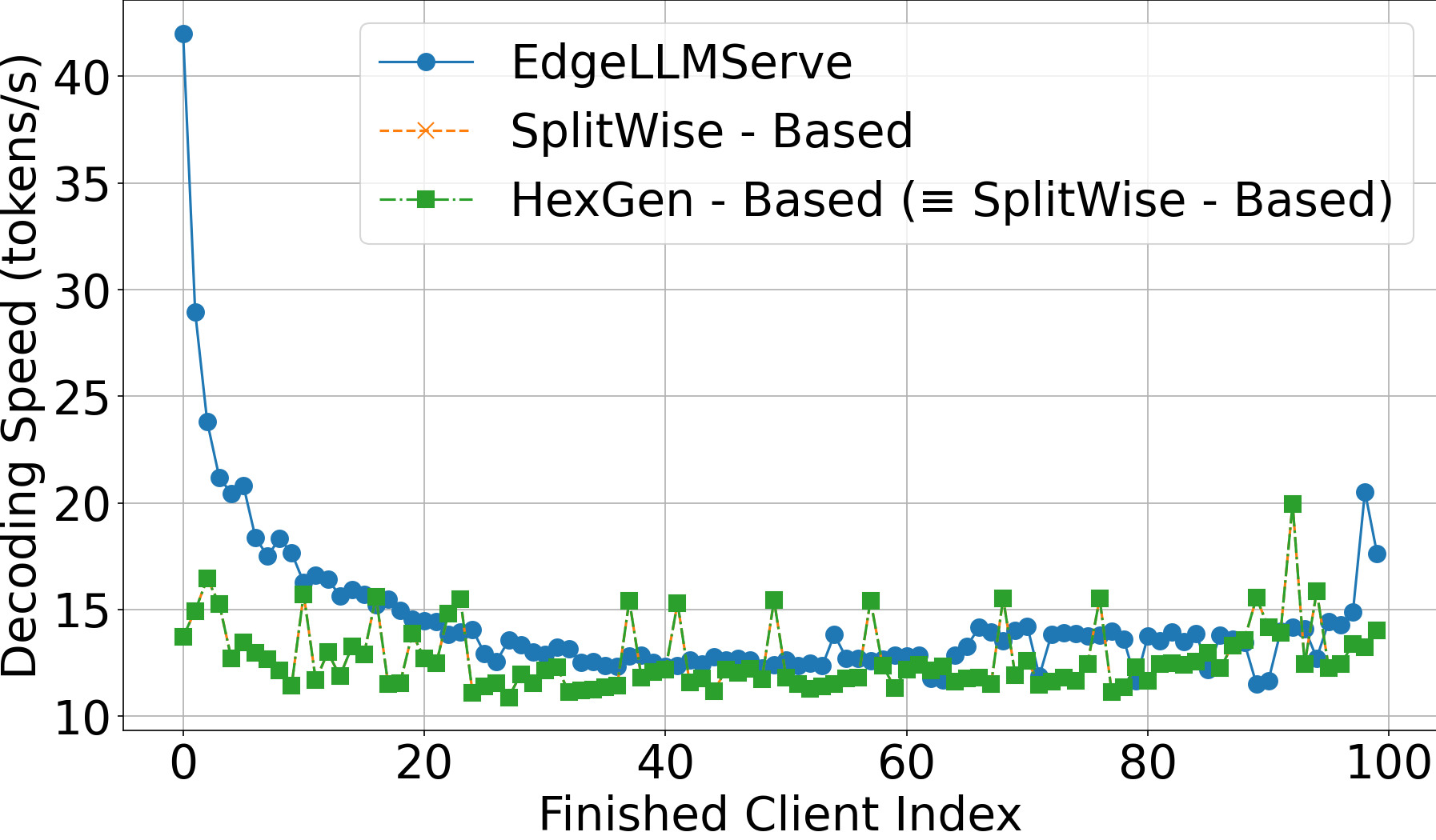}
        \caption{Decoding speed}
    \end{subfigure}\hfill
    \begin{subfigure}{\appfigscale\textwidth}
        \includegraphics[width=\linewidth]{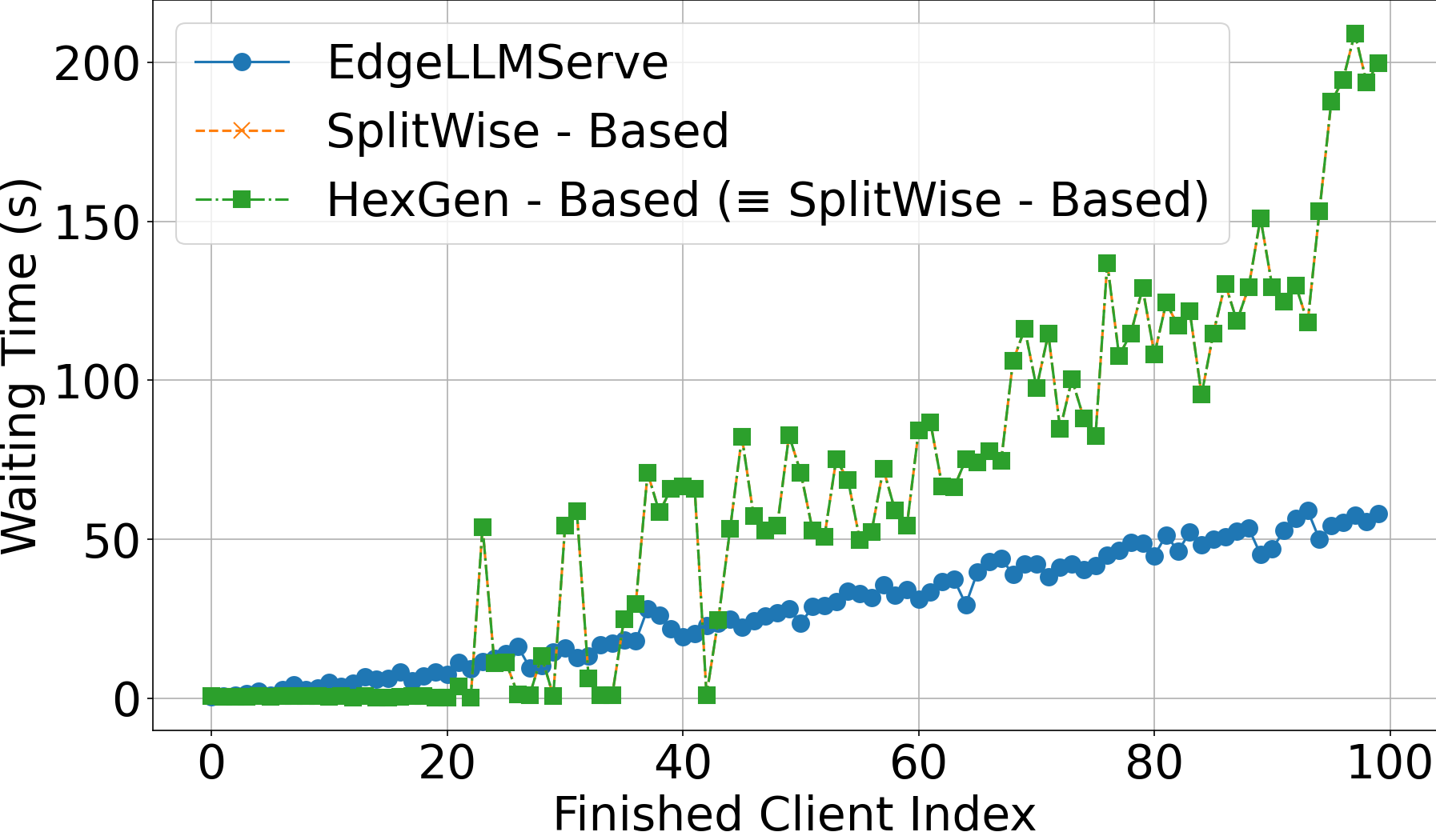}
        \caption{Waiting time}
    \end{subfigure}
    \caption{Per-request metrics --- Six devices, Gpt-Oss-20b, IELTS, arrival $3000$~ms.}
    \label{apd:f15}
\end{figure}

\paragraph{4. LZByteDance dataset. Arrival rate of 100ms}\mbox{}
\begin{table}[H]
    \centering
    \setlength{\tabcolsep}{3pt}
    \caption{Deployment plans and metrics --- Six devices, Gpt-Oss-20b, LZByteDance, arrival $100$~ms.}
    \label{apd:t16}
    \begin{subtable}[t]{0.24\textwidth}
        \centering
        \caption{HG}
        \begin{tabular}{|c|c|c|c|}
            \hline
            Mode & Dev & Slots & G--C\\\hline
            P& Dev.4$^{\star}$ & 1 & 24--0\\\hline
            \multirow{2}{*}{D} & Dev.1 & \multirow{2}{*}{32} & 14--0\\
              & Dev.7$^{\star}$ &  & 10--0\\\hline
            \multirow{3}{*}{D} & Dev.2$^{\star}$ & \multirow{3}{*}{6} & 4--6\\
              & Dev.5 &  & 7--0\\
              & Dev.6 &  & 7--0\\\hline
        \end{tabular}
    \end{subtable}\hfill
    \begin{subtable}[t]{0.24\textwidth}
        \centering
        \caption{ELS}
        \begin{tabular}{|c|c|c|c|}
            \hline
            Mode & Dev & Slots & G--C\\\hline
            P& Dev.4$^{\star}$ & 1 & 24--0\\\hline
            \multirow{2}{*}{D} & Dev.1 & \multirow{2}{*}{32} & 14--0\\
              & Dev.7$^{\star}$ &  & 10--0\\\hline
            \multirow{3}{*}{D} & Dev.2$^{\star}$ & \multirow{3}{*}{6} & 4--6\\
              & Dev.5 &  & 7--0\\
              & Dev.6 &  & 7--0\\\hline
        \end{tabular}
    \end{subtable}\hfill
    \begin{subtable}[t]{0.24\textwidth}
        \centering
        \caption{SPW}
        \begin{tabular}{|c|c|c|c|}
            \hline
            Mode & Dev & Slots & G--C\\\hline
            P& Dev.1$^{\star}$ & 1 & 20--4\\\hline
            \multirow{2}{*}{D} & Dev.2$^{\star}$ & \multirow{2}{*}{14} & 5--1\\
              & Dev.7 &  & 18--0\\\hline
            D& Dev.4$^{\star}$ & 6 & 24--0\\\hline
            \multirow{2}{*}{D} & Dev.5 & \multirow{2}{*}{2} & 15--0\\
              & Dev.6$^{\star}$ &  & 9--0\\\hline
        \end{tabular}
    \end{subtable}\hfill
    \begin{subtable}[t]{0.24\textwidth}
        \centering
        \caption{Metrics.}
        \begin{tabular}{|l|c|c|c|}
            \hline
            Method & PS & DS & W\\\hline
            HG & 737 & 17.4 & 90.0\\\hline
            \textbf{ELS} & \textbf{737} & \textbf{17.4} & \textbf{90.0}\\\hline
            SPW & 2466 & 13.6 & 114.2\\\hline
        \end{tabular}
    \end{subtable}
\end{table}
\begin{figure}[H]
    \centering
    \begin{subfigure}{\appfigscale\textwidth}
        \includegraphics[width=\linewidth]{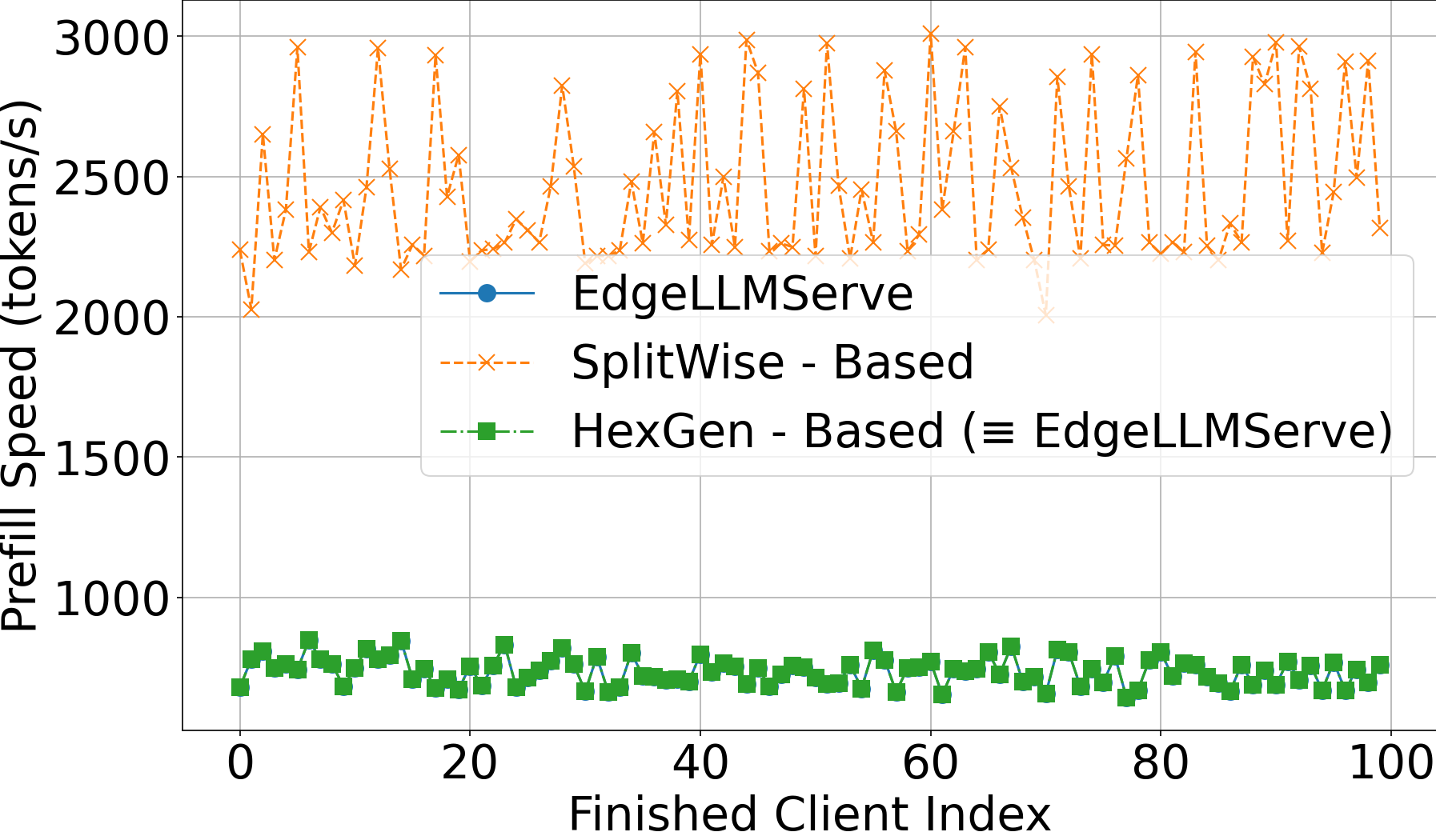}
        \caption{Prefill speed}
    \end{subfigure}\hfill
    \begin{subfigure}{\appfigscale\textwidth}
        \includegraphics[width=\linewidth]{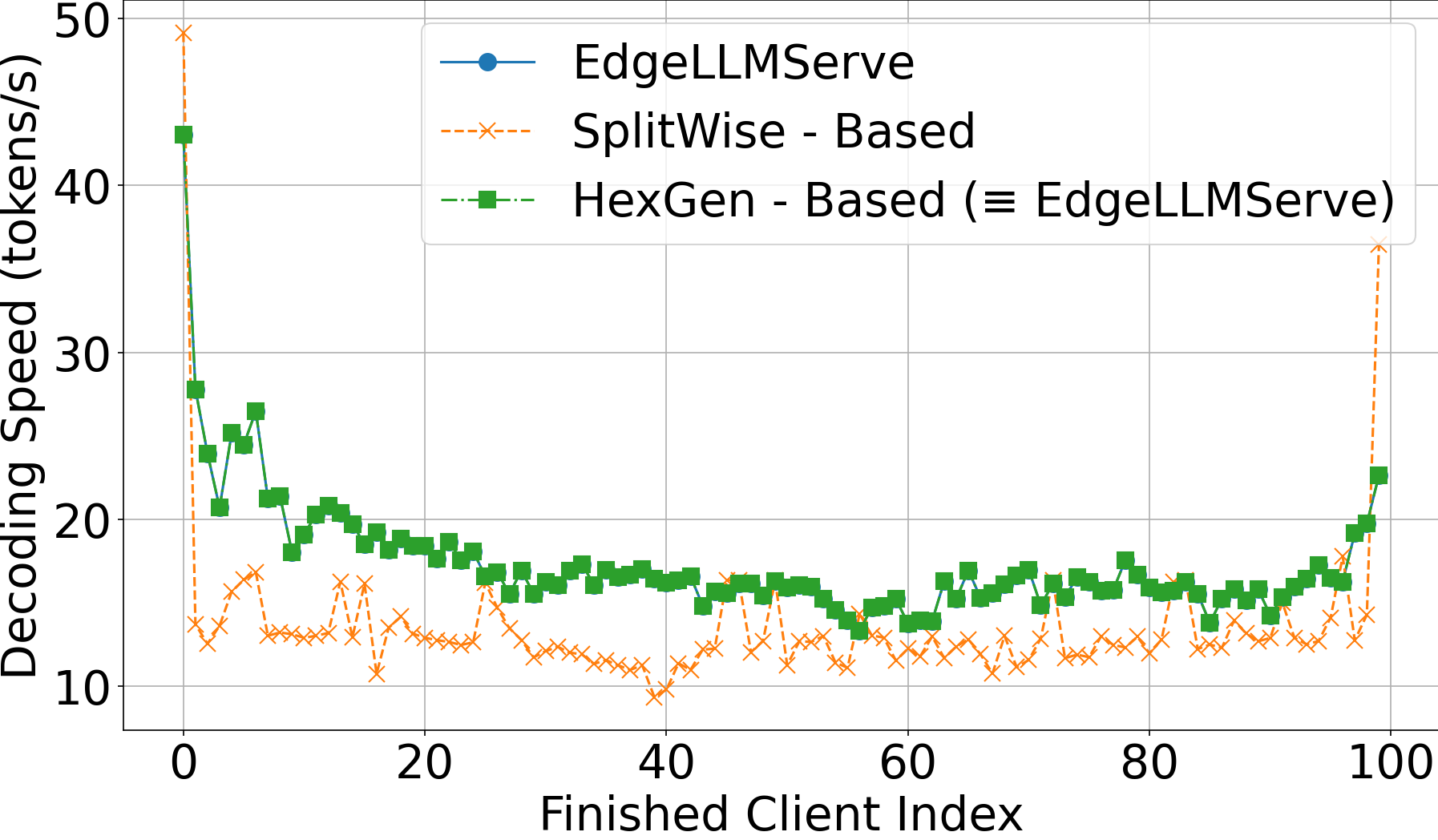}
        \caption{Decoding speed}
    \end{subfigure}\hfill
    \begin{subfigure}{\appfigscale\textwidth}
        \includegraphics[width=\linewidth]{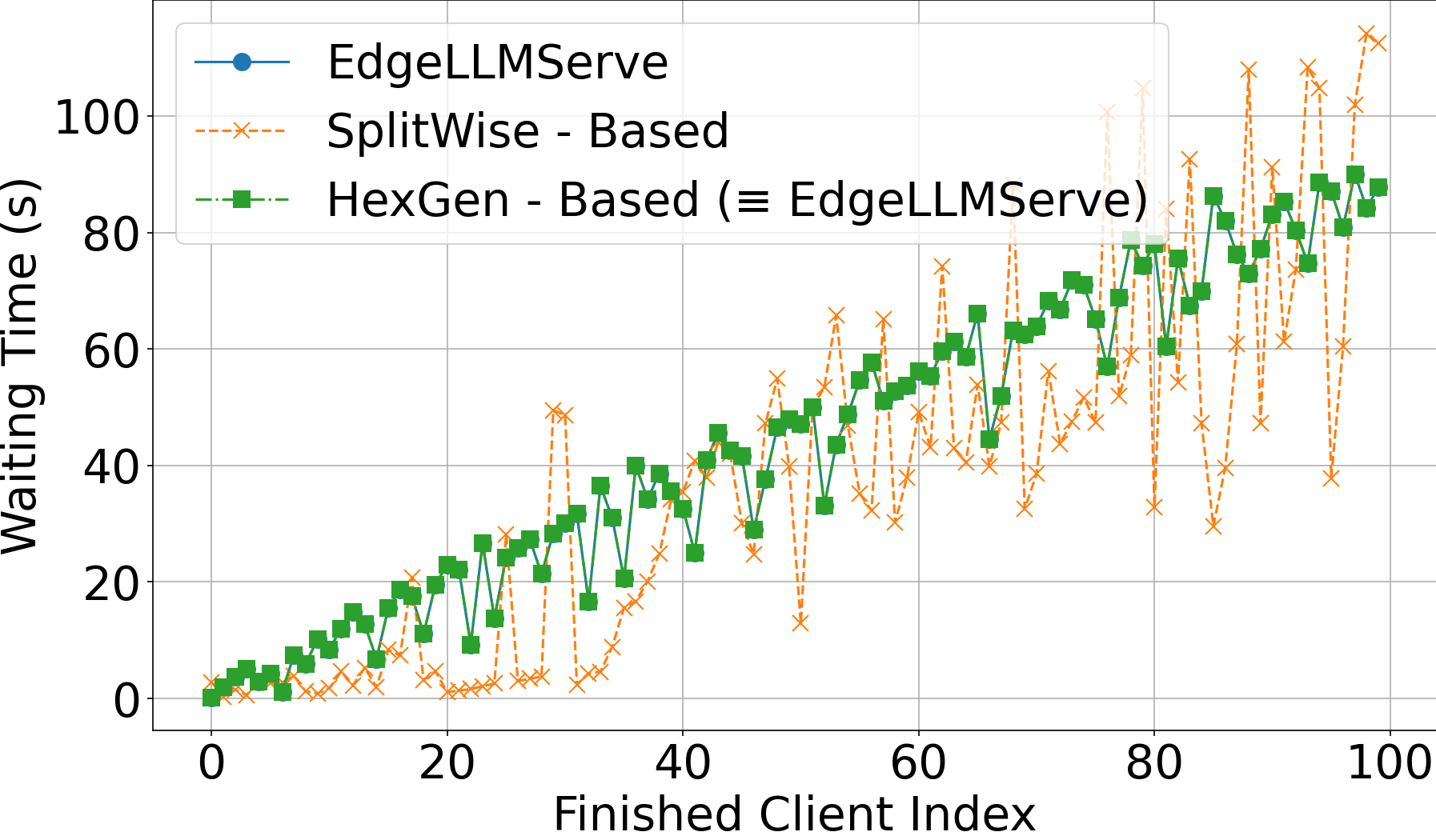}
        \caption{Waiting time}
    \end{subfigure}
    \caption{Per-request metrics --- Six devices, Gpt-Oss-20b, LZByteDance, arrival $100$~ms.}
    \label{apd:f16}
\end{figure}

\paragraph{5. LZByteDance dataset. Arrival rate of 1000ms}\mbox{}
\begin{table}[H]
    \centering
    \setlength{\tabcolsep}{3pt}
    \caption{Deployment plans and metrics --- Six devices, Gpt-Oss-20b, LZByteDance, arrival $1000$~ms.}
    \label{apd:t17}
    \begin{subtable}[t]{0.24\textwidth}
        \centering
        \caption{HG}
        \begin{tabular}{|c|c|c|c|}
            \hline
            Mode & Dev & Slots & G--C\\\hline
            P& Dev.4$^{\star}$ & 1 & 24--0\\\hline
            \multirow{2}{*}{D} & Dev.1 & \multirow{2}{*}{32} & 14--0\\
              & Dev.7$^{\star}$ &  & 10--0\\\hline
            \multirow{3}{*}{D} & Dev.2$^{\star}$ & \multirow{3}{*}{6} & 4--6\\
              & Dev.5 &  & 7--0\\
              & Dev.6 &  & 7--0\\\hline
        \end{tabular}
    \end{subtable}\hfill
    \begin{subtable}[t]{0.24\textwidth}
        \centering
        \caption{ELS}
        \begin{tabular}{|c|c|c|c|}
            \hline
            Mode & Dev & Slots & G--C\\\hline
            P& Dev.4$^{\star}$ & 1 & 24--0\\\hline
            \multirow{2}{*}{D} & Dev.1 & \multirow{2}{*}{32} & 14--0\\
              & Dev.7$^{\star}$ &  & 10--0\\\hline
            \multirow{3}{*}{D} & Dev.2$^{\star}$ & \multirow{3}{*}{6} & 4--6\\
              & Dev.5 &  & 7--0\\
              & Dev.6 &  & 7--0\\\hline
        \end{tabular}
    \end{subtable}\hfill
    \begin{subtable}[t]{0.24\textwidth}
        \centering
        \caption{SPW}
        \begin{tabular}{|c|c|c|c|}
            \hline
            Mode & Dev & Slots & G--C\\\hline
            P& Dev.1$^{\star}$ & 1 & 20--4\\\hline
            \multirow{2}{*}{D} & Dev.2$^{\star}$ & \multirow{2}{*}{14} & 5--1\\
              & Dev.7 &  & 18--0\\\hline
            D& Dev.4$^{\star}$ & 6 & 24--0\\\hline
            \multirow{2}{*}{D} & Dev.5 & \multirow{2}{*}{2} & 15--0\\
              & Dev.6$^{\star}$ &  & 9--0\\\hline
        \end{tabular}
    \end{subtable}\hfill
    \begin{subtable}[t]{0.24\textwidth}
        \centering
        \caption{Metrics.}
        \begin{tabular}{|l|c|c|c|}
            \hline
            Method & PS & DS & W\\\hline
            HG & 738 & 16.7 & 3.1\\\hline
            \textbf{ELS} & \textbf{738} & \textbf{16.7} & \textbf{3.1}\\\hline
            SPW & 2482 & 13.1 & 83.2\\\hline
        \end{tabular}
    \end{subtable}
\end{table}
\begin{figure}[H]
    \centering
    \begin{subfigure}{\appfigscale\textwidth}
        \includegraphics[width=\linewidth]{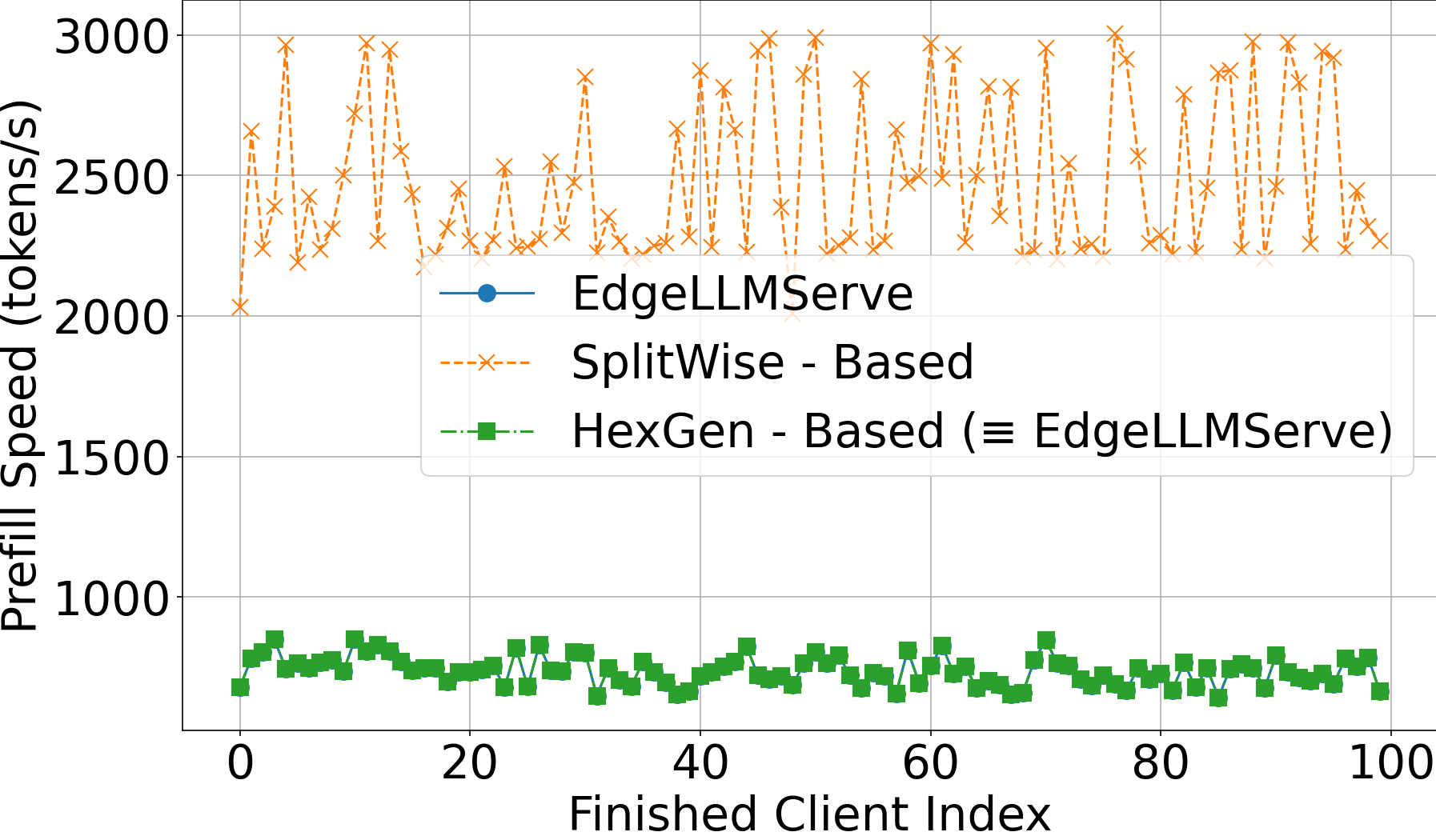}
        \caption{Prefill speed}
    \end{subfigure}\hfill
    \begin{subfigure}{\appfigscale\textwidth}
        \includegraphics[width=\linewidth]{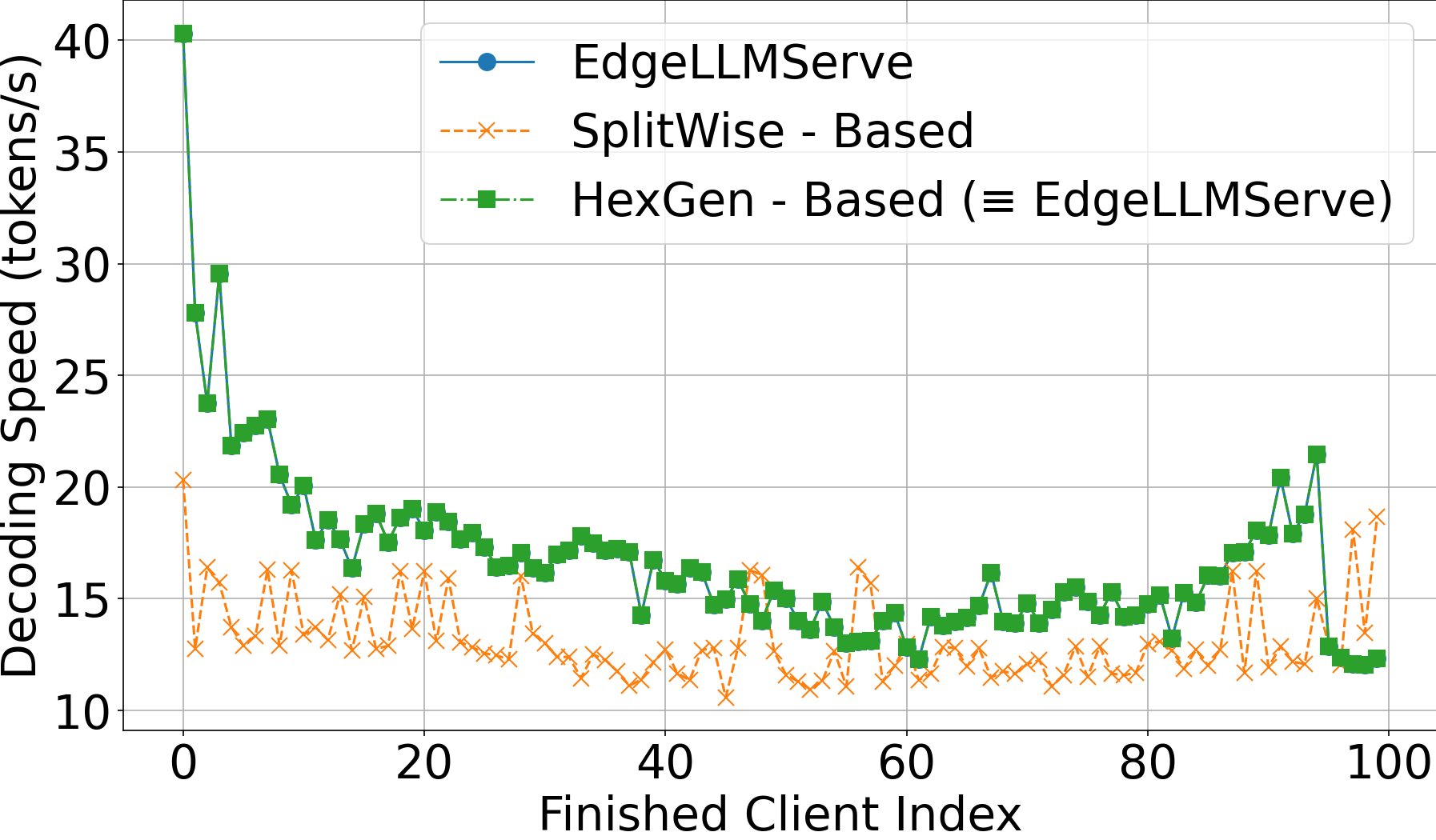}
        \caption{Decoding speed}
    \end{subfigure}\hfill
    \begin{subfigure}{\appfigscale\textwidth}
        \includegraphics[width=\linewidth]{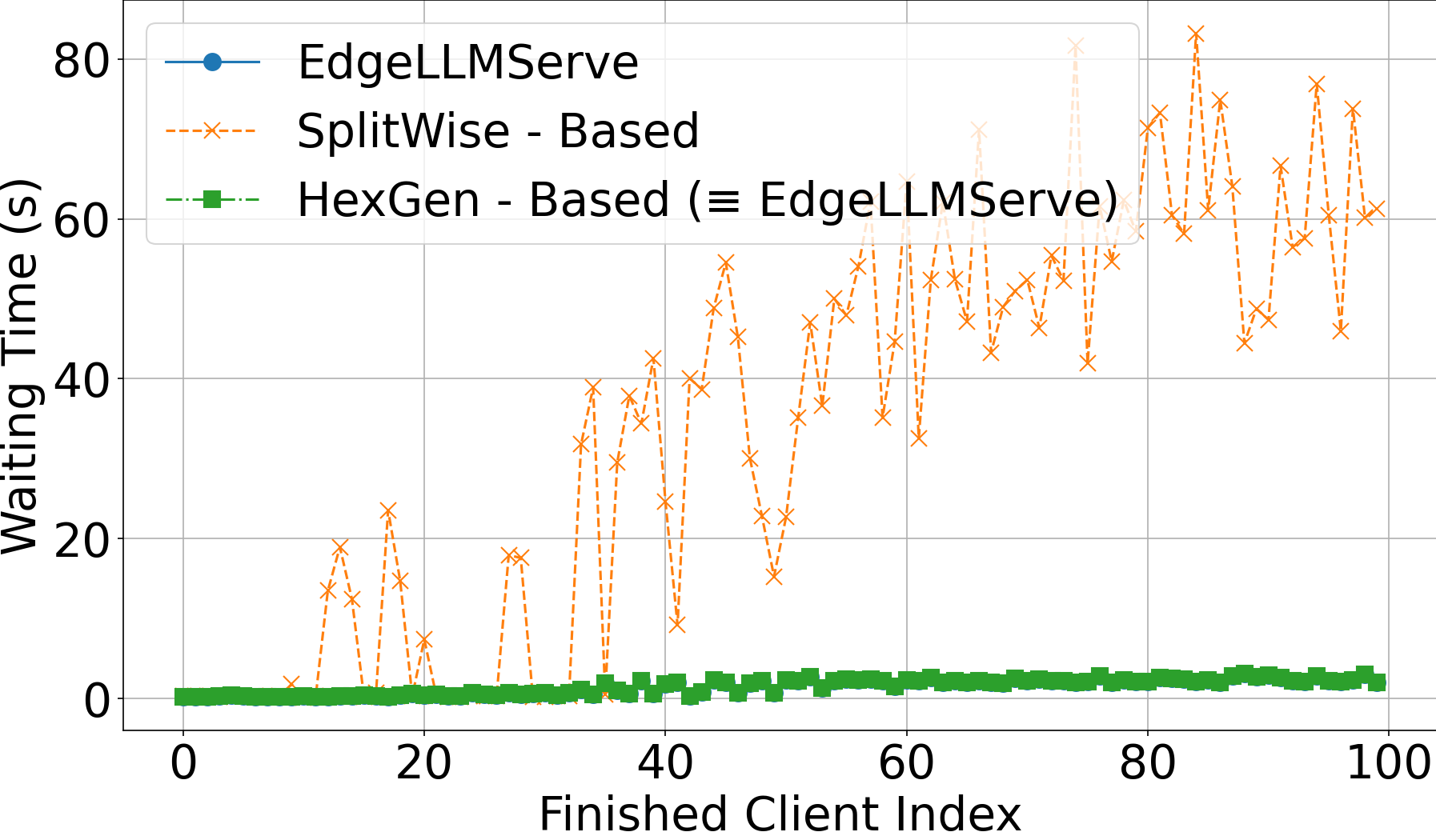}
        \caption{Waiting time}
    \end{subfigure}
    \caption{Per-request metrics --- Six devices, Gpt-Oss-20b, LZByteDance, arrival $1000$~ms.}
    \label{apd:f17}
\end{figure}

\paragraph{6. LZByteDance dataset. Arrival rate of 3000ms}\mbox{}
\begin{table}[H]
    \centering
    \setlength{\tabcolsep}{3pt}
    \caption{Deployment plans and metrics --- Six devices, Gpt-Oss-20b, LZByteDance, arrival $3000$~ms.}
    \label{apd:t18}
    \begin{subtable}[t]{0.24\textwidth}
        \centering
        \caption{HG}
        \begin{tabular}{|c|c|c|c|}
            \hline
            Mode & Dev & Slots & G--C\\\hline
            P& Dev.4$^{\star}$ & 1 & 24--0\\\hline
            \multirow{2}{*}{D} & Dev.1 & \multirow{2}{*}{32} & 14--0\\
              & Dev.7$^{\star}$ &  & 10--0\\\hline
            \multirow{3}{*}{D} & Dev.2$^{\star}$ & \multirow{3}{*}{6} & 4--6\\
              & Dev.5 &  & 7--0\\
              & Dev.6 &  & 7--0\\\hline
        \end{tabular}
    \end{subtable}\hfill
    \begin{subtable}[t]{0.24\textwidth}
        \centering
        \caption{ELS}
        \begin{tabular}{|c|c|c|c|}
            \hline
            Mode & Dev & Slots & G--C\\\hline
            P& Dev.2$^{\star}$ & 1 & 5--19\\\hline
            \multirow{2}{*}{D} & Dev.1 & \multirow{2}{*}{32} & 14--0\\
              & Dev.7$^{\star}$ &  & 10--0\\\hline
            D& Dev.4$^{\star}$ & 6 & 24--0\\\hline
            \multirow{2}{*}{D} & Dev.5 & \multirow{2}{*}{2} & 15--0\\
              & Dev.6$^{\star}$ &  & 9--0\\\hline
        \end{tabular}
    \end{subtable}\hfill
    \begin{subtable}[t]{0.24\textwidth}
        \centering
        \caption{SPW}
        \begin{tabular}{|c|c|c|c|}
            \hline
            Mode & Dev & Slots & G--C\\\hline
            P& Dev.1$^{\star}$ & 1 & 20--4\\\hline
            \multirow{2}{*}{D} & Dev.2$^{\star}$ & \multirow{2}{*}{14} & 5--1\\
              & Dev.7 &  & 18--0\\\hline
            D& Dev.4$^{\star}$ & 6 & 24--0\\\hline
            \multirow{2}{*}{D} & Dev.5 & \multirow{2}{*}{2} & 15--0\\
              & Dev.6$^{\star}$ &  & 9--0\\\hline
        \end{tabular}
    \end{subtable}\hfill
    \begin{subtable}[t]{0.24\textwidth}
        \centering
        \caption{Metrics.}
        \begin{tabular}{|l|c|c|c|}
            \hline
            Method & PS & DS & W\\\hline
            HG & 700 & 20.4 & 0.7\\\hline
            \textbf{ELS} & \textbf{294} & \textbf{19.1} & \textbf{0.8}\\\hline
            SPW & 2479 & 17.0 & 0.7\\\hline
        \end{tabular}
    \end{subtable}
\end{table}
\begin{figure}[H]
    \centering
    \begin{subfigure}{\appfigscale\textwidth}
        \includegraphics[width=\linewidth]{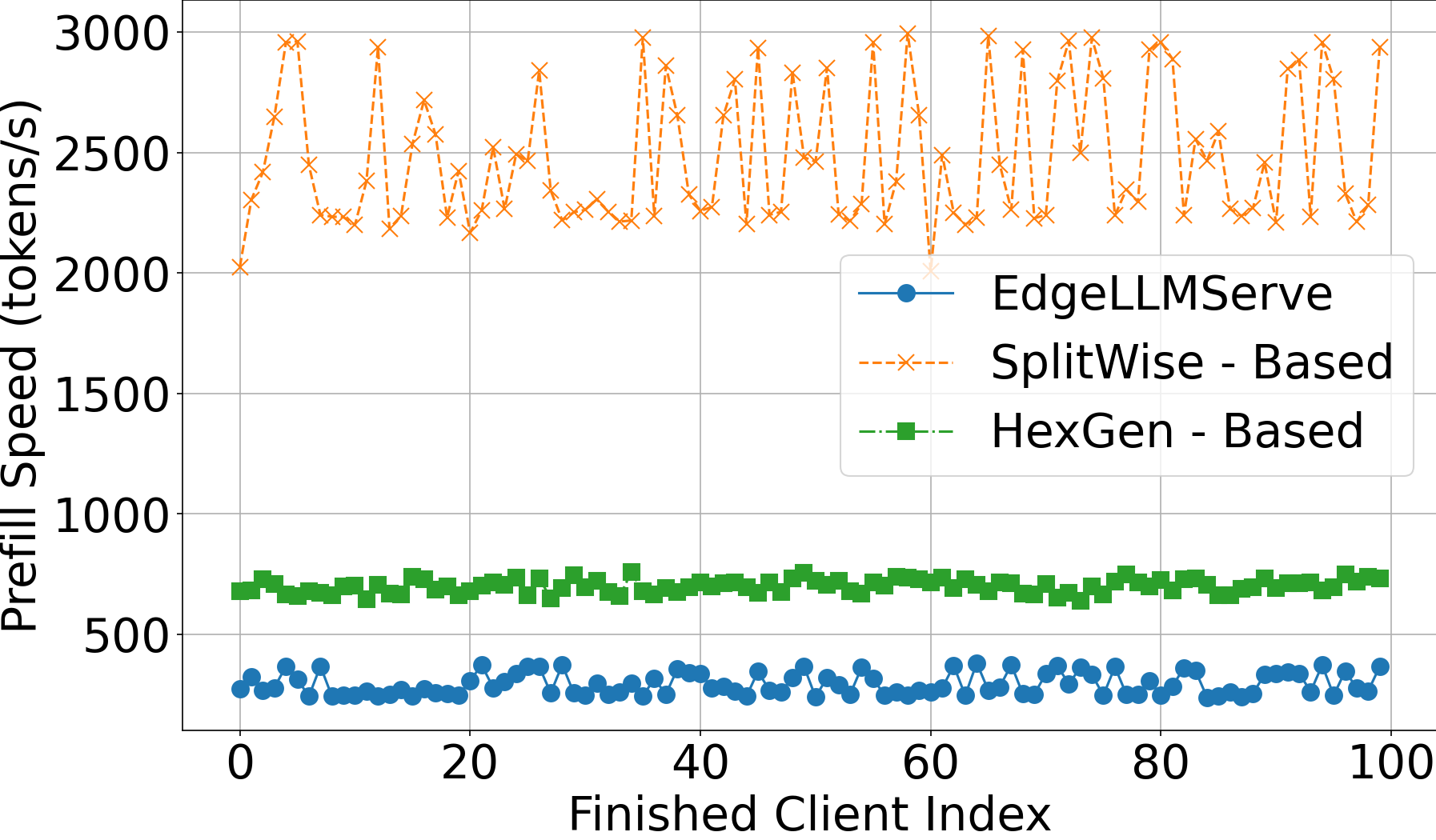}
        \caption{Prefill speed}
    \end{subfigure}\hfill
    \begin{subfigure}{\appfigscale\textwidth}
        \includegraphics[width=\linewidth]{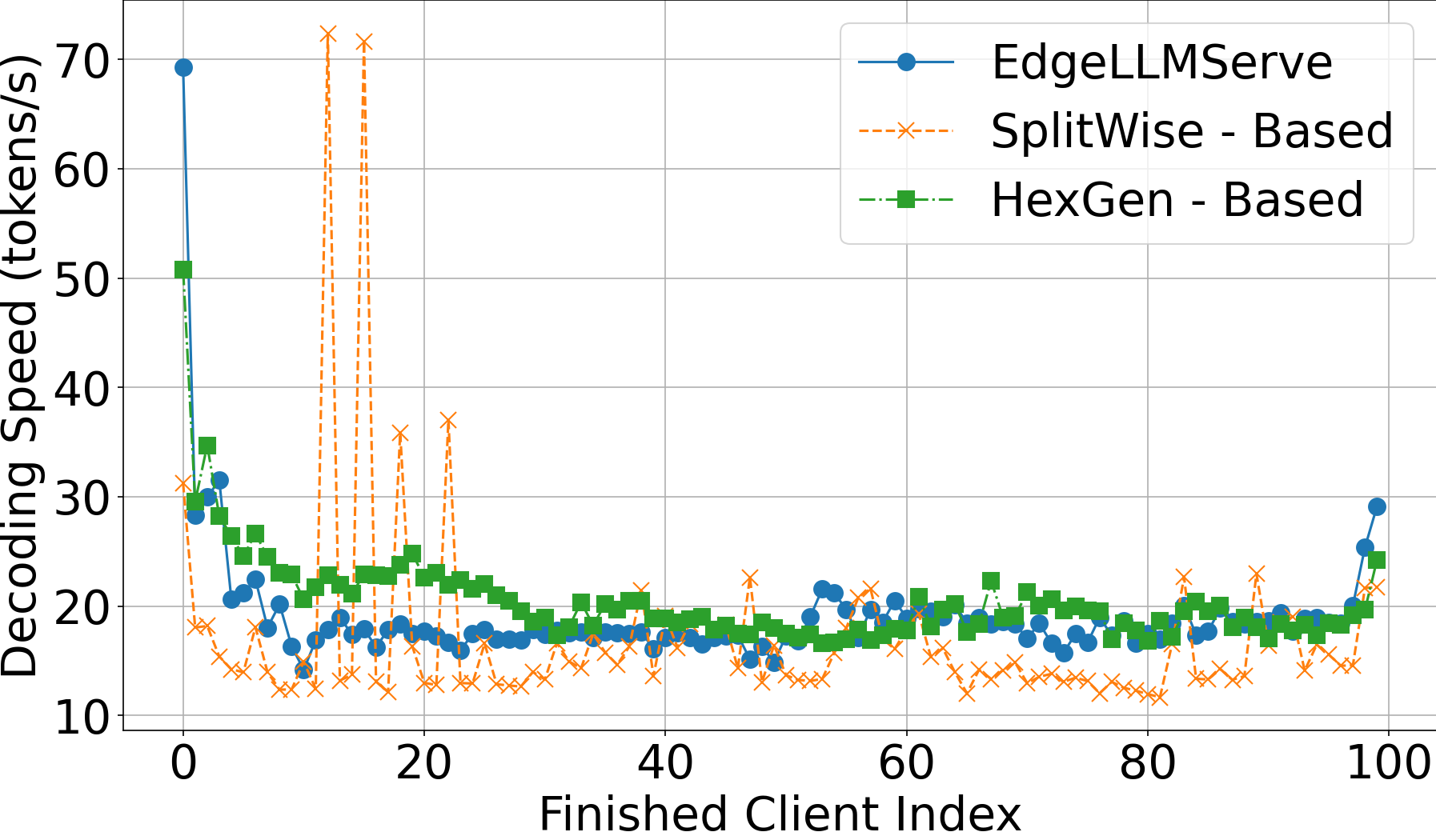}
        \caption{Decoding speed}
    \end{subfigure}\hfill
    \begin{subfigure}{\appfigscale\textwidth}
        \includegraphics[width=\linewidth]{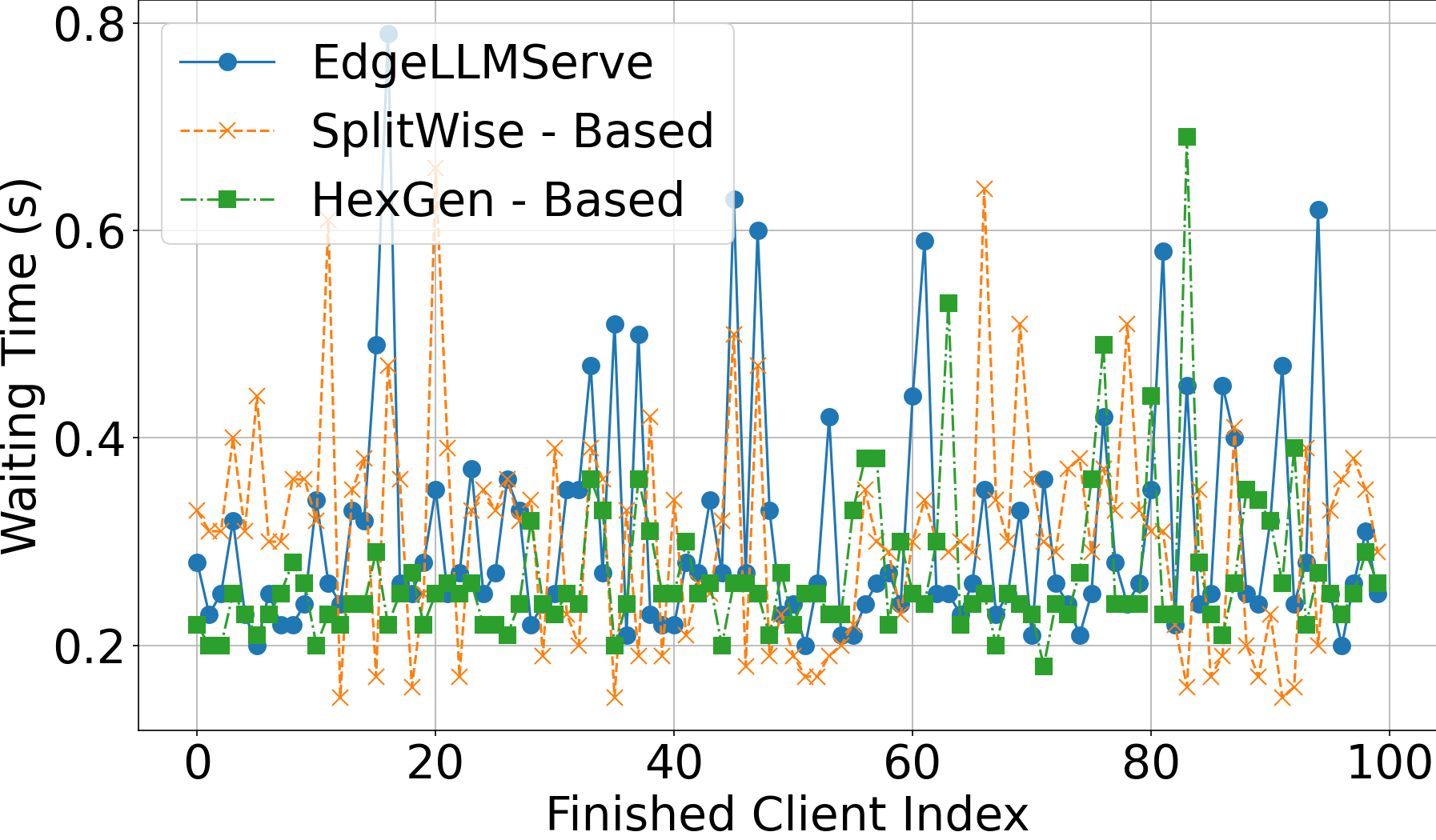}
        \caption{Waiting time}
    \end{subfigure}
    \caption{Per-request metrics --- Six devices, Gpt-Oss-20b, LZByteDance, arrival $3000$~ms.}
    \label{apd:f18}
\end{figure}

\subsubsection{Qwen3-MoE-30B}
\paragraph{1. IELTS dataset. Arrival rate of 100ms}\mbox{}
\begin{table}[H]
    \centering
    \setlength{\tabcolsep}{3pt}
    \caption{Deployment plans and metrics --- Six devices, Qwen3-MoE-30B, IELTS, arrival $100$~ms.}
    \vspace{-0.3cm}
    \label{apd:t19}
    \begin{subtable}[t]{0.24\textwidth}
        \centering
        \caption{HG}
        \begin{tabular}{|c|c|c|c|}
            \hline
            Mode & Dev & Slots & G--C\\\hline
            P& Dev.1$^{\star}$ & 1 & 26--22\\\hline
            \multirow{2}{*}{D} & Dev.2$^{\star}$ & \multirow{2}{*}{4} & 6--22\\
              & Dev.6 &  & 20--0\\\hline
            \multirow{3}{*}{D} & Dev.4$^{\star}$ & \multirow{3}{*}{12} & 23--0\\
              & Dev.5 &  & 7--0\\
              & Dev.7 &  & 18--0\\\hline
        \end{tabular}
    \end{subtable}\hfill
    \begin{subtable}[t]{0.24\textwidth}
        \centering
        \caption{ELS}
        \begin{tabular}{|c|c|c|c|}
            \hline
            Mode & Dev & Slots & G--C\\\hline
            P& Dev.1$^{\star}$ & 1 & 26--22\\\hline
            \multirow{2}{*}{D} & Dev.2$^{\star}$ & \multirow{2}{*}{4} & 6--22\\
              & Dev.6 &  & 20--0\\\hline
            \multirow{3}{*}{D} & Dev.4$^{\star}$ & \multirow{3}{*}{12} & 23--0\\
              & Dev.5 &  & 7--0\\
              & Dev.7 &  & 18--0\\\hline
        \end{tabular}
    \end{subtable}\hfill
    \begin{subtable}[t]{0.24\textwidth}
        \centering
        \caption{SPW}
        \begin{tabular}{|c|c|c|c|}
            \hline
            Mode & Dev & Slots & G--C\\\hline
            P& Dev.1$^{\star}$ & 1 & 26--22\\\hline
            \multirow{3}{*}{D} & Dev.2$^{\star}$ & \multirow{3}{*}{6} & 6--10\\
              & Dev.6 &  & 10--0\\
              & Dev.7 &  & 22--0\\\hline
            \multirow{2}{*}{D} & Dev.4$^{\star}$ & \multirow{2}{*}{8} & 36--0\\
              & Dev.5 &  & 12--0\\\hline
        \end{tabular}
    \end{subtable}\hfill
    \begin{subtable}[t]{0.24\textwidth}
        \centering
        \caption{Metrics.}
        \begin{tabular}{|l|c|c|c|}
            \hline
            Method & PS & DS & W\\\hline
            HG & 1412 & 11.0 & 880.6\\\hline
            \textbf{ELS} & \textbf{1412} & \textbf{11.0} & \textbf{880.6}\\\hline
            SPW & 1409 & 12.9 & 985.5\\\hline
        \end{tabular}
    \end{subtable}
\end{table}
\begin{figure}[H]
    \centering
    \begin{subfigure}{\appfigscale\textwidth}
        \includegraphics[width=\linewidth]{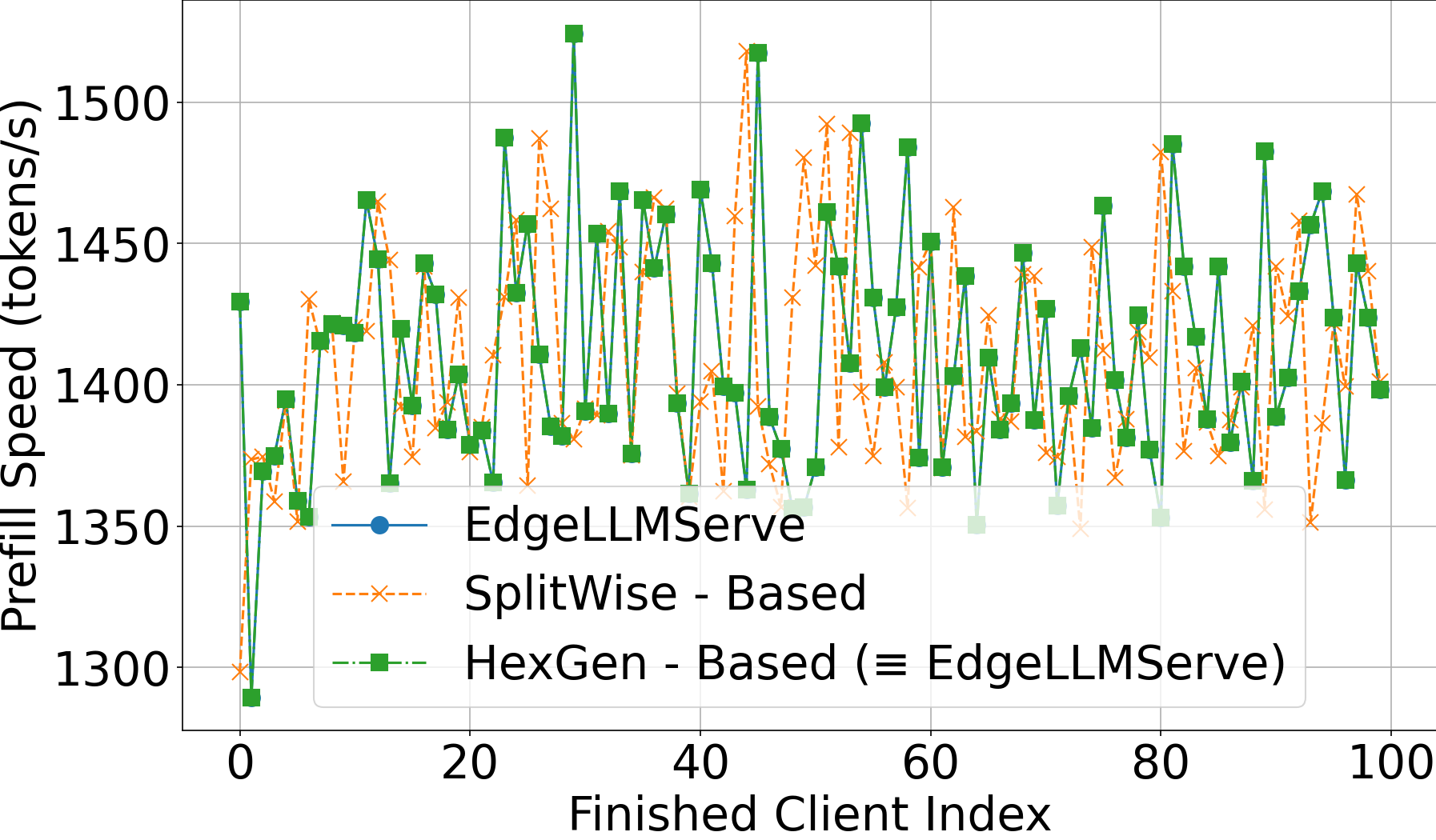}
        \caption{Prefill speed}
    \end{subfigure}\hfill
    \begin{subfigure}{\appfigscale\textwidth}
        \includegraphics[width=\linewidth]{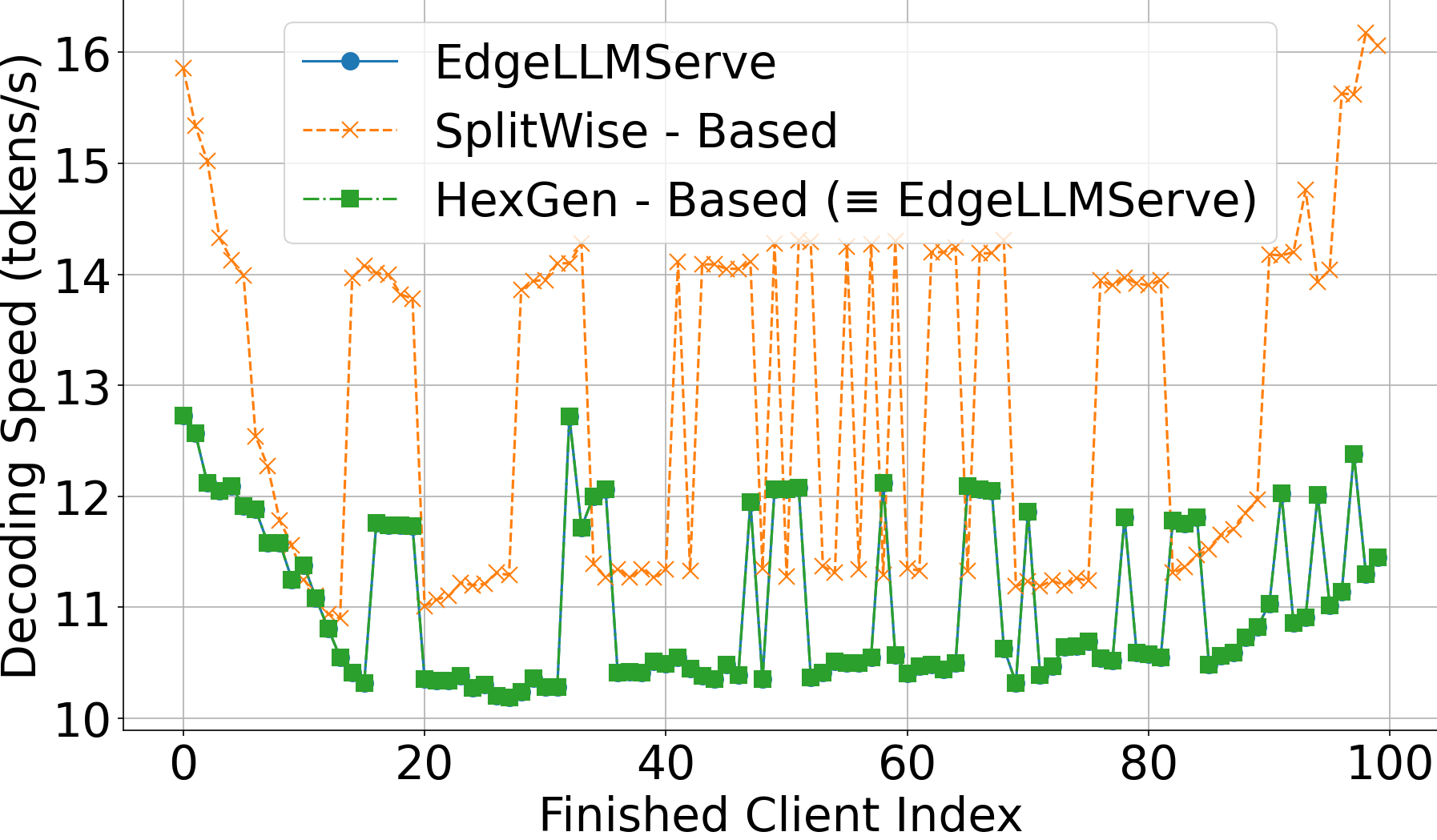}
        \caption{Decoding speed}
    \end{subfigure}\hfill
    \begin{subfigure}{\appfigscale\textwidth}
        \includegraphics[width=\linewidth]{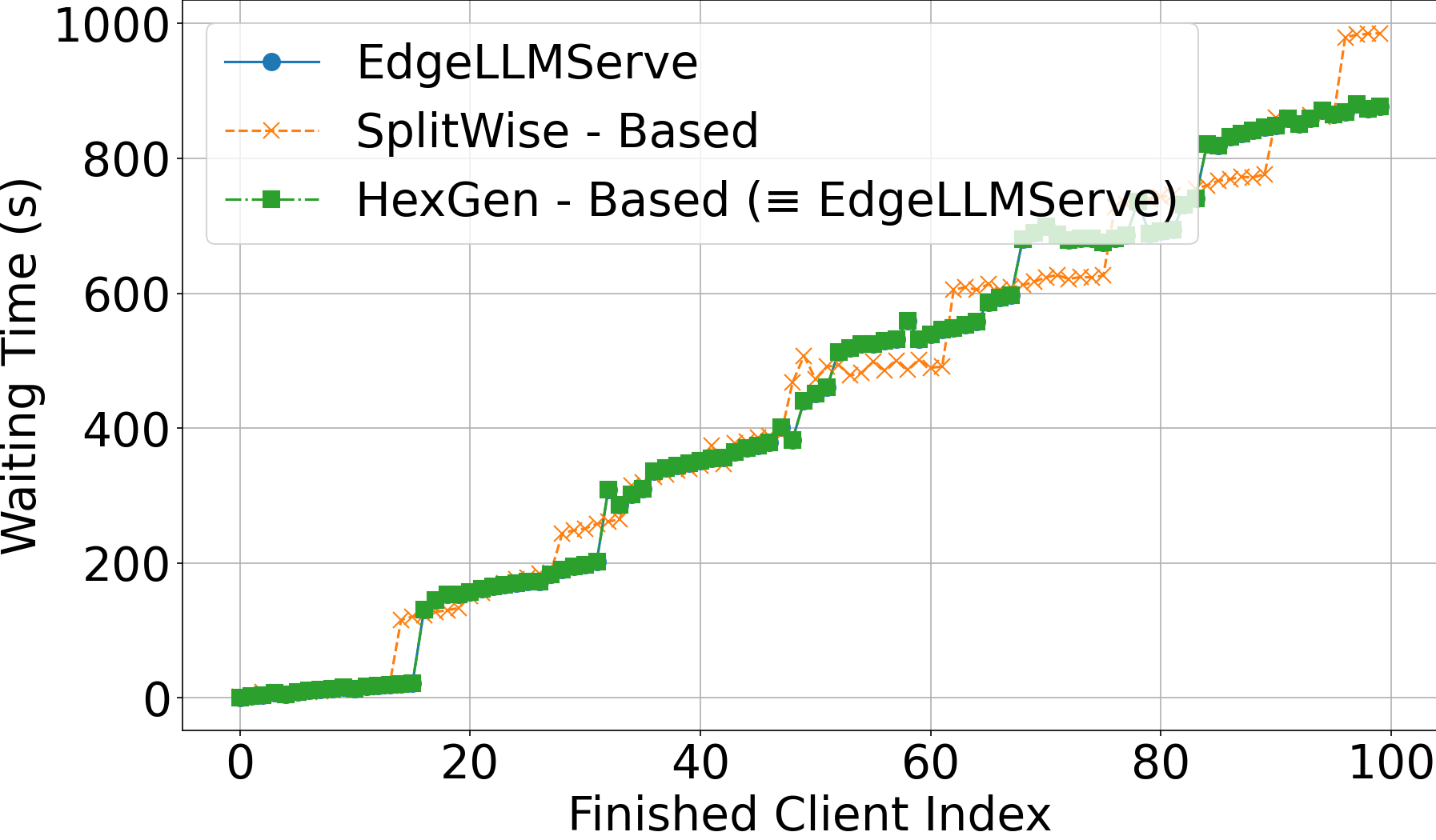}
        \caption{Waiting time}
    \end{subfigure}
    \vspace{-0.3cm}
    \caption{Per-request metrics --- Six devices, Qwen3-MoE-30B, IELTS, arrival $100$~ms.}
    \label{apd:f19}
\end{figure}

\paragraph{2. IELTS dataset. Arrival rate of 1000ms}\mbox{}
\begin{table}[H]
    \centering
    \setlength{\tabcolsep}{3pt}
    \caption{Deployment plans and metrics --- Six devices, Qwen3-MoE-30B, IELTS, arrival $1000$~ms.}
    \vspace{-0.3cm}
    \label{apd:t20}
    \begin{subtable}[t]{0.24\textwidth}
        \centering
        \caption{HG}
        \begin{tabular}{|c|c|c|c|}
            \hline
            Mode & Dev & Slots & G--C\\\hline
            P& Dev.1$^{\star}$ & 1 & 26--22\\\hline
            \multirow{2}{*}{D} & Dev.2$^{\star}$ & \multirow{2}{*}{4} & 6--22\\
              & Dev.6 &  & 20--0\\\hline
            \multirow{3}{*}{D} & Dev.4$^{\star}$ & \multirow{3}{*}{12} & 23--0\\
              & Dev.5 &  & 7--0\\
              & Dev.7 &  & 18--0\\\hline
        \end{tabular}
    \end{subtable}\hfill
    \begin{subtable}[t]{0.24\textwidth}
        \centering
        \caption{ELS}
        \begin{tabular}{|c|c|c|c|}
            \hline
            Mode & Dev & Slots & G--C\\\hline
            P& Dev.1$^{\star}$ & 1 & 26--22\\\hline
            \multirow{2}{*}{D} & Dev.2$^{\star}$ & \multirow{2}{*}{4} & 6--22\\
              & Dev.6 &  & 20--0\\\hline
            \multirow{3}{*}{D} & Dev.4$^{\star}$ & \multirow{3}{*}{12} & 23--0\\
              & Dev.5 &  & 7--0\\
              & Dev.7 &  & 18--0\\\hline
        \end{tabular}
    \end{subtable}\hfill
    \begin{subtable}[t]{0.24\textwidth}
        \centering
        \caption{SPW}
        \begin{tabular}{|c|c|c|c|}
            \hline
            Mode & Dev & Slots & G--C\\\hline
            P& Dev.1$^{\star}$ & 1 & 26--22\\\hline
            \multirow{3}{*}{D} & Dev.2$^{\star}$ & \multirow{3}{*}{6} & 6--10\\
              & Dev.6 &  & 10--0\\
              & Dev.7 &  & 22--0\\\hline
            \multirow{2}{*}{D} & Dev.4$^{\star}$ & \multirow{2}{*}{8} & 36--0\\
              & Dev.5 &  & 12--0\\\hline
        \end{tabular}
    \end{subtable}\hfill
    \begin{subtable}[t]{0.24\textwidth}
        \centering
        \caption{Metrics.}
        \begin{tabular}{|l|c|c|c|}
            \hline
            Method & PS & DS & W\\\hline
            HG & 1411 & 11.0 & 794.1\\\hline
            \textbf{ELS} & \textbf{1411} & \textbf{11.0} & \textbf{794.1}\\\hline
            SPW & 1412 & 12.9 & 898.6\\\hline
        \end{tabular}
    \end{subtable}
\end{table}
\begin{figure}[H]
    \centering
    \begin{subfigure}{\appfigscale\textwidth}
        \includegraphics[width=\linewidth]{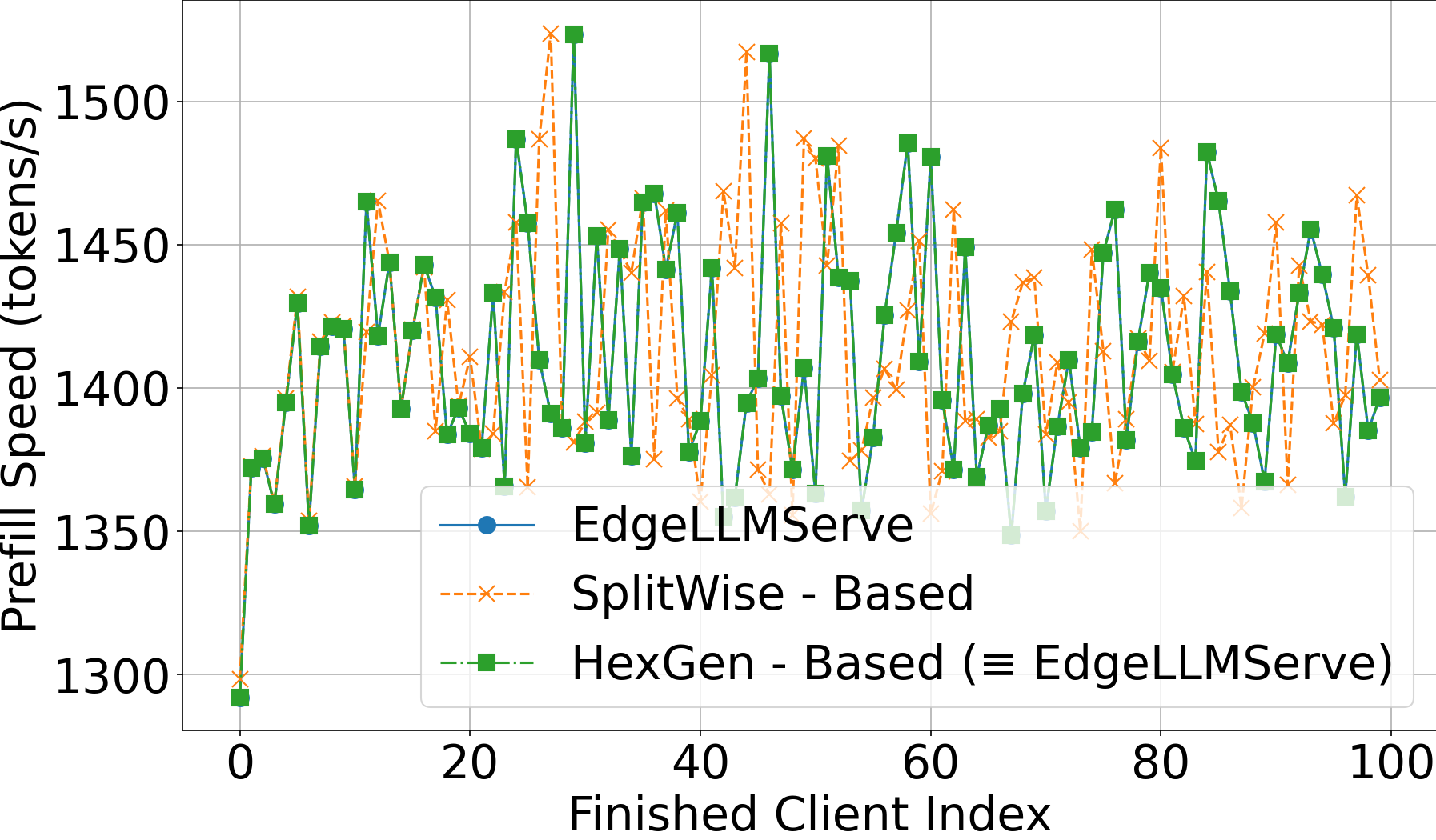}
        \caption{Prefill speed}
    \end{subfigure}\hfill
    \begin{subfigure}{\appfigscale\textwidth}
        \includegraphics[width=\linewidth]{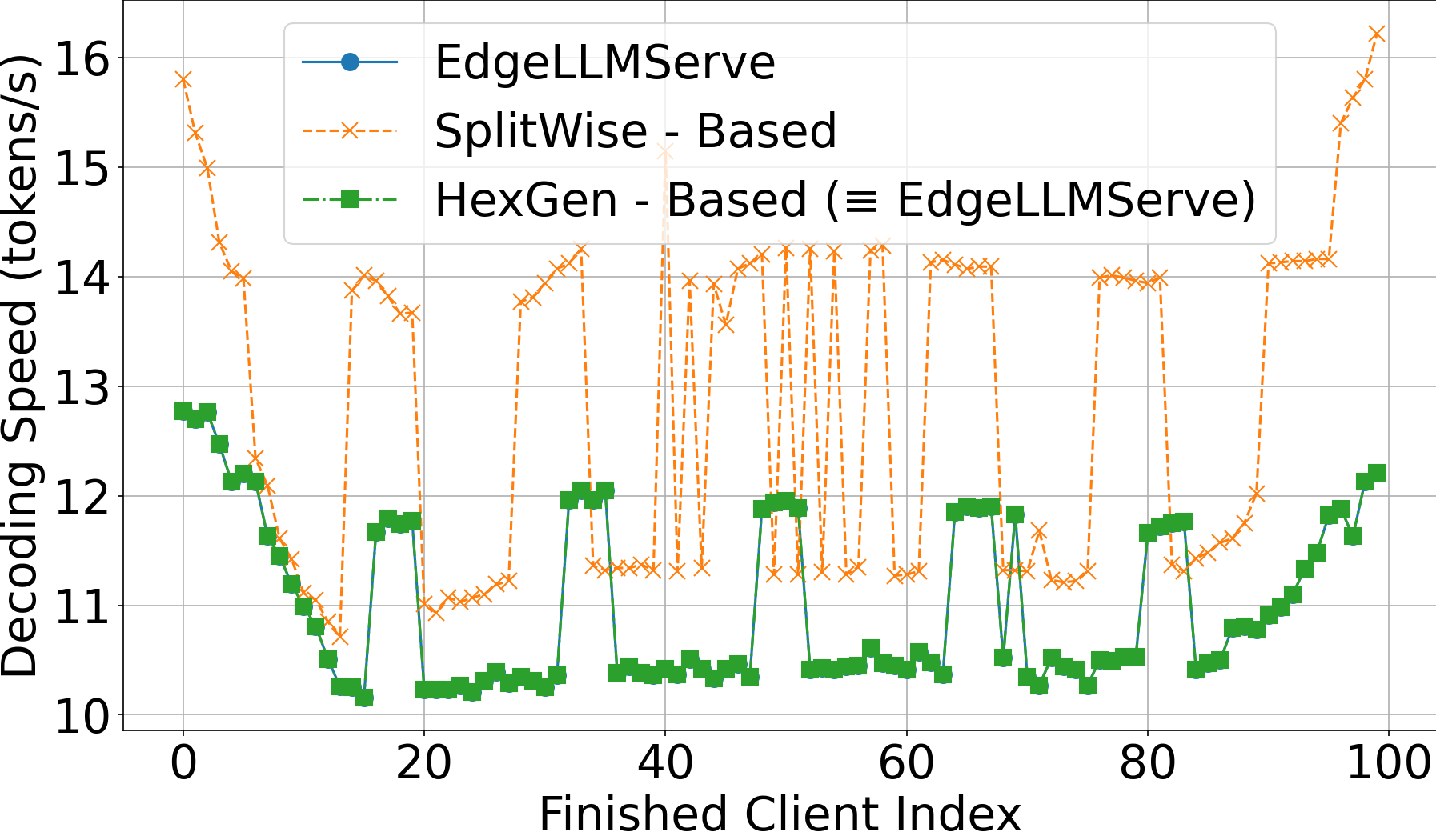}
        \caption{Decoding speed}
    \end{subfigure}\hfill
    \begin{subfigure}{\appfigscale\textwidth}
        \includegraphics[width=\linewidth]{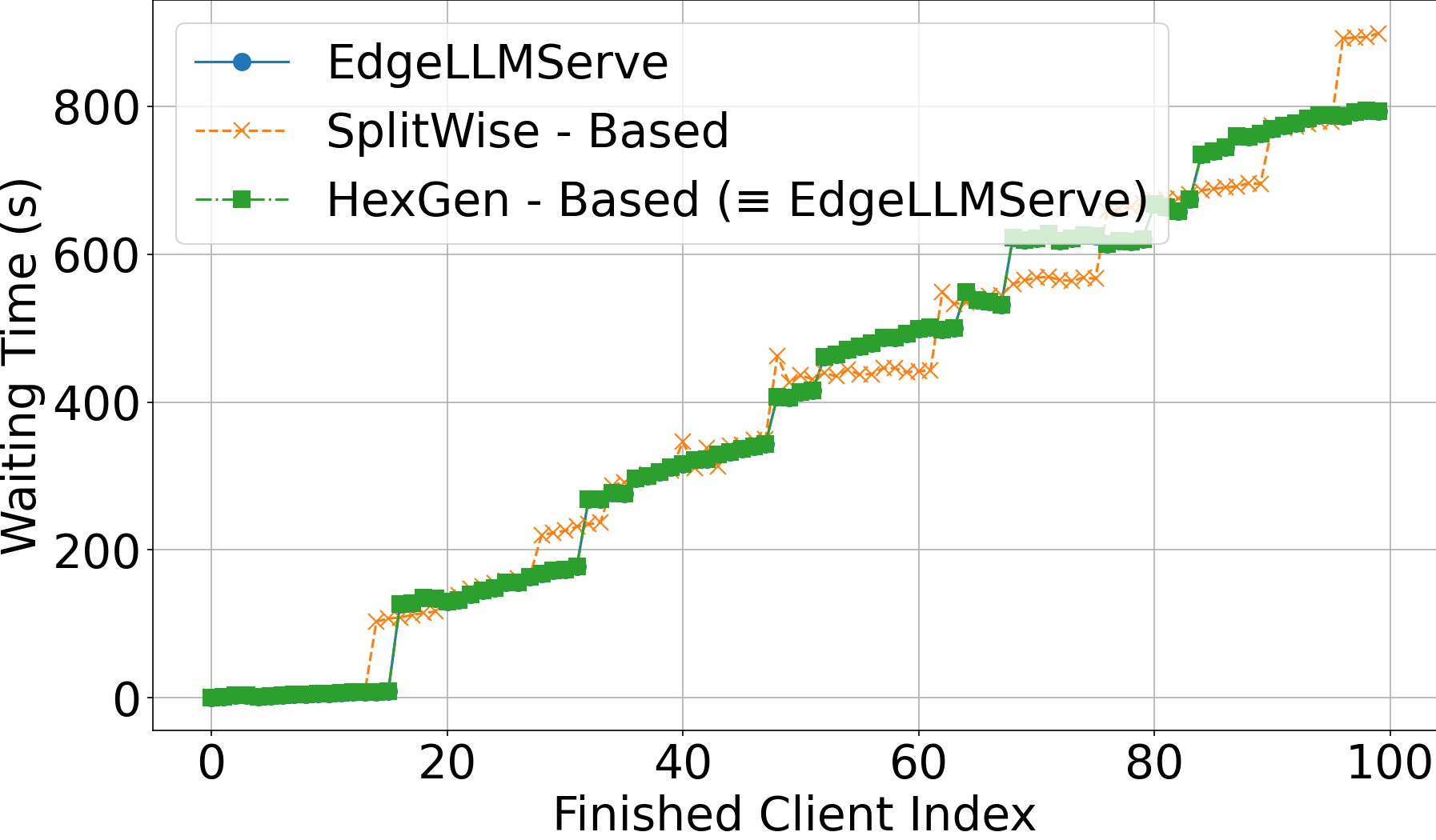}
        \caption{Waiting time}
    \end{subfigure}
    \vspace{-0.3cm}
    \caption{Per-request metrics --- Six devices, Qwen3-MoE-30B, IELTS, arrival $1000$~ms.}
    \label{apd:f20}
\end{figure}

\paragraph{3. IELTS dataset. Arrival rate of 3000ms}\mbox{}
\begin{table}[H]
    \centering
    \setlength{\tabcolsep}{3pt}
    \caption{Deployment plans and metrics --- Six devices, Qwen3-MoE-30B, IELTS, arrival $3000$~ms.}
    \label{apd:t21}
    \begin{subtable}[t]{0.24\textwidth}
        \centering
        \caption{HG}
        \begin{tabular}{|c|c|c|c|}
            \hline
            Mode & Dev & Slots & G--C\\\hline
            P& Dev.1$^{\star}$ & 1 & 26--22\\\hline
            \multirow{2}{*}{D} & Dev.2$^{\star}$ & \multirow{2}{*}{4} & 6--22\\
              & Dev.6 &  & 20--0\\\hline
            \multirow{3}{*}{D} & Dev.4$^{\star}$ & \multirow{3}{*}{12} & 23--0\\
              & Dev.5 &  & 7--0\\
              & Dev.7 &  & 18--0\\\hline
        \end{tabular}
    \end{subtable}\hfill
    \begin{subtable}[t]{0.24\textwidth}
        \centering
        \caption{ELS}
        \begin{tabular}{|c|c|c|c|}
            \hline
            Mode & Dev & Slots & G--C\\\hline
            P& Dev.1$^{\star}$ & 1 & 26--22\\\hline
            \multirow{2}{*}{D} & Dev.2$^{\star}$ & \multirow{2}{*}{4} & 6--22\\
              & Dev.6 &  & 20--0\\\hline
            \multirow{3}{*}{D} & Dev.4$^{\star}$ & \multirow{3}{*}{12} & 23--0\\
              & Dev.5 &  & 7--0\\
              & Dev.7 &  & 18--0\\\hline
        \end{tabular}
    \end{subtable}\hfill
    \begin{subtable}[t]{0.24\textwidth}
        \centering
        \caption{SPW}
        \begin{tabular}{|c|c|c|c|}
            \hline
            Mode & Dev & Slots & G--C\\\hline
            P& Dev.1$^{\star}$ & 1 & 26--22\\\hline
            \multirow{3}{*}{D} & Dev.2$^{\star}$ & \multirow{3}{*}{6} & 6--10\\
              & Dev.6 &  & 10--0\\
              & Dev.7 &  & 22--0\\\hline
            \multirow{2}{*}{D} & Dev.4$^{\star}$ & \multirow{2}{*}{8} & 36--0\\
              & Dev.5 &  & 12--0\\\hline
        \end{tabular}
    \end{subtable}\hfill
    \begin{subtable}[t]{0.24\textwidth}
        \centering
        \caption{Metrics.}
        \begin{tabular}{|l|c|c|c|}
            \hline
            Method & PS & DS & W\\\hline
            HG & 1412 & 10.9 & 742.9\\\hline
            \textbf{ELS} & \textbf{1412} & \textbf{10.9} & \textbf{742.9}\\\hline
            SPW & 1412 & 12.9 & 839.8\\\hline
        \end{tabular}
    \end{subtable}
\end{table}
\begin{figure}[H]
    \centering
    \begin{subfigure}{\appfigscale\textwidth}
        \includegraphics[width=\linewidth]{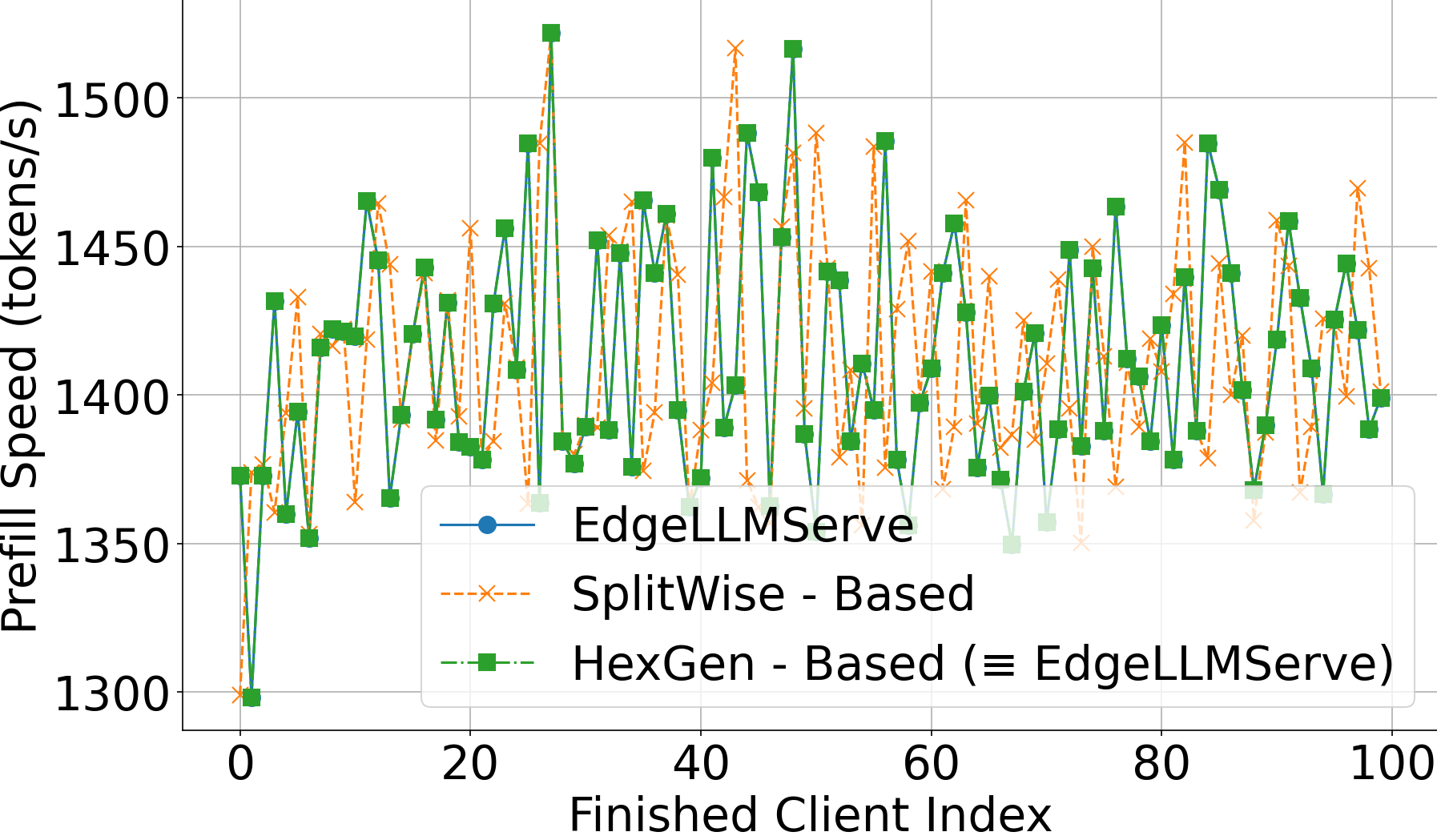}
        \caption{Prefill speed}
    \end{subfigure}\hfill
    \begin{subfigure}{\appfigscale\textwidth}
        \includegraphics[width=\linewidth]{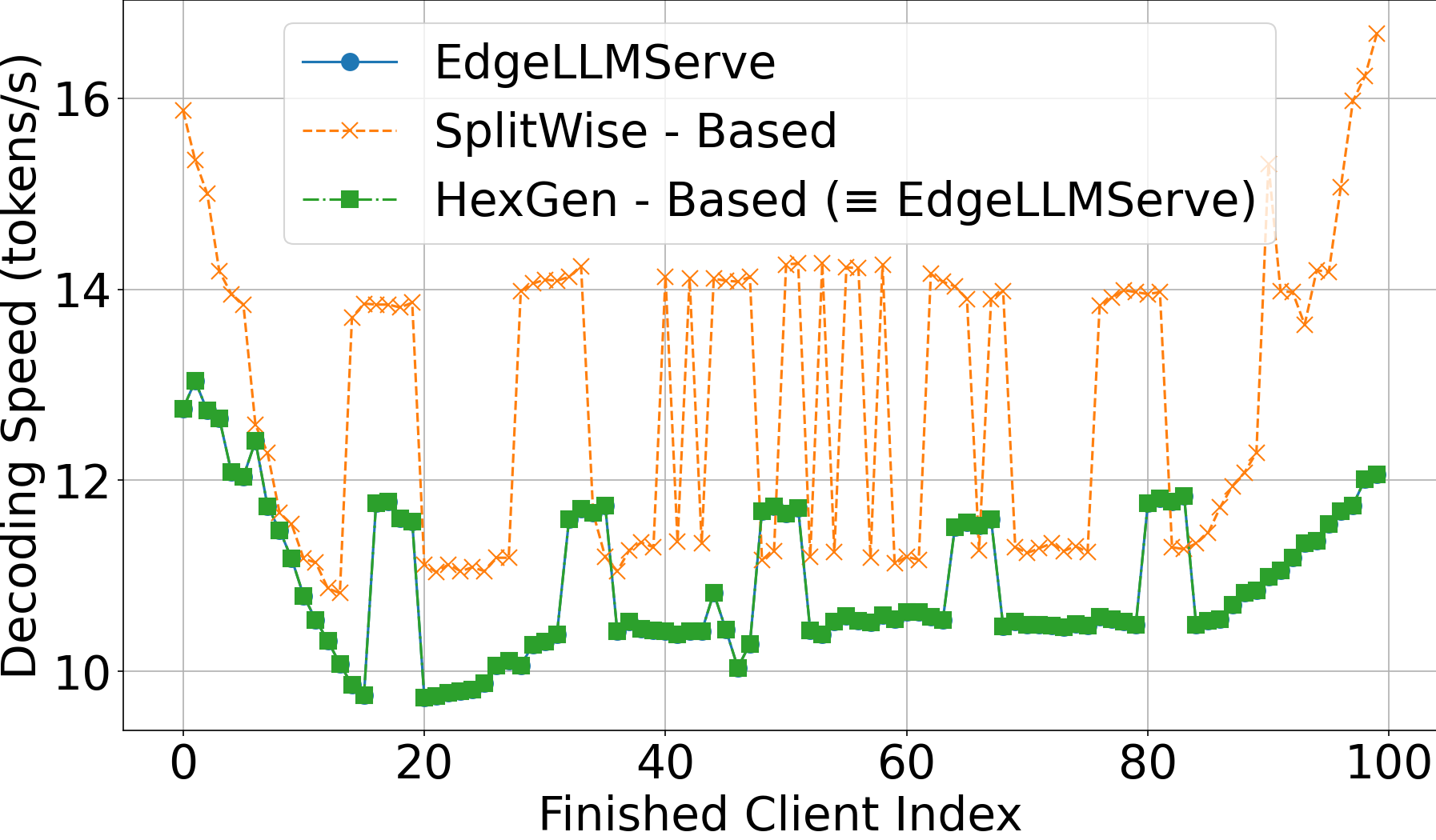}
        \caption{Decoding speed}
    \end{subfigure}\hfill
    \begin{subfigure}{\appfigscale\textwidth}
        \includegraphics[width=\linewidth]{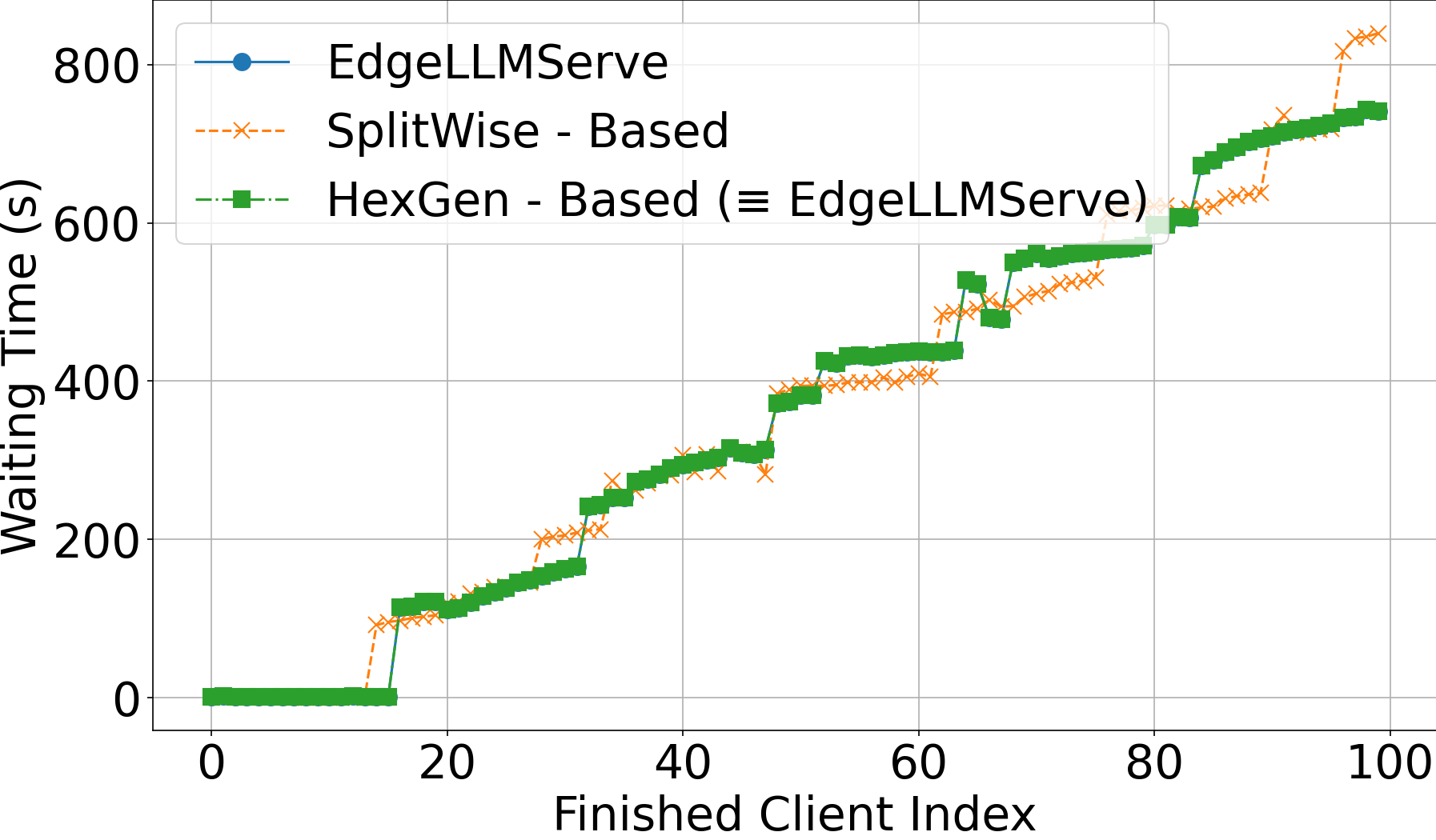}
        \caption{Waiting time}
    \end{subfigure}
    \caption{Per-request metrics --- Six devices, Qwen3-MoE-30B, IELTS, arrival $3000$~ms.}
    \label{apd:f21}
\end{figure}

\paragraph{4. LZByteDance dataset. Arrival rate of 100ms}\mbox{}
\begin{table}[H]
    \centering
    \setlength{\tabcolsep}{3pt}
    \caption{Deployment plans and metrics --- Six devices, Qwen3-MoE-30B, LZByteDance, arrival $100$~ms.}
    \label{apd:t22}
    \begin{subtable}[t]{0.24\textwidth}
        \centering
        \caption{HG}
        \begin{tabular}{|c|c|c|c|}
            \hline
            Mode & Dev & Slots & G--C\\\hline
            P& Dev.4$^{\star}$ & 1 & 48--0\\\hline
            \multirow{3}{*}{D} & Dev.1$^{\star}$ & \multirow{3}{*}{12} & 16--5\\
              & Dev.2 &  & 6--3\\
              & Dev.7 &  & 18--0\\\hline
            \multirow{2}{*}{D} & Dev.5$^{\star}$ & \multirow{2}{*}{2} & 22--0\\
              & Dev.6 &  & 26--0\\\hline
        \end{tabular}
    \end{subtable}\hfill
    \begin{subtable}[t]{0.24\textwidth}
        \centering
        \caption{ELS}
        \begin{tabular}{|c|c|c|c|}
            \hline
            Mode & Dev & Slots & G--C\\\hline
            P& Dev.4$^{\star}$ & 1 & 48--0\\\hline
            \multirow{3}{*}{D} & Dev.1$^{\star}$ & \multirow{3}{*}{12} & 16--5\\
              & Dev.2 &  & 6--3\\
              & Dev.7 &  & 18--0\\\hline
            \multirow{2}{*}{D} & Dev.5$^{\star}$ & \multirow{2}{*}{2} & 22--0\\
              & Dev.6 &  & 26--0\\\hline
        \end{tabular}
    \end{subtable}\hfill
    \begin{subtable}[t]{0.24\textwidth}
        \centering
        \caption{SPW}
        \begin{tabular}{|c|c|c|c|}
            \hline
            Mode & Dev & Slots & G--C\\\hline
            P& Dev.1$^{\star}$ & 1 & 26--22\\\hline
            \multirow{3}{*}{D} & Dev.2$^{\star}$ & \multirow{3}{*}{6} & 6--10\\
              & Dev.6 &  & 10--0\\
              & Dev.7 &  & 22--0\\\hline
            \multirow{2}{*}{D} & Dev.4$^{\star}$ & \multirow{2}{*}{8} & 36--0\\
              & Dev.5 &  & 12--0\\\hline
        \end{tabular}
    \end{subtable}\hfill
    \begin{subtable}[t]{0.24\textwidth}
        \centering
        \caption{Metrics.}
        \begin{tabular}{|l|c|c|c|}
            \hline
            Method & PS & DS & W\\\hline
            HG & 557 & 13.5 & 370.9\\\hline
            \textbf{ELS} & \textbf{557} & \textbf{13.5} & \textbf{370.9}\\\hline
            SPW & 1141 & 13.0 & 408.4\\\hline
        \end{tabular}
    \end{subtable}
\end{table}
\begin{figure}[H]
    \centering
    \begin{subfigure}{\appfigscale\textwidth}
        \includegraphics[width=\linewidth]{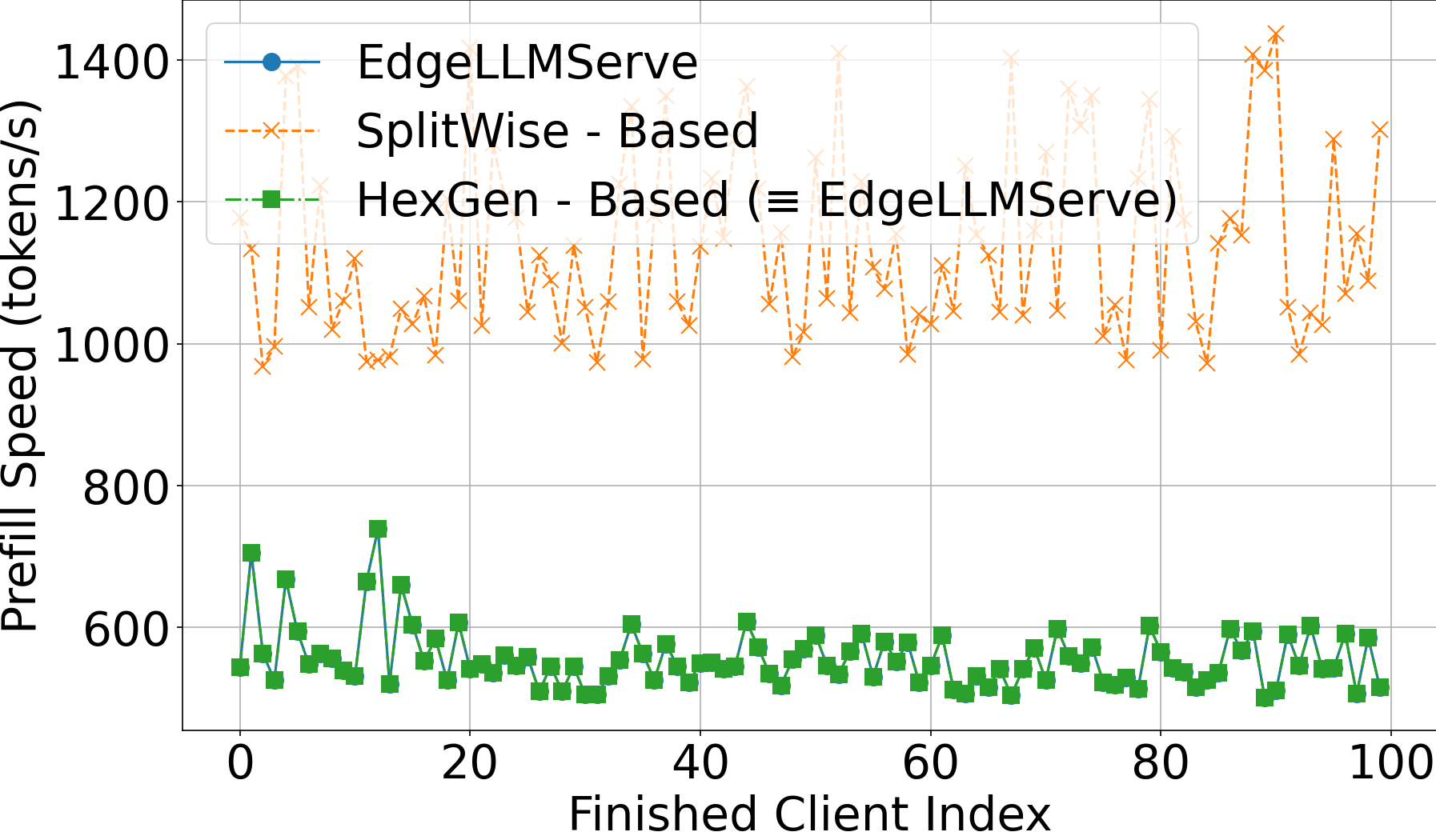}
        \caption{Prefill speed}
    \end{subfigure}\hfill
    \begin{subfigure}{\appfigscale\textwidth}
        \includegraphics[width=\linewidth]{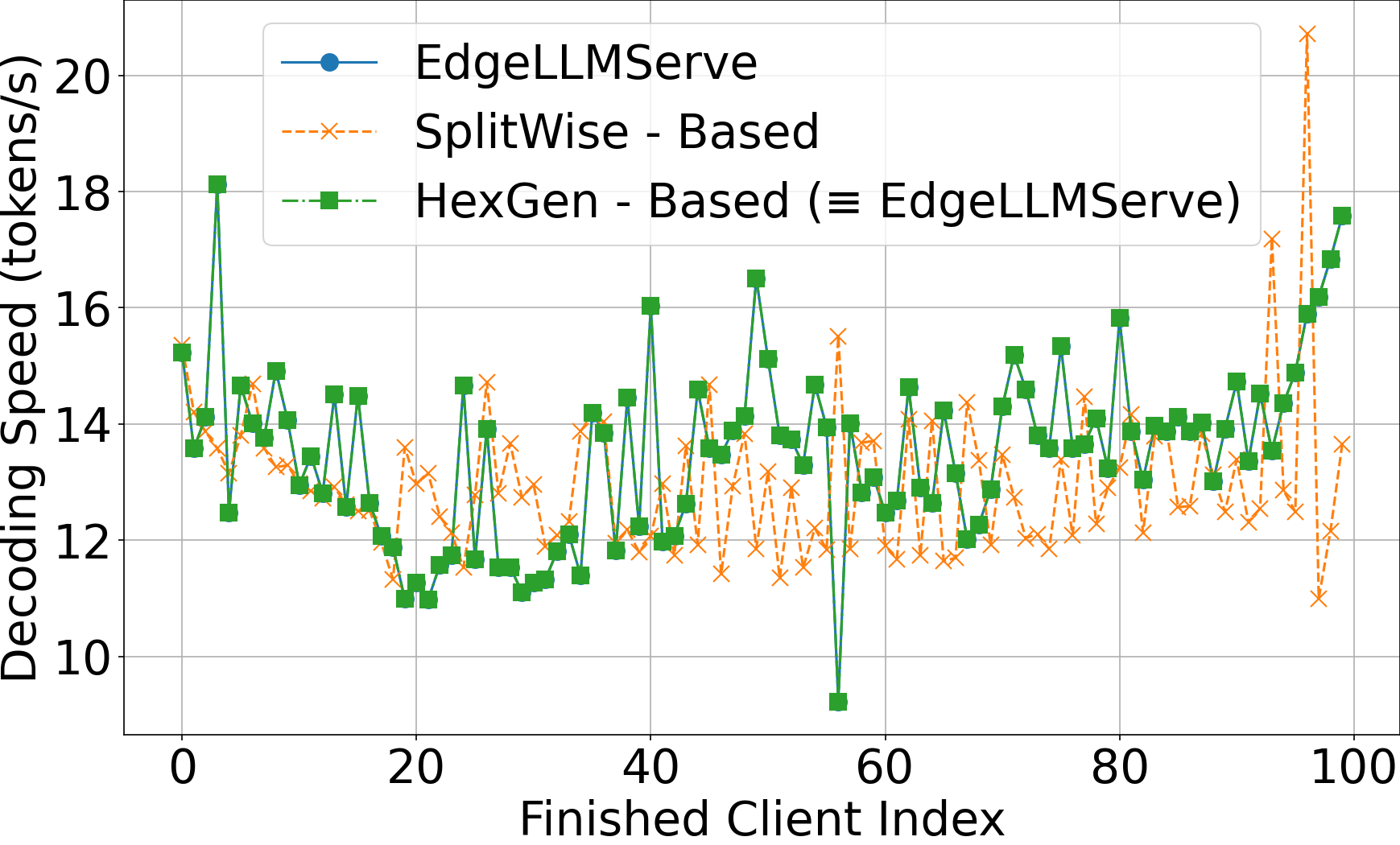}
        \caption{Decoding speed}
    \end{subfigure}\hfill
    \begin{subfigure}{\appfigscale\textwidth}
        \includegraphics[width=\linewidth]{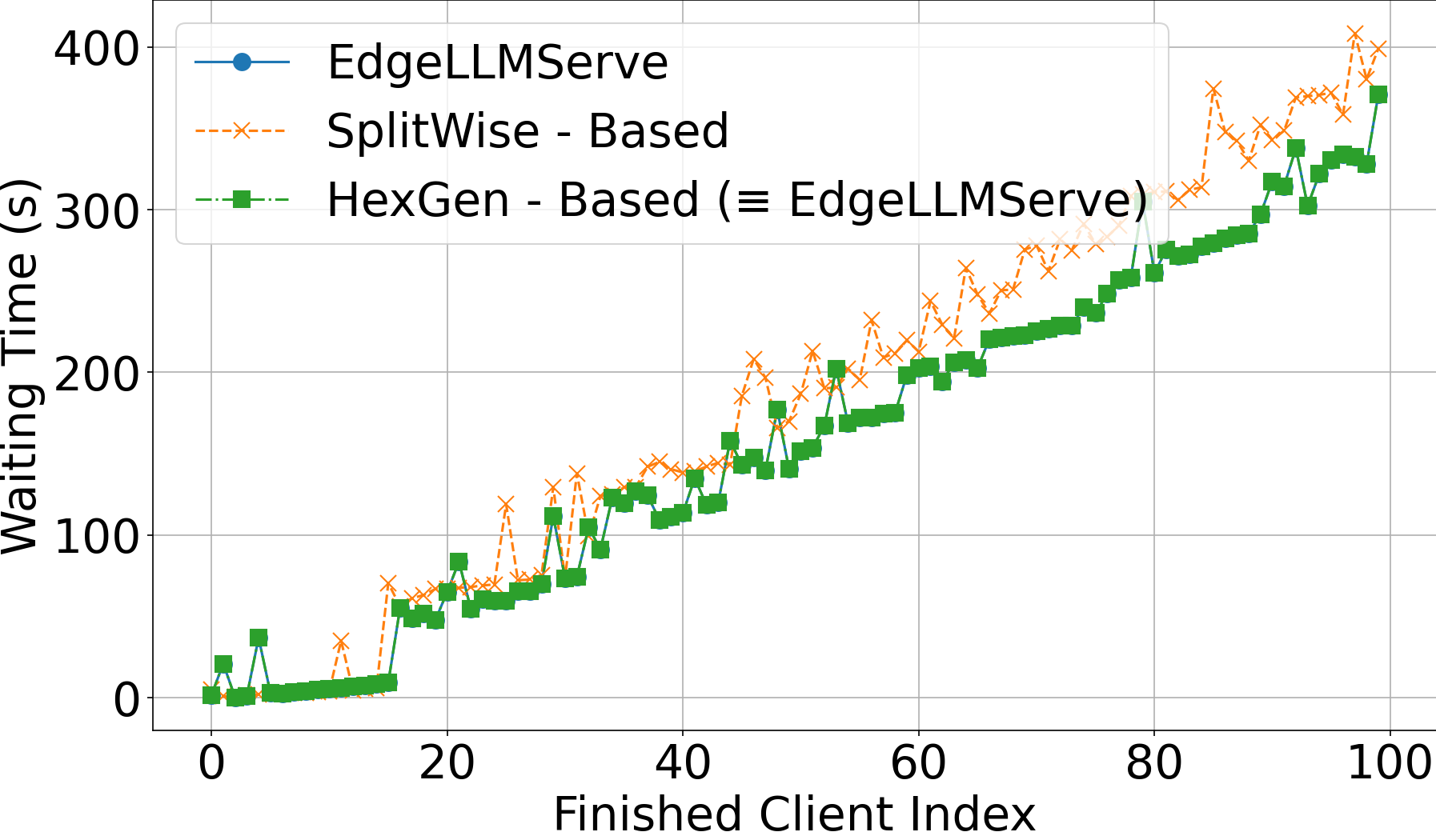}
        \caption{Waiting time}
    \end{subfigure}
    \caption{Per-request metrics --- Six devices, Qwen3-MoE-30B, LZByteDance, arrival $100$~ms.}
    \label{apd:f22}
\end{figure}

\paragraph{5. LZByteDance dataset. Arrival rate of 1000ms}\mbox{}
\begin{table}[H]
    \centering
    \setlength{\tabcolsep}{3pt}
    \caption{Deployment plans and metrics --- Six devices, Qwen3-MoE-30B, LZByteDance, arrival $1000$~ms.}
    \label{apd:t23}
    \begin{subtable}[t]{0.24\textwidth}
        \centering
        \caption{HG}
        \begin{tabular}{|c|c|c|c|}
            \hline
            Mode & Dev & Slots & G--C\\\hline
            P& Dev.4$^{\star}$ & 1 & 48--0\\\hline
            \multirow{3}{*}{D} & Dev.1$^{\star}$ & \multirow{3}{*}{12} & 16--5\\
              & Dev.2 &  & 6--3\\
              & Dev.7 &  & 18--0\\\hline
            \multirow{2}{*}{D} & Dev.5$^{\star}$ & \multirow{2}{*}{2} & 22--0\\
              & Dev.6 &  & 26--0\\\hline
        \end{tabular}
    \end{subtable}\hfill
    \begin{subtable}[t]{0.24\textwidth}
        \centering
        \caption{ELS}
        \begin{tabular}{|c|c|c|c|}
            \hline
            Mode & Dev & Slots & G--C\\\hline
            P& Dev.4$^{\star}$ & 1 & 48--0\\\hline
            \multirow{3}{*}{D} & Dev.1$^{\star}$ & \multirow{3}{*}{12} & 16--5\\
              & Dev.2 &  & 6--3\\
              & Dev.7 &  & 18--0\\\hline
            \multirow{2}{*}{D} & Dev.5$^{\star}$ & \multirow{2}{*}{2} & 22--0\\
              & Dev.6 &  & 26--0\\\hline
        \end{tabular}
    \end{subtable}\hfill
    \begin{subtable}[t]{0.24\textwidth}
        \centering
        \caption{SPW}
        \begin{tabular}{|c|c|c|c|}
            \hline
            Mode & Dev & Slots & G--C\\\hline
            P& Dev.1$^{\star}$ & 1 & 26--22\\\hline
            \multirow{3}{*}{D} & Dev.2$^{\star}$ & \multirow{3}{*}{6} & 6--10\\
              & Dev.6 &  & 10--0\\
              & Dev.7 &  & 22--0\\\hline
            \multirow{2}{*}{D} & Dev.4$^{\star}$ & \multirow{2}{*}{8} & 36--0\\
              & Dev.5 &  & 12--0\\\hline
        \end{tabular}
    \end{subtable}\hfill
    \begin{subtable}[t]{0.24\textwidth}
        \centering
        \caption{Metrics.}
        \begin{tabular}{|l|c|c|c|}
            \hline
            Method & PS & DS & W\\\hline
            HG & 562 & 13.7 & 228.2\\\hline
            \textbf{ELS} & \textbf{562} & \textbf{13.7} & \textbf{228.2}\\\hline
            SPW & 1147 & 12.9 & 269.2\\\hline
        \end{tabular}
    \end{subtable}
\end{table}
\begin{figure}[H]
    \centering
    \begin{subfigure}{\appfigscale\textwidth}
        \includegraphics[width=\linewidth]{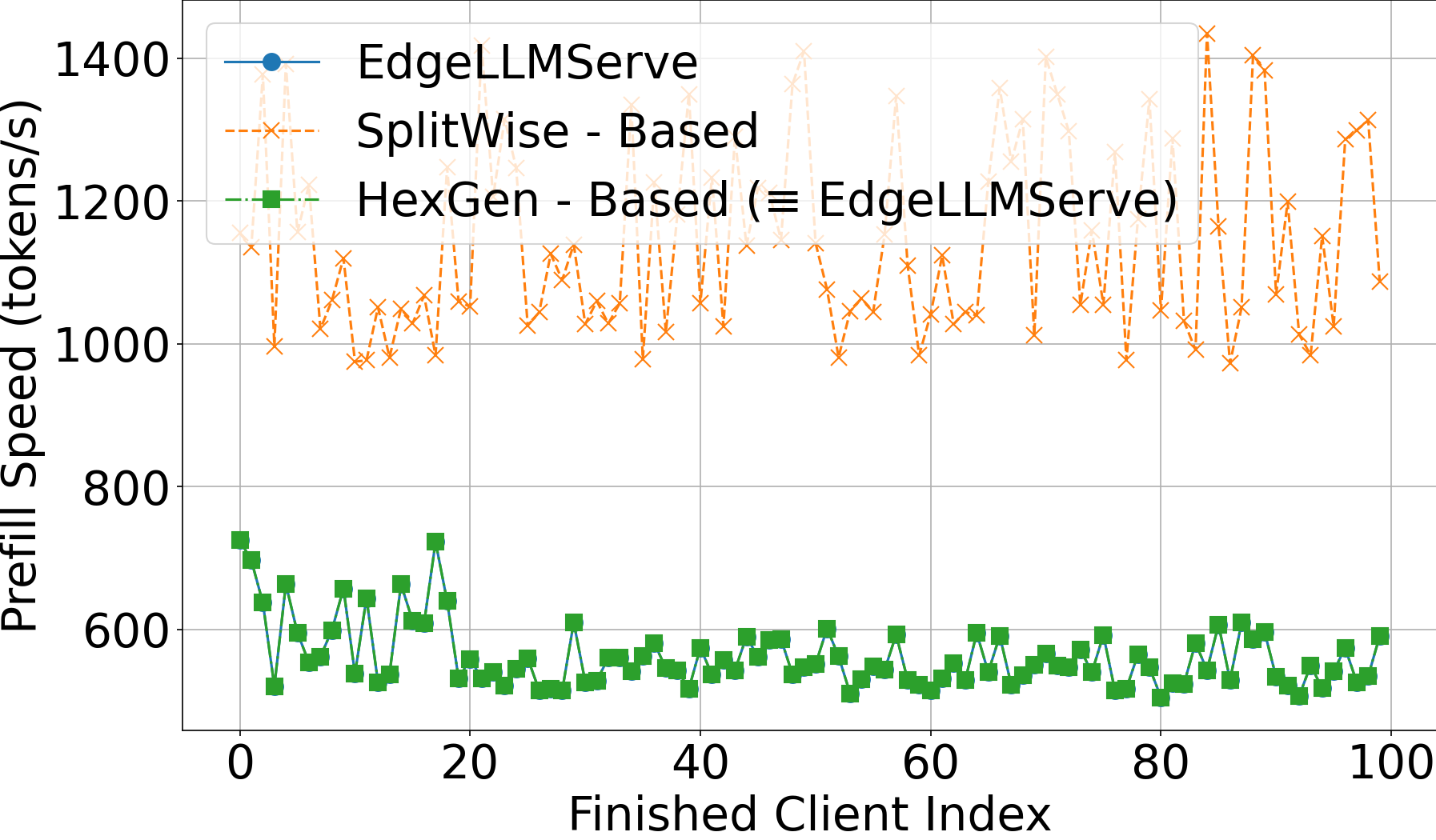}
        \caption{Prefill speed}
    \end{subfigure}\hfill
    \begin{subfigure}{\appfigscale\textwidth}
        \includegraphics[width=\linewidth]{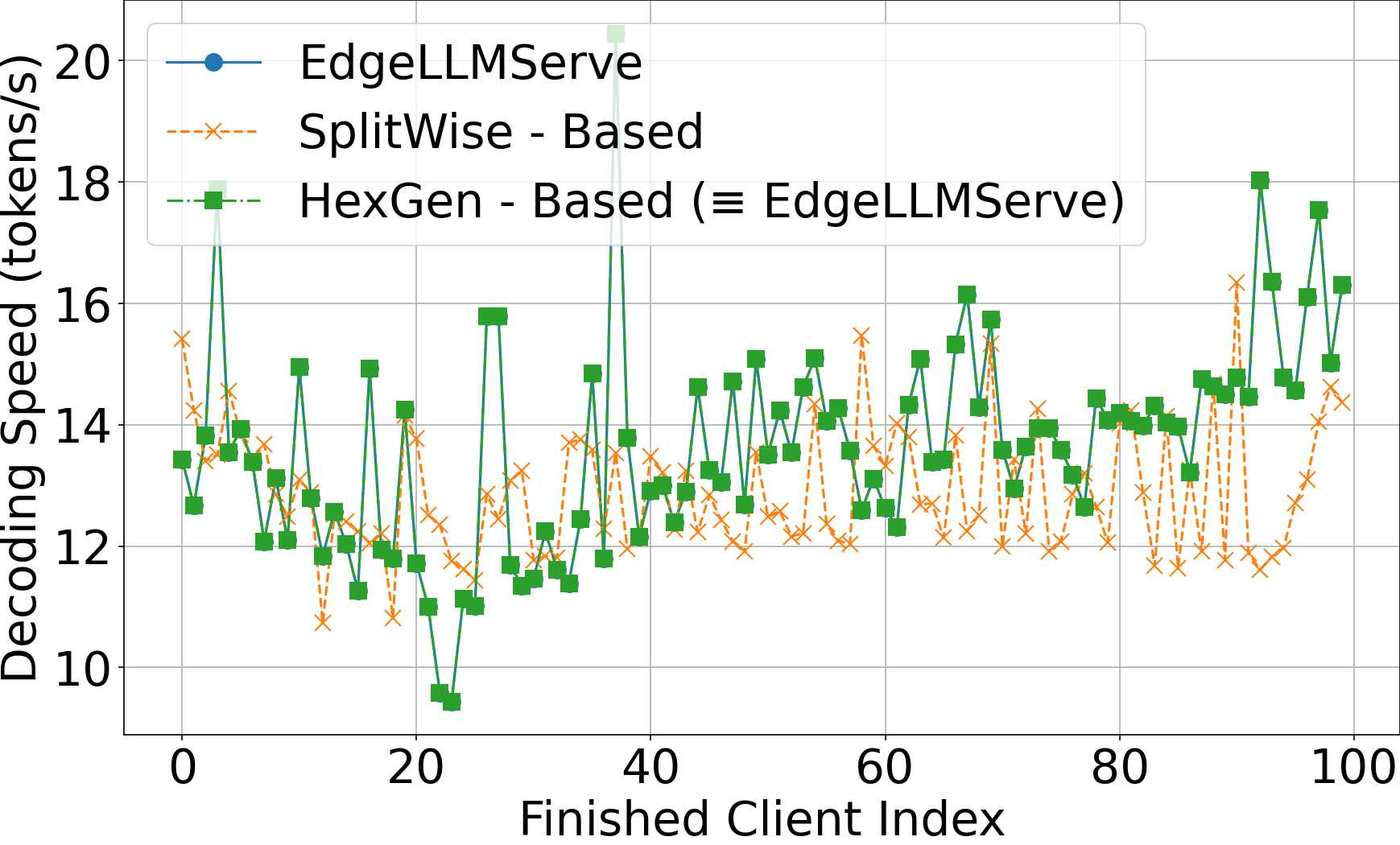}
        \caption{Decoding speed}
    \end{subfigure}\hfill
    \begin{subfigure}{\appfigscale\textwidth}
        \includegraphics[width=\linewidth]{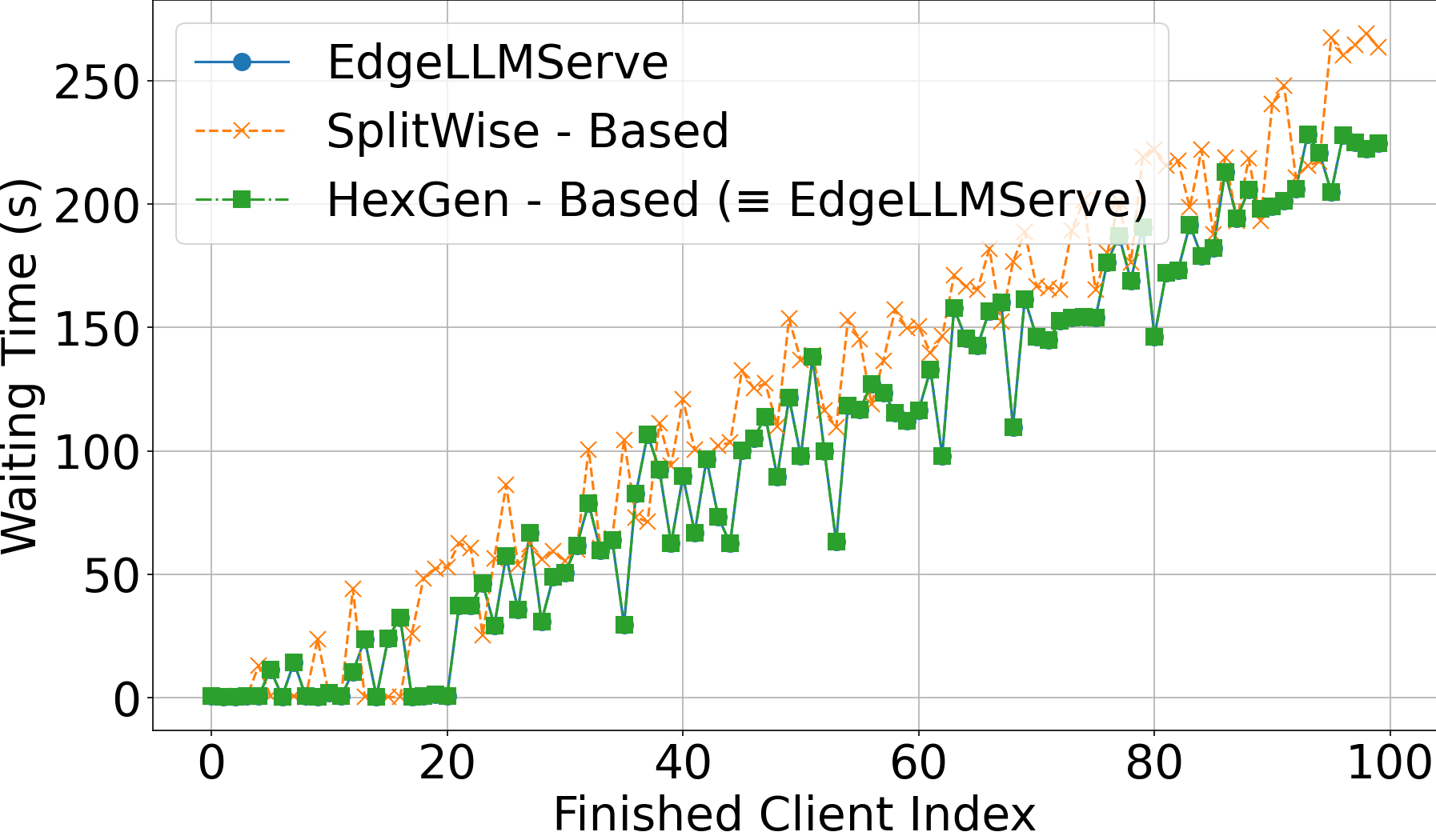}
        \caption{Waiting time}
    \end{subfigure}
    \caption{Per-request metrics --- Six devices, Qwen3-MoE-30B, LZByteDance, arrival $1000$~ms.}
    \label{apd:f23}
\end{figure}

\paragraph{6. LZByteDance dataset. Arrival rate of 3000ms}\mbox{}
\begin{table}[H]
    \centering
    \setlength{\tabcolsep}{3pt}
    \caption{Deployment plans and metrics --- Six devices, Qwen3-MoE-30B, LZByteDance, arrival $3000$~ms.}
    \label{apd:t24}
    \begin{subtable}[t]{0.24\textwidth}
        \centering
        \caption{HG}
        \begin{tabular}{|c|c|c|c|}
            \hline
            Mode & Dev & Slots & G--C\\\hline
            P& Dev.4$^{\star}$ & 1 & 48--0\\\hline
            \multirow{3}{*}{D} & Dev.1$^{\star}$ & \multirow{3}{*}{12} & 16--5\\
              & Dev.2 &  & 6--3\\
              & Dev.7 &  & 18--0\\\hline
            \multirow{2}{*}{D} & Dev.5$^{\star}$ & \multirow{2}{*}{2} & 22--0\\
              & Dev.6 &  & 26--0\\\hline
        \end{tabular}
    \end{subtable}\hfill
    \begin{subtable}[t]{0.24\textwidth}
        \centering
        \caption{ELS}
        \begin{tabular}{|c|c|c|c|}
            \hline
            Mode & Dev & Slots & G--C\\\hline
            P& Dev.2$^{\star}$ & 1 & 9--39\\\hline
            \multirow{3}{*}{D} & Dev.1$^{\star}$ & \multirow{3}{*}{15} & 14--6\\
              & Dev.4 &  & 11--0\\
              & Dev.7 &  & 17--0\\\hline
            \multirow{2}{*}{D} & Dev.5$^{\star}$ & \multirow{2}{*}{2} & 22--0\\
              & Dev.6 &  & 26--0\\\hline
        \end{tabular}
    \end{subtable}\hfill
    \begin{subtable}[t]{0.24\textwidth}
        \centering
        \caption{SPW}
        \begin{tabular}{|c|c|c|c|}
            \hline
            Mode & Dev & Slots & G--C\\\hline
            P& Dev.1$^{\star}$ & 1 & 26--22\\\hline
            \multirow{3}{*}{D} & Dev.2$^{\star}$ & \multirow{3}{*}{6} & 6--10\\
              & Dev.6 &  & 10--0\\
              & Dev.7 &  & 22--0\\\hline
            \multirow{2}{*}{D} & Dev.4$^{\star}$ & \multirow{2}{*}{8} & 36--0\\
              & Dev.5 &  & 12--0\\\hline
        \end{tabular}
    \end{subtable}\hfill
    \begin{subtable}[t]{0.24\textwidth}
        \centering
        \caption{Metrics.}
        \begin{tabular}{|l|c|c|c|}
            \hline
            Method & PS & DS & W\\\hline
            HG & 563 & 14.4 & 102.2\\\hline
            \textbf{ELS} & \textbf{206} & \textbf{12.2} & \textbf{99.8}\\\hline
            SPW & 1147 & 13.0 & 129.6\\\hline
        \end{tabular}
    \end{subtable}
\end{table}
\begin{figure}[H]
    \centering
    \begin{subfigure}{\appfigscale\textwidth}
        \includegraphics[width=\linewidth]{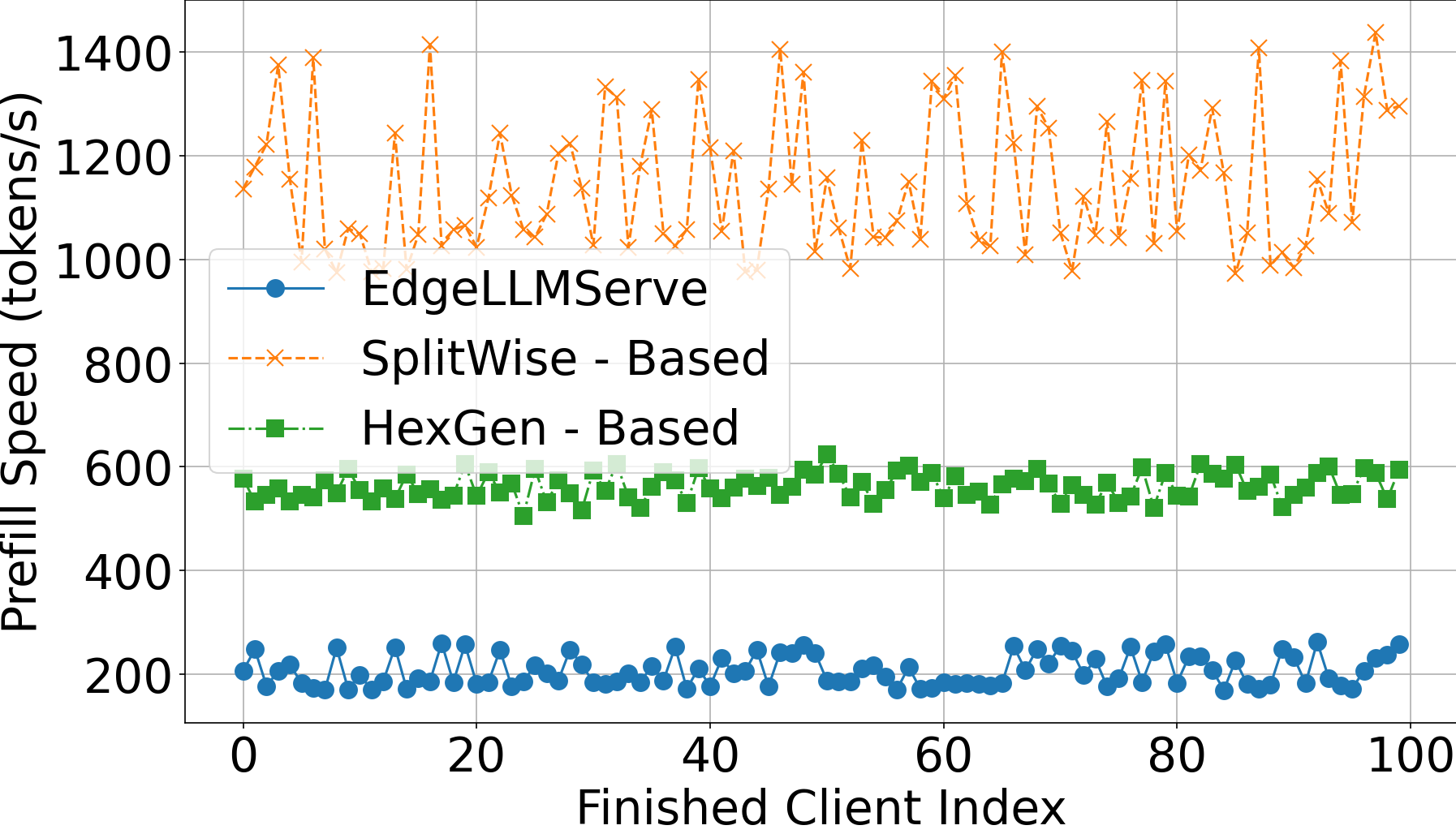}
        \caption{Prefill speed}
    \end{subfigure}\hfill
    \begin{subfigure}{\appfigscale\textwidth}
        \includegraphics[width=\linewidth]{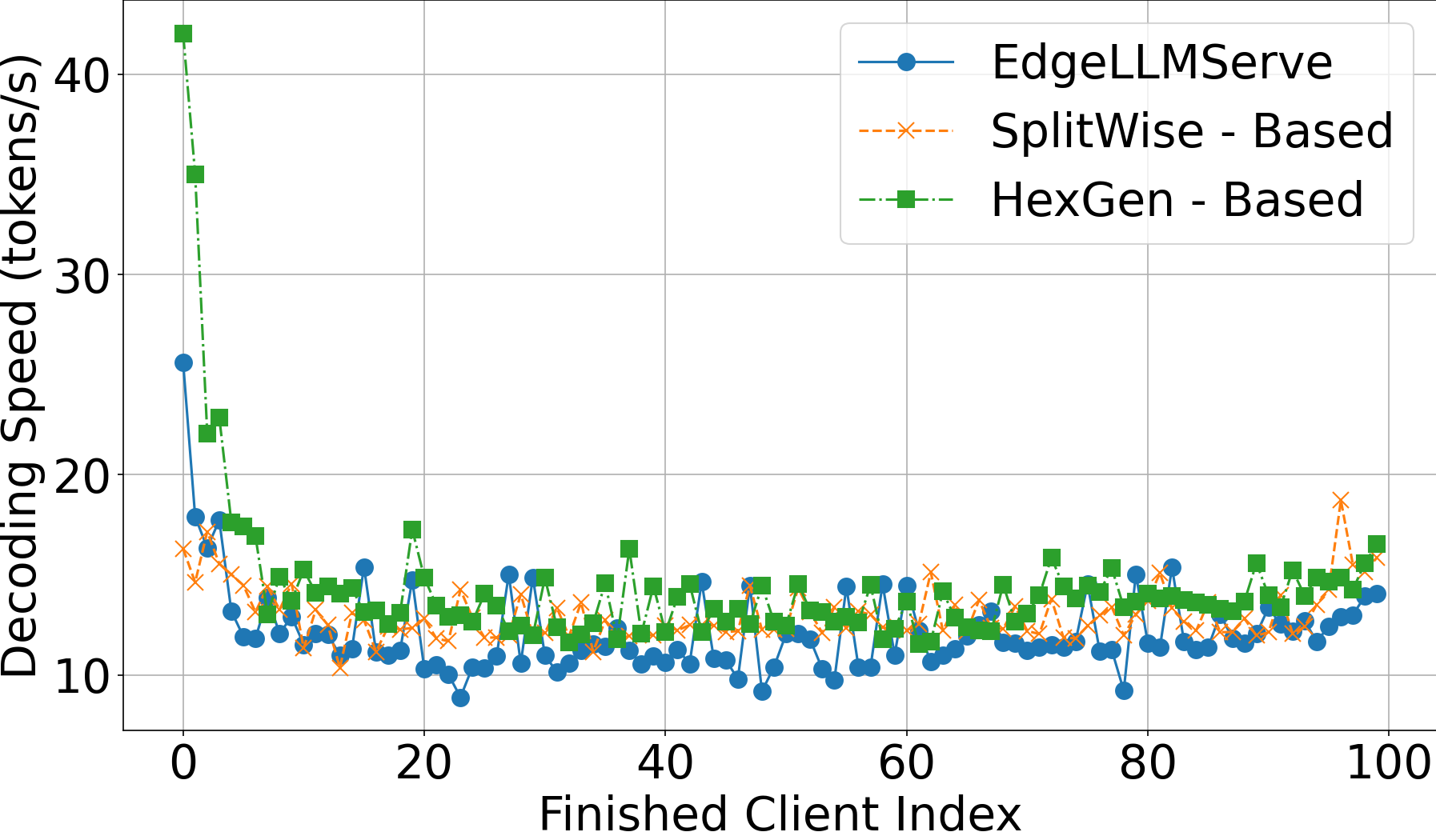}
        \caption{Decoding speed}
    \end{subfigure}\hfill
    \begin{subfigure}{\appfigscale\textwidth}
        \includegraphics[width=\linewidth]{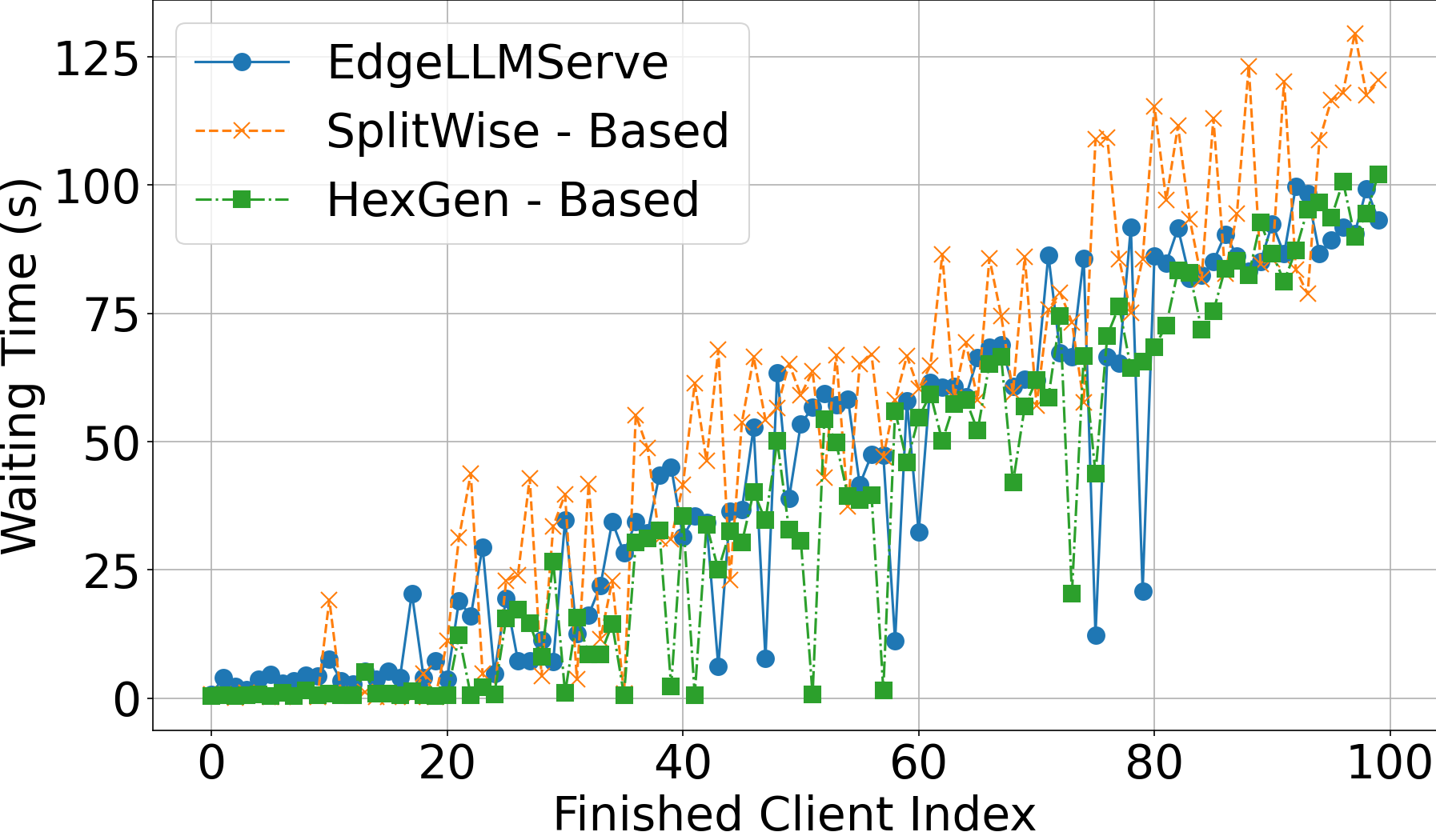}
        \caption{Waiting time}
    \end{subfigure}
    \caption{Per-request metrics --- Six devices, Qwen3-MoE-30B, LZByteDance, arrival $3000$~ms.}
    \label{apd:f24}
\end{figure}

\subsection{Eight devices}
\subsubsection{Gpt-Oss-20b}
\paragraph{1. IELTS dataset. Arrival rate of 100ms}\mbox{}
\vspace{-0.3cm}
\begin{table}[H]
    \centering
    \setlength{\tabcolsep}{3pt}
    \caption{Deployment plans and metrics --- Eight devices, Gpt-Oss-20b, IELTS, arrival $100$~ms.}
    \vspace{-0.3cm}
    \label{apd:t25}
    \begin{subtable}[t]{0.24\textwidth}
        \centering
        \caption{HG}
        \begin{tabular}{|c|c|c|c|}
            \hline
            Mode & Dev & Slots & G--C\\\hline
            P& Dev.1$^{\star}$ & 1 & 20--4\\\hline
            \multirow{2}{*}{D} & Dev.2$^{\star}$ & \multirow{2}{*}{8} & 5--4\\
              & Dev.4 &  & 15--0\\\hline
            D& Dev.3$^{\star}$ & 5 & 24--0\\\hline
            \multirow{2}{*}{D} & Dev.5 & \multirow{2}{*}{2} & 15--0\\
              & Dev.6$^{\star}$ &  & 9--0\\\hline
            \multirow{2}{*}{D} & Dev.7 & \multirow{2}{*}{30} & 13--0\\
              & Dev.8$^{\star}$ &  & 11--0\\\hline
        \end{tabular}
    \end{subtable}\hfill
    \begin{subtable}[t]{0.24\textwidth}
        \centering
        \caption{ELS}
        \begin{tabular}{|c|c|c|c|}
            \hline
            Mode & Dev & Slots & G--C\\\hline
            P& Dev.1$^{\star}$ & 1 & 20--4\\\hline
            \multirow{2}{*}{D} & Dev.2$^{\star}$ & \multirow{2}{*}{8} & 5--4\\
              & Dev.4 &  & 15--0\\\hline
            D& Dev.3$^{\star}$ & 5 & 24--0\\\hline
            \multirow{2}{*}{D} & Dev.5 & \multirow{2}{*}{2} & 15--0\\
              & Dev.6$^{\star}$ &  & 9--0\\\hline
            \multirow{2}{*}{D} & Dev.7 & \multirow{2}{*}{30} & 13--0\\
              & Dev.8$^{\star}$ &  & 11--0\\\hline
        \end{tabular}
    \end{subtable}\hfill
    \begin{subtable}[t]{0.24\textwidth}
        \centering
        \caption{SPW}
        \begin{tabular}{|c|c|c|c|}
            \hline
            Mode & Dev & Slots & G--C\\\hline
            P& Dev.1$^{\star}$ & 1 & 20--4\\\hline
            \multirow{2}{*}{D} & Dev.2$^{\star}$ & \multirow{2}{*}{8} & 5--4\\
              & Dev.4 &  & 15--0\\\hline
            D& Dev.3$^{\star}$ & 5 & 24--0\\\hline
            \multirow{2}{*}{D} & Dev.5 & \multirow{2}{*}{2} & 15--0\\
              & Dev.6$^{\star}$ &  & 9--0\\\hline
            \multirow{2}{*}{D} & Dev.7 & \multirow{2}{*}{30} & 13--0\\
              & Dev.8$^{\star}$ &  & 11--0\\\hline
        \end{tabular}
    \end{subtable}\hfill
    \begin{subtable}[t]{0.24\textwidth}
        \centering
        \caption{Metrics.}
        \begin{tabular}{|l|c|c|c|}
            \hline
            Method & PS & DS & W\\\hline
            HG & 2913 & 11.5 & 200.9\\\hline
            \textbf{ELS} & \textbf{2913} & \textbf{11.5} & \textbf{200.9}\\\hline
            SPW & 2913 & 11.5 & 200.9\\\hline
        \end{tabular}
    \end{subtable}
\end{table}
\vspace{-0.3cm}
\begin{figure}[H]
    \centering
    \begin{subfigure}{\appfigscale\textwidth}
        \includegraphics[width=\linewidth]{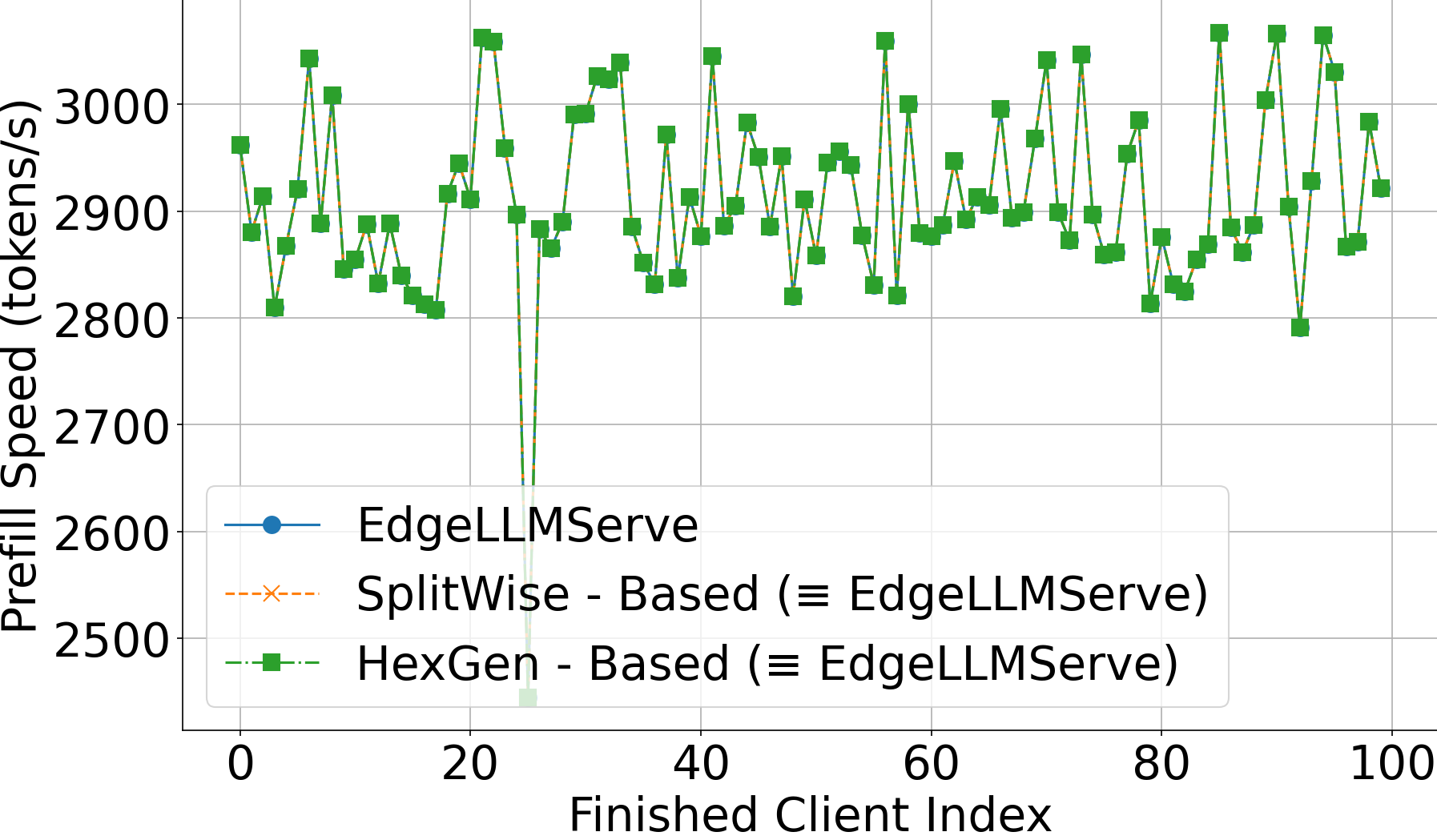}
        \caption{Prefill speed}
    \end{subfigure}\hfill
    \begin{subfigure}{\appfigscale\textwidth}
        \includegraphics[width=\linewidth]{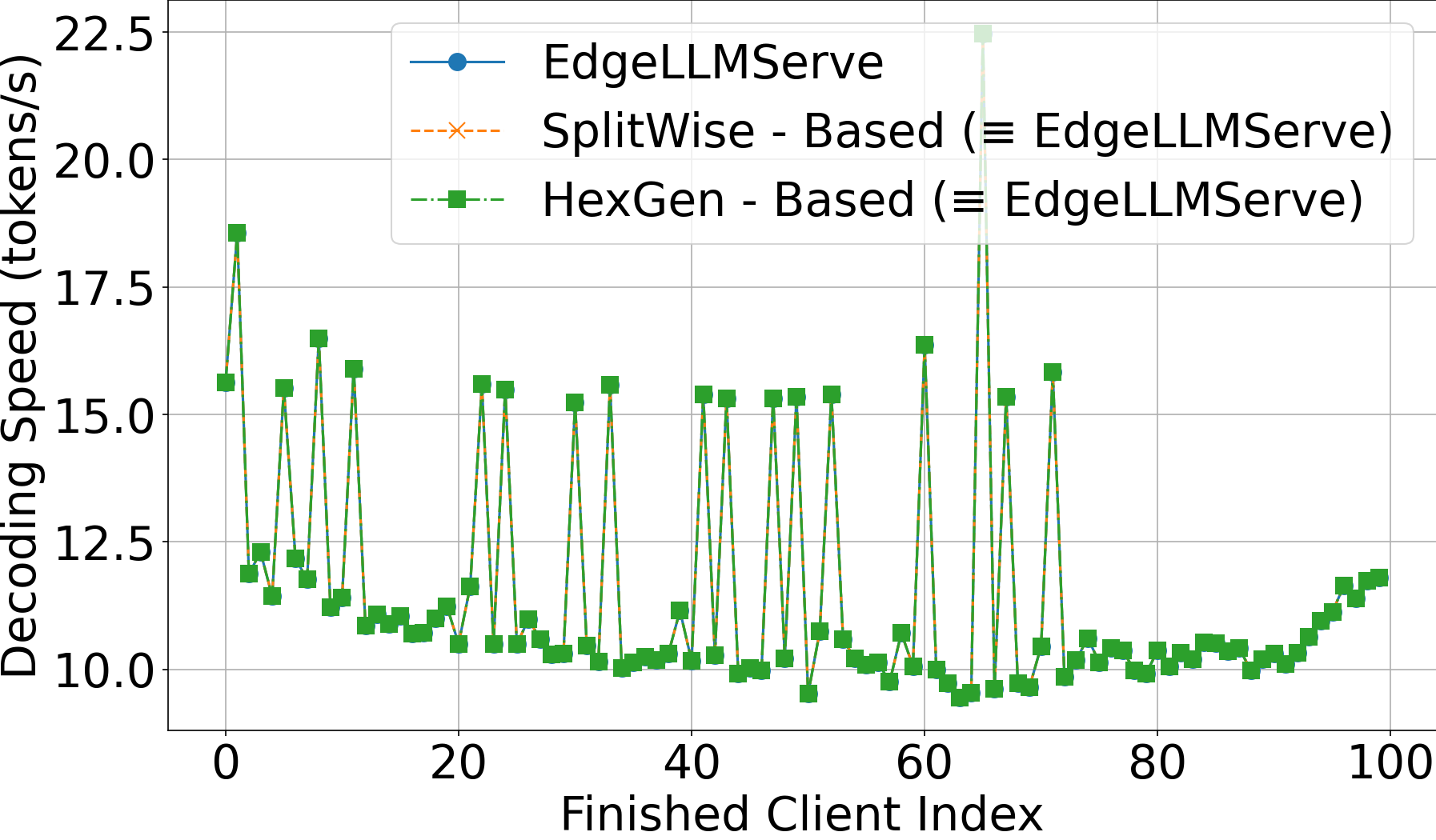}
        \caption{Decoding speed}
    \end{subfigure}\hfill
    \begin{subfigure}{\appfigscale\textwidth}
        \includegraphics[width=\linewidth]{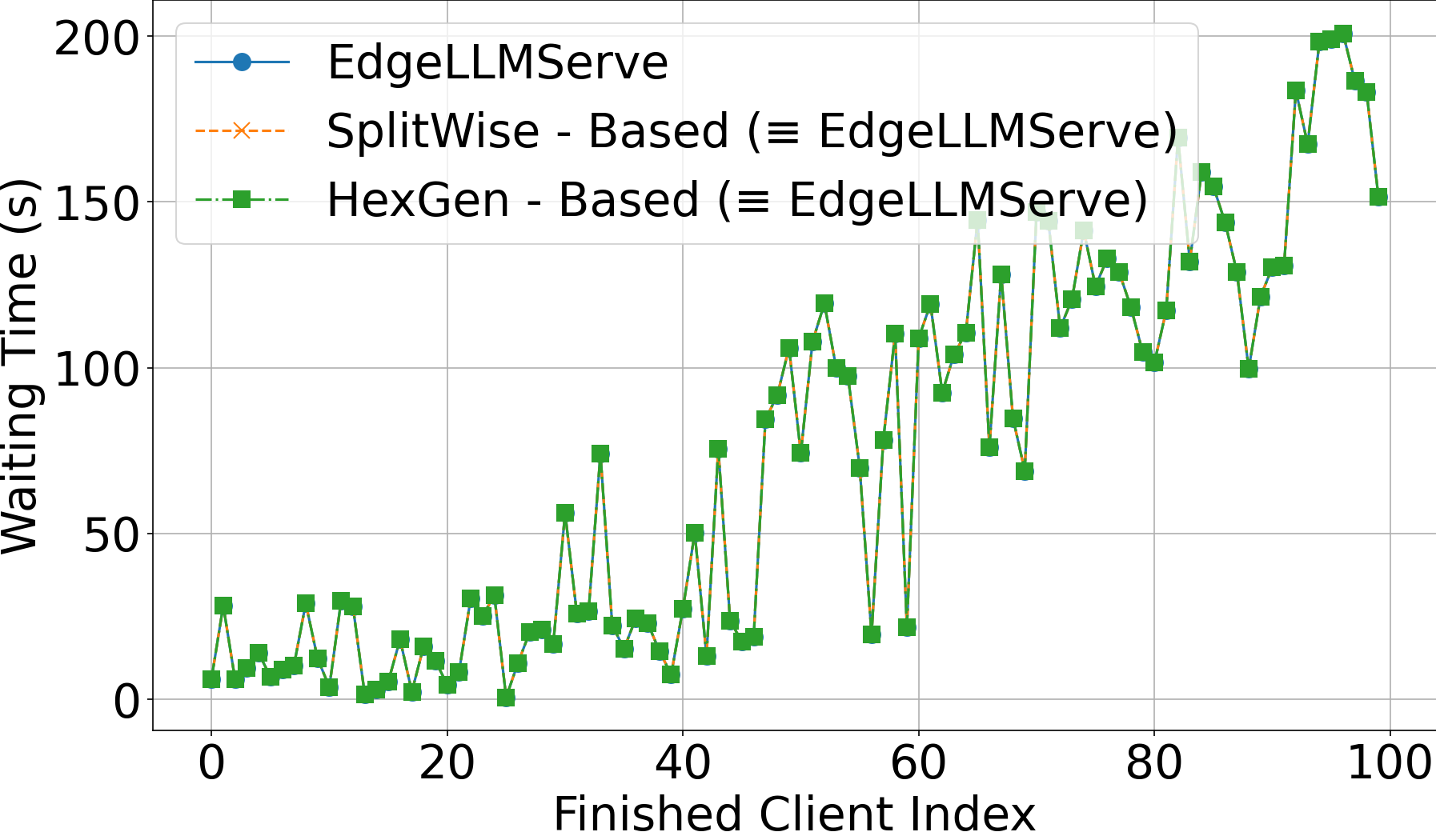}
        \caption{Waiting time}
    \end{subfigure}
    \vspace{-0.3cm}
    \caption{Per-request metrics --- Eight devices, Gpt-Oss-20b, IELTS, arrival $100$~ms.}
    \label{apd:f25}
\end{figure}

\vspace{-0.3cm}
\paragraph{2. IELTS dataset. Arrival rate of 1000ms}\mbox{}
\vspace{-0.3cm}
\begin{table}[H]
    \centering
    \setlength{\tabcolsep}{3pt}
    \caption{Deployment plans and metrics --- Eight devices, Gpt-Oss-20b, IELTS, arrival $1000$~ms.}
    \vspace{-0.3cm}
    \label{apd:t26}
    \begin{subtable}[t]{0.24\textwidth}
        \centering
        \caption{HG}
        \begin{tabular}{|c|c|c|c|}
            \hline
            Mode & Dev & Slots & G--C\\\hline
            P& Dev.1$^{\star}$ & 1 & 20--4\\\hline
            \multirow{2}{*}{D} & Dev.2$^{\star}$ & \multirow{2}{*}{8} & 5--4\\
              & Dev.4 &  & 15--0\\\hline
            D& Dev.3$^{\star}$ & 5 & 24--0\\\hline
            \multirow{2}{*}{D} & Dev.5 & \multirow{2}{*}{2} & 15--0\\
              & Dev.6$^{\star}$ &  & 9--0\\\hline
            \multirow{2}{*}{D} & Dev.7 & \multirow{2}{*}{30} & 13--0\\
              & Dev.8$^{\star}$ &  & 11--0\\\hline
        \end{tabular}
    \end{subtable}\hfill
    \begin{subtable}[t]{0.24\textwidth}
        \centering
        \caption{ELS}
        \begin{tabular}{|c|c|c|c|}
            \hline
            Mode & Dev & Slots & G--C\\\hline
            P& Dev.1$^{\star}$ & 1 & 20--4\\\hline
            \multirow{2}{*}{D} & Dev.2$^{\star}$ & \multirow{2}{*}{8} & 5--4\\
              & Dev.4 &  & 15--0\\\hline
            D& Dev.3$^{\star}$ & 5 & 24--0\\\hline
            \multirow{2}{*}{D} & Dev.5 & \multirow{2}{*}{2} & 15--0\\
              & Dev.6$^{\star}$ &  & 9--0\\\hline
            \multirow{2}{*}{D} & Dev.7 & \multirow{2}{*}{30} & 13--0\\
              & Dev.8$^{\star}$ &  & 11--0\\\hline
        \end{tabular}
    \end{subtable}\hfill
    \begin{subtable}[t]{0.24\textwidth}
        \centering
        \caption{SPW}
        \begin{tabular}{|c|c|c|c|}
            \hline
            Mode & Dev & Slots & G--C\\\hline
            P& Dev.1$^{\star}$ & 1 & 20--4\\\hline
            \multirow{2}{*}{D} & Dev.2$^{\star}$ & \multirow{2}{*}{8} & 5--4\\
              & Dev.4 &  & 15--0\\\hline
            D& Dev.3$^{\star}$ & 5 & 24--0\\\hline
            \multirow{2}{*}{D} & Dev.5 & \multirow{2}{*}{2} & 15--0\\
              & Dev.6$^{\star}$ &  & 9--0\\\hline
            \multirow{2}{*}{D} & Dev.7 & \multirow{2}{*}{30} & 13--0\\
              & Dev.8$^{\star}$ &  & 11--0\\\hline
        \end{tabular}
    \end{subtable}\hfill
    \begin{subtable}[t]{0.24\textwidth}
        \centering
        \caption{Metrics.}
        \begin{tabular}{|l|c|c|c|}
            \hline
            Method & PS & DS & W\\\hline
            HG & 2915 & 11.8 & 130.4\\\hline
            \textbf{ELS} & \textbf{2915} & \textbf{11.8} & \textbf{130.4}\\\hline
            SPW & 2915 & 11.8 & 130.4\\\hline
        \end{tabular}
    \end{subtable}
\end{table}
\vspace{-0.3cm}
\begin{figure}[H]
    \centering
    \begin{subfigure}{\appfigscale\textwidth}
        \includegraphics[width=\linewidth]{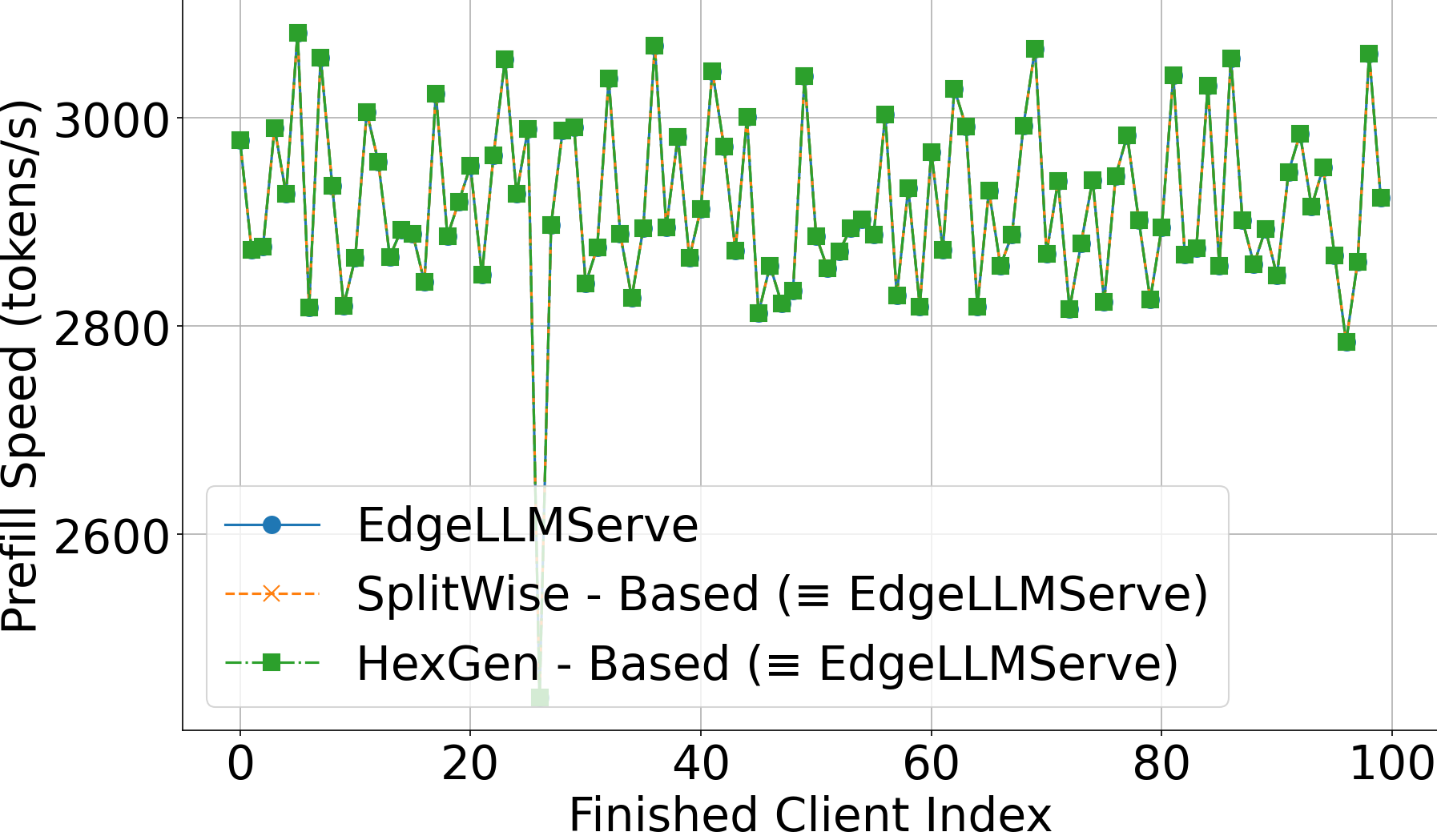}
        \caption{Prefill speed}
    \end{subfigure}\hfill
    \begin{subfigure}{\appfigscale\textwidth}
        \includegraphics[width=\linewidth]{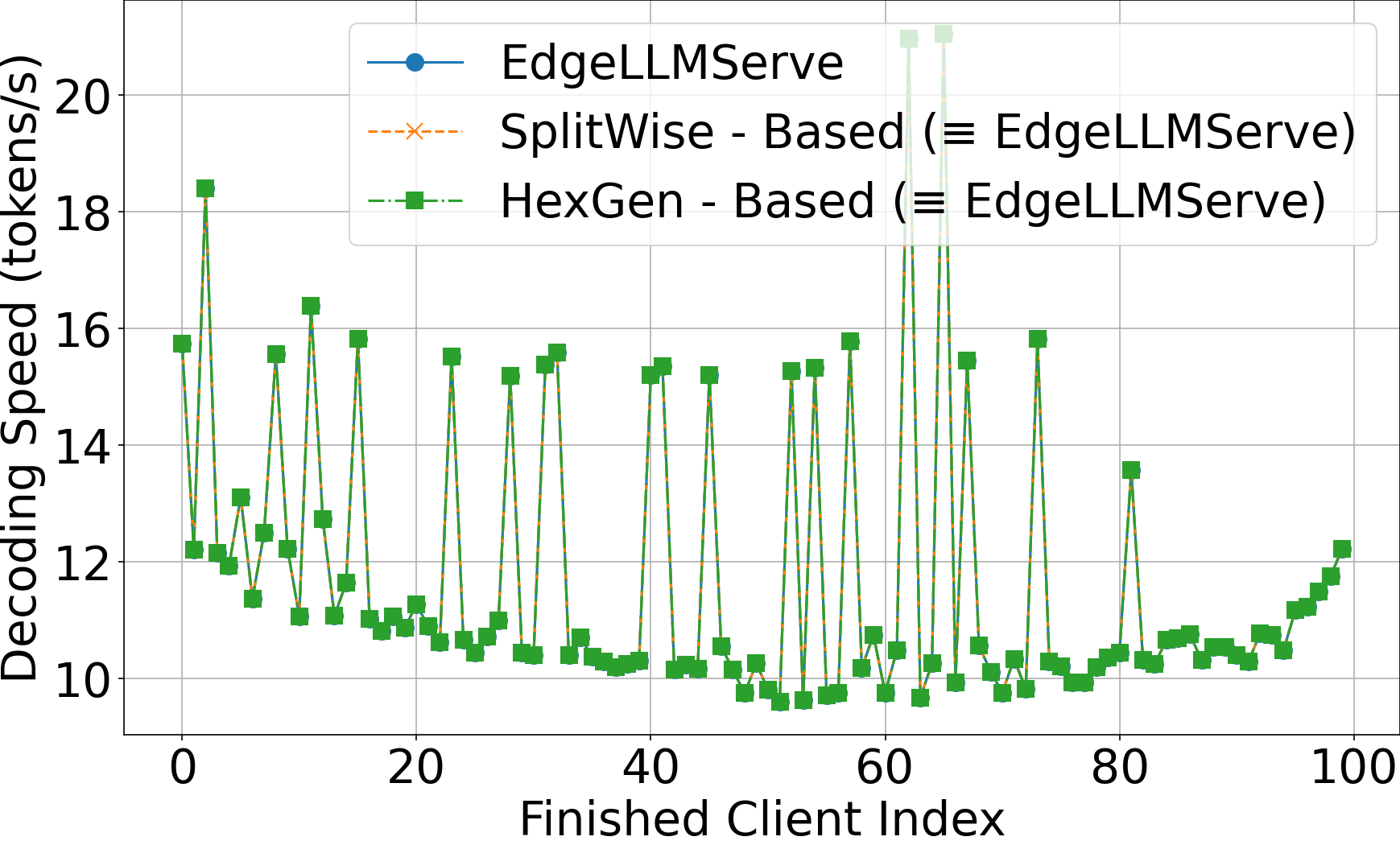}
        \caption{Decoding speed}
    \end{subfigure}\hfill
    \begin{subfigure}{\appfigscale\textwidth}
        \includegraphics[width=\linewidth]{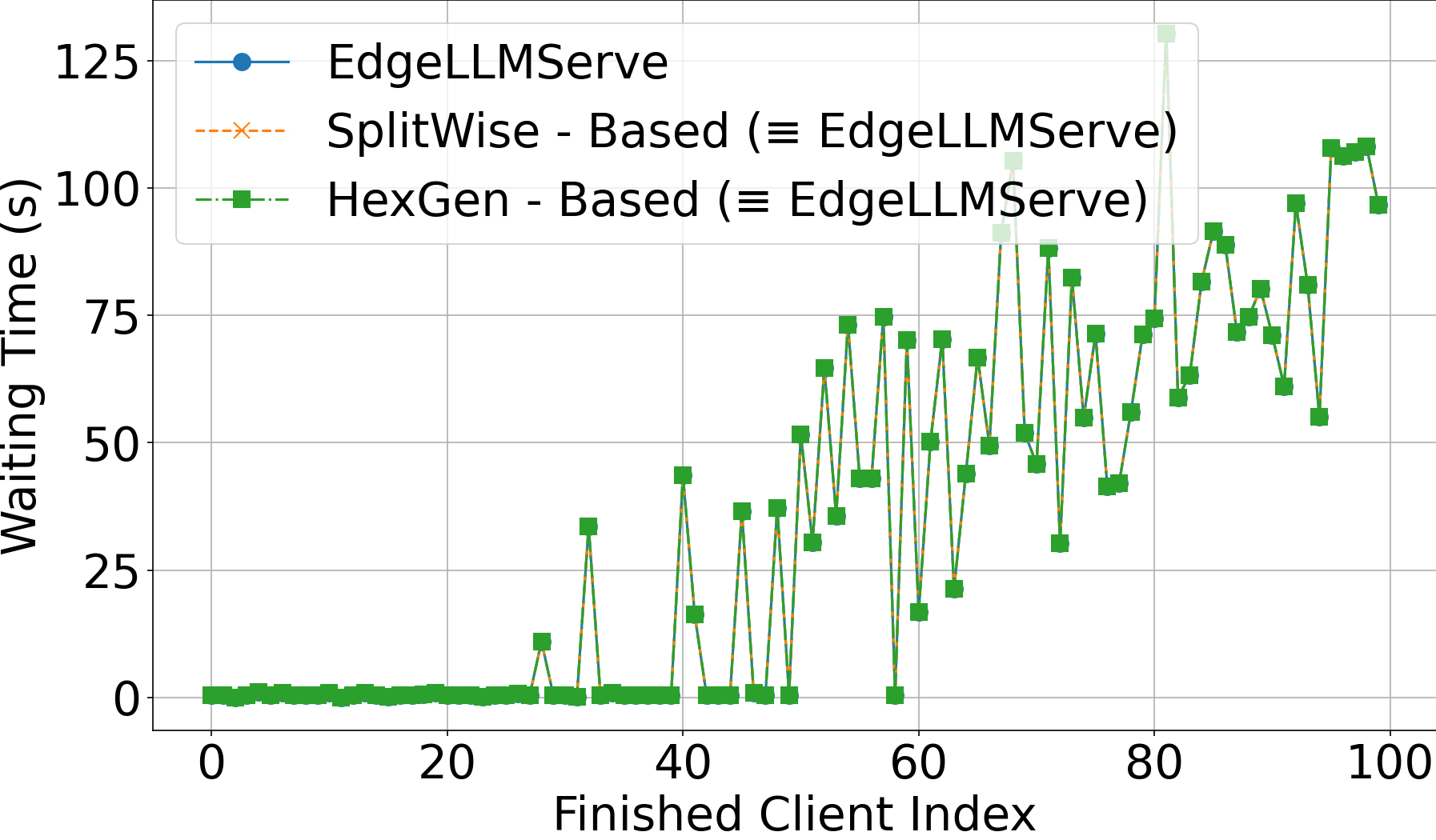}
        \caption{Waiting time}
    \end{subfigure}
    \vspace{-0.3cm}
    \caption{Per-request metrics --- Eight devices, Gpt-Oss-20b, IELTS, arrival $1000$~ms.}
    \label{apd:f26}
\end{figure}

\paragraph{3. IELTS dataset. Arrival rate of 3000ms}\mbox{}
\vspace{-0.3cm}
\begin{table}[H]
    \centering
    \setlength{\tabcolsep}{3pt}
    \caption{Deployment plans and metrics --- Eight devices, Gpt-Oss-20b, IELTS, arrival $3000$~ms.}
    \vspace{-0.3cm}
    \label{apd:t27}
    \begin{subtable}[t]{0.24\textwidth}
        \centering
        \caption{HG}
        \begin{tabular}{|c|c|c|c|}
            \hline
            Mode & Dev & Slots & G--C\\\hline
            P& Dev.1$^{\star}$ & 1 & 20--4\\\hline
            \multirow{2}{*}{D} & Dev.2$^{\star}$ & \multirow{2}{*}{8} & 5--4\\
              & Dev.4 &  & 15--0\\\hline
            D& Dev.3$^{\star}$ & 5 & 24--0\\\hline
            \multirow{2}{*}{D} & Dev.5 & \multirow{2}{*}{2} & 15--0\\
              & Dev.6$^{\star}$ &  & 9--0\\\hline
            \multirow{2}{*}{D} & Dev.7 & \multirow{2}{*}{30} & 13--0\\
              & Dev.8$^{\star}$ &  & 11--0\\\hline
        \end{tabular}
    \end{subtable}\hfill
    \begin{subtable}[t]{0.24\textwidth}
        \centering
        \caption{ELS}
        \begin{tabular}{|c|c|c|c|}
            \hline
            Mode & Dev & Slots & G--C\\\hline
            P& Dev.3$^{\star}$ & 1 & 24--0\\\hline
            \multirow{2}{*}{D} & Dev.1 & \multirow{2}{*}{14} & 17--1\\
              & Dev.2$^{\star}$ &  & 5--1\\\hline
            D& Dev.4$^{\star}$ & 6 & 24--0\\\hline
            \multirow{2}{*}{D} & Dev.5 & \multirow{2}{*}{2} & 15--0\\
              & Dev.6$^{\star}$ &  & 9--0\\\hline
            \multirow{2}{*}{D} & Dev.7 & \multirow{2}{*}{30} & 13--0\\
              & Dev.8$^{\star}$ &  & 11--0\\\hline
        \end{tabular}
    \end{subtable}\hfill
    \begin{subtable}[t]{0.24\textwidth}
        \centering
        \caption{SPW}
        \begin{tabular}{|c|c|c|c|}
            \hline
            Mode & Dev & Slots & G--C\\\hline
            P& Dev.1$^{\star}$ & 1 & 20--4\\\hline
            \multirow{2}{*}{D} & Dev.2$^{\star}$ & \multirow{2}{*}{8} & 5--4\\
              & Dev.4 &  & 15--0\\\hline
            D& Dev.3$^{\star}$ & 5 & 24--0\\\hline
            \multirow{2}{*}{D} & Dev.5 & \multirow{2}{*}{2} & 15--0\\
              & Dev.6$^{\star}$ &  & 9--0\\\hline
            \multirow{2}{*}{D} & Dev.7 & \multirow{2}{*}{30} & 13--0\\
              & Dev.8$^{\star}$ &  & 11--0\\\hline
        \end{tabular}
    \end{subtable}\hfill
    \begin{subtable}[t]{0.24\textwidth}
        \centering
        \caption{Metrics.}
        \begin{tabular}{|l|c|c|c|}
            \hline
            Method & PS & DS & W\\\hline
            HG & 2924 & 12.5 & 2.1\\\hline
            \textbf{ELS} & \textbf{768} & \textbf{15.7} & \textbf{1.9}\\\hline
            SPW & 2924 & 12.5 & 2.1\\\hline
        \end{tabular}
    \end{subtable}
\end{table}
\vspace{-0.3cm}
\begin{figure}[H]
    \centering
    \begin{subfigure}{\appfigscale\textwidth}
        \includegraphics[width=\linewidth]{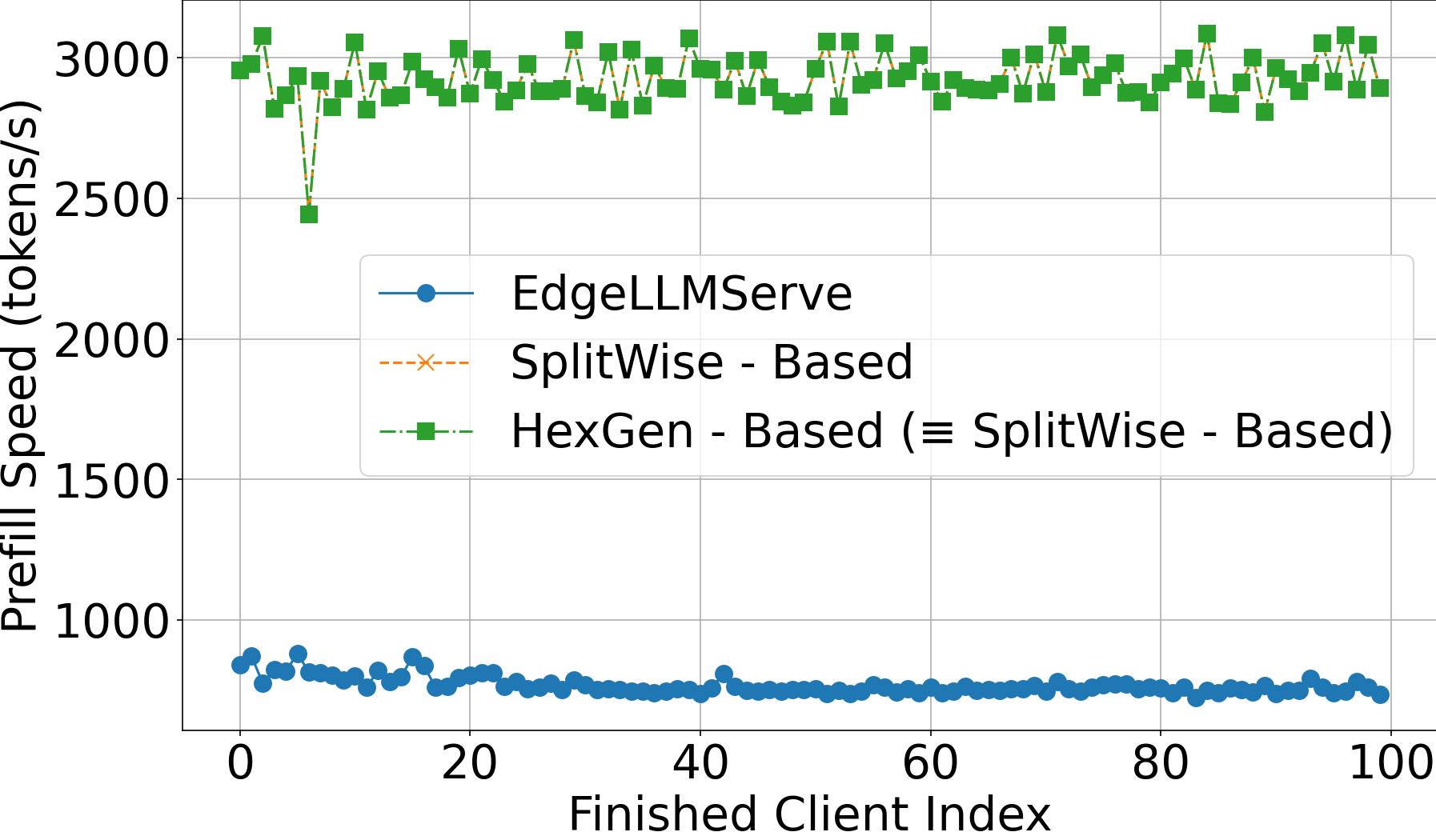}
        \caption{Prefill speed}
    \end{subfigure}\hfill
    \begin{subfigure}{\appfigscale\textwidth}
        \includegraphics[width=\linewidth]{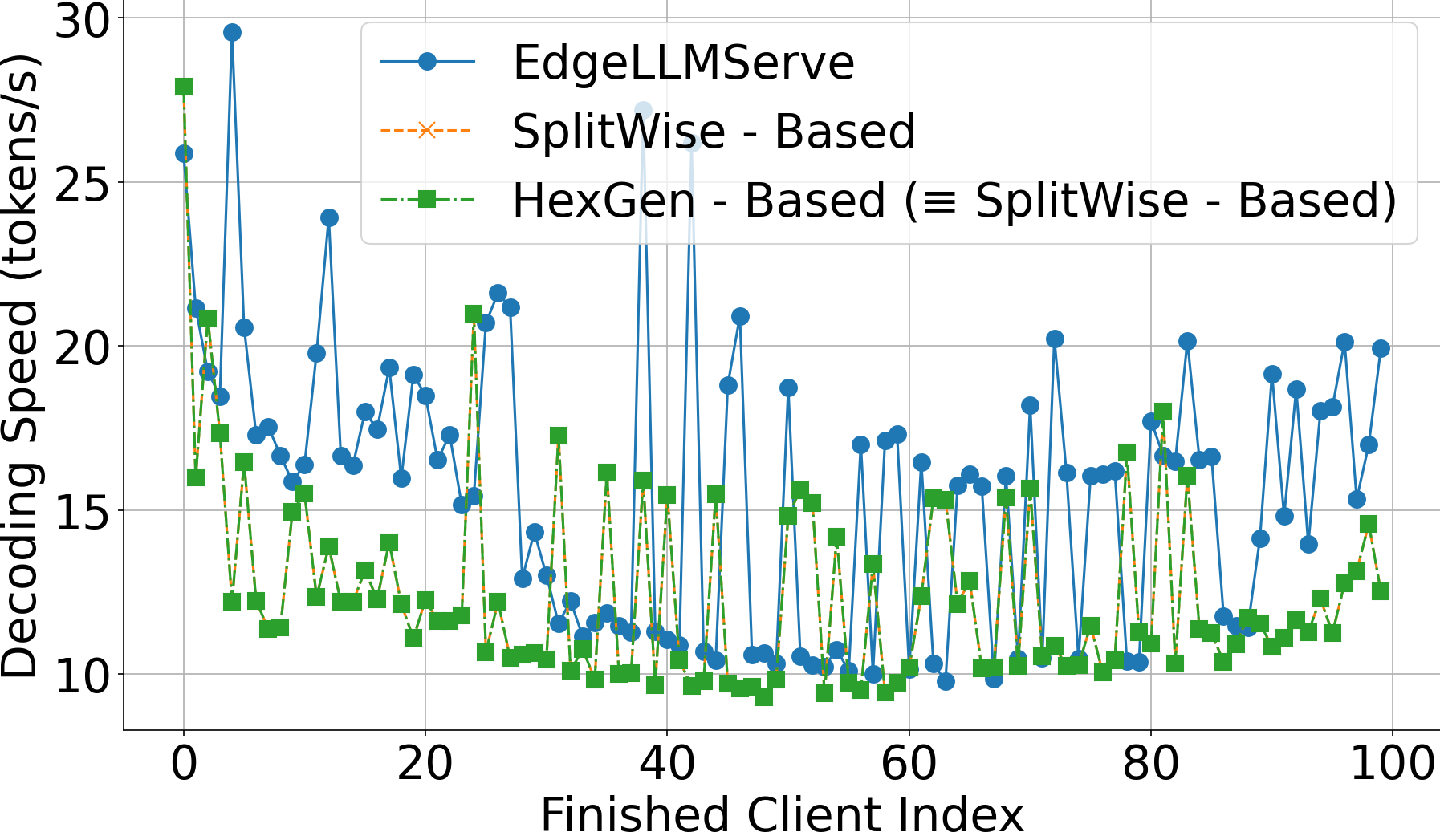}
        \caption{Decoding speed}
    \end{subfigure}\hfill
    \begin{subfigure}{\appfigscale\textwidth}
        \includegraphics[width=\linewidth]{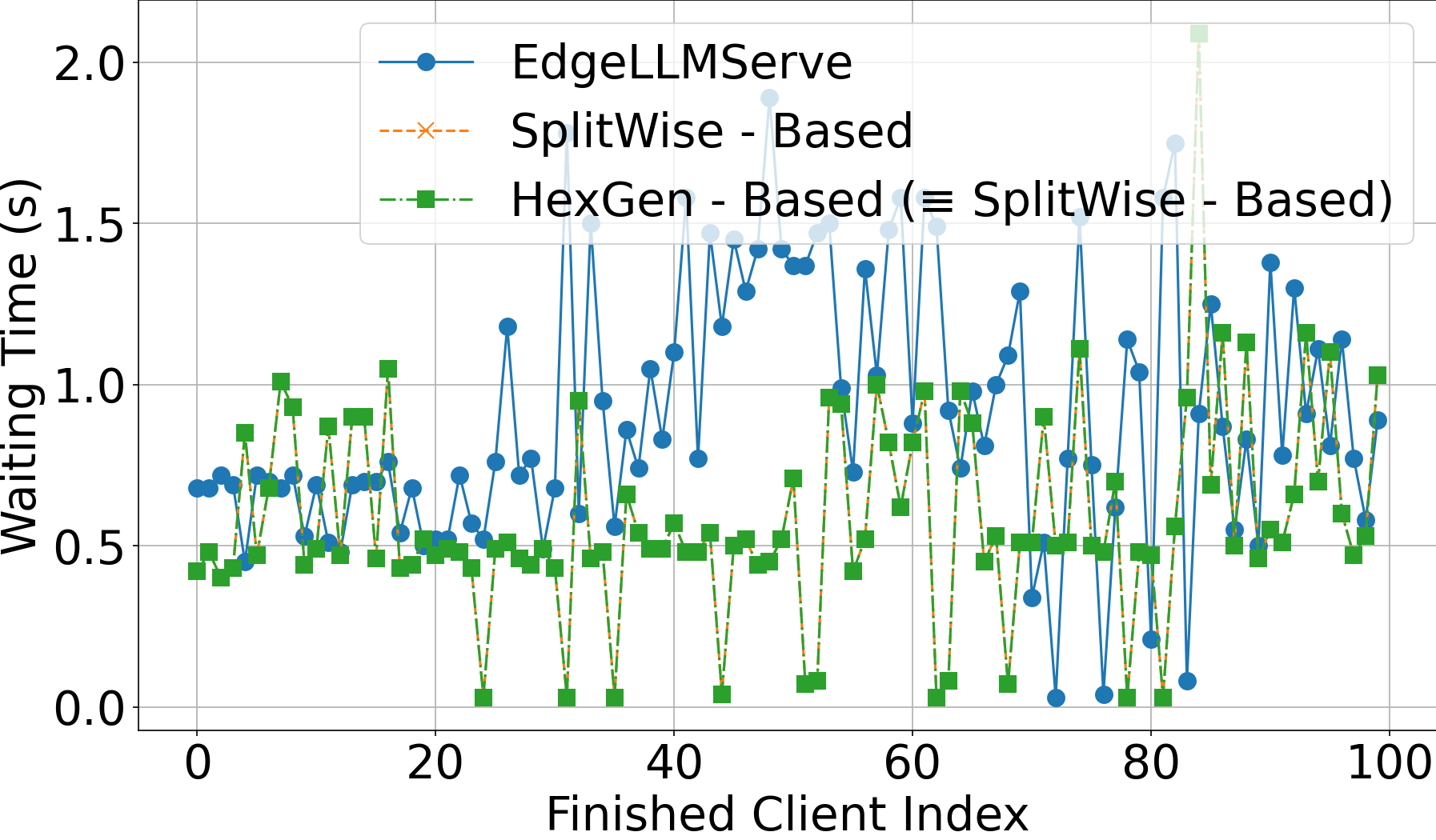}
        \caption{Waiting time}
    \end{subfigure}
    \vspace{-0.3cm}
    \caption{Per-request metrics --- Eight devices, Gpt-Oss-20b, IELTS, arrival $3000$~ms.}
    \label{apd:f27}
\end{figure}

\vspace{-0.3cm}
\paragraph{4. LZByteDance dataset. Arrival rate of 100ms}\mbox{}
\vspace{-0.3cm}
\begin{table}[H]
    \centering
    \setlength{\tabcolsep}{3pt}
    \caption{Deployment plans and metrics --- Eight devices, Gpt-Oss-20b, LZByteDance, arrival $100$~ms.}
    \vspace{-0.3cm}
    \label{apd:t28}
    \begin{subtable}[t]{0.24\textwidth}
        \centering
        \caption{HG}
        \begin{tabular}{|c|c|c|c|}
            \hline
            Mode & Dev & Slots & G--C\\\hline
            P& Dev.1$^{\star}$ & 1 & 20--4\\\hline
            \multirow{2}{*}{D} & Dev.2$^{\star}$ & \multirow{2}{*}{8} & 5--4\\
              & Dev.4 &  & 15--0\\\hline
            D& Dev.3$^{\star}$ & 5 & 24--0\\\hline
            \multirow{2}{*}{D} & Dev.5 & \multirow{2}{*}{2} & 15--0\\
              & Dev.6$^{\star}$ &  & 9--0\\\hline
            \multirow{2}{*}{D} & Dev.7 & \multirow{2}{*}{30} & 13--0\\
              & Dev.8$^{\star}$ &  & 11--0\\\hline
        \end{tabular}
    \end{subtable}\hfill
    \begin{subtable}[t]{0.24\textwidth}
        \centering
        \caption{ELS}
        \begin{tabular}{|c|c|c|c|}
            \hline
            Mode & Dev & Slots & G--C\\\hline
            P& Dev.1$^{\star}$ & 1 & 20--4\\\hline
            \multirow{2}{*}{D} & Dev.2$^{\star}$ & \multirow{2}{*}{8} & 5--4\\
              & Dev.4 &  & 15--0\\\hline
            D& Dev.3$^{\star}$ & 5 & 24--0\\\hline
            \multirow{2}{*}{D} & Dev.5 & \multirow{2}{*}{2} & 15--0\\
              & Dev.6$^{\star}$ &  & 9--0\\\hline
            \multirow{2}{*}{D} & Dev.7 & \multirow{2}{*}{30} & 13--0\\
              & Dev.8$^{\star}$ &  & 11--0\\\hline
        \end{tabular}
    \end{subtable}\hfill
    \begin{subtable}[t]{0.24\textwidth}
        \centering
        \caption{SPW}
        \begin{tabular}{|c|c|c|c|}
            \hline
            Mode & Dev & Slots & G--C\\\hline
            P& Dev.1$^{\star}$ & 1 & 20--4\\\hline
            \multirow{2}{*}{D} & Dev.2$^{\star}$ & \multirow{2}{*}{8} & 5--4\\
              & Dev.4 &  & 15--0\\\hline
            D& Dev.3$^{\star}$ & 5 & 24--0\\\hline
            \multirow{2}{*}{D} & Dev.5 & \multirow{2}{*}{2} & 15--0\\
              & Dev.6$^{\star}$ &  & 9--0\\\hline
            \multirow{2}{*}{D} & Dev.7 & \multirow{2}{*}{30} & 13--0\\
              & Dev.8$^{\star}$ &  & 11--0\\\hline
        \end{tabular}
    \end{subtable}\hfill
    \begin{subtable}[t]{0.24\textwidth}
        \centering
        \caption{Metrics.}
        \begin{tabular}{|l|c|c|c|}
            \hline
            Method & PS & DS & W\\\hline
            HG & 2464 & 13.1 & 66.6\\\hline
            \textbf{ELS} & \textbf{2464} & \textbf{13.1} & \textbf{66.6}\\\hline
            SPW & 2464 & 13.1 & 66.6\\\hline
        \end{tabular}
    \end{subtable}
\end{table}
\vspace{-0.3cm}
\begin{figure}[H]
    \centering
    \begin{subfigure}{\appfigscale\textwidth}
        \includegraphics[width=\linewidth]{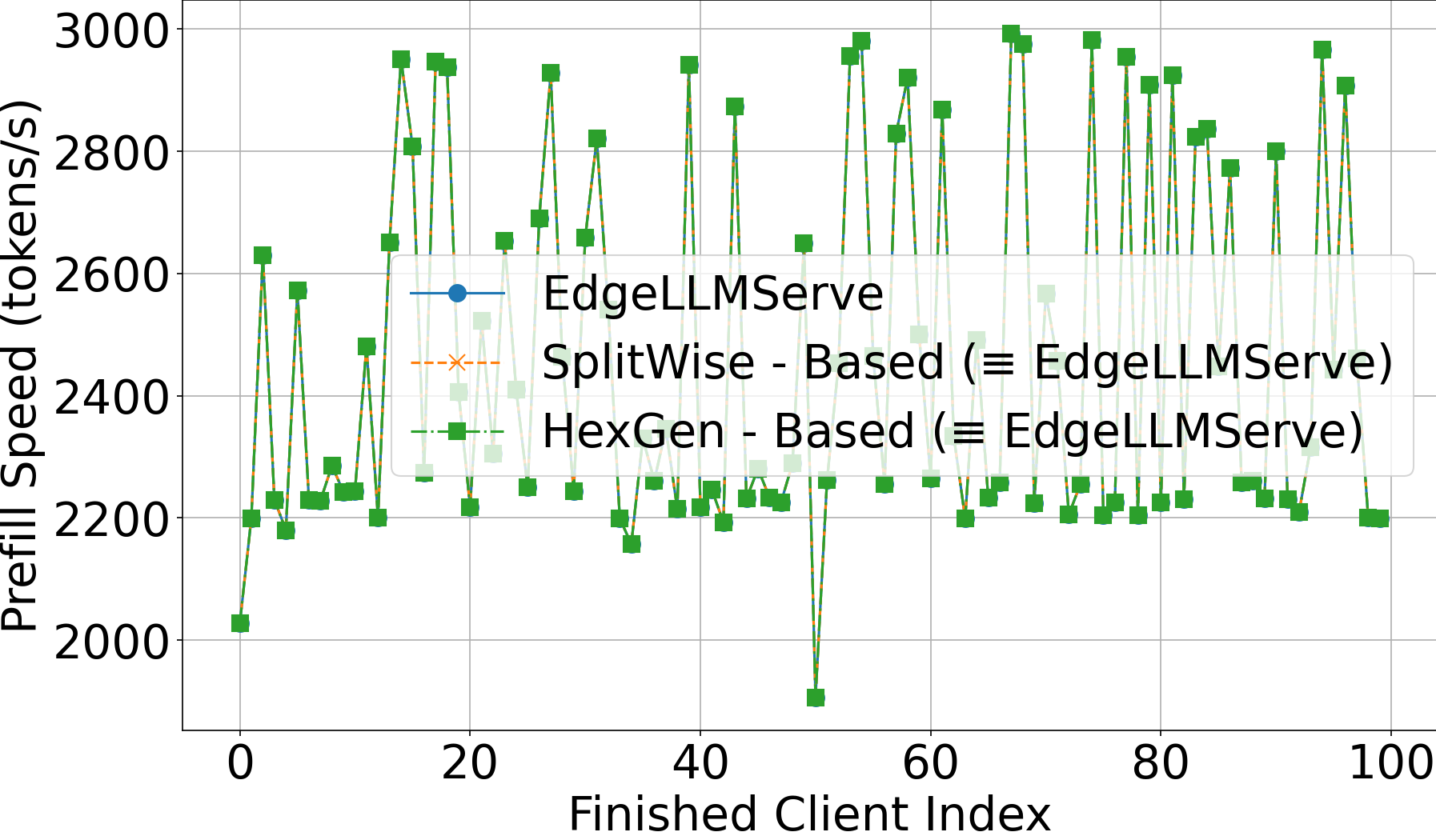}
        \caption{Prefill speed}
    \end{subfigure}\hfill
    \begin{subfigure}{\appfigscale\textwidth}
        \includegraphics[width=\linewidth]{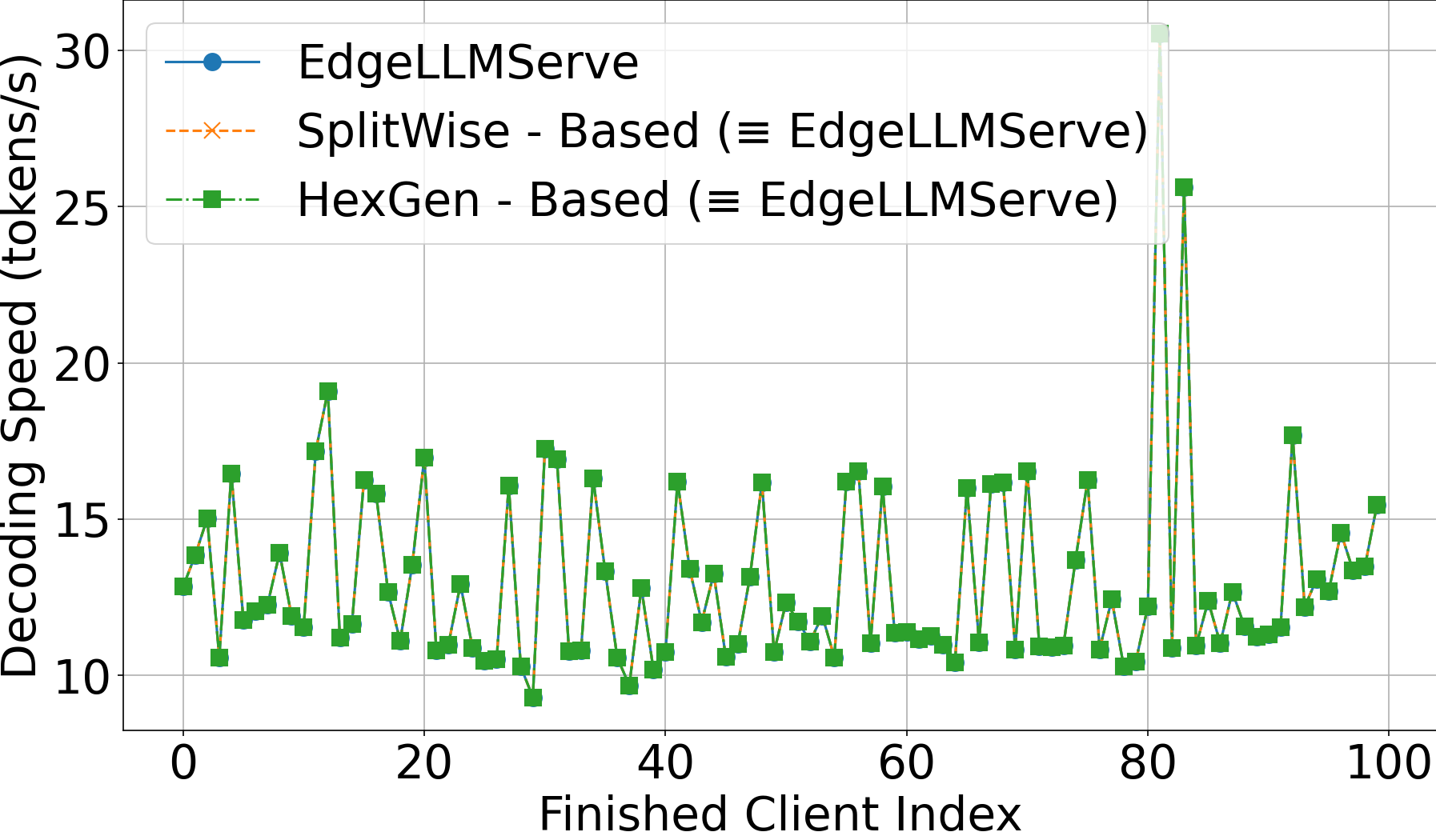}
        \caption{Decoding speed}
    \end{subfigure}\hfill
    \begin{subfigure}{\appfigscale\textwidth}
        \includegraphics[width=\linewidth]{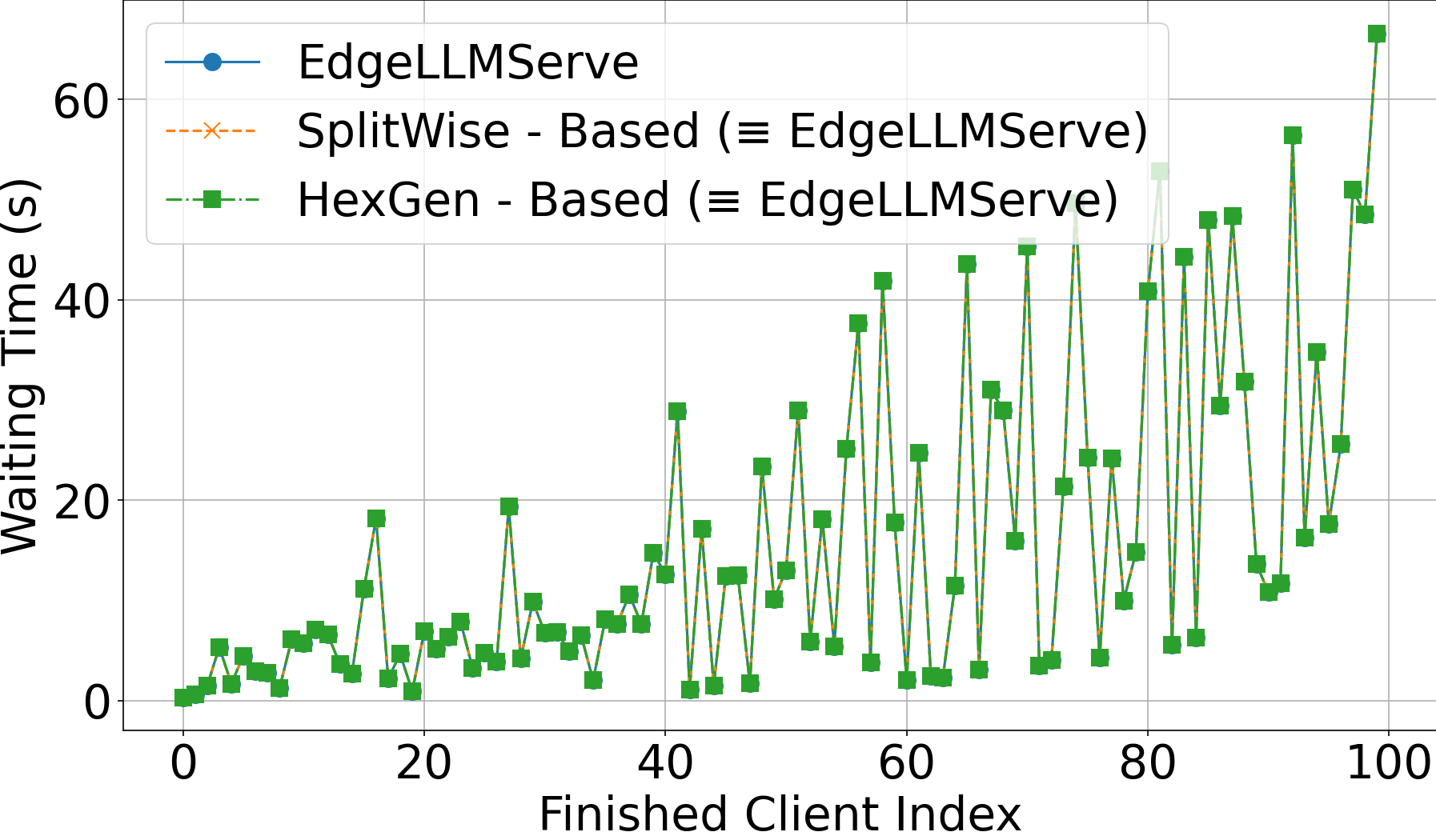}
        \caption{Waiting time}
    \end{subfigure}
    \vspace{-0.3cm}
    \caption{Per-request metrics --- Eight devices, Gpt-Oss-20b, LZByteDance, arrival $100$~ms.}
    \label{apd:f28}
\end{figure}

\paragraph{5. LZByteDance dataset. Arrival rate of 1000ms}\mbox{}
\vspace{-0.3cm}
\begin{table}[H]
    \centering
    \setlength{\tabcolsep}{3pt}
    \caption{Deployment plans and metrics --- Eight devices, Gpt-Oss-20b, LZByteDance, arrival $1000$~ms.}
    \vspace{-0.3cm}
    \label{apd:t29}
    \begin{subtable}[t]{0.24\textwidth}
        \centering
        \caption{HG}
        \begin{tabular}{|c|c|c|c|}
            \hline
            Mode & Dev & Slots & G--C\\\hline
            P& Dev.1$^{\star}$ & 1 & 20--4\\\hline
            \multirow{2}{*}{D} & Dev.2$^{\star}$ & \multirow{2}{*}{8} & 5--4\\
              & Dev.4 &  & 15--0\\\hline
            D& Dev.3$^{\star}$ & 5 & 24--0\\\hline
            \multirow{2}{*}{D} & Dev.5 & \multirow{2}{*}{2} & 15--0\\
              & Dev.6$^{\star}$ &  & 9--0\\\hline
            \multirow{2}{*}{D} & Dev.7 & \multirow{2}{*}{30} & 13--0\\
              & Dev.8$^{\star}$ &  & 11--0\\\hline
        \end{tabular}
    \end{subtable}\hfill
    \begin{subtable}[t]{0.24\textwidth}
        \centering
        \caption{ELS}
        \begin{tabular}{|c|c|c|c|}
            \hline
            Mode & Dev & Slots & G--C\\\hline
            P& Dev.3$^{\star}$ & 1 & 24--0\\\hline
            \multirow{2}{*}{D} & Dev.1 & \multirow{2}{*}{14} & 17--1\\
              & Dev.2$^{\star}$ &  & 5--1\\\hline
            D& Dev.4$^{\star}$ & 6 & 24--0\\\hline
            \multirow{2}{*}{D} & Dev.5 & \multirow{2}{*}{2} & 15--0\\
              & Dev.6$^{\star}$ &  & 9--0\\\hline
            \multirow{2}{*}{D} & Dev.7 & \multirow{2}{*}{30} & 13--0\\
              & Dev.8$^{\star}$ &  & 11--0\\\hline
        \end{tabular}
    \end{subtable}\hfill
    \begin{subtable}[t]{0.24\textwidth}
        \centering
        \caption{SPW}
        \begin{tabular}{|c|c|c|c|}
            \hline
            Mode & Dev & Slots & G--C\\\hline
            P& Dev.1$^{\star}$ & 1 & 20--4\\\hline
            \multirow{2}{*}{D} & Dev.2$^{\star}$ & \multirow{2}{*}{8} & 5--4\\
              & Dev.4 &  & 15--0\\\hline
            D& Dev.3$^{\star}$ & 5 & 24--0\\\hline
            \multirow{2}{*}{D} & Dev.5 & \multirow{2}{*}{2} & 15--0\\
              & Dev.6$^{\star}$ &  & 9--0\\\hline
            \multirow{2}{*}{D} & Dev.7 & \multirow{2}{*}{30} & 13--0\\
              & Dev.8$^{\star}$ &  & 11--0\\\hline
        \end{tabular}
    \end{subtable}\hfill
    \begin{subtable}[t]{0.24\textwidth}
        \centering
        \caption{Metrics.}
        \begin{tabular}{|l|c|c|c|}
            \hline
            Method & PS & DS & W\\\hline
            HG & 2480 & 13.0 & 31.5\\\hline
            \textbf{ELS} & \textbf{754} & \textbf{16.2} & \textbf{0.7}\\\hline
            SPW & 2480 & 13.0 & 31.5\\\hline
        \end{tabular}
    \end{subtable}
\end{table}
\vspace{-0.3cm}
\begin{figure}[H]
    \centering
    \begin{subfigure}{\appfigscale\textwidth}
        \includegraphics[width=\linewidth]{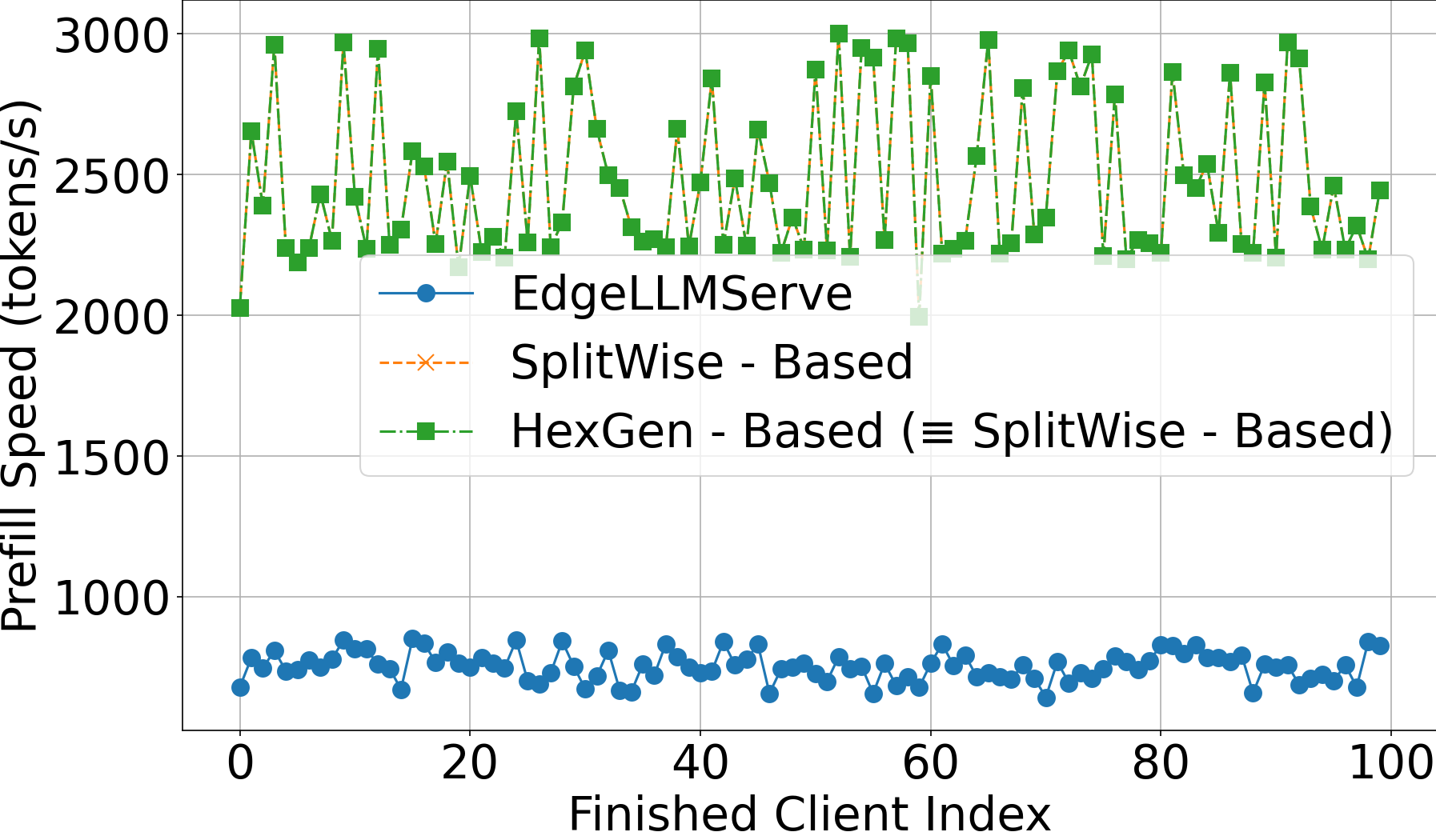}
        \caption{Prefill speed}
    \end{subfigure}\hfill
    \begin{subfigure}{\appfigscale\textwidth}
        \includegraphics[width=\linewidth]{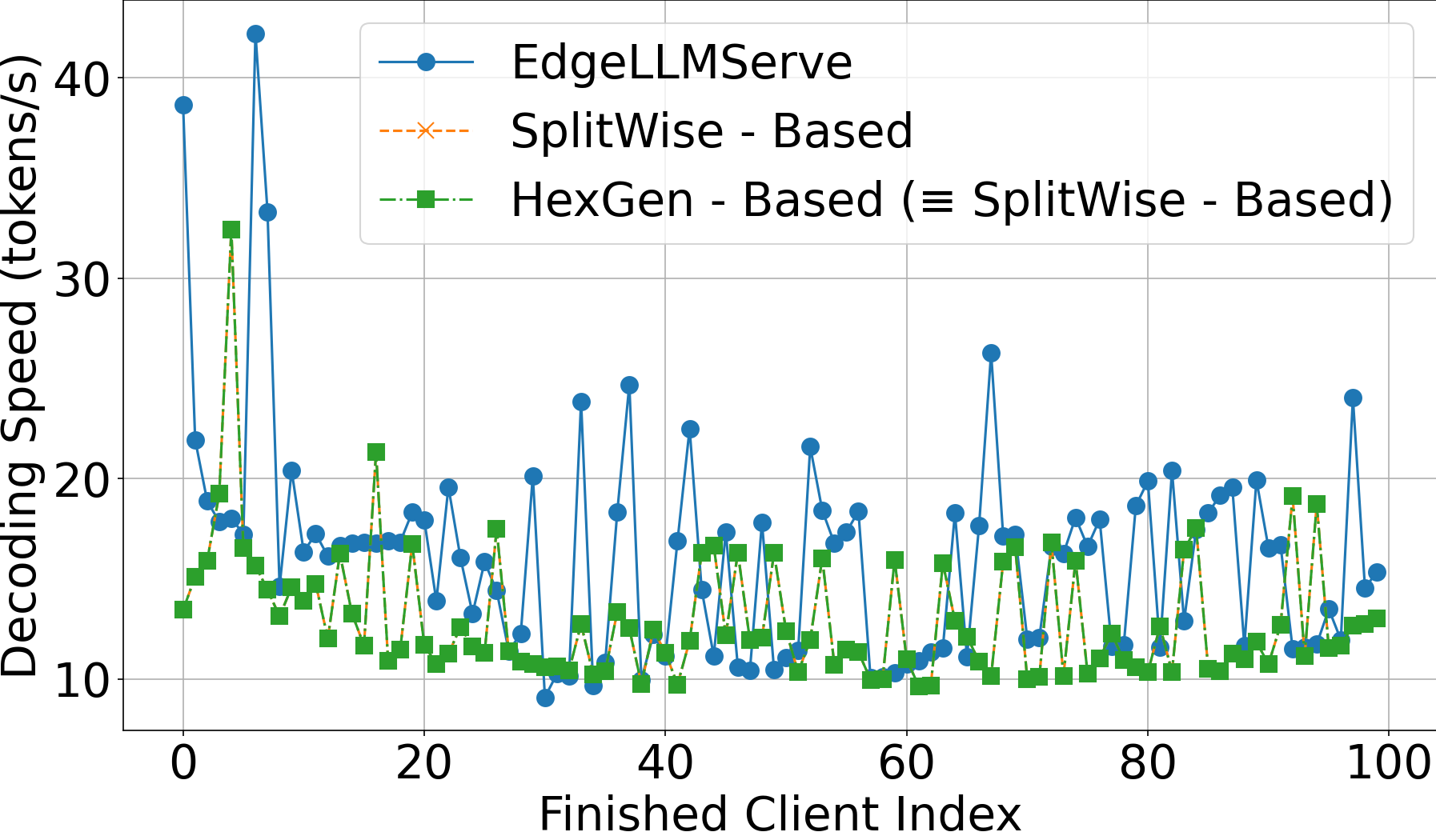}
        \caption{Decoding speed}
    \end{subfigure}\hfill
    \begin{subfigure}{\appfigscale\textwidth}
        \includegraphics[width=\linewidth]{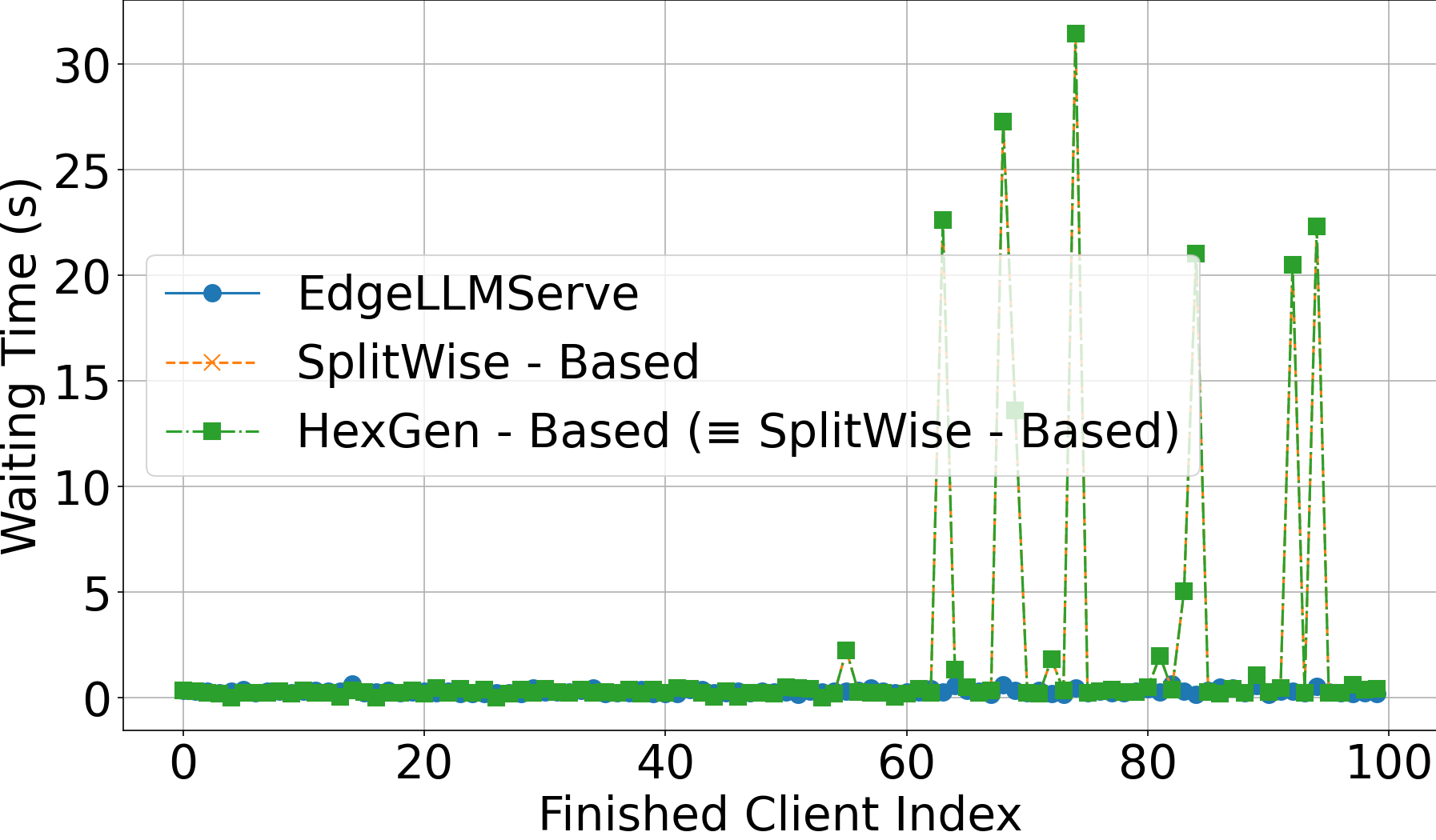}
        \caption{Waiting time}
    \end{subfigure}
\vspace{-0.3cm}
    \caption{Per-request metrics --- Eight devices, Gpt-Oss-20b, LZByteDance, arrival $1000$~ms.}
    \label{apd:f29}
\end{figure}

\vspace{-0.3cm}
\paragraph{6. LZByteDance dataset. Arrival rate of 3000ms}\mbox{}
\vspace{-0.3cm}
\begin{table}[H]
    \centering
    \setlength{\tabcolsep}{3pt}
    \caption{Deployment plans and metrics --- Eight devices, Gpt-Oss-20b, LZByteDance, arrival $3000$~ms.}
    \vspace{-0.3cm}
    \label{apd:t30}
    \begin{subtable}[t]{0.24\textwidth}
        \centering
        \caption{HG}
        \begin{tabular}{|c|c|c|c|}
            \hline
            Mode & Dev & Slots & G--C\\\hline
            P& Dev.1$^{\star}$ & 1 & 20--4\\\hline
            \multirow{2}{*}{D} & Dev.2$^{\star}$ & \multirow{2}{*}{8} & 5--4\\
              & Dev.4 &  & 15--0\\\hline
            D& Dev.3$^{\star}$ & 5 & 24--0\\\hline
            \multirow{2}{*}{D} & Dev.5 & \multirow{2}{*}{2} & 15--0\\
              & Dev.6$^{\star}$ &  & 9--0\\\hline
            \multirow{2}{*}{D} & Dev.7 & \multirow{2}{*}{30} & 13--0\\
              & Dev.8$^{\star}$ &  & 11--0\\\hline
        \end{tabular}
    \end{subtable}\hfill
    \begin{subtable}[t]{0.24\textwidth}
        \centering
        \caption{ELS}
        \begin{tabular}{|c|c|c|c|}
            \hline
            Mode & Dev & Slots & G--C\\\hline
            P& Dev.2$^{\star}$ & 1 & 5--19\\\hline
            \multirow{2}{*}{D} & Dev.1 & \multirow{2}{*}{32} & 14--0\\
              & Dev.7$^{\star}$ &  & 10--0\\\hline
            \multirow{3}{*}{D} & Dev.3$^{\star}$ & \multirow{3}{*}{6} & 14--0\\
              & Dev.5 &  & 5--0\\
              & Dev.6 &  & 5--0\\\hline
            \multirow{2}{*}{D} & Dev.4$^{\star}$ & \multirow{2}{*}{18} & 8--0\\
              & Dev.8 &  & 16--0\\\hline
        \end{tabular}
    \end{subtable}\hfill
    \begin{subtable}[t]{0.24\textwidth}
        \centering
        \caption{SPW}
        \begin{tabular}{|c|c|c|c|}
            \hline
            Mode & Dev & Slots & G--C\\\hline
            P& Dev.1$^{\star}$ & 1 & 20--4\\\hline
            \multirow{2}{*}{D} & Dev.2$^{\star}$ & \multirow{2}{*}{8} & 5--4\\
              & Dev.4 &  & 15--0\\\hline
            D& Dev.3$^{\star}$ & 5 & 24--0\\\hline
            \multirow{2}{*}{D} & Dev.5 & \multirow{2}{*}{2} & 15--0\\
              & Dev.6$^{\star}$ &  & 9--0\\\hline
            \multirow{2}{*}{D} & Dev.7 & \multirow{2}{*}{30} & 13--0\\
              & Dev.8$^{\star}$ &  & 11--0\\\hline
        \end{tabular}
    \end{subtable}\hfill
    \begin{subtable}[t]{0.24\textwidth}
        \centering
        \caption{Metrics.}
        \begin{tabular}{|l|c|c|c|}
            \hline
            Method & PS & DS & W\\\hline
            HG & 2481 & 18.9 & 0.6\\\hline
            \textbf{ELS} & \textbf{294} & \textbf{21.5} & \textbf{0.6}\\\hline
            SPW & 2481 & 18.9 & 0.6\\\hline
        \end{tabular}
    \end{subtable}
\end{table}
\begin{figure}[H]
    \centering
    \begin{subfigure}{\appfigscale\textwidth}
        \includegraphics[width=\linewidth]{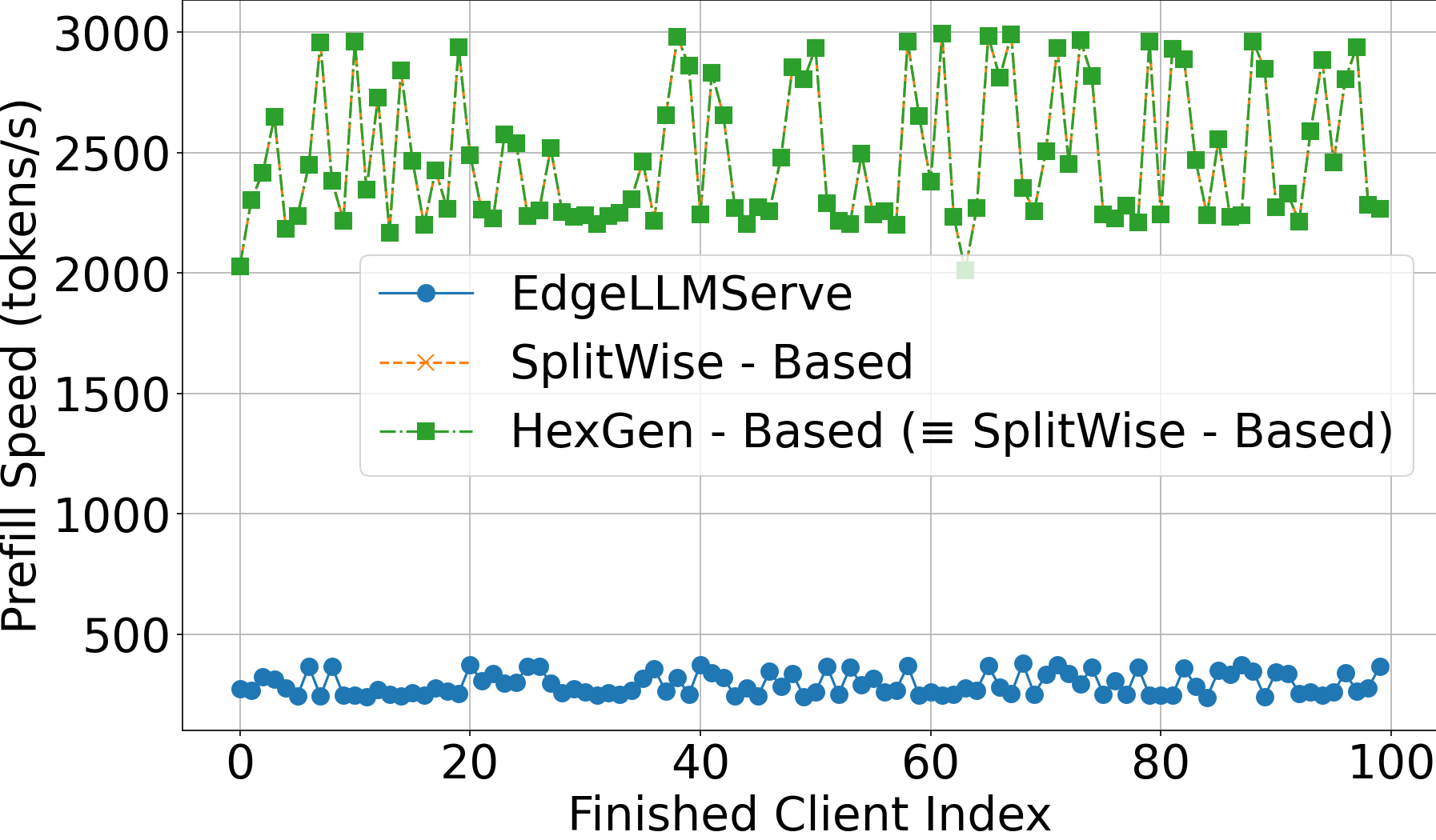}
        \caption{Prefill speed}
    \end{subfigure}\hfill
    \begin{subfigure}{\appfigscale\textwidth}
        \includegraphics[width=\linewidth]{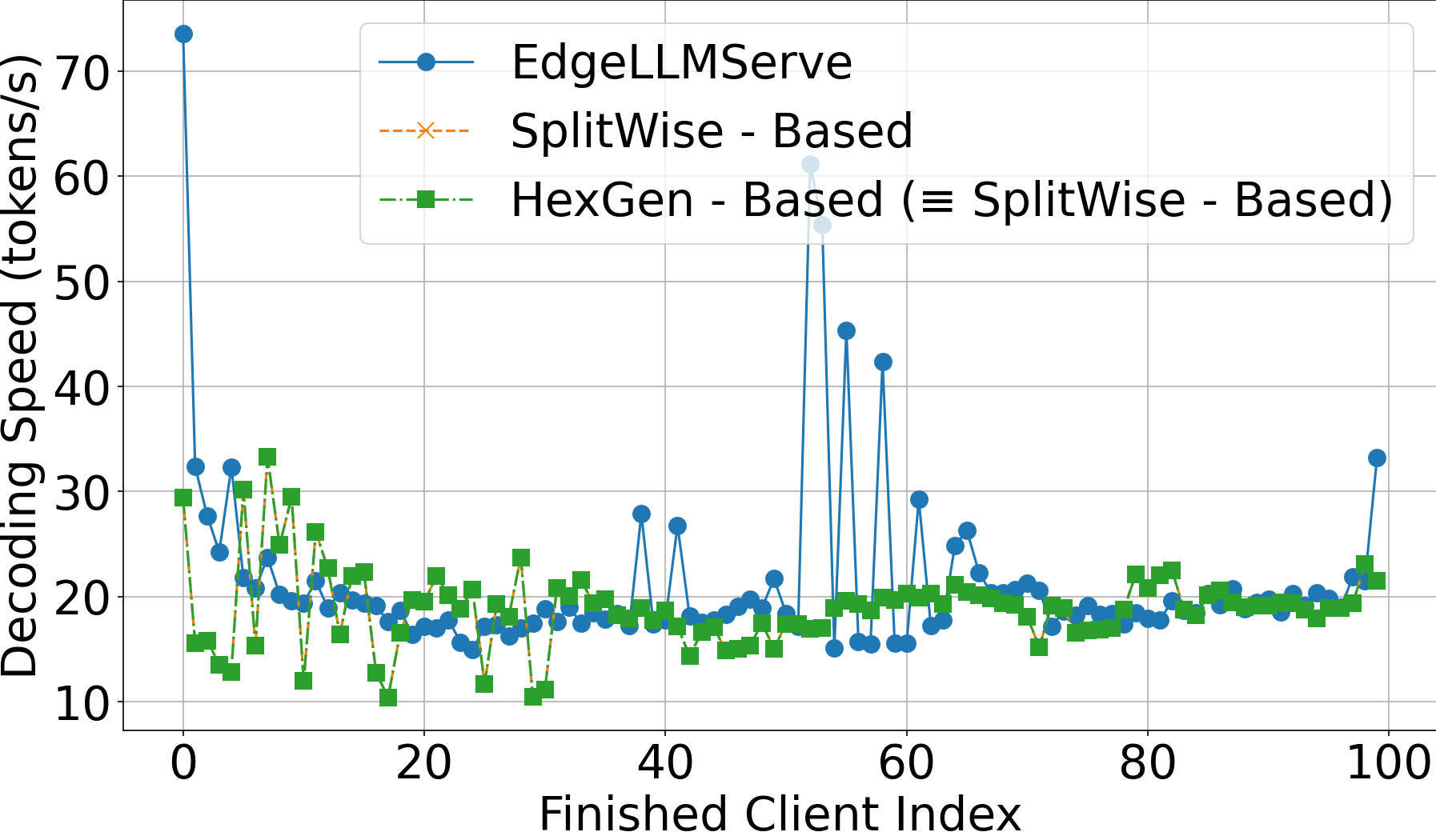}
        \caption{Decoding speed}
    \end{subfigure}\hfill
    \begin{subfigure}{\appfigscale\textwidth}
        \includegraphics[width=\linewidth]{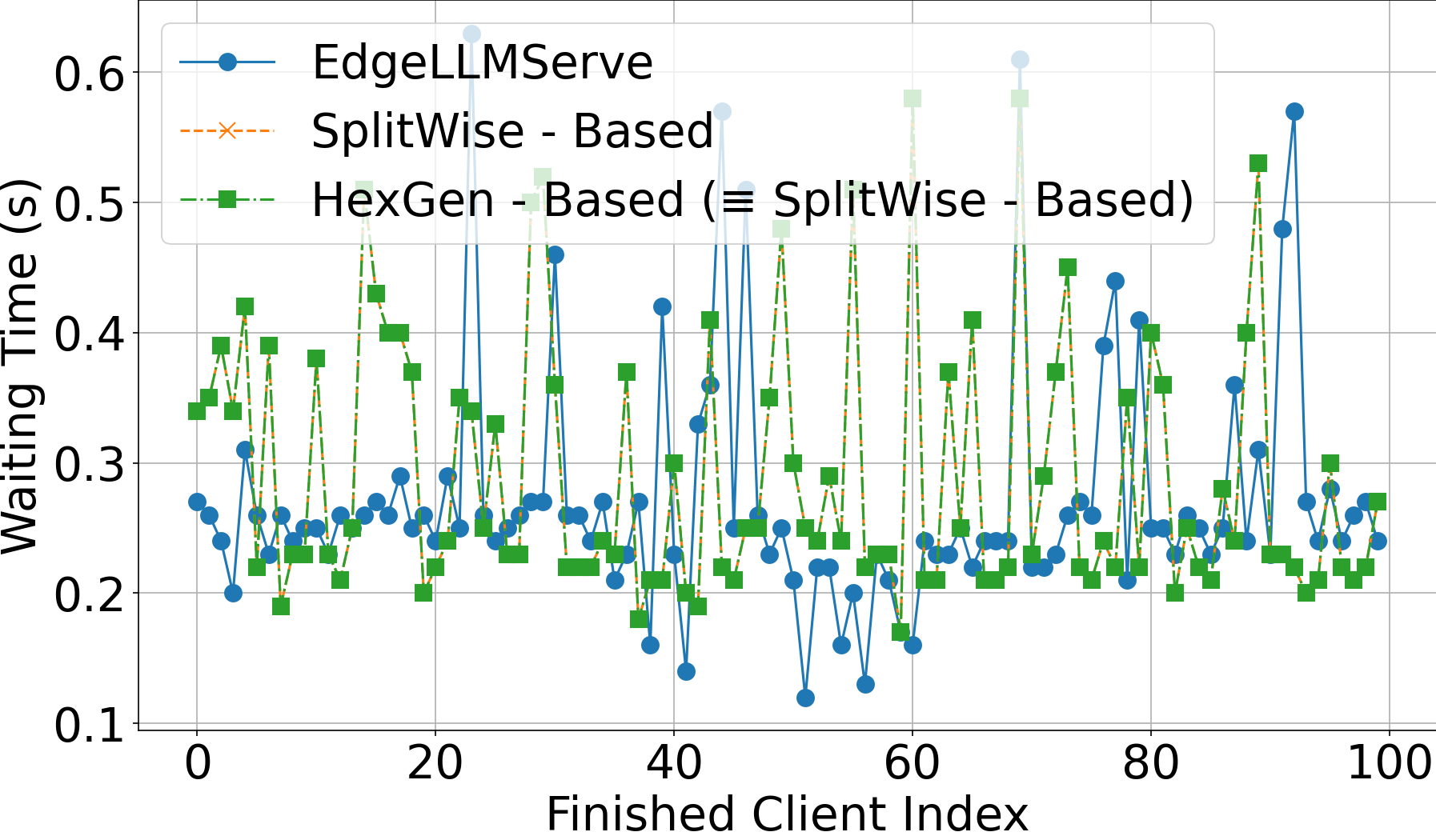}
        \caption{Waiting time}
    \end{subfigure}
    \vspace{-0.3cm}
    \caption{Per-request metrics --- Eight devices, Gpt-Oss-20b, LZByteDance, arrival $3000$~ms.}
    \label{apd:f30}
\end{figure}

\subsubsection{Qwen3-MoE-30B}
\paragraph{1. IELTS dataset. Arrival rate of 100ms}\mbox{}
\vspace{-0.3cm}
\begin{table}[H]
    \centering
    \setlength{\tabcolsep}{3pt}
    \caption{Deployment plans and metrics --- Eight devices, Qwen3-MoE-30B, IELTS, arrival $100$~ms.}
    \label{apd:t31}
    \vspace{-0.3cm}
    \begin{subtable}[t]{0.24\textwidth}
        \centering
        \caption{HG}
        \begin{tabular}{|c|c|c|c|}
            \hline
            Mode & Dev & Slots & G--C\\\hline
            P& Dev.1$^{\star}$ & 1 & 26--22\\\hline
            \multirow{3}{*}{D} & Dev.2 & \multirow{3}{*}{9} & 7--3\\
              & Dev.7 &  & 19--0\\
              & Dev.8$^{\star}$ &  & 19--0\\\hline
            \multirow{2}{*}{D} & Dev.3$^{\star}$ & \multirow{2}{*}{8} & 36--0\\
              & Dev.6 &  & 12--0\\\hline
            \multirow{2}{*}{D} & Dev.4$^{\star}$ & \multirow{2}{*}{8} & 36--0\\
              & Dev.5 &  & 12--0\\\hline
        \end{tabular}
    \end{subtable}\hfill
    \begin{subtable}[t]{0.24\textwidth}
        \centering
        \caption{ELS}
        \begin{tabular}{|c|c|c|c|}
            \hline
            Mode & Dev & Slots & G--C\\\hline
            P& Dev.1$^{\star}$ & 1 & 26--22\\\hline
            \multirow{3}{*}{D} & Dev.2 & \multirow{3}{*}{9} & 7--3\\
              & Dev.7 &  & 19--0\\
              & Dev.8$^{\star}$ &  & 19--0\\\hline
            \multirow{2}{*}{D} & Dev.3$^{\star}$ & \multirow{2}{*}{8} & 36--0\\
              & Dev.6 &  & 12--0\\\hline
            \multirow{2}{*}{D} & Dev.4$^{\star}$ & \multirow{2}{*}{8} & 36--0\\
              & Dev.5 &  & 12--0\\\hline
        \end{tabular}
    \end{subtable}\hfill
    \begin{subtable}[t]{0.24\textwidth}
        \centering
        \caption{SPW}
        \begin{tabular}{|c|c|c|c|}
            \hline
            Mode & Dev & Slots & G--C\\\hline
            \multirow{2}{*}{P} & Dev.7$^{\star}$ & \multirow{2}{*}{2} & 26--0\\
              & Dev.8 &  & 22--0\\\hline
            \multirow{2}{*}{D} & Dev.1$^{\star}$ & \multirow{2}{*}{10} & 13--19\\
              & Dev.2 &  & 5--11\\\hline
            \multirow{2}{*}{D} & Dev.3$^{\star}$ & \multirow{2}{*}{8} & 36--0\\
              & Dev.6 &  & 12--0\\\hline
            \multirow{2}{*}{D} & Dev.4$^{\star}$ & \multirow{2}{*}{8} & 36--0\\
              & Dev.5 &  & 12--0\\\hline
        \end{tabular}
    \end{subtable}\hfill
    \begin{subtable}[t]{0.24\textwidth}
        \centering
        \caption{Metrics.}
        \begin{tabular}{|l|c|c|c|}
            \hline
            Method & PS & DS & W\\\hline
            HG & 1412 & 11.9 & 440.8\\\hline
            \textbf{ELS} & \textbf{1412} & \textbf{11.9} & \textbf{440.8}\\\hline
            SPW & 1373 & 11.8 & 516.8\\\hline
        \end{tabular}
    \end{subtable}
\end{table}
\vspace{-0.3cm}
\begin{figure}[H]
    \centering
    \begin{subfigure}{\appfigscale\textwidth}
        \includegraphics[width=\linewidth]{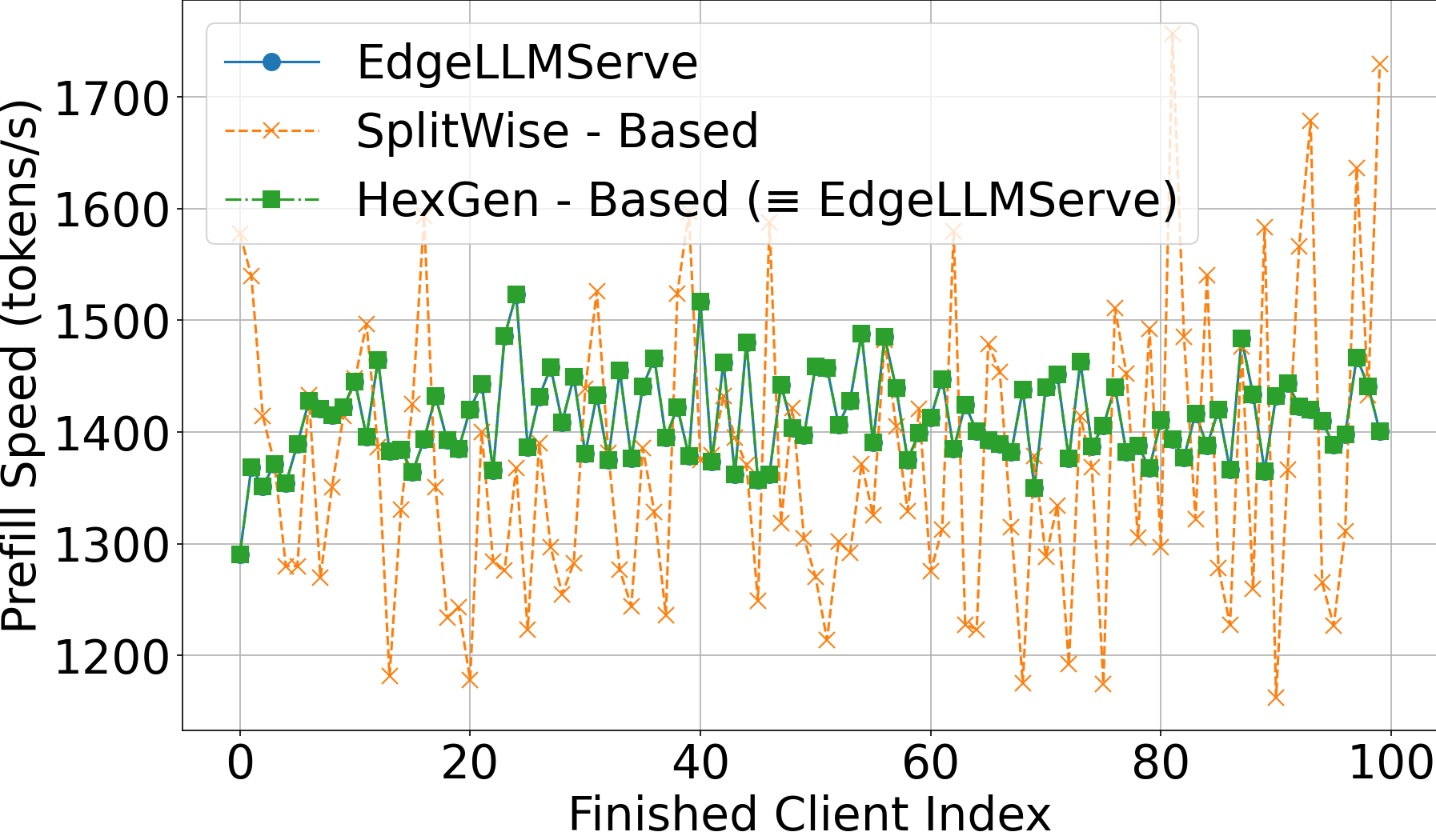}
        \caption{Prefill speed}
    \end{subfigure}\hfill
    \begin{subfigure}{\appfigscale\textwidth}
        \includegraphics[width=\linewidth]{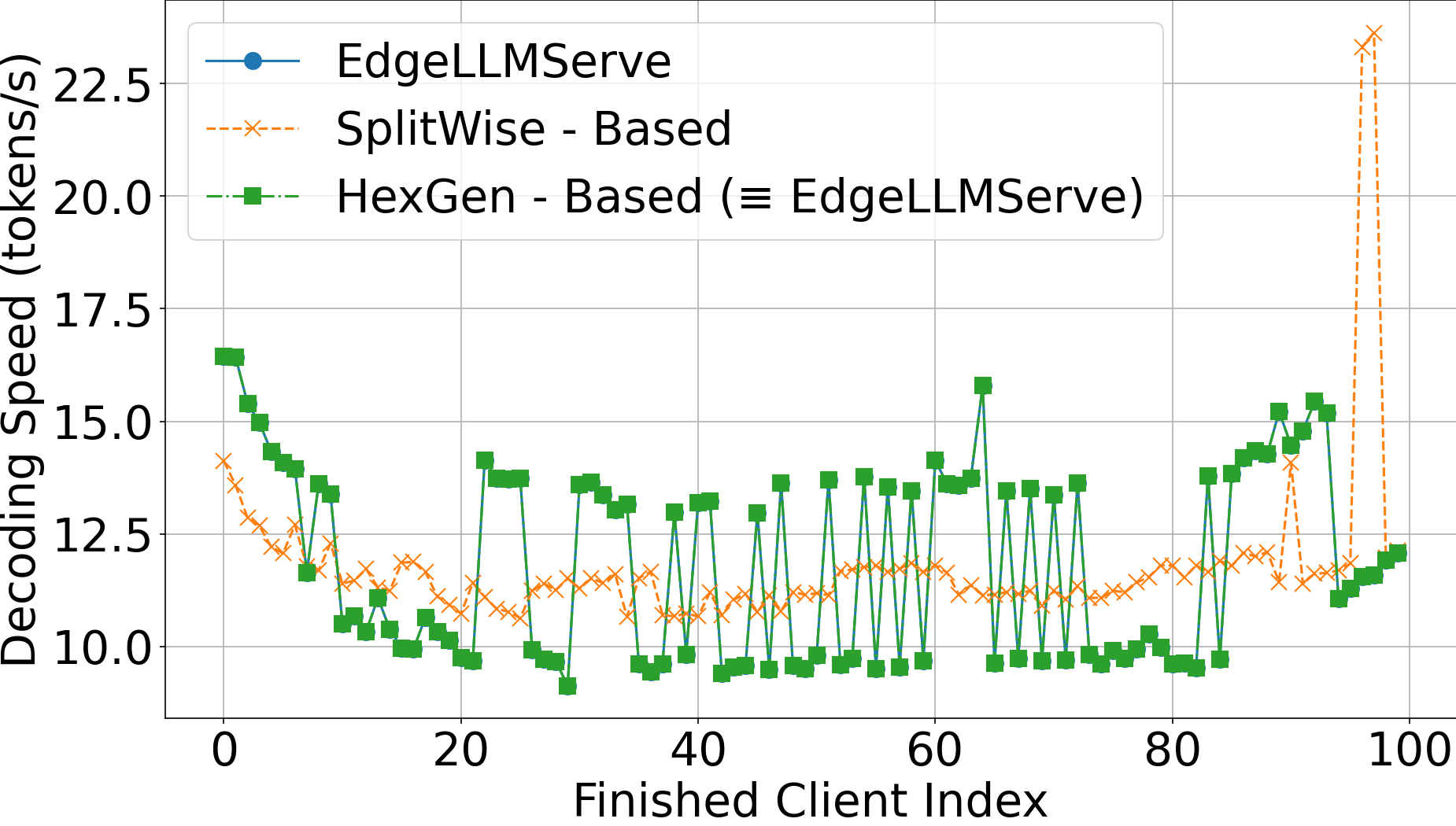}
        \caption{Decoding speed}
    \end{subfigure}\hfill
    \begin{subfigure}{\appfigscale\textwidth}
        \includegraphics[width=\linewidth]{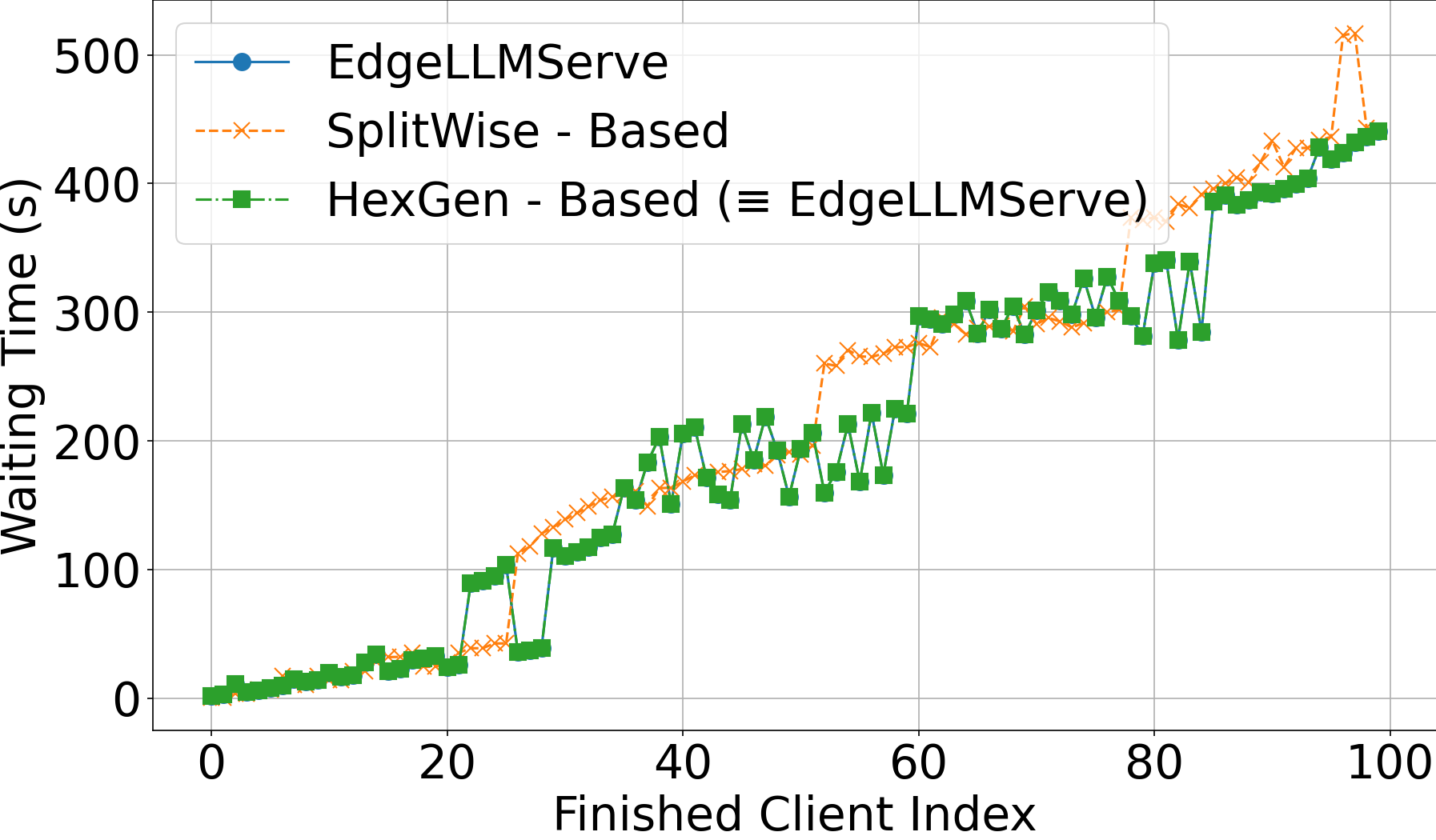}
        \caption{Waiting time}
    \end{subfigure}
    \vspace{-0.3cm}
    \caption{Per-request metrics --- Eight devices, Qwen3-MoE-30B, IELTS, arrival $100$~ms.}
    \label{apd:f31}
\end{figure}

\vspace{-0.3cm}
\paragraph{2. IELTS dataset. Arrival rate of 1000ms}\mbox{}
\vspace{-0.3cm}
\begin{table}[H]
    \centering
    \setlength{\tabcolsep}{3pt}
    \caption{Deployment plans and metrics --- Eight devices, Qwen3-MoE-30B, IELTS, arrival $1000$~ms.}
    \vspace{-0.3cm}
    \label{apd:t32}
    \begin{subtable}[t]{0.24\textwidth}
        \centering
        \caption{HG}
        \begin{tabular}{|c|c|c|c|}
            \hline
            Mode & Dev & Slots & G--C\\\hline
            P& Dev.1$^{\star}$ & 1 & 26--22\\\hline
            \multirow{3}{*}{D} & Dev.2 & \multirow{3}{*}{9} & 7--3\\
              & Dev.7 &  & 19--0\\
              & Dev.8$^{\star}$ &  & 19--0\\\hline
            \multirow{2}{*}{D} & Dev.3$^{\star}$ & \multirow{2}{*}{8} & 36--0\\
              & Dev.6 &  & 12--0\\\hline
            \multirow{2}{*}{D} & Dev.4$^{\star}$ & \multirow{2}{*}{8} & 36--0\\
              & Dev.5 &  & 12--0\\\hline
        \end{tabular}
    \end{subtable}\hfill
    \begin{subtable}[t]{0.24\textwidth}
        \centering
        \caption{ELS}
        \begin{tabular}{|c|c|c|c|}
            \hline
            Mode & Dev & Slots & G--C\\\hline
            P& Dev.1$^{\star}$ & 1 & 26--22\\\hline
            \multirow{3}{*}{D} & Dev.2 & \multirow{3}{*}{9} & 7--3\\
              & Dev.7 &  & 19--0\\
              & Dev.8$^{\star}$ &  & 19--0\\\hline
            \multirow{2}{*}{D} & Dev.3$^{\star}$ & \multirow{2}{*}{8} & 36--0\\
              & Dev.6 &  & 12--0\\\hline
            \multirow{2}{*}{D} & Dev.4$^{\star}$ & \multirow{2}{*}{8} & 36--0\\
              & Dev.5 &  & 12--0\\\hline
        \end{tabular}
    \end{subtable}\hfill
    \begin{subtable}[t]{0.24\textwidth}
        \centering
        \caption{SPW}
        \begin{tabular}{|c|c|c|c|}
            \hline
            Mode & Dev & Slots & G--C\\\hline
            \multirow{2}{*}{P} & Dev.7$^{\star}$ & \multirow{2}{*}{2} & 26--0\\
              & Dev.8 &  & 22--0\\\hline
            \multirow{2}{*}{D} & Dev.1$^{\star}$ & \multirow{2}{*}{10} & 13--19\\
              & Dev.2 &  & 5--11\\\hline
            \multirow{2}{*}{D} & Dev.3$^{\star}$ & \multirow{2}{*}{8} & 36--0\\
              & Dev.6 &  & 12--0\\\hline
            \multirow{2}{*}{D} & Dev.4$^{\star}$ & \multirow{2}{*}{8} & 36--0\\
              & Dev.5 &  & 12--0\\\hline
        \end{tabular}
    \end{subtable}\hfill
    \begin{subtable}[t]{0.24\textwidth}
        \centering
        \caption{Metrics.}
        \begin{tabular}{|l|c|c|c|}
            \hline
            Method & PS & DS & W\\\hline
            HG & 1412 & 12.1 & 329.9\\\hline
            \textbf{ELS} & \textbf{1412} & \textbf{12.1} & \textbf{329.9}\\\hline
            SPW & 1385 & 12.0 & 431.0\\\hline
        \end{tabular}
    \end{subtable}
\end{table}
\vspace{-0.3cm}
\begin{figure}[H]
    \centering
    \begin{subfigure}{\appfigscale\textwidth}
        \includegraphics[width=\linewidth]{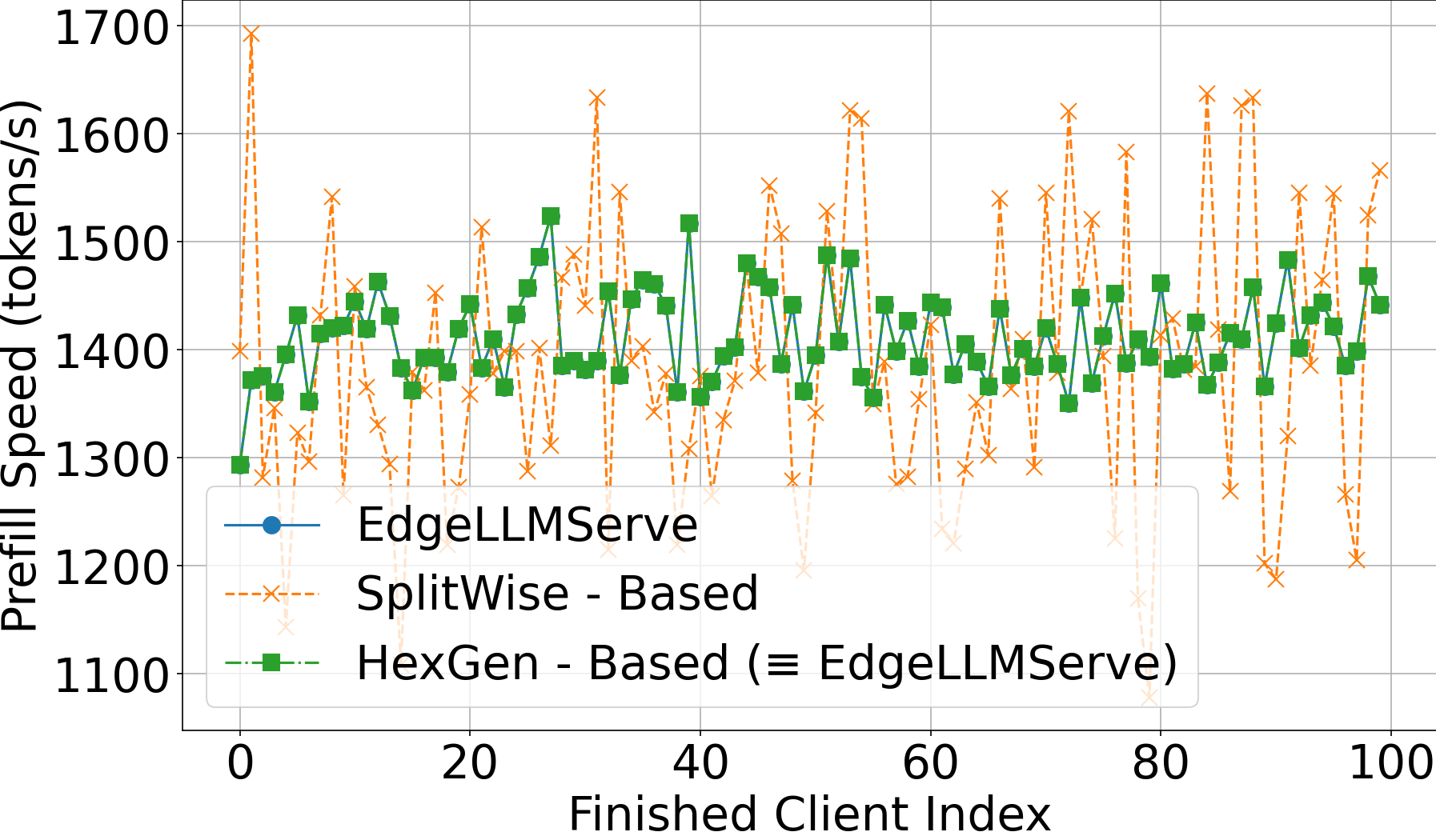}
        \caption{Prefill speed}
    \end{subfigure}\hfill
    \begin{subfigure}{\appfigscale\textwidth}
        \includegraphics[width=\linewidth]{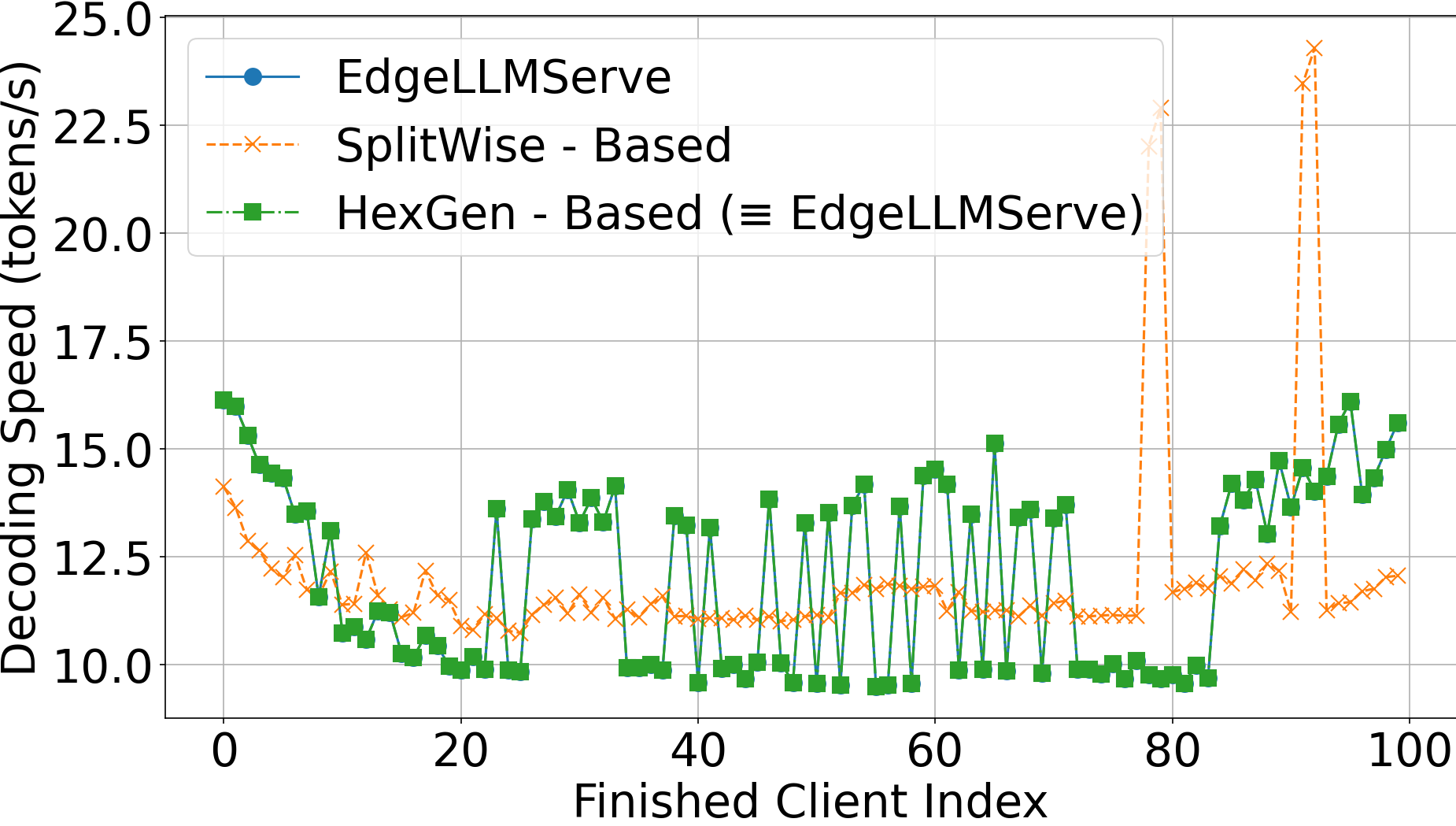}
        \caption{Decoding speed}
    \end{subfigure}\hfill
    \begin{subfigure}{\appfigscale\textwidth}
        \includegraphics[width=\linewidth]{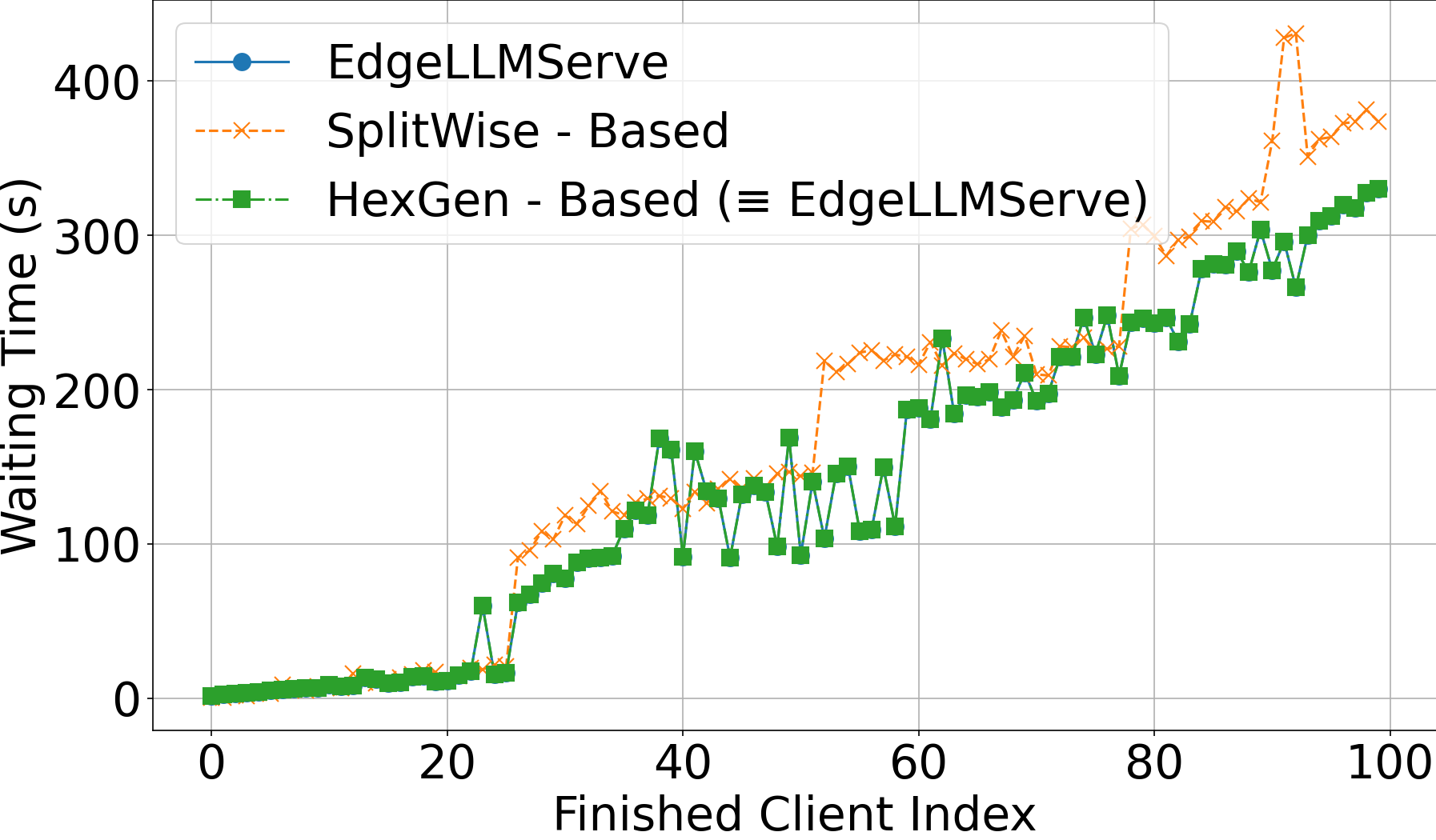}
        \caption{Waiting time}
    \end{subfigure}
    \vspace{-0.3cm}
    \caption{Per-request metrics --- Eight devices, Qwen3-MoE-30B, IELTS, arrival $1000$~ms.}
    \label{apd:f32}
\end{figure}

\paragraph{3. IELTS dataset. Arrival rate of 3000ms}\mbox{}
\vspace{-0.3cm}
\begin{table}[H]
    \centering
    \setlength{\tabcolsep}{3pt}
    \caption{Deployment plans and metrics --- Eight devices, Qwen3-MoE-30B, IELTS, arrival $3000$~ms.}
    \vspace{-0.3cm}
    \label{apd:t33}
    \begin{subtable}[t]{0.24\textwidth}
        \centering
        \caption{HG}
        \begin{tabular}{|c|c|c|c|}
            \hline
            Mode & Dev & Slots & G--C\\\hline
            P& Dev.1$^{\star}$ & 1 & 26--22\\\hline
            \multirow{3}{*}{D} & Dev.2 & \multirow{3}{*}{9} & 7--3\\
              & Dev.7 &  & 19--0\\
              & Dev.8$^{\star}$ &  & 19--0\\\hline
            \multirow{2}{*}{D} & Dev.3$^{\star}$ & \multirow{2}{*}{8} & 36--0\\
              & Dev.6 &  & 12--0\\\hline
            \multirow{2}{*}{D} & Dev.4$^{\star}$ & \multirow{2}{*}{8} & 36--0\\
              & Dev.5 &  & 12--0\\\hline
        \end{tabular}
    \end{subtable}\hfill
    \begin{subtable}[t]{0.24\textwidth}
        \centering
        \caption{ELS}
        \begin{tabular}{|c|c|c|c|}
            \hline
            Mode & Dev & Slots & G--C\\\hline
            P& Dev.1$^{\star}$ & 1 & 26--22\\\hline
            \multirow{3}{*}{D} & Dev.2 & \multirow{3}{*}{9} & 7--3\\
              & Dev.7 &  & 19--0\\
              & Dev.8$^{\star}$ &  & 19--0\\\hline
            \multirow{2}{*}{D} & Dev.3$^{\star}$ & \multirow{2}{*}{8} & 36--0\\
              & Dev.6 &  & 12--0\\\hline
            \multirow{2}{*}{D} & Dev.4$^{\star}$ & \multirow{2}{*}{8} & 36--0\\
              & Dev.5 &  & 12--0\\\hline
        \end{tabular}
    \end{subtable}\hfill
    \begin{subtable}[t]{0.24\textwidth}
        \centering
        \caption{SPW}
        \begin{tabular}{|c|c|c|c|}
            \hline
            Mode & Dev & Slots & G--C\\\hline
            \multirow{2}{*}{P} & Dev.7$^{\star}$ & \multirow{2}{*}{2} & 26--0\\
              & Dev.8 &  & 22--0\\\hline
            \multirow{2}{*}{D} & Dev.1$^{\star}$ & \multirow{2}{*}{10} & 13--19\\
              & Dev.2 &  & 5--11\\\hline
            \multirow{2}{*}{D} & Dev.3$^{\star}$ & \multirow{2}{*}{8} & 36--0\\
              & Dev.6 &  & 12--0\\\hline
            \multirow{2}{*}{D} & Dev.4$^{\star}$ & \multirow{2}{*}{8} & 36--0\\
              & Dev.5 &  & 12--0\\\hline
        \end{tabular}
    \end{subtable}\hfill
    \begin{subtable}[t]{0.24\textwidth}
        \centering
        \caption{Metrics.}
        \begin{tabular}{|l|c|c|c|}
            \hline
            Method & PS & DS & W\\\hline
            HG & 1412 & 11.9 & 260.5\\\hline
            \textbf{ELS} & \textbf{1412} & \textbf{11.9} & \textbf{260.5}\\\hline
            SPW & 1776 & 11.8 & 342.9\\\hline
        \end{tabular}
    \end{subtable}
\end{table}
\vspace{-0.3cm}
\begin{figure}[H]
    \centering
    \begin{subfigure}{\appfigscale\textwidth}
        \includegraphics[width=\linewidth]{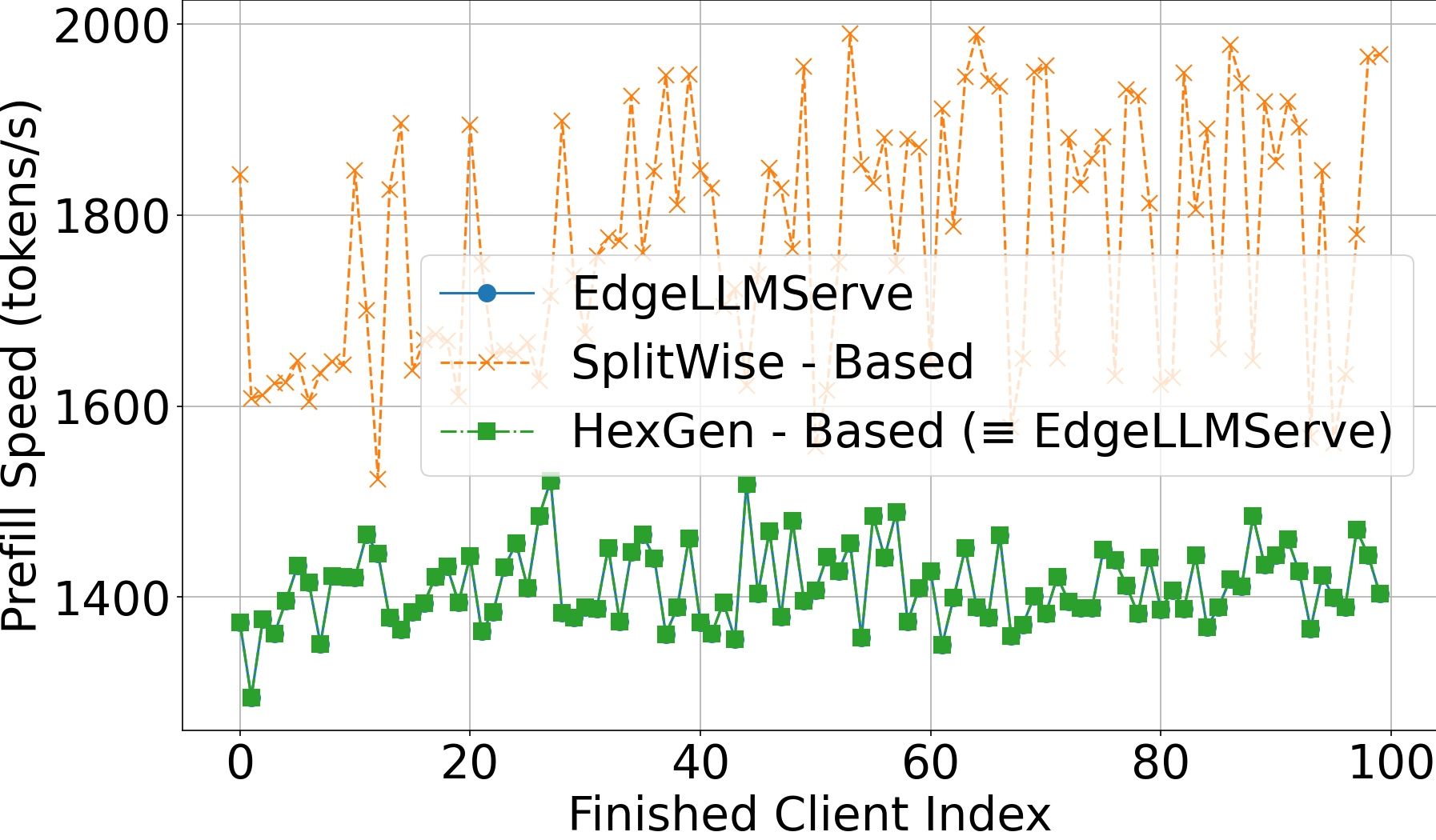}
        \caption{Prefill speed}
    \end{subfigure}\hfill
    \begin{subfigure}{\appfigscale\textwidth}
        \includegraphics[width=\linewidth]{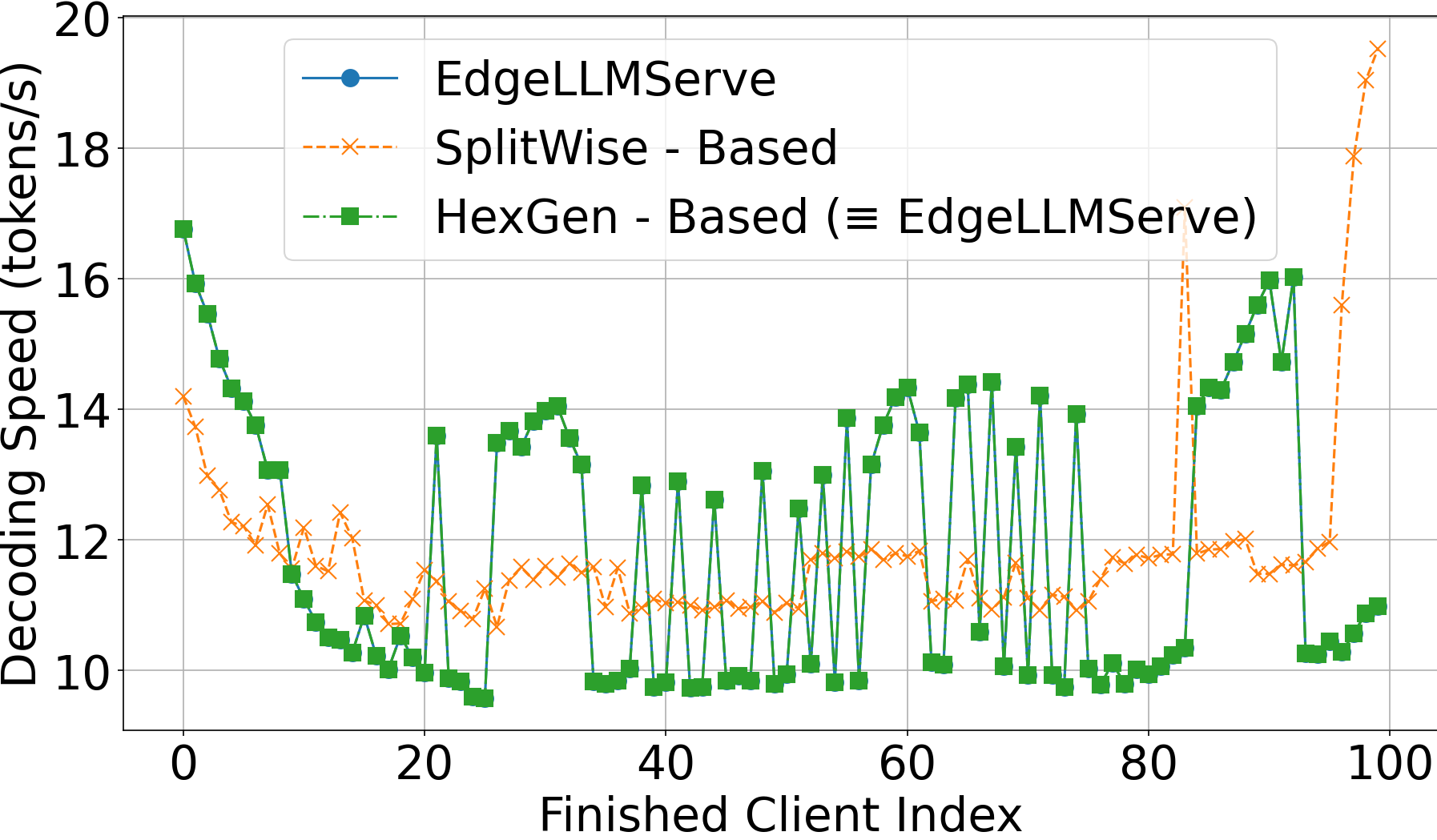}
        \caption{Decoding speed}
    \end{subfigure}\hfill
    \begin{subfigure}{\appfigscale\textwidth}
        \includegraphics[width=\linewidth]{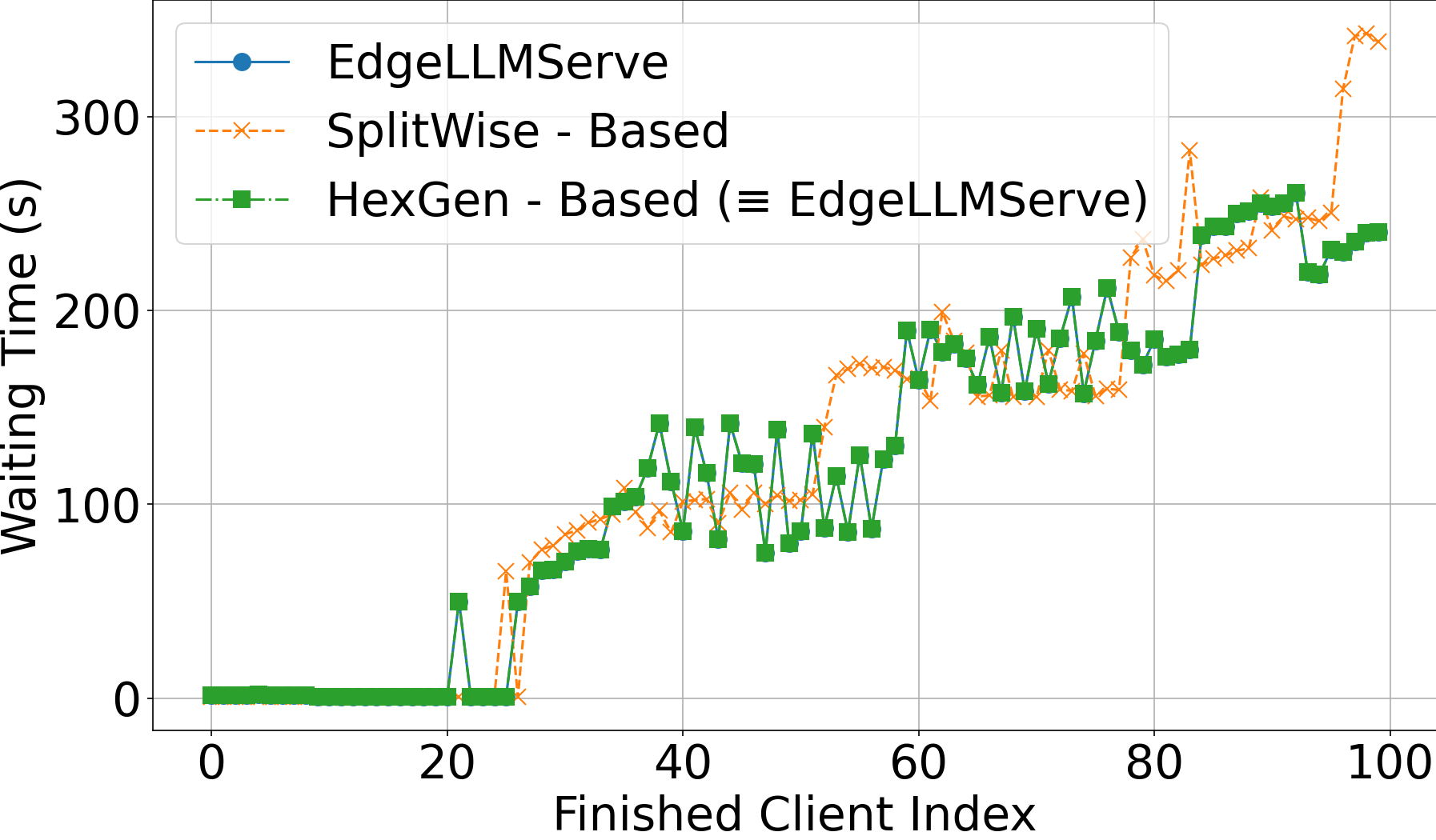}
        \caption{Waiting time}
    \end{subfigure}
    \vspace{-0.3cm}
    \caption{Per-request metrics --- Eight devices, Qwen3-MoE-30B, IELTS, arrival $3000$~ms.}
    \label{apd:f33}
\end{figure}

\vspace{-0.3cm}
\paragraph{4. LZByteDance dataset. Arrival rate of 100ms}\mbox{}
\vspace{-0.3cm}
\begin{table}[H]
    \centering
    \setlength{\tabcolsep}{3pt}
    \caption{Deployment plans and metrics --- Eight devices, Qwen3-MoE-30B, LZByteDance, arrival $100$~ms.}
    \vspace{-0.3cm}
    \label{apd:t34}
    \begin{subtable}[t]{0.24\textwidth}
        \centering
        \caption{HG}
        \begin{tabular}{|c|c|c|c|}
            \hline
            Mode & Dev & Slots & G--C\\\hline
            P& Dev.1$^{\star}$ & 1 & 26--22\\\hline
            \multirow{3}{*}{D} & Dev.2 & \multirow{3}{*}{9} & 7--3\\
              & Dev.7 &  & 19--0\\
              & Dev.8$^{\star}$ &  & 19--0\\\hline
            \multirow{2}{*}{D} & Dev.3$^{\star}$ & \multirow{2}{*}{8} & 36--0\\
              & Dev.6 &  & 12--0\\\hline
            \multirow{2}{*}{D} & Dev.4$^{\star}$ & \multirow{2}{*}{8} & 36--0\\
              & Dev.5 &  & 12--0\\\hline
        \end{tabular}
    \end{subtable}\hfill
    \begin{subtable}[t]{0.24\textwidth}
        \centering
        \caption{ELS}
        \begin{tabular}{|c|c|c|c|}
            \hline
            Mode & Dev & Slots & G--C\\\hline
            P& Dev.1$^{\star}$ & 1 & 26--22\\\hline
            \multirow{3}{*}{D} & Dev.2 & \multirow{3}{*}{9} & 7--3\\
              & Dev.7 &  & 19--0\\
              & Dev.8$^{\star}$ &  & 19--0\\\hline
            \multirow{2}{*}{D} & Dev.3$^{\star}$ & \multirow{2}{*}{8} & 36--0\\
              & Dev.6 &  & 12--0\\\hline
            \multirow{2}{*}{D} & Dev.4$^{\star}$ & \multirow{2}{*}{8} & 36--0\\
              & Dev.5 &  & 12--0\\\hline
        \end{tabular}
    \end{subtable}\hfill
    \begin{subtable}[t]{0.24\textwidth}
        \centering
        \caption{SPW}
        \begin{tabular}{|c|c|c|c|}
            \hline
            Mode & Dev & Slots & G--C\\\hline
            \multirow{2}{*}{P} & Dev.7$^{\star}$ & \multirow{2}{*}{2} & 26--0\\
              & Dev.8 &  & 22--0\\\hline
            \multirow{2}{*}{D} & Dev.1$^{\star}$ & \multirow{2}{*}{10} & 13--19\\
              & Dev.2 &  & 5--11\\\hline
            \multirow{2}{*}{D} & Dev.3$^{\star}$ & \multirow{2}{*}{8} & 36--0\\
              & Dev.6 &  & 12--0\\\hline
            \multirow{2}{*}{D} & Dev.4$^{\star}$ & \multirow{2}{*}{8} & 36--0\\
              & Dev.5 &  & 12--0\\\hline
        \end{tabular}
    \end{subtable}\hfill
    \begin{subtable}[t]{0.24\textwidth}
        \centering
        \caption{Metrics.}
        \begin{tabular}{|l|c|c|c|}
            \hline
            Method & PS & DS & W\\\hline
            HG & 1146 & 13.5 & 109.6\\\hline
            \textbf{ELS} & \textbf{1146} & \textbf{13.5} & \textbf{109.6}\\\hline
            SPW & 1175 & 13.2 & 256.9\\\hline
        \end{tabular}
    \end{subtable}
\end{table}
\vspace{-0.3cm}
\begin{figure}[H]
    \centering
    \begin{subfigure}{\appfigscale\textwidth}
        \includegraphics[width=\linewidth]{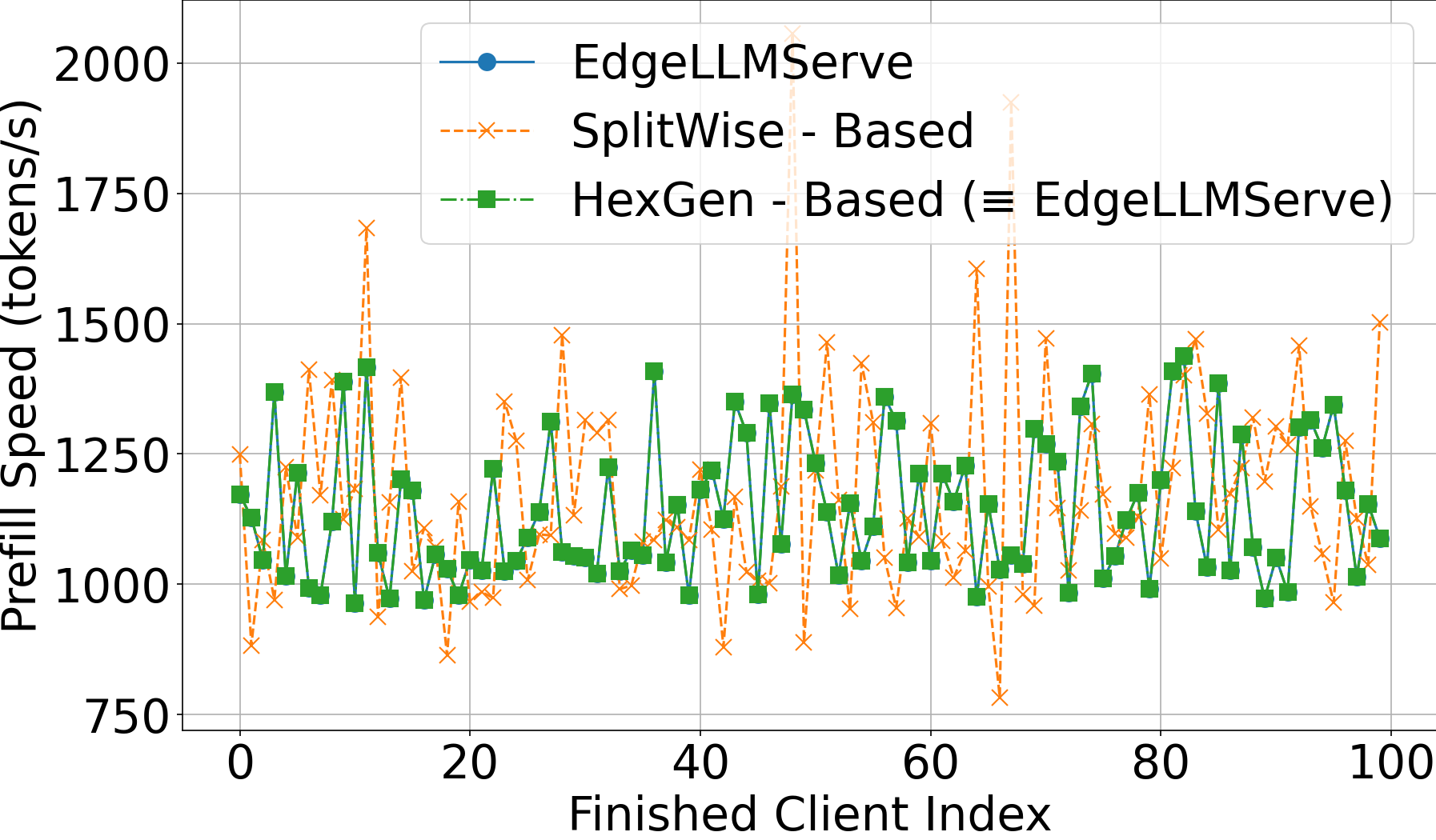}
        \caption{Prefill speed}
    \end{subfigure}\hfill
    \begin{subfigure}{\appfigscale\textwidth}
        \includegraphics[width=\linewidth]{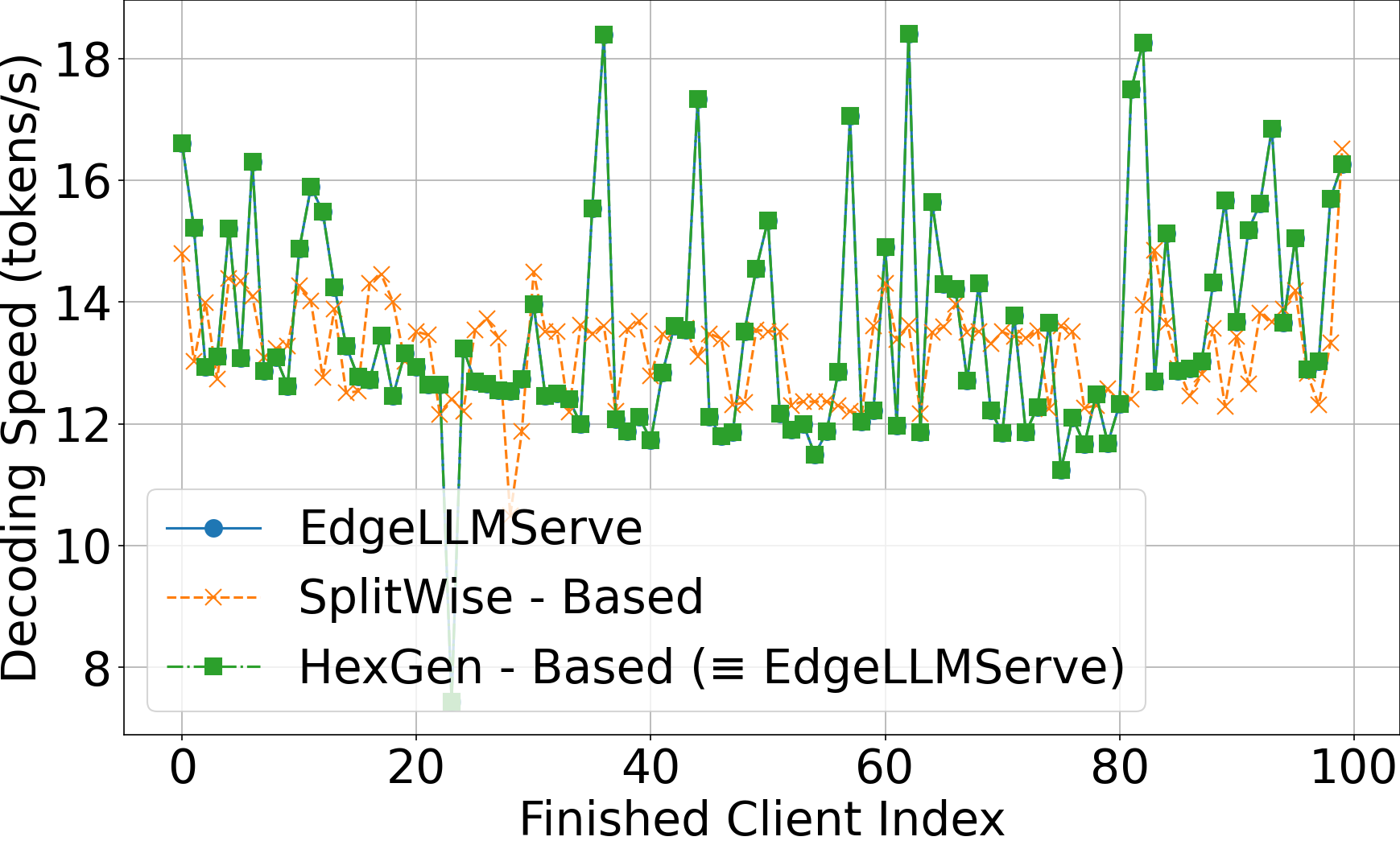}
        \caption{Decoding speed}
    \end{subfigure}\hfill
    \begin{subfigure}{\appfigscale\textwidth}
        \includegraphics[width=\linewidth]{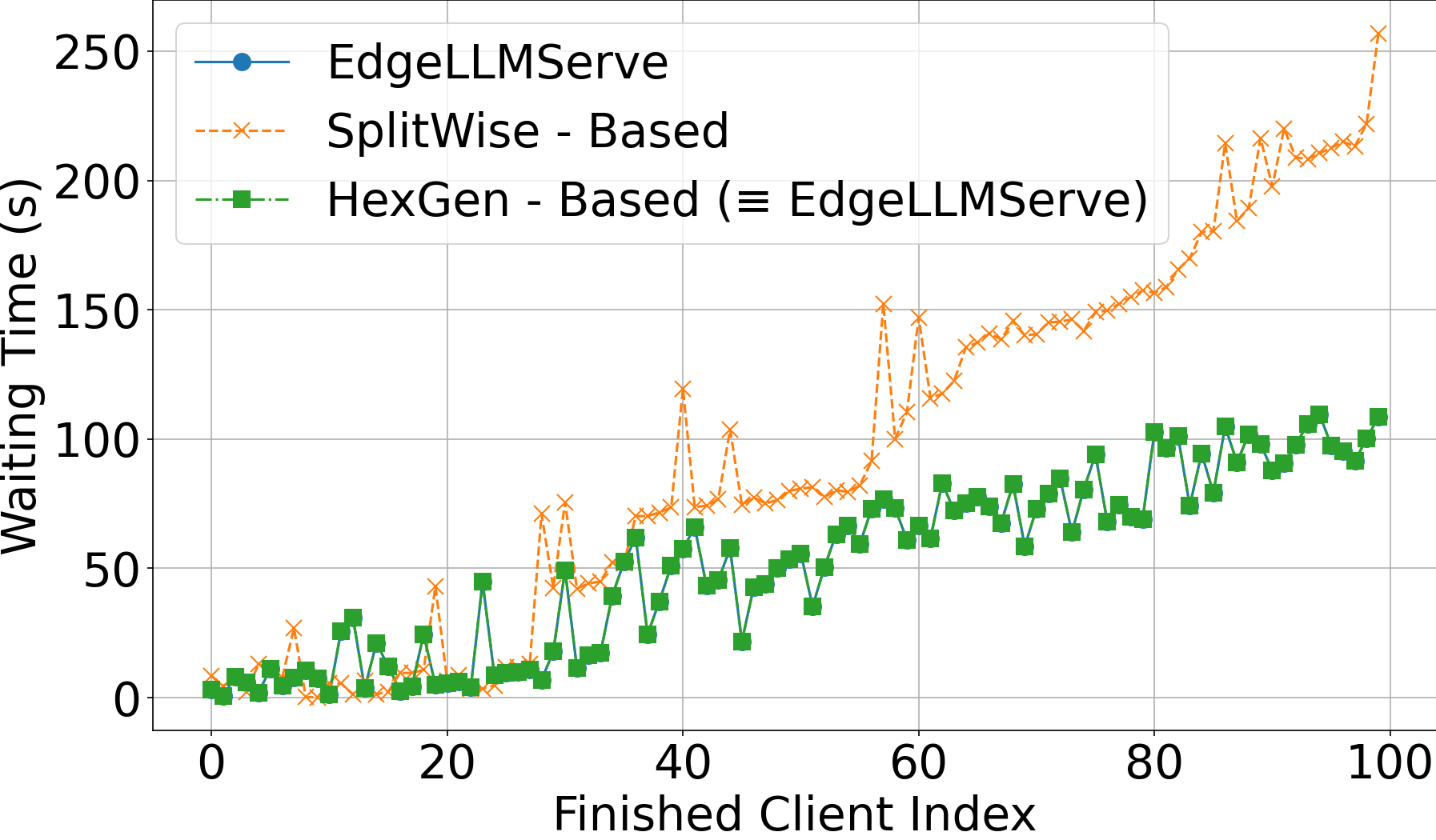}
        \caption{Waiting time}
    \end{subfigure}
    \vspace{-0.3cm}
    \caption{Per-request metrics --- Eight devices, Qwen3-MoE-30B, LZByteDance, arrival $100$~ms.}
    \label{apd:f34}
\end{figure}

\paragraph{5. LZByteDance dataset. Arrival rate of 2000ms}\mbox{}
\vspace{-0.3cm}
\begin{table}[H]
    \centering
    \setlength{\tabcolsep}{3pt}
    \caption{Deployment plans and metrics --- Eight devices, Qwen3-MoE-30B, LZByteDance, arrival $2000$~ms.}
    \vspace{-0.3cm}
    \label{apd:t35}
    \begin{subtable}[t]{0.24\textwidth}
        \centering
        \caption{HG}
        \begin{tabular}{|c|c|c|c|}
            \hline
            Mode & Dev & Slots & G--C\\\hline
            P& Dev.1$^{\star}$ & 1 & 26--22\\\hline
            \multirow{3}{*}{D} & Dev.2 & \multirow{3}{*}{9} & 7--3\\
              & Dev.7 &  & 19--0\\
              & Dev.8$^{\star}$ &  & 19--0\\\hline
            \multirow{2}{*}{D} & Dev.3$^{\star}$ & \multirow{2}{*}{8} & 36--0\\
              & Dev.6 &  & 12--0\\\hline
            \multirow{2}{*}{D} & Dev.4$^{\star}$ & \multirow{2}{*}{8} & 36--0\\
              & Dev.5 &  & 12--0\\\hline
        \end{tabular}
    \end{subtable}\hfill
    \begin{subtable}[t]{0.24\textwidth}
        \centering
        \caption{ELS}
        \begin{tabular}{|c|c|c|c|}
            \hline
            Mode & Dev & Slots & G--C\\\hline
            P& Dev.3$^{\star}$ & 1 & 48--0\\\hline
            \multirow{3}{*}{D} & Dev.1$^{\star}$ & \multirow{3}{*}{18} & 13--5\\
              & Dev.7 &  & 15--0\\
              & Dev.8 &  & 15--0\\\hline
            \multirow{2}{*}{D} & Dev.2$^{\star}$ & \multirow{2}{*}{4} & 6--22\\
              & Dev.6 &  & 20--0\\\hline
            \multirow{2}{*}{D} & Dev.4$^{\star}$ & \multirow{2}{*}{8} & 36--0\\
              & Dev.5 &  & 12--0\\\hline
        \end{tabular}
    \end{subtable}\hfill
    \begin{subtable}[t]{0.24\textwidth}
        \centering
        \caption{SPW}
        \begin{tabular}{|c|c|c|c|}
            \hline
            Mode & Dev & Slots & G--C\\\hline
            \multirow{2}{*}{P} & Dev.1$^{\star}$ & \multirow{2}{*}{2} & 25--12\\
              & Dev.2 &  & 11--0\\\hline
            \multirow{3}{*}{D} & Dev.3$^{\star}$ & \multirow{3}{*}{12} & 12--0\\
              & Dev.7 &  & 18--0\\
              & Dev.8 &  & 18--0\\\hline
            \multirow{3}{*}{D} & Dev.4$^{\star}$ & \multirow{3}{*}{6} & 30--0\\
              & Dev.5 &  & 9--0\\
              & Dev.6 &  & 9--0\\\hline
        \end{tabular}
    \end{subtable}\hfill
    \begin{subtable}[t]{0.24\textwidth}
        \centering
        \caption{Metrics.}
        \begin{tabular}{|l|c|c|c|}
            \hline
            Method & PS & DS & W\\\hline
            HG & 1146 & 13.8 & 136.6\\\hline
            \textbf{ELS} & \textbf{622} & \textbf{12.3} & \textbf{73.3}\\\hline
            SPW & 1282 & 15.0 & 174.0\\\hline
        \end{tabular}
    \end{subtable}
\end{table}
\vspace{-0.3cm}
\begin{figure}[H]
    \centering
    \begin{subfigure}{\appfigscale\textwidth}
        \includegraphics[width=\linewidth]{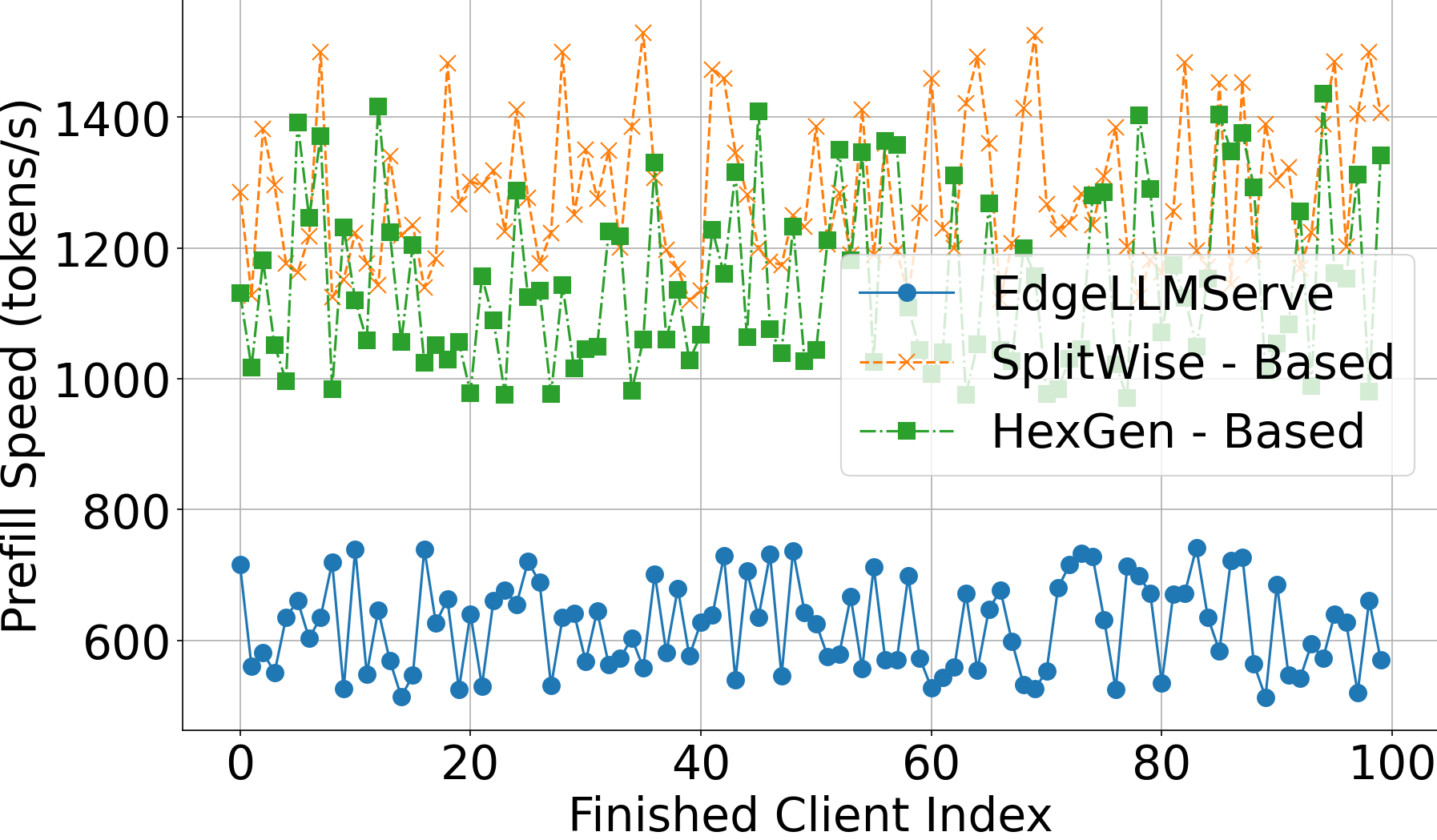}
        \caption{Prefill speed}
    \end{subfigure}\hfill
    \begin{subfigure}{\appfigscale\textwidth}
        \includegraphics[width=\linewidth]{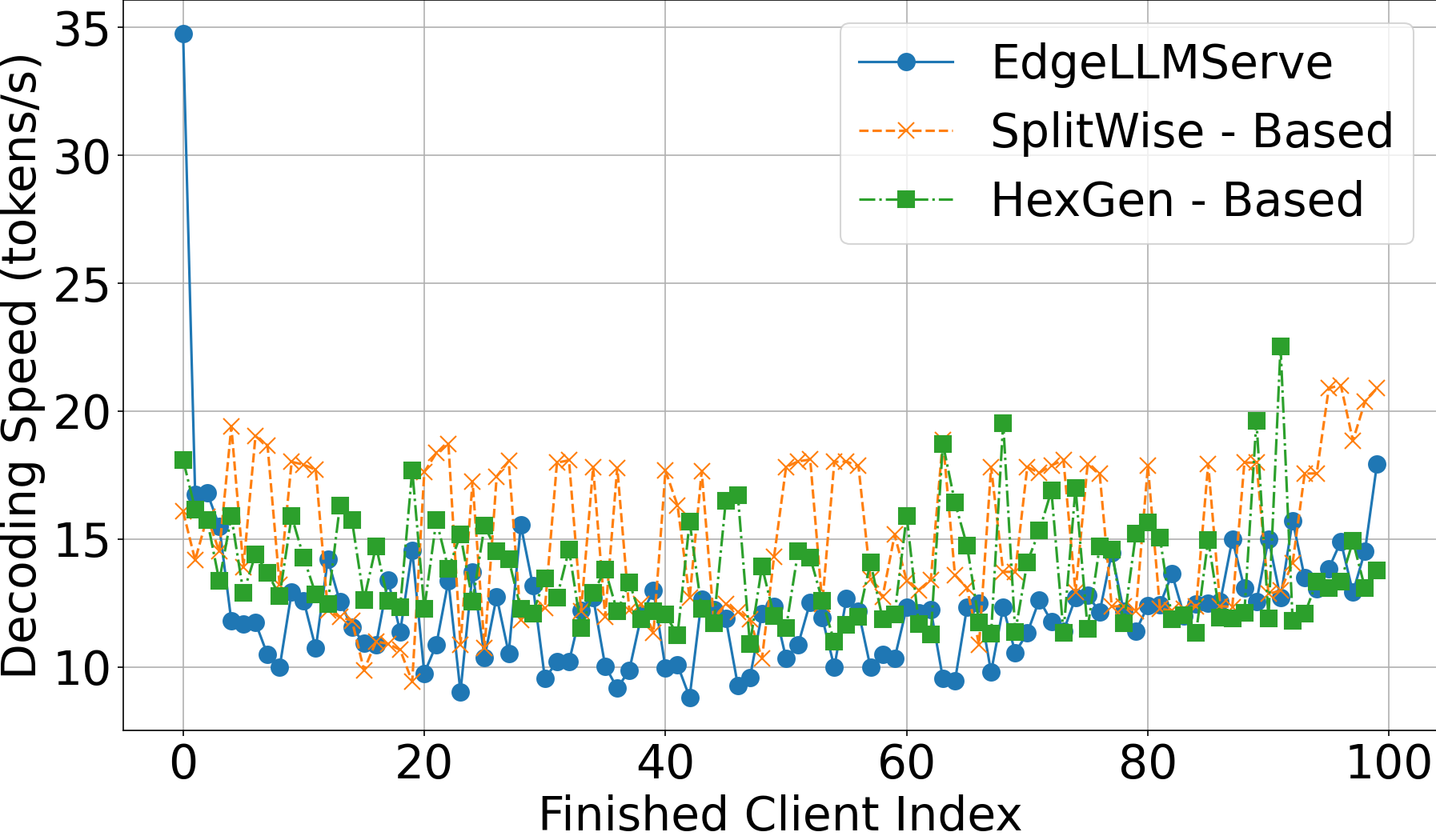}
        \caption{Decoding speed}
    \end{subfigure}\hfill
    \begin{subfigure}{\appfigscale\textwidth}
        \includegraphics[width=\linewidth]{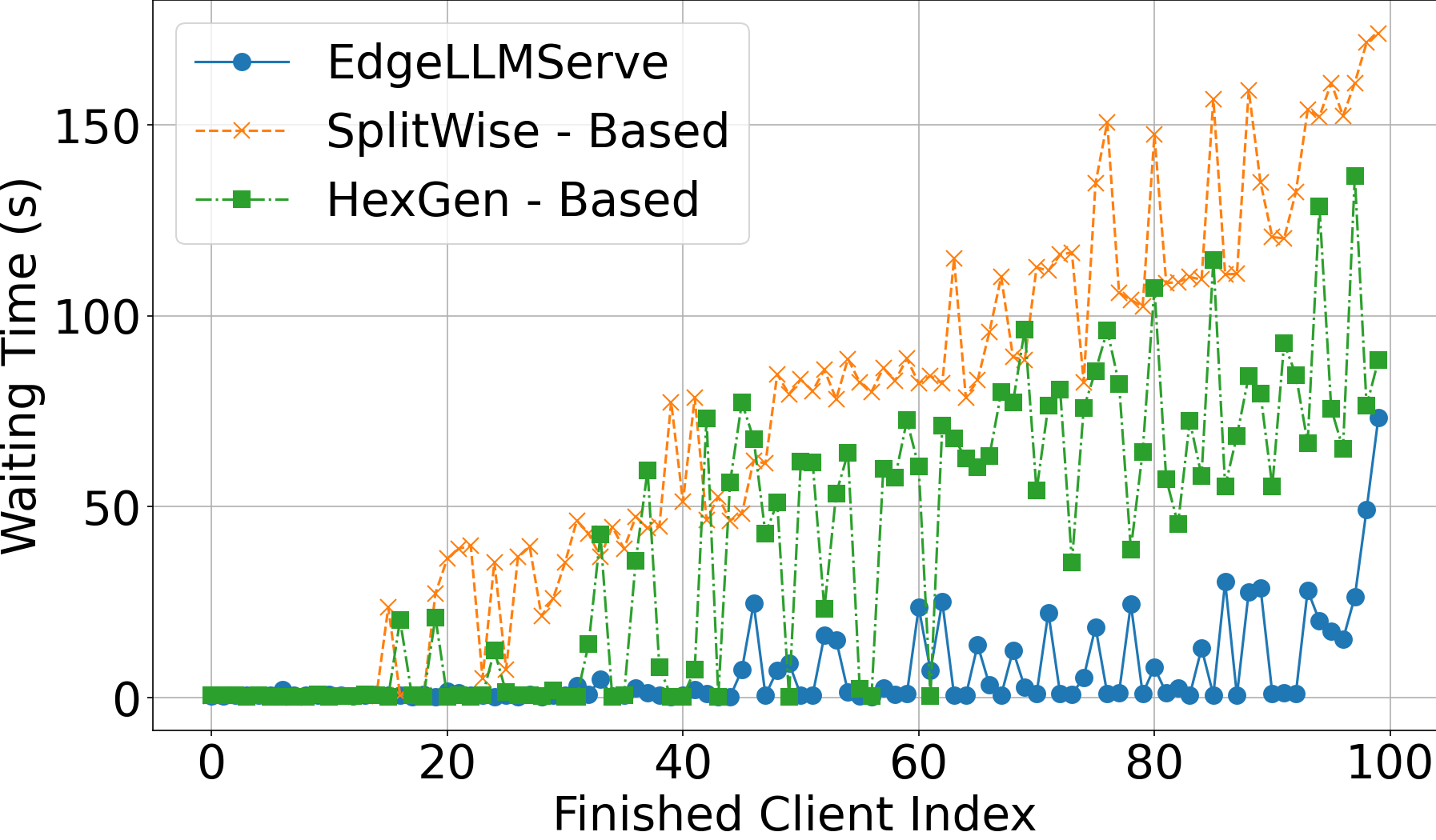}
        \caption{Waiting time}
    \end{subfigure}
    \vspace{-0.3cm}
    \caption{Per-request metrics --- Eight devices, Qwen3-MoE-30B, LZByteDance, arrival $2000$~ms.}
    \label{apd:f35}
\end{figure}

\vspace{-0.3cm}
\paragraph{6. LZByteDance dataset. Arrival rate of 3000ms}\mbox{}
\vspace{-0.3cm}
\begin{table}[H]
    \centering
    \setlength{\tabcolsep}{3pt}
    \caption{Deployment plans and metrics --- Eight devices, Qwen3-MoE-30B, LZByteDance, arrival $3000$~ms.}
    \vspace{-0.3cm}
    \label{apd:t36}
    \begin{subtable}[t]{0.24\textwidth}
        \centering
        \caption{HG}
        \begin{tabular}{|c|c|c|c|}
            \hline
            Mode & Dev & Slots & G--C\\\hline
            P& Dev.1$^{\star}$ & 1 & 26--22\\\hline
            \multirow{3}{*}{D} & Dev.2 & \multirow{3}{*}{9} & 7--3\\
              & Dev.7 &  & 19--0\\
              & Dev.8$^{\star}$ &  & 19--0\\\hline
            \multirow{2}{*}{D} & Dev.3$^{\star}$ & \multirow{2}{*}{8} & 36--0\\
              & Dev.6 &  & 12--0\\\hline
            \multirow{2}{*}{D} & Dev.4$^{\star}$ & \multirow{2}{*}{8} & 36--0\\
              & Dev.5 &  & 12--0\\\hline
        \end{tabular}
    \end{subtable}\hfill
    \begin{subtable}[t]{0.24\textwidth}
        \centering
        \caption{ELS}
        \begin{tabular}{|c|c|c|c|}
            \hline
            Mode & Dev & Slots & G--C\\\hline
            \multirow{2}{*}{P} & Dev.2$^{\star}$ & \multirow{2}{*}{2} & 8--20\\
              & Dev.6 &  & 20--0\\\hline
            \multirow{3}{*}{D} & Dev.1$^{\star}$ & \multirow{3}{*}{18} & 13--5\\
              & Dev.7 &  & 15--0\\
              & Dev.8 &  & 15--0\\\hline
            \multirow{2}{*}{D} & Dev.3$^{\star}$ & \multirow{2}{*}{8} & 36--0\\
              & Dev.5 &  & 12--0\\\hline
            D& Dev.4$^{\star}$ & 4 & 48--0\\\hline
        \end{tabular}
    \end{subtable}\hfill
    \begin{subtable}[t]{0.24\textwidth}
        \centering
        \caption{SPW}
        \begin{tabular}{|c|c|c|c|}
            \hline
            Mode & Dev & Slots & G--C\\\hline
            \multirow{2}{*}{P} & Dev.1$^{\star}$ & \multirow{2}{*}{2} & 25--12\\
              & Dev.2 &  & 11--0\\\hline
            \multirow{3}{*}{D} & Dev.3$^{\star}$ & \multirow{3}{*}{12} & 12--0\\
              & Dev.7 &  & 18--0\\
              & Dev.8 &  & 18--0\\\hline
            \multirow{3}{*}{D} & Dev.4$^{\star}$ & \multirow{3}{*}{6} & 30--0\\
              & Dev.5 &  & 9--0\\
              & Dev.6 &  & 9--0\\\hline
        \end{tabular}
    \end{subtable}\hfill
    \begin{subtable}[t]{0.24\textwidth}
        \centering
        \caption{Metrics.}
        \begin{tabular}{|l|c|c|c|}
            \hline
            Method & PS & DS & W\\\hline
            HG & 1147 & 13.9 & 66.3\\\hline
            \textbf{ELS} & \textbf{188} & \textbf{21.2} & \textbf{1.8}\\\hline
            SPW & 954 & 12.8 & 57.7\\\hline
        \end{tabular}
    \end{subtable}
\end{table}
\vspace{-0.3cm}
\begin{figure}[H]
    \centering
    \begin{subfigure}{\appfigscale\textwidth}
        \includegraphics[width=\linewidth]{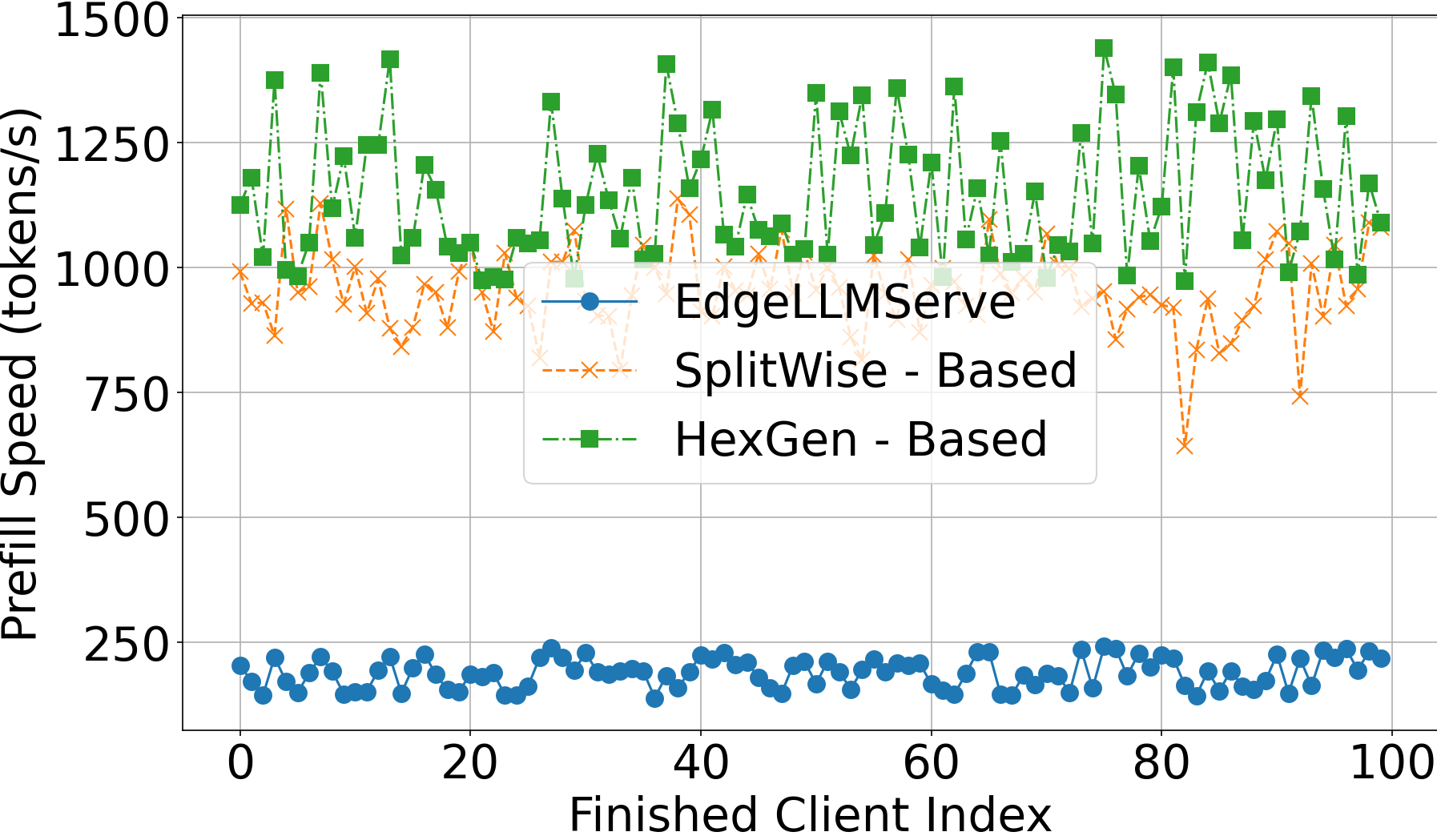}
        \caption{Prefill speed}
    \end{subfigure}\hfill
    \begin{subfigure}{\appfigscale\textwidth}
        \includegraphics[width=\linewidth]{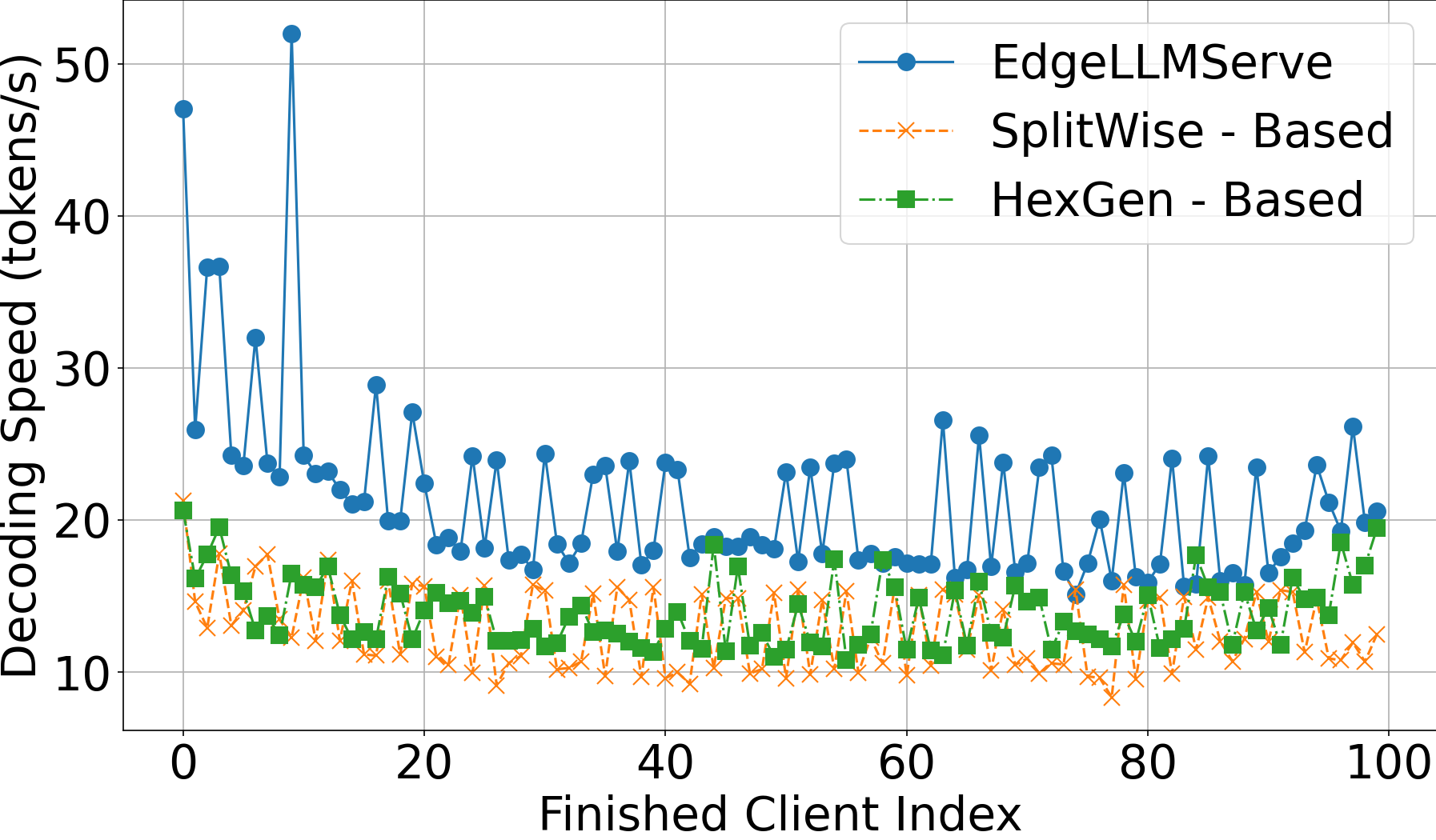}
        \caption{Decoding speed}
    \end{subfigure}\hfill
    \begin{subfigure}{\appfigscale\textwidth}
        \includegraphics[width=\linewidth]{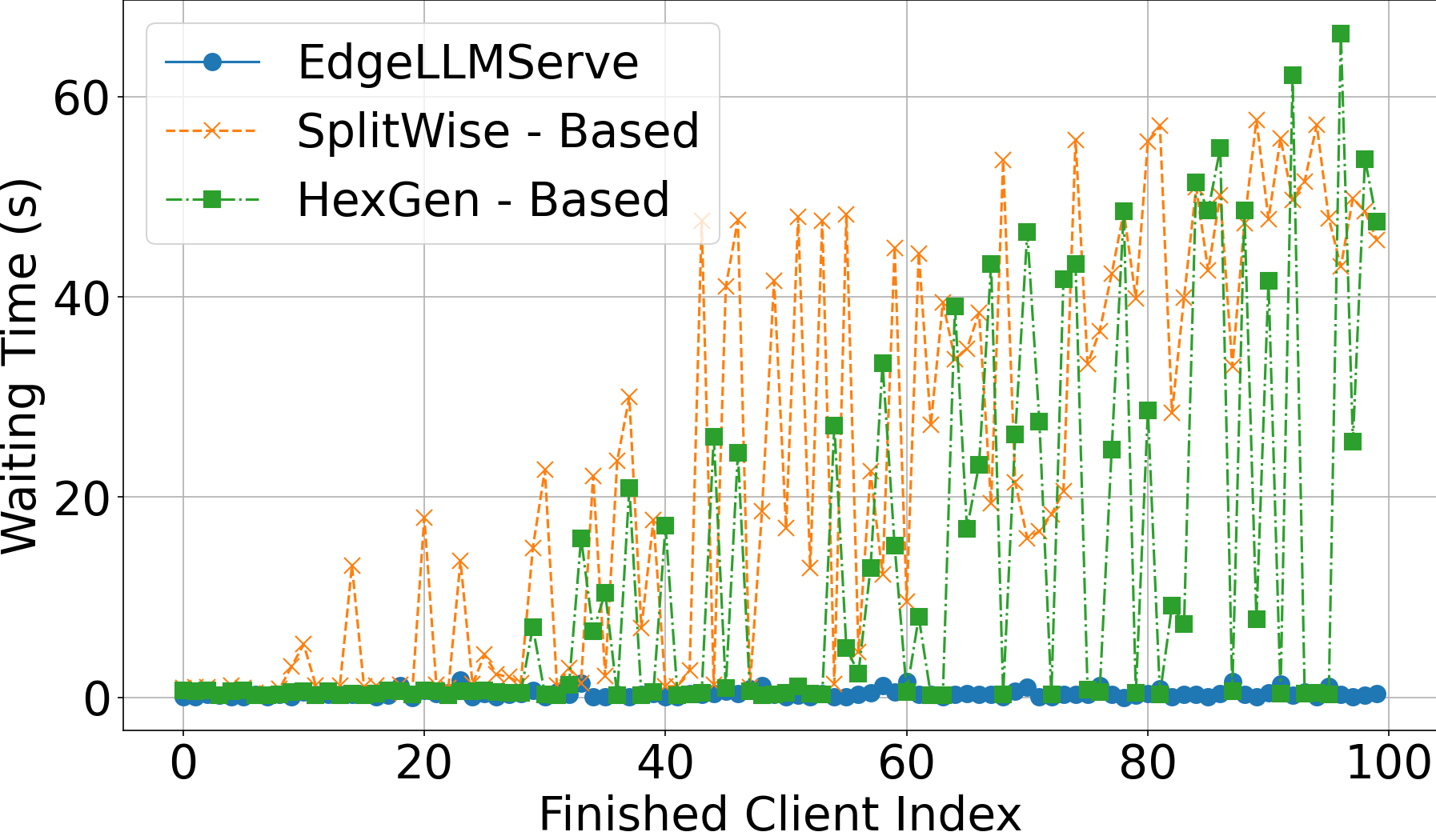}
        \caption{Waiting time}
    \end{subfigure}
    \vspace{-0.3cm}
    \caption{Per-request metrics --- Eight devices, Qwen3-MoE-30B, LZByteDance, arrival $3000$~ms.}
    \label{apd:f36}
\end{figure}

\end{document}